%% file: hermesTMDs.tex
\documentclass[11pt,a4paper]{article}

\input{./latex/packages}
\input{./latex/acronyms}                 
\input{./latex/macros}                   
\input{./latex/mathematics}              
\input{./latex/particles}                
\input{./latex/physics}                  

\usepackage{slantsc}
\usepackage[T1]{fontenc}

\usepackage{enumitem}

\allowdisplaybreaks

\newcommand{\slim}{\mskip 1.5mu}              

\newcommand{\lf}{\left}
\newcommand{\rg}{\right}

\long\def\symbolfootnote[#1]#2{\begingroup%
\def\thefootnote{\fnsymbol{footnote}}\footnote[#1]{#2}\endgroup}
\def\hermesauthor[#1]#2{{#2}$^{\, #1}$}
\def\hermesinstitute[#1]#2{$^{#1\,}$ {#2}\\}
\renewcommand{\thefootnote}{\alph{footnote}}
\def\nowat[#1]#2{\(^,\)\footnote[#1]{#2}}

\begin{document}

\graphicspath{{./plots/}}

\input{./sections/title}


\newpage

\input{./sections/introduction}

\input{./sections/theory}
\input{./sections/measurement}

\input{./sections/interpretation}

\input{./sections/conclusion}
\input{./sections/appendices}


\bibliographystyle{./latex/JHEP-e}
\bibliography{hermesTMDs}

\end{document}

%% file: latex/packages.tex
\usepackage{color}            
\usepackage[pdftex]{graphicx} 
\usepackage{ifthen}           
\usepackage{./latex/jheppub}  
\usepackage{./latex/lineno}   
\usepackage{xspace}           

%% file: latex/acronyms.tex
\newcommand{\dis}[1][long]{%
 \ifthenelse{\equal{#1}{long}}{deep-inelastic scattering\xspace}%
 {\textsc{DIS}\xspace}%
}

\newcommand{\sia}{\ensuremath{e^{+}e^{-}} annihilation\xspace}

\newcommand{\ssa}[1][small]{%
\ifthenelse{\equal{#1}{long}}{single-spin asymmetries\xspace}%
{\textsc{SSA}\xspace}%
}

\newcommand{\lssa}[1][long]{%
\ifthenelse{\equal{#1}{long}}{longitudinal single-spin asymmetries\xspace}%
{\textsc{LSSA}\xspace}%
}

\newcommand{\tssa}[1][long]{%
\ifthenelse{\equal{#1}{long}}{transverse single-spin asymmetries\xspace}%
{\textsc{TSSA}\xspace}%
}

\newcommand{\dsa}[1][small]{%
\ifthenelse{\equal{#1}{long}}{double-spin asymmetries\xspace}%
{\textsc{DSA}\xspace}%
}

\newcommand{\sfa}[1][small]{%
\ifthenelse{\equal{#1}{long}}{structure-function asymmetries\xspace}%
{\textsc{SFA}\xspace}%
}

\newcommand{\csa}[1][small]{%
\ifthenelse{\equal{#1}{long}}{cross-section asymmetries\xspace}%
{\textsc{CSA}\xspace}%
}

\newcommand{\Teven}[1][small]{%
\ifthenelse{\equal{#1}{long}}{naive time reversal even\xspace}%
{naive-\(T\)-even\xspace}%
}
\newcommand{\Todd}[1][small]{%
\ifthenelse{\equal{#1}{long}}{naive time reversal odd\xspace}%
{naive-\(T\)-odd\xspace}%
}

\newcommand{\QCD}[1][small]{%
\ifthenelse{\equal{#1}{long}}{quantum chromodynamics\xspace}%
{\textsc{QCD}\xspace}%
}
\newcommand{\QED}{\textsc{QED}\xspace} 

\newcommand{\pdf}[1][small]{%
\ifthenelse{\equal{#1}{long}}{parton distribution functions\xspace}%
{\textsc{PDF}\xspace}%
}

\newcommand{\tmd}[1][small]{%
\ifthenelse{\equal{#1}{long}}{transverse-momentum-dependent parton distribution functions\xspace}%
{\textsc{TMD}\xspace}%
}
\newcommand{\Tmd}[1][small]{%
\ifthenelse{\equal{#1}{long}}{Transverse-momentum-dependent parton distribution functions\xspace}%
{\textsc{TMD}\xspace}%
}

\newcommand{\clas}{\textsc{CLAS}\xspace}
\newcommand{\compass}{\textsc{COMPASS}\xspace}

\newcommand{\desy}{\textsc{DESY}\xspace}
 
\newcommand{\hera}{\textsc{HERA}\xspace} 
\newcommand{\hermes}{\textsc{HERMES}\xspace}

\newcommand{\rich}[1][small]{%
\ifthenelse{\equal{#1}{long}}{ring-imaging Cherenkov detector\xspace}%
{\textsc{RICH}\xspace}}

\newcommand{\pid}{\textsc{PID}\xspace}

\newcommand{\geant}{\textsc{Geant3}\xspace}

\newcommand{\minuit}{\textsc{Minuit}\xspace}
\newcommand{\pythia}{\textsc{Pythia6.2}\xspace}
\newcommand{\radgen}{\textsc{RadGen}\xspace}

%% file: latex/macros.tex
\newcommand{\rangeho}[2]{\ensuremath{\left]#1;\,#2\right]}\xspace}

\newcommand{\tableenv}[4]{
\begin{table}[#1]
\centering%
{#2}
\caption{#3}\label{#4}
\end{table}}

\newcommand{\ToDo}[2][short]{%
\ifthenelse{\equal{#1}{long}}%
{\begin{center}
\begin{tabular}{|p{12cm}|}\arrayrulecolor{darkred}\hline
\textcolor{darkred}{\textbf{ToDo:}}\\\hline
\textcolor{black}{#2}\\\hline
\end {tabular}
\end{center}}
{\footnote{\textcolor{darkred}{\textbf{ToDo:} #2}}} }

%% file: latex/mathematics.tex
\newcommand{\abs}[1]{\ensuremath{\left|#1\right|}}
\newcommand{\mean}[1]{\ensuremath{\left<#1\right>}\xspace}

\newcommand{\euvec}[1]{\ensuremath{\mathbf{#1}}}
\newcommand{\unitvec}[1]{\ensuremath{\mathbf{\hat{#1}}}}



%% file: latex/particles.tex
\newcommand{\proton}{\ensuremath{p}\xspace}
\newcommand{\pbar}{\ensuremath{\bar{p}}\xspace}
\newcommand{\piplus}{\ensuremath{\pi^{\,+}}\xspace}
\newcommand{\pizero}{\ensuremath{\pi^{\,0}}\xspace}
\newcommand{\piminus}{\ensuremath{\pi^{\,-}}\xspace}
\newcommand{\kplus}{\ensuremath{K^{\,+}}\xspace}
\newcommand{\kminus}{\ensuremath{K^{\,-}}\xspace}



%% file: latex/physics.tex
\newcommand{\conv}[3]{\ensuremath{ {\cal C} \bigl[ {#3} \; {#1} \; {#2} \bigr]}}

\newcommand{\aup}[1]{\ensuremath{A_{\text{U}\perp}^{#1}}\xspace}

\newcommand{\yield}[1]{%
\ifthenelse{\equal{#1}{up}}%
{\ensuremath{N^h_\Uparrow}\xspace}
{\ensuremath{N^h_\Downarrow}\xspace}}

\newcommand{\alp}[1]{\ensuremath{A_{\text{L}\perp}^{#1}}\xspace}

\newcommand{\sigmauu}[1]{\ensuremath{\sigma_{\,\text{UU}}^{\,#1}}\xspace}

\newcommand{\sigmaut}[2]{%
\ifthenelse{\equal{#2}{T}}%
{\ensuremath{\sigma_{\,\text{UT}}^{\,#1}}\xspace}
{\ensuremath{\sigma_{\,\text{U}{\,#2}}^{\,#1}}\xspace}}

\newcommand{\sigmalt}[2]{%
\ifthenelse{\equal{#2}{T}}%
{\ensuremath{\sigma_{\,\text{LT}}^{\,#1}}\xspace}
{\ensuremath{\sigma_{\,\text{L}{\,#2}}^{\,#1}}\xspace}}

\newcommand{\sinemodulation}[1]{\ensuremath{\sin{\left(#1\right)}}}
\newcommand{\cosinemodulation}[1]{\ensuremath{\cos{\left(#1\right)}}}

\newcommand{\structure}[1]{\ensuremath{F_{\,\text{#1}}}}
\newcommand{\structureh}[1]{\ensuremath{F_{\,\text{#1}}^{{\,h}}}}
\newcommand{\structuresin}[2]{\ensuremath{F_{\,\text{#1}}^{\,\sinemodulation{#2}}}}
\newcommand{\structuresinh}[2]{\ensuremath{F_{\,\text{#1}}^{\,\sinemodulation{#2},h}}}
\newcommand{\structurecos}[2]{\ensuremath{F_{\,\text{#1}}^{\,\cosinemodulation{#2}}}}

\newcommand{\convolution}[1]{\ensuremath{\mathcal{C}\left[{#1}\right]}}

\newcommand{\pdff}[1]{\ensuremath{f_{\,1}^{\,{#1}}}\xspace}
\newcommand{\pdfg}[1]{\ensuremath{g_{\,1}^{\,{#1}}}\xspace}
\newcommand{\pdfgtwo}[1]{\ensuremath{g_{\,2}^{\,{#1}}}\xspace}
\newcommand{\pdfh}[1]{\ensuremath{h_{\,1}^{\,{#1}}}\xspace}

\newcommand{\pdffpt}[1]{\ensuremath{f_{\,1}^{\,{#1}}\left(\x,\ptsqr\right)}\xspace}
\newcommand{\pdfgpt}[1]{\ensuremath{g_{\,1}^{\,{#1}}\left(\x,\ptsqr\right)}\xspace}
\newcommand{\pdfhpt}[1]{\ensuremath{h_{\,1}^{\,{#1}}\left(\x,\ptsqr\right)}\xspace}

\newcommand{\tmdsivers}[1]{\ensuremath{f_{\,1\text{T}}^{\,\perp,{#1}}}\xspace}

\newcommand{\tmdsiverspt}[1]{\ensuremath{f_{\,1\text{T}}^{\,\perp,{#1}}\left(\x,\ptsqr\right)}\xspace}
\newcommand{\tmdbm}[1]{\ensuremath{h_{\,1}^{\,\perp,{#1}}}\xspace}

\newcommand{\tmdbmpt}[1]{\ensuremath{h_{\,1}^{\,\perp,{#1}}\left(\x,\ptsqr\right)}\xspace}
\newcommand{\tmdgt}[1]{\ensuremath{g_{\,1\text{T}}^{\,{#1}}}\xspace}
\newcommand{\tmdonegt}[1]{\ensuremath{g_{\,1\text{T}}^{\,(1),{#1}}}\xspace}
\newcommand{\tmdgtpt}[1]{\ensuremath{g_{\,1\text{T}}^{\,{#1}}\left(\x,\ptsqr\right)}\xspace}
\newcommand{\tmdhl}[1]{\ensuremath{h_{\,1\text{L}}^{\,\perp,{#1}}}\xspace}
\newcommand{\tmdonehl}[1]{\ensuremath{h_{\,1\text{L}}^{\,\perp(1),{#1}}}\xspace}
\newcommand{\tmdhlpt}[1]{\ensuremath{h_{\,1\text{L}}^{\,\perp,{#1}}\left(\x,\ptsqr\right)}\xspace}
\newcommand{\tmdht}[1]{\ensuremath{h_{\,1\text{T}}^{\,\perp,{#1}}}\xspace}

\newcommand{\tmdhtpt}[1]{\ensuremath{h_{\,1\text{T}}^{\,\perp,{#1}}\left(\x,\ptsqr\right)}\xspace}

\newcommand{\twistet}[1]{\ensuremath{e_{\,\text{T}}^{\,{#1}}}\xspace}
\newcommand{\twistetperp}[1]{\ensuremath{e_{\,\text{T}}^{\,\perp,{#1}}}\xspace}
\newcommand{\twistft}[1]{\ensuremath{f_{\,\text{T}}^{\,{#1}}}\xspace}
\newcommand{\twistftperp}[1]{\ensuremath{f_{\,\text{T}}^{\,\perp,{#1}}}\xspace}
\newcommand{\twistgt}[1]{\ensuremath{g_{\,\text{T}}^{\,{#1}}}\xspace}
\newcommand{\twistgtperp}[1]{\ensuremath{g_{\,\text{T}}^{\,\perp,{#1}}}\xspace}
\newcommand{\twisthl}[1]{\ensuremath{h_{\,\text{L}}^{\,{#1}}}\xspace}

\newcommand{\twistht}[1]{\ensuremath{h_{\,\text{T}}^{\,{#1}}}\xspace}
\newcommand{\twisthtperp}[1]{\ensuremath{h_{\,\text{T}}^{\,\perp,{#1}}}\xspace}

\newcommand{\ffd}[1]{\ensuremath{D_{\,1}^{\,{#1}}\left(\z\right)}\xspace}
\newcommand{\ffdkt}[1]{\ensuremath{D_{\,1}^{\,{#1}}\left(\z,z^2\ktsqr\right)}\xspace}
\newcommand{\ffdmod}[1]{\ensuremath{D_{\,1}^{\,{#1}}}\xspace}
\newcommand{\ffcollinskt}[1]{\ensuremath{H_{\,1}^{\,\perp,{#1}}\left(\z,\z^2\ktsqr\right)}\xspace}
\newcommand{\ffcollins}[1]{\ensuremath{H_{\,1}^{\,\perp,{#1}}\left(\z\right)}\xspace}
\newcommand{\ffcollinsmod}[1]{\ensuremath{H_{\,1}^{\,\perp,{#1}}}\xspace}

\newcommand{\collins}[1]{\ensuremath{2\left<\sin{\left(\phi+\phis\right)}\right>_{\text{UT}}^{#1}}\xspace}
\newcommand{\collinsexp}[1]{\ensuremath{2\left<\sin{\left(\phi+\phis\right)}\right>_{\text{U}\perp}^{#1}}\xspace}
\newcommand{\sivers}[1]{\ensuremath{2\left<\sin{\left(\phi-\phis\right)}\right>_{\text{UT}}^{#1}}\xspace}
\newcommand{\siversexp}[1]{\ensuremath{2\left<\sin{\left(\phi-\phis\right)}\right>_{\text{U}\perp}^{#1}}\xspace}

\newcommand{\sinthreephiexp}[1]{\ensuremath{2\left<\sin{\left(3\phi-\phis\right)}\right>_{\text{U}\perp}^{#1}}\xspace}
\newcommand{\sintwophi}[1]{\ensuremath{2\left<\sin{\left(2\phi-\phis\right)}\right>_{\text{UT}}^{#1}}\xspace}
\newcommand{\sintwophiexp}[1]{\ensuremath{2\left<\sin{\left(2\phi-\phis\right)}\right>_{\text{U}\perp}^{#1}}\xspace}
\newcommand{\sinphis}[1]{\ensuremath{2\left<\sin{\left(\phis\right)}\right>_{\text{UT}}^{#1}}\xspace}
\newcommand{\sinphisexp}[1]{\ensuremath{2\left<\sin{\left(\phis\right)}\right>_{\text{U}\perp}^{#1}}\xspace}
\newcommand{\sintwophilexp}[1]{\ensuremath{2\left<\sin{\left(2\phi+\phis\right)}\right>_{\text{U}\perp}^{#1}}\xspace}
\newcommand{\cosphi}[1]{\ensuremath{2\left<\cos{\left(\phih-\phis\right)}\right>_{\text{LT}}^{#1}}\xspace}
\newcommand{\cosphiexp}[1]{\ensuremath{2\left<\cos{\left(\phih-\phis\right)}\right>_{\text{L}\perp}^{#1}}\xspace}
\newcommand{\cosphilexp}[1]{\ensuremath{2\left<\cos{\left(\phih+\phis\right)}\right>_{\text{L}\perp}^{#1}}\xspace}
\newcommand{\cosphis}[1]{\ensuremath{2\left<\cos{\left(\phis\right)}\right>_{\text{LT}}^{#1}}\xspace}
\newcommand{\cosphisexp}[1]{\ensuremath{2\left<\cos{\left(\phis\right)}\right>_{\text{L}\perp}^{#1}}\xspace}
\newcommand{\costwophi}[1]{\ensuremath{2\left<\cos{\left(2\phih-\phis\right)}\right>_{\text{LT}}^{#1}}\xspace}
\newcommand{\costwophiexp}[1]{\ensuremath{2\left<\cos{\left(2\phih-\phis\right)}\right>_{\text{L}\perp}^{#1}}\xspace}
\newcommand{\unpolcosphi}[1]{\ensuremath{2\left<\cos{\left(\phih\right)}\right>_{\text{UU}}^{#1}}\xspace}
\newcommand{\unpolcostwophi}[1]{\ensuremath{2\left<\cos{\left(2\phih\right)}\right>_{\text{UU}}^{#1}}\xspace}
\newcommand{\sinphi}[1]{\ensuremath{2\left<\sin{\left(\phih\right)}\right>_{\text{UL}}^{#1}}\xspace}
\newcommand{\sinphiexp}[1]{\ensuremath{2\left<\sin{\left(\phih\right)}\right>_{\text{U}\parallel}^{#1}}\xspace}
\newcommand{\sintwophiul}[1]{\ensuremath{2\left<\sin{\left(2\phih\right)}\right>_{\text{UL}}^{#1}}\xspace}
\newcommand{\sintwophiulexp}[1]{\ensuremath{2\left<\sin{\left(2\phih\right)}\right>_{\text{U}\parallel}^{#1}}\xspace}
\newcommand{\sinthreephiulexp}[1]{\ensuremath{2\left<\sin{\left(3\phih\right)}\right>_{\text{U}\parallel}^{#1}}\xspace}

\newcommand{\coszerophillexp}[1]{\ensuremath{2\left<\cos{\left(0\phih\right)}\right>_{\text{L}\parallel}^{#1}}\xspace}
\newcommand{\cosphill}[1]{\ensuremath{2\left<\cos{\left(\phih\right)}\right>_{\text{LL}}^{#1}}\xspace}
\newcommand{\cosphillexp}[1]{\ensuremath{2\left<\cos{\left(\phih\right)}\right>_{\text{L}\parallel}^{#1}}\xspace}
\newcommand{\costwophillexp}[1]{\ensuremath{2\left<\cos{\left(2\phih\right)}\right>_{\text{L}\parallel}^{#1}}\xspace}

\newcommand{\collinsexpSFA}[1]{\ensuremath{2\langle\sin{\left(\phi+\phis\right)} / \epsilon \, \rangle_{\text{U}\perp}^{#1}}\xspace}

\newcommand{\siversexpSFA}[1]{\ensuremath{2\langle\sin{\left(\phi-\phis\right)}\, \rangle_{\text{U}\perp}^{#1}}\xspace}

\newcommand{\sinthreephiexpSFA}[1]{\ensuremath{2\langle\sin{\left(3\phi-\phis\right)} / \epsilon \, \rangle_{\text{U}\perp}^{#1}}\xspace}

\newcommand{\sintwophiexpSFA}[1]{\ensuremath{2\langle\sin{\left(2\phi-\phis\right)} / \sqrt{2\epsilon(1+\epsilon)} \, \rangle_{\text{U}\perp}^{#1}}\xspace}

\newcommand{\sinphisexpSFA}[1]{\ensuremath{2\langle\sin{\left(\phis\right)} / \sqrt{2\epsilon(1+\epsilon)} \, \rangle_{\text{U}\perp}^{#1}}\xspace}
\newcommand{\sintwophilexpSFA}[1]{\ensuremath{2\langle\sin{\left(2\phi+\phis\right)} / \epsilon \, \rangle_{\text{U}\perp}^{#1}}\xspace}

\newcommand{\cosphiexpSFA}[1]{\ensuremath{2\langle\cos{\left(\phih-\phis\right)} / \sqrt{1-\epsilon^{2}} \, \rangle_{\text{L}\perp}^{#1}}\xspace}
\newcommand{\cosphilexpSFA}[1]{\ensuremath{2\langle\cos{\left(\phih+\phis\right)} / \sqrt{2\epsilon(1-\epsilon)} \, \rangle_{\text{L}\perp}^{#1}}\xspace}

\newcommand{\cosphisexpSFA}[1]{\ensuremath{2\langle\cos{\left(\phis\right)} / \sqrt{2\epsilon(1-\epsilon)} \, \rangle_{\text{L}\perp}^{#1}}\xspace}

\newcommand{\costwophiexpSFA}[1]{\ensuremath{2\langle\cos{\left(2\phih-\phis\right)} / \sqrt{2\epsilon(1-\epsilon)} \, \rangle_{\text{L}\perp}^{#1}}\xspace}

\newcommand{\q}{\ensuremath{\euvec{q}}\xspace}
\newcommand{\qt}{\ensuremath{\euvec{q}_T}\xspace}
\newcommand{\qtsqr}{\ensuremath{\euvec{q}_T^2}\xspace}
\newcommand{\pt}{\ensuremath{\euvec{p}_T}\xspace}
\newcommand{\ptsqr}{\ensuremath{\euvec{p}_T^2}\xspace}
\newcommand{\kt}{\ensuremath{\euvec{k}_T}\xspace}
\newcommand{\ktsqr}{\ensuremath{\euvec{k}_T^2}\xspace}

\newcommand{\Ph}{\ensuremath{\euvec{P}_h}\xspace}
\newcommand{\Phabs}{\ensuremath{\abs{\euvec{P}_h}}\xspace}
\newcommand{\Phperp}{\ensuremath{\euvec{P}_{h\perp}}\xspace}
\newcommand{\Phperpsqr}{\ensuremath{\Phperpabs^{2}}\xspace}
\newcommand{\Phperpabs}{\ensuremath{P_{h\perp}}\xspace}
\newcommand{\Q}{\ensuremath{Q^{\,2}}\xspace}

\newcommand{\W}{\ensuremath{W^2}\xspace}
\newcommand{\x}{\ensuremath{x}\xspace}

\newcommand{\xb}{\ensuremath{x}\xspace} 
\newcommand{\y}{\ensuremath{y}\xspace}
\newcommand{\z}{\ensuremath{z}\xspace}

\newcommand{\xf}{\ensuremath{x_F}\xspace}

\newcommand{\phis}{\ensuremath{\phi_{S}}\xspace}
\newcommand{\phih}{\ensuremath{\phi}\xspace}

\newcommand{\unit}[1]{\ensuremath{\,\text{\texttt{#1}}}}
\newcommand{\mrad}{\unit{mrad}}

\newcommand{\gev}{\unit{GeV}}
\newcommand{\GeV}{\gev}
\newcommand{\Tm}{\unit{Tm}}

%% file: sections/title.tex

\subheader{DESY Report 20-119}

\title{Azimuthal single- and double-spin asymmetries in semi-inclusive deep-inelastic lepton scattering by transversely polarized protons}

\abstract{A comprehensive set of azimuthal single-spin and double-spin asymmetries in semi-inclusive leptoproduction of pions, charged kaons, protons, and antiprotons 
from transversely polarized protons is presented. These asymmetries include the previously published \hermes results on Collins and Sivers asymmetries, the analysis of which has been extended to include protons and antiprotons and also to an extraction in a three-dimensional kinematic binning and enlarged phase space. They are complemented by corresponding results for the remaining four single-spin and four double-spin asymmetries allowed in the one-photon-exchange approximation of the semi-inclusive deep-inelastic scattering process for target-polarization orientation perpendicular to the direction of the incoming lepton beam.
Among those results, significant non-vanishing \cosinemodulation{\phih-\phis} modulations provide evidence for a sizable worm-gear (II) distribution, \tmdgtpt{q}. Most of the other modulations are found to be consistent with zero with the notable exception of large \sinemodulation{\phis} modulations for charged pions and  \kplus.
}

\input{./sections/authorsJHEP-hermesTMDs}

\collaboration{The HERMES Collaboration}

\keywords{Lepton-nucleon scattering, fixed-target experiments, QCD, polarization}


\maketitle

%% file: sections/authorsJHEP-hermesTMDs.tex
\author[13,16]{A.~Airapetian}
\author[26]{N.~Akopov}
\author[6]{Z.~Akopov}
\author[7]{E.C.~Aschenauer}
\author[25]{W.~Augustyniak}
\author[26,a]{R.~Avakian\note{Deceased.}}
\author[21]{A.~Bacchetta}
\author[19,a]{S.~Belostotski}
\author[20]{V.~Bryzgalov}
\author[11]{G.P.~Capitani}
\author[22]{E.~Cisbani}
\author[10]{G.~Ciullo}
\author[10]{M.~Contalbrigo}
\author[6]{W.~Deconinck}
\author[2]{R.~De~Leo}
\author[11]{E.~De~Sanctis}
\author[9]{M.~Diefenthaler}
\author[11]{P.~Di~Nezza}
\author[13]{M.~D\"uren}
\author[26]{G.~Elbakian}
\author[5]{F.~Ellinghaus}
\author[11]{A.~Fantoni}
\author[23]{L.~Felawka}
\author[6,19,23]{G.~Gavrilov}
\author[26]{V.~Gharibyan}
\author[6]{Y.~Holler}
\author[20]{A.~Ivanilov}
\author[1,a]{H.E.~Jackson}
\author[12]{S.~Joosten}
\author[14]{R.~Kaiser}
\author[6,26]{G.~Karyan}
\author[5]{E.~Kinney}
\author[19]{A.~Kisselev}
\author[17]{V.~Kozlov}
\author[9,19]{P.~Kravchenko}
\author[2]{L.~Lagamba}
\author[18]{L.~Lapik\'as}
\author[10]{P.~Lenisa}
\author[16]{W.~Lorenzon}
\author[19]{S.I.~Manaenkov}
\author[25,a]{B.~Marianski}
\author[26]{H.~Marukyan}
\author[24]{Y.~Miyachi}
\author[10,26]{A.~Movsisyan}
\author[11]{V.~Muccifora}
\author[19]{Y.~Naryshkin}
\author[9]{A.~Nass}
\author[26]{G.~Nazaryan}
\author[7]{W.-D.~Nowak}
\author[10]{L.L.~Pappalardo}
\author[1]{P.E.~Reimer}
\author[11]{A.R.~Reolon}
\author[7,15]{C.~Riedl}
\author[9]{K.~Rith}
\author[14]{G.~Rosner}
\author[6]{A.~Rostomyan}
\author[15]{J.~Rubin}
\author[12]{D.~Ryckbosch}
\author[21]{A.~Sch\"afer}
\author[3,4,12]{G.~Schnell}
\author[14]{B.~Seitz}
\author[24]{T.-A.~Shibata}
\author[8]{V.~Shutov}
\author[10]{M.~Statera}
\author[17]{A.~Terkulov}
\author[12]{M.~Tytgat}
\author[12]{Y.~Van~Haarlem}
\author[12]{C.~Van~Hulse}
\author[3,19]{D.~Veretennikov}
\author[2]{I.~Vilardi}
\author[9]{S.~Yaschenko}
\author[9]{D.~Zeiler}
\author[6]{B.~Zihlmann}
\author[25]{P.~Zupranski}

\affiliation[1]{Physics Division, Argonne National Laboratory, Argonne, Illinois 60439-4843, USA}
\affiliation[2]{Istituto Nazionale di Fisica Nucleare, Sezione di Bari, 70124 Bari, Italy}
\affiliation[3]{Department of Theoretical Physics, University of the Basque Country UPV/EHU, 48080 Bilbao, Spain}
\affiliation[4]{IKERBASQUE, Basque Foundation for Science, 48013 Bilbao, Spain}
\affiliation[5]{Nuclear Physics Laboratory, University of Colorado, Boulder, Colorado 80309-0390, USA}
\affiliation[6]{DESY, 22603 Hamburg, Germany}
\affiliation[7]{DESY, 15738 Zeuthen, Germany}
\affiliation[8]{Joint Institute for Nuclear Research, 141980 Dubna, Russia}
\affiliation[9]{Physikalisches Institut, Universit\"at Erlangen-N\"urnberg, 91058 Erlangen, Germany}
\affiliation[10]{Istituto Nazionale di Fisica Nucleare, Sezione di Ferrara, and Dipartimento di Fisica e Scienze della Terra, Universit\`a di Ferrara, 44122 Ferrara, Italy}
\affiliation[11]{Istituto Nazionale di Fisica Nucleare, Laboratori Nazionali di Frascati, 00044 Frascati, Italy}
\affiliation[12]{Department of Physics and Astronomy, Ghent University, 9000 Gent, Belgium}
\affiliation[13]{II. Physikalisches Institut, Justus-Liebig Universit\"at Gie{\ss}en, 35392 Gie{\ss}en, Germany}
\affiliation[14]{SUPA, School of Physics and Astronomy, University of Glasgow, Glasgow G12 8QQ, United Kingdom}
\affiliation[15]{Department of Physics, University of Illinois, Urbana, Illinois 61801-3080, USA}
\affiliation[16]{Randall Laboratory of Physics, University of Michigan, Ann Arbor, Michigan 48109-1040, USA }
\affiliation[17]{Lebedev Physical Institute, 117924 Moscow, Russia}
\affiliation[18]{National Institute for Subatomic Physics (Nikhef), 1009 DB Amsterdam, The Netherlands}
\affiliation[19]{Petersburg Nuclear Physics Institute, National Research Center Kurchatov Institute, Gatchina, 188300 Leningrad Region, Russia}
\affiliation[20]{Institute for High Energy Physics, National Research Center Kurchatov Institute, Protvino, 142281 Moscow Region, Russia}
\affiliation[21]{Institut f\"ur Theoretische Physik, Universit\"at Regensburg, 93040 Regensburg, Germany}
\affiliation[22]{Istituto Nazionale di Fisica Nucleare, Sezione di Roma, Gruppo Collegato Sanit\`a, and Istituto Superiore di Sanit\`a, 00161 Roma, Italy}
\affiliation[23]{TRIUMF, Vancouver, British Columbia V6T 2A3, Canada}
\affiliation[24]{Department of Physics, Tokyo Institute of Technology, Tokyo 152, Japan}
\affiliation[25]{National Centre for Nuclear Research, 00-689 Warsaw, Poland}
\affiliation[26]{Yerevan Physics Institute, 375036 Yerevan, Armenia}

%% file: sections/introduction.tex
\section{Introduction}\label{sec-introduction}

The present knowledge of the internal structure of the nucleon has emerged from
half a century of increasingly precise experimental investigation, in
particular of deep-inelastic scattering (DIS) of leptons (see, e.g., refs.~\cite{EPJA52,Gao:2017yyd}).
This process is traditionally interpreted in the collinear approximation of the quark-parton model,
where the main variable represents the longitudinal momentum of the quark expressed as
a fraction \x of that of the nucleon, in a frame in which the latter is
very large (``infinite-momentum frame'').\footnote{More formally, \x is the fraction of the nucleon's light-cone ``\(+\)'' momentum carried by the quark.}
One reason for this field to continue flourishing
is the intrinsic richness of the subject~\cite{Collins:2011zzd}.  
Technological advances in polarized beams and targets applied to the \dis process
make it possible to reveal correlations between
the spins of both partons and parent nucleon and 
the longitudinal and transverse components of the momentum of the partons.
The key aspects are control of polarizations in the initial state
without excessive penalty in luminosity,
as well as substantial acceptance
permitting detection of 
not only the scattered leptons but also
identified hadrons in the final state.
The distribution of these hadrons carries information about the struck quark's
transverse momentum, \pt, combined with transverse momentum acquired in the
fragmentation process, and the type of hadron 
provides information about the struck quark's flavor.

All parton distribution functions (PDFs) evolve with the hard scale
represented in \dis by \Q, where $-$\Q is the square of the 
four-momentum of the exchanged virtual photon.\footnote{For 
brevity, this dependence will be often omitted in the notation used here.}
More important in the context of the work presented here is that all PDFs can depend 
not only on \x but also on \pt. If the full dependence on these two variables
is retained, they are referred to as transverse-momentum dependent (TMD)
PDFs.

At leading twist\footnote{Following the ``working definition'' of
  Jaffe~\cite{Jaffe:1996zw}, twist \(t\) denotes the order  \(2-t\) of power
  suppression in the hard scale of the process under study, leading twist
  corresponding to twist 2 in this context.}, there are eight TMD PDFs.
Only three of them survive integration over \pt and therefore have a
corresponding standard collinear PDF:
the polarization-averaged or `unpolarized' distribution \pdffpt{q}, the quark 
helicity distribution \pdfgpt{q}, and the transversity distribution \pdfhpt{q}. 
While some information is available on the \pt dependence of \pdffpt{q}, 
very little is known about the \pt dependence of the other two.

The five leading-twist TMD PDFs that do {\em not} survive 
integration over \pt typically describe a correlation between \pt and
the spin direction of the parent nucleon and/or the ejected quark (and always 
implicitly \x as well). 
Three of these TMD PDFs are chiral odd\footnote{The definition of a quark PDF contains two quark fields:
  chiral-odd functions change sign if the chirality of the field operators is reversed~\cite{Jaffe:1996zw}.}
like the transversity distribution,
being related to transverse polarization of the struck quark.
This property excludes them from influencing any 
inclusive-DIS observable, at least
neglecting mass-suppressed effects.  
Chiral-odd PDFs appear only in observables 
involving two chiral-odd partners.  
Examples of such partnerships are two chiral-odd PDFs in the Drell--Yan process,
or a chiral-odd PDF with a chiral-odd fragmentation function (FF)
describing production of hadrons in semi-inclusive \dis.

Two TMD PDFs, the Sivers distribution \tmdsiverspt{q} and the Boer--Mulders distribution \tmdbmpt{q}
(see section \ref{theory-siversandboermulders}),
are rather intriguing because they are odd under naive time reversal (\Todd),
meaning that they describe a dependence on a triple product of two momenta and a spin
vector, which changes sign upon inverting all 
three-momenta and angular momenta.
As will be discussed below, the first observation
of a non-zero value for a \Todd TMD PDF led to the realization that this property
challenges the traditional concepts of factorization and universality of PDFs.
Furthermore, the \Todd property of TMD PDFs
provides a mechanism to explain the otherwise puzzling observation of
single-spin asymmetries (SSAs) in either hadron-hadron collisions or  \dis.

There are now indications that a substantial contribution to the helicity sum rule
for the nucleon comes from parton orbital angular 
momentum (cf.~refs.~\cite{Engelhardt:2019lyy,Alexandrou:2020sml}).
A tantalizing aspect of TMD PDFs is that some of them are related 
to the orbital angular momentum of quarks.
Non-zero values of these TMD PDFs require the
presence of nucleon wave function components with different orbital angular
momenta. However, no quantitative relationship between a TMD PDF and orbital 
angular momentum has yet been identified.

\begin{figure}[t]
\centering
\includegraphics[width=0.5\textwidth,keepaspectratio]{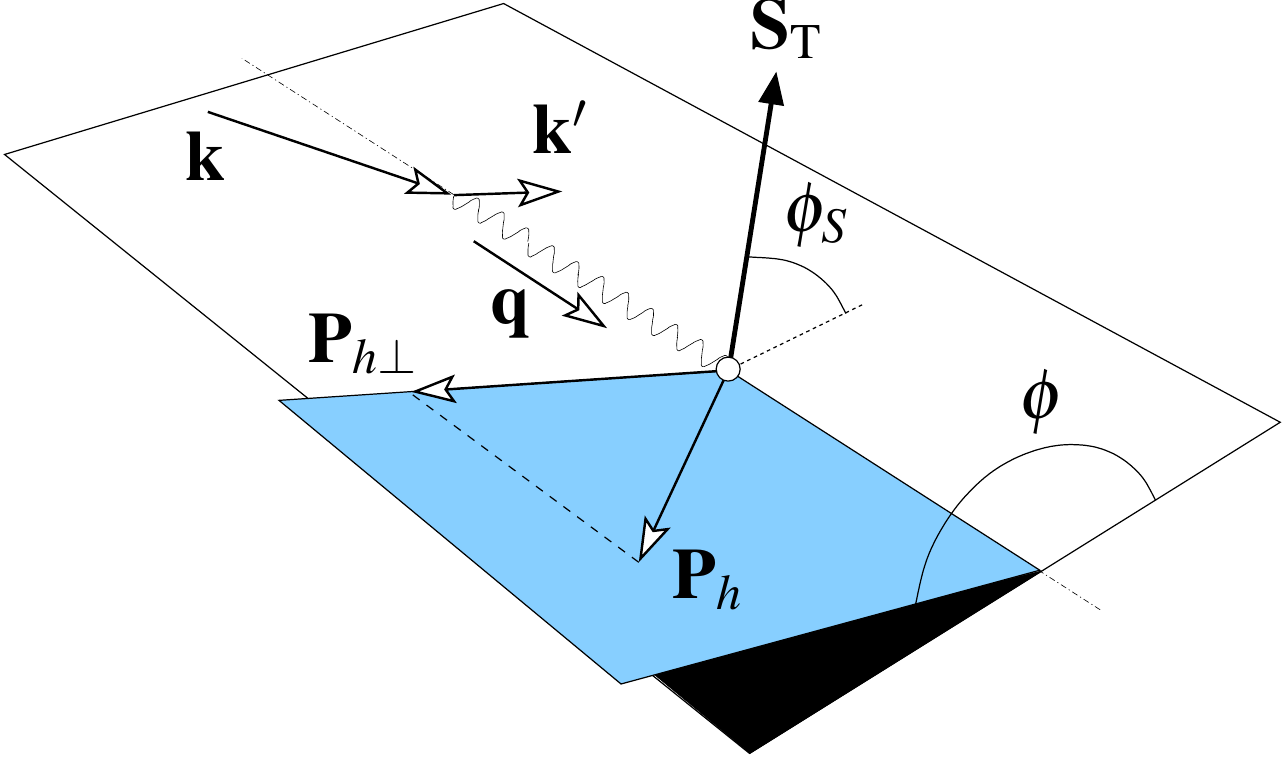}
  \caption{Following the {\em Trento conventions}~\cite{Bacchetta04}, \phih is defined to be the angle between the lepton scattering plane and the plane spanned by the virtual-photon momentum \( \text{\textbf{q}} \equiv \text{\textbf{k}}' - \text{\textbf{k}}\) (the difference of the momenta of the outgoing and incoming lepton) and \textbf{P}\(_{h}\), the momentum vector of the observed hadron, about the virtual-photon axis. Likewise, the angle \phis is defined as the angle between the lepton scattering plane and the target-polarization vector \textbf{S}\(_{T}\) of the transversely polarized nucleon.}
 \label{fig:angles}
\end{figure}

TMD PDFs can be experimentally constrained in semi-inclusive \dis by measurements of azimuthal distributions 
of the scattered lepton and produced hadrons about the direction of the exchanged virtual photon.
The Fourier harmonics of those distributions relate to specific structure functions.
The involved angles with respect to the lepton scattering plane are the azimuthal angle \phih
of the detected hadron and --- when target
polarization is involved --- the azimuthal angle \phis
of the polarization component orthogonal to the
direction of the virtual photon, as depicted in figure~\ref{fig:angles}.

At small transverse momentum,
factorization theorems make it possible to express the structure functions as
convolutions over quark transverse momentum of a TMD PDF and a
TMD FF~\cite{Collins:2011zzd}. 
TMD PDFs and TMD FFs will collectively be denoted as TMDs, when needed.
As final-state polarizations are not measured in the present work,
only two leading-twist TMD FFs are available to couple
to the leading-twist TMD PDFs in the structure functions.
The chiral-even TMD PDFs are convoluted with the polarization-averaged TMD FF \ffdkt{q\rightarrow h}, 
while the chiral-odd TMD PDFs, such as the transversity distribution, 
are convoluted with the chiral-odd \Todd Collins TMD FF \ffcollinskt{q\rightarrow h}, representing a correlation between the transverse polarization
of the fragmenting quark and the transverse momentum $\z\kt$ of the produced hadron carrying
the fraction \z of the energy of the virtual photon in the target-rest frame.  
Thus, the Collins fragmentation function acts as a `quark polarimeter'.
Table \ref{theory-tmdsummary} summarizes some properties of the leading-twist TMDs.

\tableenv{t}{
\begin{tabular}{|lccc|}
\hline
\textbf{Name} & \hspace*{0.4cm}\textbf{\tmd PDF/FF} \hspace*{0.25cm}& \hspace*{0.3cm}\textbf{Chirality}\hspace*{0.3cm} & \textbf{Naive time reversal} \\
\hline
Polarization-averaged & \pdff{q} & even & even \\
Helicity   & \pdfg{q} & even & even \\
Transversity & \pdfh{q} & odd & even  \\
Sivers & \tmdsivers{q} & even & odd  \\
Boer--Mulders & \tmdbm{q} & odd & odd  \\
Pretzelosity & \tmdht{q} & odd & even  \\
Worm-gear (I) & \tmdhl{q} & odd & even  \\
Worm-gear (II) & \tmdgt{q} & even & even  \\ \hline
Polarization-averaged & \ffdmod{q\rightarrow h} & even & even \\
Collins & \ffcollinsmod{q\rightarrow h} & odd & odd \\
\hline
\end{tabular}}{Leading-twist TMD distribution and fragmentation functions 
and their key symmetry properties. 
Only the first three TMD PDFs and \ffdmod{q\rightarrow h} survive integration over transverse momentum.
}{theory-tmdsummary}

In this work, azimuthal asymmetries in the yield of pions and charged kaons
are extracted from semi-inclusive \dis data recorded with a transversely polarized 
hydrogen target at the \hermes experiment.  They are supplemented with the first such 
measurements for proton and antiproton electroproduction. 
Fourier amplitudes of single-spin asymmetries
are presented that arise from the transversity \pdfhpt{q},
the Sivers \tmdsiverspt{q}, and the pretzelosity 
\tmdhtpt{q} distributions.  Also, a Fourier amplitude 
related to the worm-gear distribution \tmdgtpt{q} is extracted from the double-spin
asymmetry (DSA) requiring longitudinally polarized beams.
Besides these leading-twist contributions,
kinematically suppressed Fourier amplitudes are also reported, e.g., those related to the other worm-gear distribution \tmdhlpt{q}
due to a small longitudinal component of the target-polarization vector (cf.~appendix~\ref{sec-app-longpol}), or those involving twist-3 TMDs.
All Fourier amplitudes for charged mesons and for protons are extracted in a 
three-dimensional binning in the kinematic variables \xb, \z, as well as the transverse hadron momentum,
which will greatly facilitate disentangling the underlying dynamics of the 
partonic nucleon structure and of the fragmentation process. 
The Fourier amplitudes are extracted also in one-dimensional binning in those variables. 
Due to insufficient yields, the Fourier amplitudes for neutral pions and for antiprotons are provided in only the one-dimensional binning.

%% file: sections/theory.tex
\section{TMDs in semi-inclusive \dis}\label{sec-theory}

\subsection{Structure functions in the semi-inclusive DIS cross section}\label{theory-observables}

The observables of interest in this work are Fourier
amplitudes of the semi-inclusive DIS cross section, selected in all cases
by the polarization direction of the target nucleon with respect to the direction of the virtual photon, 
and --- in some cases --- also by the helicity of the beam lepton.
The azimuthal dependence of the cross section for leptoproduction of hadrons on a nucleon $N$ can be decomposed in the
one-photon-exchange approximation in terms of semi-inclusive DIS structure functions as~\cite{Bacchetta:2006tn}
\begin{align}
\lefteqn{\frac{\text{d} \sigma^{lN\to lhX}}{\text{d} \xb \, \text{d} y\, \text{d} \phis \,\text{d} z\, \text{d} \phih\, \text{d} \Phperpabs^2}
\propto \Biggl\{
 \structure{UU,T} + \epsilon\slim \structure{UU,L}  }  \nonumber \\ 
 & \quad   
 + \sqrt{2\,\epsilon (1+\epsilon)} \,\cosinemodulation{\phih}\, \structurecos{UU}{\phih}
 + \epsilon \, \cosinemodulation{2\phih}  \structurecos{UU}{2\phih}
 + \lambda_l\, \sqrt{2\,\epsilon (1-\epsilon)}\, \sinemodulation{\phih}\, \structuresin{LU}{\phih}
      \phantom{\Bigg[ \Bigg] }
      \nonumber \\  
 & \quad 
 + S_L\, \Bigg[ 
      \sqrt{2\, \epsilon (1+\epsilon)}\,  \sinemodulation{\phih}\, \structuresin{UL}{\phih}
      +  \epsilon \, \sinemodulation{2\phih}\, \structuresin{UL}{2\phih}
	\Bigg]
	\nonumber \\  
 & \quad 
 + S_L \lambda_l\, \Bigg[ \,
      \sqrt{1-\epsilon^2}\, \structure{LL}
      +\sqrt{2\,\epsilon (1-\epsilon)}\,  \cosinemodulation{\phih}\, \structurecos{LL}{\phih}
	\Bigg]
	\nonumber \\  
 & \quad 
 + S_T \, \Bigg[
      \sinemodulation{\phih-\phis} \, \Bigl( \structuresin{UT,T}{\phih -\phis} + \epsilon\, \structuresin{UT,L}{\phih -\phis} \Bigr)
      \label{theory-fourier}  \\  
 & \quad  \qquad \qquad 
  + \epsilon\, \sinemodulation{\phih+\phis} \, \structuresin{UT}{\phih +\phis}
  + \epsilon\, \sinemodulation{3\phih-\phis} \, \structuresin{UT}{3\phih -\phis}
	\phantom{\Bigg[ \Bigg] }
	\nonumber \\  
 & \quad \qquad \qquad 
  + \sqrt{2\,\epsilon (1+\epsilon)}\, \sinemodulation{\phis} \, \structuresin{UT}{\phis}
  + \sqrt{2\,\epsilon (1+\epsilon)}\, \sinemodulation{2\phih-\phis} \, \structuresin{UT}{2\phih -\phis}
	\Bigg]
	\nonumber \\  
 & \quad  
  + S_T \lambda_l \, \Bigg[
    \sqrt{1-\epsilon^2}\,\cosinemodulation{\phih-\phis} \, \structurecos{LT}{\phih -\phis}
	\nonumber \\  
 & \quad \qquad \qquad 
  + \sqrt{2\,\epsilon (1-\epsilon)}\, \cosinemodulation{\phis} \, \structurecos{LT}{\phis}
  + \sqrt{2\,\epsilon (1-\epsilon)}\, \cosinemodulation{2\phih-\phis} \, \structurecos{LT}{2\phih -\phis}
	\Bigg] 
 \Biggr\},\nonumber
\end{align}
where \(\xb\equiv \Q/(2P\cdot q)\),\footnote{While the right-hand side of this equation corresponds to the Bjorken variable, it coincides with the light-cone momentum fraction introduced in section~\ref{sec-introduction} in the Bjorken limit.}
\(y\equiv (P\cdot q)/(P\cdot k)\), 
\(z\equiv(P\cdot P_h)/(P\cdot q)\), 
with \(q\), \(P\), $k$, $k'$ and $P_h$ representing the four-momenta of the 
exchanged virtual photon, initial-state target nucleon, incident and outgoing lepton, and produced hadron \(h\), respectively.
Furthermore, 
\begin{equation}
   \Phperpabs \equiv \left| \Ph - \frac{(\Ph \cdot \q)\q}{|\q|^2} \right| 
\label{eq:Phperp} 
\end{equation}
is the magnitude of the hadron's transverse momentum, 
\(\lambda_{l} =+1\) (\(\lambda_{l} =-1\)) denotes right-handed (left-handed) beam leptons in the lepton-nucleon center-of-mass system, 
and the ``photon polarization parameter'' $\epsilon \equiv \frac{1-y-\frac{1}{4}\gamma^{2}y^2}{1-y+\frac{1}{4}y^{2}(\gamma^2+2)}$ 
is the ratio of longitudinal to transverse photon flux, where $\gamma \equiv 2M\xb/Q$ with $M$ the mass of the target nucleon.

The structure functions \(F\) depend in general on \x, \z, \Phperpabs and \Q.
The first subscript U (L) on the structure functions represents unpolarized (longitudinally polarized)
beam, while the second subscript T (L) denotes transverse (longitudinal) target polarization \(S_T\) (\(S_L\)). 
When present, the third subscript T (L) denotes transverse (longitudinal) virtual photons.
In principle, all structure functions have a dependence on the hadron type, although the hadron label \(h\) is omitted for compactness.

As will be discussed in more detail in section~\ref{sec:qpm}, the transverse-polarization-dependent azimuthal modulations
appearing in the fifth, sixth, and eighth line of eq.~\eqref{theory-fourier} arise 
as convolutions of leading-twist (twist-2) TMDs,
while the remaining transverse-polarization dependent terms involve twist-3 TMDs.

\subsection{Connection between structure functions and TMDs}\label{sec:qpm}

According to factorization theorems (see, e.g.,
refs.~\cite{Collins:1981uw,Collins:1984kg,Collins:1989gx,Ji:2004wu,Collins:2011zzd,Echevarria:2012js,Rogers:2015sqa,Collins:2017oxh}
and references therein),
at small transverse momentum\footnote{See ref.~\cite{Boglione:2019nwk} and appendix \ref{app:TMDfactorization}
  for a discussion on the limits of applicability of the TMD formalism.} 
the structure functions in eq.~\eqref{theory-fourier} 
can be written as convolutions in transverse-momentum space of a TMD PDF and
a TMD FF, possibly accompanied by a weighting factor \(w(\pt,\kt)\), i.e.,
\begin{equation}
  F  \left( \xb,\z,\Phperpabs,\Q \right) = {\cal C} \bigl[ w f^{q} D^{q\to h}
    \bigr]~,
\label{eq:strucfunc-generic}
\end{equation}
where the notation ${\cal C}$ indicates the convolution
\begin{equation}
  \begin{split}
    {\cal C} \bigl[ w f^{q} D^{q\to h} \bigr] \equiv
   {\sum_q} e_q^2 H(\Q) \! \int &\text{d}^2 \pt \text{d}^2 \kt \, \delta^{(2)}
    \biggl(\pt - \kt -\frac{\Phperp}{\z} \biggr)
\\
   & w(\pt,\kt) f^{q}(\x,\ptsqr;\Q) D^{q\to h}(\z,\ktsqr;\Q) .
\label{eq:convolution}
  \end{split}
\end{equation}
Here, \(e_q\) are the quark electric charges in units of the
elementary charge, \(H\) is a hard function that can be computed perturbatively
as a power expansion in the strong coupling constant \(\alpha_S\)~\cite{Collins:2017oxh}.
The TMDs included in the convolution depend on \Q in a way dictated by TMD evolution
equations~\cite{Collins:1981uk,Aybat:2011zv,Echevarria:2012pw}.\footnote{TMDs depend on two scales, usually denoted as $\mu^2$ and
  $\zeta$, but for simplicity we set them both to be equal to the hard scale \Q.} 
At variance with collinear PDFs, TMD evolution contains a
universal, flavor- and spin-independent nonperturbative component, which has to
be fixed from data~\cite{Bacchetta:2017gcc,Bacchetta:2019sam,Scimemi:2019cmh} 
or computed in lattice QCD~\cite{Ji:2019sxk,Vladimirov:2020ofp,Shanahan:2020zxr}.
At parton-model level, the hard function reduces to
unity, the TMDs become independent of \Q and the convolutions correspond to
the definition in, e.g., ref.~\cite{Bacchetta:2006tn}.

Table~\ref{theory-tmdsummary2} summarizes the correspondence between the 
leading-twist azimuthal modulations defined in eq.~\eqref{theory-fourier} and the 
TMDs appearing in the structure-function expressions.
Further details are provided below.

\tableenv{t}{
\begin{tabular}{|lccc|}
\hline
\textbf{Name}\hspace*{1.8cm} 	& \hspace*{.6cm} \textbf{\tmd PDF}  \hspace*{.6cm} 	& \hspace*{.6cm} \textbf{\tmd FF} \hspace*{.6cm} 	& \textbf{Structure function} \\
\hline
Transversity 	& \pdfh{q} 		& \ffcollinsmod{q\to h} 	& \structuresin{UT}{\phih+\phis} 	\\
Sivers 		& \tmdsivers{q} & \ffdmod{q\to h} 		& \structuresin{UT}{\phih-\phis}  	\\
Boer--Mulders 	& \tmdbm{q} 	& \ffcollinsmod{q\to h} 	& \structurecos{UU}{2\phih}		\\
Pretzelosity 	& \tmdht{q} 	& \ffcollinsmod{q\to h} 	& \structuresin{UT}{3\phih+\phis} 	\\
Worm-gear (I) 	& \tmdhl{q} 	& \ffcollinsmod{q\to h} 	& \structuresin{UL}{2\phih} 		\\
Worm-gear (II) 	& \tmdgt{q} 	& \ffdmod{q\to h} 		& \structurecos{LT}{\phih-\phis} 		\\
\hline
\end{tabular}}{Leading-twist TMD PDFs that do not survive integration over \pt, 
together with the TMD FFs with which they appear in their associated leading semi-inclusive DIS structure functions.
}{theory-tmdsummary2}

\subsubsection{The transversity distribution}\label{theory-transversity}

The transversity distribution has the probabilistic interpretation as 
the difference in number densities of quarks with transverse 
polarization parallel and anti-parallel to the transverse polarization of
the parent nucleon~\cite{Jaffe:1991kp}.  
Among the three leading-twist PDFs surviving integration 
over \pt, it is the only one that involves transverse quark polarization
and is thereby chiral-odd.  
Unlike the polarization-averaged and the quark-helicity distributions, QCD evolution of the transversity in a 
spin-$\frac{1}{2}$ hadron does not mix quarks with gluons because 
of helicity conservation~\cite{Artru:1989zv}.

The transversity distribution \pdfh{q} appears together with the Collins
fragmentation function \ffcollinsmod{q\rightarrow h} in the structure function
\begin{equation}
  \structuresin{UT}{\phih+\phis} \left(\xb,z,\Phperpabs,\Q \right)
   = \conv{\pdfh{q}}{\ffcollinsmod{q\rightarrow h}}{- \frac{\unitvec{h} \cdot \kt}{M_h}} ~,
 \label{eq:QPM-collins}
\end{equation}
where 
$M_h$ is the mass of the produced hadron  and
$\unitvec{h}=\Phperp/|\Phperp|$.

Azimuthal asymmetries related to \structuresin{UT}{\phih+\phis}
as a function of single kinematic variables have been published by the
\hermes Collaboration for charged pions~\cite{Airapetian:2004tw} and later
for pions, charged kaons, as well as the pion charge-difference~\cite{Airapetian:2010ds},
all from a transversely polarized hydrogen target.  
In the present work, the three-dimensional dependences of 
the so-called Collins asymmetries go beyond the original works of refs.~\cite{Airapetian:2004tw,Airapetian:2010ds}, 
which concentrated on one-dimensional binning in either the kinematic variable \x, \z, or \Phperpabs.
In addition, results for protons and antiprotons obtained here for the first time are included.

\subsubsection{The Sivers distribution}\label{theory-siversandboermulders}

The Sivers and Boer--Mulders functions are the only TMDs that are \Todd.
The chiral-even Sivers function \tmdsivers{q}~\cite{Sivers:1989cc} 
has the probabilistic interpretation as
the dependence of the number density of quarks on the orientation of \pt 
with respect
to the transverse polarization of the parent nucleon, while the chiral-odd
Boer--Mulders function \tmdbm{q}~\cite{Boer:1997nt} relates \pt to the transverse polarization of the
struck quark in an unpolarized nucleon.
The Boer--Mulders function is not further discussed in this paper, but relevant measurements and discussions are reported in
refs.~\cite{Airapetian:2012yg,Adolph:2014pwc} and the references therein.

Among the TMDs that do not survive integration over \pt,
these \Todd functions have thus far
received the most attention, both experimentally and theoretically.  
The possible existence of the Sivers 
function was proposed already three decades ago~\cite{Sivers:1989cc} in an effort 
to explain the unexpected 
single-spin asymmetries that had appeared in the production of pions from the 
collision of unpolarized with transversely polarized protons~\cite{Antille:1980th}.
That interpretation came under doubt when the \Todd Collins fragmentation function 
was proposed as an alternative mechanism, and it was demonstrated that
the existence of such \Todd TMD PDFs would violate the
fundamental time reversal symmetry~\cite{Collins:2002kn}.

A flurry of theoretical activity was
inspired by a seminal model calculation~\cite{Brodsky:2002cx} showing how the Sivers function
could legitimately arise through overlap integrals of quark wave functions with different orbital angular momenta,
together with a final-state interaction of the ejected quark with the 
target remnant. This soon led to the realization~\cite{Collins:2002kn,
Ji:2002aa, Belitsky:2002sm} that the aforementioned demonstration applied
only to \pt-integrated PDFs, in the definition of which a gauge link in the final
state could legitimately be neglected.  The gauge-invariant definition of TMD PDFs
requires this gauge link, which then provides the phase necessary for the interference
associated with the \Todd property.  The link can be interpreted as a final-state
interaction of the ejected quark with the color field of the target remnant.  This 
interaction can be considered to be embodied in the TMD PDF itself, with \pt
representing the transverse momentum {\em following} 
the interaction~\cite{Burkardt:2003uw}.

Incorporation of the gauge link into factorization proofs had a profound impact.
The concept of universality of leading-twist distribution functions had to be 
generalized to
allow for specific interaction dependences.  In the case of the Sivers function,
and in fact for all \Todd TMDs,
they are predicted to appear with the opposite sign in the expressions for \dis and 
Drell--Yan cross sections~\cite{Collins:2002kn}, reflecting the appearance of the embodied 
interaction in the final or initial state, respectively.\footnote{In the context 
of the present work, these distributions should therefore in principle appear with the label `DIS'.}
While the existence of a nonzero Sivers function was finally firmly established by data for semi-inclusive \dis of leptons
with transversely polarized targets~\cite{Airapetian:2004tw,Airapetian:2009ae,Alekseev:2010rw}, the
experimental verification of this direct prediction of QCD is eagerly awaited. 
Recent measurements of transverse-spin asymmetries in weak-boson production and the Drell--Yan process~\cite{Adamczyk:2015gyk,Aghasyan:2017jop}, 
albeit not sufficiently precise, are consistent with the sign change predicted. 

Much of the interest in the Sivers function arises from the evidence linking it
to orbital angular momentum of quarks.  Model calculations have found quark
wave function components with differing orbital angular momenta to be necessary
for a non-zero Sivers function.
The same statement can be made for relativistic theories of the anomalous magnetic moment $\kappa$ of the nucleon.
In fact, the same wave function components appear in both cases~\cite{Burkardt:2005km}.  
Under certain plausible assumptions, such as an attractive final-state interaction,
the sign of the Sivers function for each quark flavor is related to the sign
of the contribution of this flavor to $\kappa$~\cite{Burkardt:2002ks}. The predicted
relationship is consistent with experiment~\cite{Airapetian:2009ae}.
A quantitative estimate of orbital angular
momentum based on the Sivers function was attempted~\cite{Bacchetta:2011gx},
but it was based on restrictive assumptions~\cite{Pasquini:2019evu}.

In semi-inclusive \dis, the Sivers function appears convoluted with the unpolarized fragmentation function in the structure function~\cite{Boer:1997nt}
\begin{equation}
   \structuresin{UT}{\phih-\phis} \left(\xb,z,\Phperpabs,\Q \right) 
    = \convolution{ -\frac{\unitvec{h}\cdot \pt}{M} \, \tmdsivers{q} \, D_1^{q\rightarrow h} } ~.
\label{eq:QPM-sivers}
\end{equation}
It should be noted that the \(\sin(\phi-\phis)\) modulation of the semi-inclusive DIS cross section is the only one,
besides the azimuthally uniform denominator of all the \ssa and \dsa amplitudes,
that can in principle receive contributions from longitudinally polarized photons; these contributions, however, are vanishing
at leading and subleading twist in the region of low transverse momentum.

The \hermes Collaboration presented results for closely related asymmetries
for identified pions and charged kaons, as well as for the pion charge-difference asymmetry from a 
transversely polarized hydrogen target~\cite{Airapetian:2004tw,Airapetian:2009ae}.
In the present work, the three-dimensional dependences 
go again beyond those original works, which concentrated on
one-dimensional kinematic binning in either \x, \z, or \Phperpabs.
Furthermore, results for protons and antiprotons are presented here for the first time.

\subsubsection{The pretzelosity distribution}\label{theory-pretzelosity}

The \Teven chiral-odd pretzelosity TMD \tmdht{q},
introduced for the first time by Mulders and Tangerman~\cite{Tangerman:1995hw},
has the probabilistic interpretation
as the dependence of the number density of quarks on the relative orientation of \pt
and the transverse polarizations of both the quark and parent nucleon.  
In a helicity basis, this tensor structure corresponds to a flip of the quark helicity and
nucleon helicity in opposite directions.
The struck quark therefore has to absorb two units of orbital
angular momentum $L_z$, requiring either the presence of $s-d$
interference in the nucleon wave function, or matrix elements that are 
quadratic in a $p$ wave component. 
Other properties of the pretzelosity distribution are given, e.g., in ref.~\cite{Avakian:2008dz}.
In various models, such as bag or spectator models, the pretzelosity distribution appears
as the difference between helicity and transversity distributions, and hence
can be interpreted as representing relativistic effects in the nucleon structure.\footnote{For a non-relativistic system, where boosts and rotations commute, the transversity and helicity distributions would coincide (cf.~ref.~\cite{Jaffe:1996zw}).}
The name pretzelosity is loosely connected to the fact that this TMD is related to
a quadrupolar distortion of the quark
density~\cite{Miller:2007ae,Burkardt:2007rv}.

Being chiral-odd, 
pretzelosity appears in semi-inclusive DIS convoluted with the Collins fragmentation function
leading to a \(\sin\lf(3\phih - \phis \rg)\) modulation of the cross section~\cite{Tangerman:1995hw,Kotzinian:1994dv}:
\begin{align}
 \lefteqn{\structuresin{UT}{3\phih-\phis} \left( \xb,z,\Phperpabs,\Q \right) =} \nonumber \\
& \qquad \qquad  \convolution{\frac{2(\unitvec{h}\cdot\pt ) (\pt\cdot\kt )+\ptsqr(\unitvec{h}\cdot\kt )-4(\unitvec{h}\cdot\pt )^2 (\unitvec{h}\cdot\kt)}{2M^2M_h} \,\tmdht{q} \, \ffcollinsmod{q\rightarrow h}} ~.
\label{eq:sinthreephi}
\end{align}

The only existing measurement of this asymmetry comes from the Jefferson Lab Hall A Collaboration~\cite{Zhang:2013dow};
a transversely polarized $^{3}$He target was used, effectively a target of transversely polarized neutrons. 
The resulting asymmetry amplitudes are consistent with zero, both for $\pi^{+}$ and $\pi^{-}$. 
The measurements presented here for pions, charged kaons as well as for protons and antiprotons are the first of their kind for 
scattering off transversely polarized protons.

\subsubsection{The worm-gear distributions}\label{theory-wormgear}

The TMD PDFs \tmdgtpt{q} and
\tmdhlpt{q}~\cite{Ralston:1979ys,Tangerman:1995hw,Kotzinian:1995cz}
respectively describe
the number density of longitudinally polarized quarks in a transversely
polarized nucleon and of transversely polarized quarks in a longitudinally
polarized nucleon. 
The name ``worm gear'' refers to the orthogonal orientation of
the spins of quarks and nucleons. 
Both distributions are \Teven, and \tmdgtpt{q} is chiral-even while
\tmdhlpt{q} is chiral-odd.

A feature that distinguishes the two worm-gear distributions from all other TMDs is
that, in light-cone quark models, the corresponding generalized parton 
distributions (GPDs) vanish~\cite{Diehl:2005jf}. Furthermore,
model calculations~\cite{Boffi:2009sh} find that the two
distributions are closely related: $\tmdgtpt{q} = -\tmdhlpt{q}$.
However, this cannot be generally true at all scales due to the different evolution of 
chiral-even versus chiral-odd distributions (cf.~ref.~\cite{Bacchetta:2013pqa}).

In the Wandzura--Wilczek-type approximation (see, e.g.,
\cite{Avakian:2007mv}),
relations can be established between the worm-gear distributions \tmdgtpt{q} and \tmdhlpt{q} 
and the helicity and transversity distributions, respectively\footnote{For the adaptation of the original Wandzura--Wilczek
  approximation~\cite{Wandzura:1977qf} to semi-inclusive DIS see~\cite{Bastami:2018xqd} and references therein.} 
\begin{align}
\tmdonegt{q} \equiv \int \! \text{d}\ptsqr\, \frac{\ptsqr}{2M^{2}} \, \tmdgtpt{q} 
  &\stackrel{\text{\tiny WW}}{\approx} x \int\limits_x^1\frac{\text{d}\xi}{\xi}\, g_{\,1}^{\,q}\left(\xi \right)\, \stackrel{\text{\tiny WW}}{\approx}  x \twistgt{q} \, , \\
\tmdonehl{q} \equiv \int \! \text{d}\ptsqr\, \frac{\ptsqr}{2M^{2}} \,  \tmdhlpt{q} 
  &\stackrel{\text{\tiny WW}}{\approx}  - x^2 \int\limits_x^1\frac{\text{d}\xi}{\xi^2}\, h_{\,1}^{\,q}\left(\xi \right) \stackrel{\text{\tiny WW}}{\approx}  -\frac{1}{2} x \twisthl{q} \, ,
\end{align}
where all approximate signs involve Wandzura--Wilczek-type approximations and
the neglect of mass terms.
Experimental tests of the relations between the \ptsqr-moments of the worm-gear and the particular moments of the collinear helicity and transversity distributions would thus provide indications whether or not the relevant genuine twist-3 contributions to \twistgt{} and \twisthl{} are significant (cf.~discussion in ref.~\cite{Accardi:2009au}).

The structure function \structurecos{LT}{\phih-\phis} of the target-spin and beam-helicity dependent cross section 
provides a leading-twist signal for the worm-gear (II) distribution \tmdgtpt{q} in conjunction with the polarization-averaged fragmentation function \ffdkt{q\to h}:
\begin{equation}
 \structurecos{LT}{\phih-\phis} \left(\xb,z,\Phperpabs,\Q \right)
  = \convolution{ \frac{\unitvec{h}\cdot\pt}{M} \,\tmdgt{q} \, \ffdmod{q\rightarrow h} } \, .
\label{eq:cosphi}
\end{equation}

The Jefferson Lab Hall A Collaboration published related results for charged pions 
produced in semi-inclusive \dis off transversely polarized $^{3}$He and used these data 
to extract the corresponding Fourier amplitude for transversely polarized neutrons~\cite{Huang:2011bc}. 
While the results for positive pions are consistent with zero, the ones for negative pions provide first evidence for a non-vanishing \tmdgtpt{q}.
The measurements presented here for pions, charged kaons as well
as for protons and antiprotons are the first of their kind for transversely polarized protons.

The chiral-odd worm-gear (I) distribution \tmdhlpt{q} couples to the chiral-odd Collins 
fragmentation function. In semi-inclusive \dis from longitudinally polarized nucleons this combination gives rise to~\cite{Tangerman:1995hw,Kotzinian:1994dv}
\begin{equation}
 \structuresin{UL}{2\phih} \left(\xb,z,\Phperpabs,\Q \right) 
 = \convolution{ -\frac{2( \unitvec{h}\cdot\kt ) (\unitvec{h}\cdot\pt ) -\kt\cdot\pt}{M M_h} \, \tmdhl{q} \, \ffcollinsmod{q\rightarrow h} }\, .
\end{equation}
The primary choice for studying \structuresin{UL}{2\phih} 
is scattering off a longitudinally polarized target (cf.~ref.~\cite{Airapetian:1999tv}), as such it would not normally be included in the present measurement.
However, due to the small but non-vanishing target-spin component that is longitudinal to the virtual-photon direction
in measurements on targets polarized perpendicular to the incident-beam direction (cf.~appendix~\ref{sec-app-longpol}), 
the worm-gear (II) distribution \tmdhlpt{q} can in principle be constrained also by these data. This will be further discussed in the corresponding section~\ref{sec:ALU_ALL}.

\subsubsection{The subleading-twist amplitudes}\label{theory-subleadingtwist}

Each structure function in both the antepenultimate and the ultimate lines of eq.~\eqref{theory-fourier} 
is given by a sum of several terms, each of which contains a twist-3 TMD  
convoluted with a twist-2 TMD. 
The twist-3 objects have no probabilistic interpretation and contain
{\em interaction-dependent} terms, i.e., they involve quark-gluon correlations in the nucleon wave function.
All these terms are suppressed by the factor $(M/Q)$, and hence become negligible in the Bjorken limit. 
Nevertheless, evidences for substantial twist-3 contributions to single-spin asymmetries 
have already been found in the \hermes kinematic 
region~\cite{Airapetian:1999tv,Airapetian:2001eg,Airapetian:2002mf,Airapetian:2005jc,Airapetian:2006rx, Airapetian:2019mov}.
The \(\sin\phih\) Fourier amplitude of the \(\pi^{+}\) leptoproduction cross section
for longitudinally polarized protons was found to have magnitudes as large as 
about 5\% of the polarization-averaged cross section,
which are typical of the more sizable leading-twist Fourier amplitudes among those mentioned above.
Hence, it is of interest to also extract here the non-leading single-spin and double-spin asymmetries
for transverse target polarization.

The \structuresin{UT}{2\phih-\phis} structure function is given by
\begin{equation}
\begin{split}
 \structuresin{UT}{2\phih-\phis} 
 & \left(\xb,z,\Phperpabs,\Q \right) = \\
 &    \frac{2M}{Q}\mathcal{C} \Bigg[ 
  \frac{2(\unitvec{h} \cdot \pt)^2 -\ptsqr}{2 M^2} \left( \xb
    \twistftperp{q} \, \ffdmod{q\to h}
    - \frac{M_h}{zM} \, \tmdht{q}  \, \widetilde{H}^{\,q\to h} \right)\\
   & \qquad
   - \frac{2(\unitvec{h} \cdot \kt) (\unitvec{h} \cdot \pt) - \pt
     \cdot \kt}{2 M M_h} \, \times \\
& 
\left( \xb \twistht{q}  \, \ffcollinsmod{q\to h}  + \frac{M_h}{zM} \, \tmdgt{q} \, \widetilde{G}^{\,\perp,q\to h} 
   \; + 
  \xb  \twisthtperp{q} \,  \ffcollinsmod{q\to h}  - \frac{M_h}{zM} \, \tmdsivers{q} \, \widetilde{D}^{\,\perp,q\to h}
   \right) \Bigg].
\end{split}
\label{eq:sintwophi}
\end{equation}
The interaction-dependent fragmentation functions are indicated by a tilde. 
Similarly, the \structuresin{UT}{\phis} structure function is given by
\begin{equation}\label{theory-sinphis}
\begin{split}
 \structuresin{UT}{\phis} 
 & \left(\xb,z,\Phperpabs,\Q \right) =
  \frac{2M}{Q}\mathcal{C} \Bigg[ 
  \xb \, \twistft{q} \, \ffdmod{q\to h} - \frac{M_h}{zM} \pdfh{q} \, \widetilde{H}^{\,q\to h} \; - \frac{\pt \cdot \kt}{2 M M_h} \, \times  \qquad \qquad \qquad \qquad \qquad \\
& \qquad
 \left(  \xb  \twistht{q} \,  \ffcollinsmod{q\to h}
   + \frac{M_h}{zM} \tmdgt{q} \, \widetilde{G}^{\,\perp,q\to h} \; 
    - 
     \xb  \twisthtperp{q} \,  \ffcollinsmod{q\to h}
   + \frac{M_h}{zM} \tmdsivers{q} \, \widetilde{D}^{\,\perp,q\to h}  \right)
      \Bigg] .
\end{split}
\end{equation}

The two structure functions involve rather similar combinations of twist-2 and twist-3 distribution and fragmentation functions. 
In Wandzura--Wilczek-type approximations,
the chiral-even \Todd twist-3 distributions \twistft{q} and \twistftperp{q} are related to the Sivers function, 
while the difference (sum) of the chiral-odd \Teven twist-3 distributions \twistht{q} and \twisthtperp{q}
are related to the transversity (pretzelosity)~\cite{Bacchetta:2006tn}.  
In general, the interaction-dependent fragmentation functions disappear in the Wandzura--Wilczek-type approximation. 
The expressions for these two structure functions thus simplify significantly in such an approach~\cite{Bastami:2018xqd}.

A unique feature of the partial cross section given by eq.~\eqref{theory-sinphis}
is that it is the only contribution to the cross section
\sigmaut{h}{T} that survives integration over transverse hadron momentum~\cite{Mulders:1995dh,Bacchetta:2006tn}:
\begin{equation}
 \int \! d^2 \Phperp \, \structuresin{UT}{\phis} \left(\xb,z,\Phperpabs,\Q \right)
 = -\xb \frac{2M_h}{Q} \sum\limits_q e_q^2 \, \pdfh{q} (\xb)
 \frac{\widetilde{H}^{\,q\to h}\left(z\right)}{z} \, .
\end{equation}
It thus provides sensitivity to the transversity distribution without
involving a convolution over intrinsic transverse momenta. 
Nonetheless, due to time-reversal invariance, this modulation must vanish in the one-photon-exchange approximation in the inclusive limit~\cite{Christ:1966zz},
i.e., summing over all final-state hadrons and integrating over \z, which has indeed been demonstrated in the kinematic regime of this measurement in ref.~\cite{Airapetian:2009ab}.

Interest in \( \tilde{H}^{\,q\to h}\left(z\right) \) has grown significantly in the past years due to its connection to the single-spin asymmetries observed in \( p^\uparrow p \to \pi\, X \). 
Using Lorentz-invariance relations as well as QCD equations of motion, it was shown that both \( \tilde{H}^{\,q\to h}\left(z\right) \) and the Collins function arise from the same underlying dynamical correlator~\cite{Gamberg:2017gle}. 
As a consequence, it would be very surprising if this function vanished. 
Besides being a candidate for explaining single-spin asymmetries observed in \( p^\uparrow p \to \pi\, X \) (cf.~ref.~\cite{Aschenauer:2015ndk} and references therein),
it also contributes to transverse target single-spin asymmetries 
in inclusive electroproduction of hadrons~\cite{Gamberg:2017gle} as measured, e.g., at HERMES~\cite{Airapetian:2013bim}.

Finally, the subleading structure functions contributing to the cross section \sigmalt{h}{T}
are given by
\begin{align}\label{theory-cos2phi-phis}
 \structurecos{LT}{2\phih-\phis} 
 & \left(\xb,z,\Phperpabs,\Q \right) = \nonumber \\
 & \frac{2M}{Q}\mathcal{C} \Bigg[ 
  -\frac{2(\unitvec{h} \cdot \pt)^2 -\ptsqr}{2 M^2} \left( \xb
    \twistgtperp{q} \, \ffdmod{q\to h}
    + \frac{M_h}{zM}\tmdht{q}  \, \widetilde{E}^{\,q\to h} \right) \nonumber \\
   & \qquad + \frac{2(\unitvec{h} \cdot \kt) (\unitvec{h} \cdot \pt) - \pt
     \cdot \kt}{2 M M_h} \times \nonumber \\
 & 
  \left( \xb  \twistet{q}  \, \ffcollinsmod{q\to h}
   - \frac{M_h}{zM} \tmdgt{q} \,  \widetilde{D}^{\,\perp,q\to h}
   \; - 
  \xb  \twistetperp{q} \, \ffcollinsmod{q\to h}
   - \frac{M_h}{zM} \tmdsivers{q} \, \widetilde{G}^{\,\perp,q\to h} 
   \right) \Bigg], 
\end{align}
and
\begin{equation}\label{eq:theory-cosphis}
\begin{split}
 \structurecos{LT}{\phis} 
& \left(\xb,z,\Phperpabs,\Q \right) = 
 \frac{2M}{Q}\mathcal{C} \Bigg[ - \xb   \twistgt{q} \, \ffdmod{q\to h}
    - \frac{M_h}{zM}\pdfh{q} \, \widetilde{E}^{\,q\to h} + \frac{\pt \cdot \kt}{2 M M_h} \times \qquad \qquad \qquad \qquad \qquad \\
& \qquad \left(
  \xb  \twistet{q} \, \ffcollinsmod{q\to h}
   - \frac{M_h}{zM} \tmdgt{q} \, \widetilde{D}^{\,\perp,q\to h} \; 
    + 
 \xb  \twistetperp{q} \, \ffcollinsmod{q\to h}
   + \frac{M_h}{zM} \tmdsivers{q} \, \widetilde{G}^{\,\perp,q\to h} 
   \right) \Bigg].
\end{split}
\end{equation}

Also here, the two structure functions involve rather similar combinations of twist-2 and twist-3 distribution and fragmentation functions.
However, the expressions simplify even more in Wandzura--Wilczek-type approximations as in addition to the 
interaction-dependent fragmentation functions also the chiral-odd \Todd twist-3 distributions \twistet{q} and \twistetperp{q}
vanish, thus leaving only the contribution from the chiral-even \Teven twist-3 distributions \twistgt{q} and \twistgtperp{q}~\cite{Bastami:2018xqd}.

As is the case for the  \structuresin{UT}{\phis} structure function, the partial cross section given by eq.~\eqref{eq:theory-cosphis} 
is the only contribution to the cross section \sigmalt{h}{T} that survives integration over transverse hadron momentum~\cite{Jaffe:1993xb,Mulders:1995dh,Bacchetta:2006tn}:
\begin{multline}\label{eq:cosphisintegrated}
\int \! d^2 \Phperp \,  \structurecos{LT}{\phis} \left(\xb,z,\Phperpabs,\Q \right)
= \\
-\xb \frac{2M}{Q} \sum\limits_q e_q^2 \, 
\left( \xb   \twistgt{q} (\xb)  \ffd{q\to h}  + \frac{M_{h}}{zM} \pdfh{q}(\xb)  \widetilde{E}^{\,q\to h}\left(z\right)  \right) .
\end{multline}
Already in the early 1990s it was pointed out that this modulation provides collinear access to transversity in semi-inclusive \dis~\cite{Jaffe:1993xb},
complementary to that using dihadron fragmentation~\cite{Efremov:1992pe, Collins:1993kq}.
The challenge is to disentangle the transversity contribution from that of \twistgt{q}, in particular as the latter appears with the dominant \( \ffd{q\to h} \) fragmentation function.

In the inclusive limit, only the term in eq.~\eqref{eq:cosphisintegrated} involving \twistgt{q} can contribute. It is related to the virtual-photon--absorption asymmetries $A_{\, 2}(\xb)$, used to extract information on the inclusive-DIS structure function $\pdfgtwo{}(\xb)$:
\begin{equation}\label{eq:g2gT}
\pdfg{}(\xb)+\pdfgtwo{}(\xb) = \frac{1}{2} \sum\limits_q e_q^2 \; \xb \,  \twistgt{q} (\xb) \, .
\end{equation}
Measurements of \pdfgtwo{} of the proton have been published by several experiments~\cite{Airapetian:2011wu, Adams:1997tq, Abe:1998wq, Anthony:2002hy, Armstrong:2018xgk},
which could be used together with measurements of the helicity distributions to put constraints on the \twistgt{q} contribution to eq.~\eqref{eq:cosphisintegrated}.

There is also special interest in \twistgt{q} itself through its dependence on the interaction-dependent function \( \bar{g}_{2} \); 
this function is related to the transverse color Lorentz force
the struck quark experiences from the spectator at the moment just after it is struck by the virtual photon~\cite{Burkardt:2008ps,Aslan:2019jis}. 
That is in contrast to the Sivers function, which integrates the transverse force over the length of the struck-quark's trajectory.

None of the four twist-3 Fourier amplitudes has so far been measured in semi-inclusive \dis.

%% file: sections/measurement.tex
\section{Measurement and analysis}\label{section-measurement}

The Fourier analysis of the azimuthal transverse-target-polarization 
dependence of the semi-inclusive \dis cross section
follows closely the approach in the earlier \hermes publications 
on the Sivers and Collins effects for pions and charged 
kaons~\cite{Airapetian:2009ae,Airapetian:2010ds}. 
The relevant aspects of the \hermes experiment and the  
general analysis framework are described below, 
while the differences between this analysis and that of the previous publications are 
listed in section~\ref{sec:analysis-differences}.

\subsection{The \hermes experiment}\label{sec:experiment}

The data to be presented were collected using the \hermes spectrometer~\cite{Ackerstaff:1998av} 
at the \hera lepton storage ring during the 2002--2005 running period. 
A longitudinally polarized positron beam (electrons in 2005) with a momentum of \(27.6\GeV\) 
traversed a transversely polarized hydrogen target.

A nuclear-polarized pure-hydrogen gas target~\cite{Airapetian:2004yf} internal to the \hera lepton storage ring was
used, providing highly polarized target samples without dilution from
unpolarized target material or background arising from unwanted
scattering from the target-material container. Furthermore, this
technique included rapid reversals of target-spin orientations, 
with the sign randomly chosen at 1-3 min time intervals.
This provided a substantial reduction of time-dependent
systematic uncertainties. For the 2002-2005 running, an average degree
of polarization, perpendicular to the lepton-beam direction, of \(0.725\pm0.053\) was achieved.

The \(27.6\GeV\) electron or positron beam of \hera  
became self-polarized in the transverse direction
due to a tiny spin-flip asymmetry in the
emission of synchrotron radiation (Sokolov--Ternov effect) \cite{Sokolov:1963zn}. 
Longitudinal beam polarization was then obtained through
spin rotators installed up- and down-stream of the \hermes interaction region. 
Every few months, the longitudinal beam polarization was reversed to allow balancing of
data for the two helicity states. 
For the data presented, the typical beam-polarization values
are between 30\% and 40\% in magnitude, 
with a negligible net polarization when averaged over the whole data-taking period.

Scattered leptons and charged hadrons produced in the forward direction were 
detected within an angular acceptance of about \(\pm 170\mrad\) horizontally 
and about \(\pm(40\)--\(140) \mrad\) vertically.
Charged-particle tracks were reconstructed using a set of drift chambers 
in front of and behind the \(1.6\Tm\) dipole magnet and corrected for the bending within the target magnetic field, 
resulting in an average momentum and 
angular resolution of about 1.5\% each. 

The particle-identification system 
consisted of a dual-radiator ring-imaging Cherenkov (\rich) detector, a transition-radiation
detector, a pre-shower scintillation counter and an electromagnetic
calorimeter. The \pid system provided a lepton
identification with an efficiency of \(98\%\) and a hadron
contamination of less than \(1\%\). In the momentum range \(2\GeV<
\Phabs <15\GeV\), charged pions, kaons, and protons\footnote{
The momentum range for (anti)protons is later restricted to \(4\GeV<
\Phabs <15\GeV\) in order to avoid the low-momentum region of large 
meson contamination due to inefficiencies of the \rich.}
are identified by using the \rich detector \cite{Akopov:2000qi}, for which
a hadron-identification algorithm is applied that takes into account the event
topology~\cite{Airapetian:2012yg}.

The electromagnetic calorimeter and the pre-shower scintillation counter 
were also employed in detecting photons with an energy above 1\GeV, 
which are used here in reconstructing neutral pions. Unaffected by the magnetic 
fields of both the target and the spectrometer magnet, photons were accepted 
in the horizontal and vertical angular ranges of \(\pm 175 \mrad \) and \(\pm (43\)--\(147)\mrad \), respectively.

\begin{figure}[t]
\centering
\includegraphics[width=0.65\textwidth,keepaspectratio]{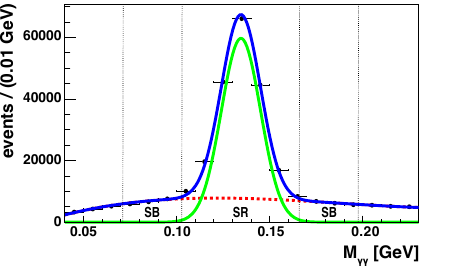}
  \caption{%
  	The two-photon invariant-mass distribution in the region of the \pizero mass for the overall data sample.
	The sum (blue line) of a Gaussian for the \pizero signal (green line) and a third-order Chebyshev polynomial for the combinatorial background (red dashed curve) are fit to data.
	The signal range used in the analysis, indicated as ``SR'', spans a \( \pm 3\sigma \) region around the \pizero peak position.
	Events for the background subtraction are selected from the sidebands denoted by ``SB''. The signal-region and sideband boundaries are indicated by vertical dotted lines.
	}
 \label{fig:pi0spectrum}
\end{figure}

Neutral pions are reconstructed using their dominant decay into two photons. 
The decay length of the \pizero is negligible compared to the resolution of the 
spectrometer, hence the decay vertex is assumed to coincide with the lepton-scattering vertex.
The photon pairs produced within the acceptance of the spectrometer generate electromagnetic 
showers in the calorimeter, a fraction of the photons starting a shower already in the lead sheet 
of the pre-shower detector, which is taken into account in the energy determination of the photon. 
For each \dis event with more than one photon detected in the calorimeter, 
the invariant mass of all possible photon-pair combinations is calculated under the assumption that the photon-pair 
originated from the lepton-scattering vertex. The resulting two-photon invariant-mass 
distribution for the overall data sample is shown in figure~\ref{fig:pi0spectrum}.
In each kinematic bin, the signal range is determined by a $\pm 3\sigma$ 
window around the \pizero peak position of the invariant-mass distribution, where $1\sigma$ 
reflects the energy resolution of the calorimeter. 
For the subtraction of the combinatorial background, events from sidebands to the left and right of the peak were used, 
appropriately weighted to reflect the amount of background in the signal region.

\subsection{Data selection}\label{measurement-kinematicrequirements}

Identified leptons were subject to various kinematic requirements in order to select a ``\dis sample'':
\begin{enumerate}[label=(\roman*)]
\item Final-state electrons and positrons are kept (including leptons with charge opposite to the beam leptons)
    in order to apply a correction for background contributions from pair-production processes.
\item The hard scattering scale of
the \dis process is constrained to \(\Q > 1\GeV^2\). 
\item Based on the chosen scale and the limited angular acceptance of the spectrometer,
the Bjorken scaling variable is required to be in the range \(0.023 < \xb < 0.6\). 
\item Scattering events originating from the excitation of
nucleon resonances and their subsequent strong decays are excluded by
the requirement \(\W>10\GeV^2\) on the squared invariant mass of the photon-nucleon system \(\W \equiv (q+P)^{2}\). 
\item The upper limit on \y is implied only
 by the calorimeter threshold of \(1.4\GeV\) (\( \y < 0.95 \)).
 The lower limit on \y is dictated by the \W constraint, resulting in a minimum \y of 0.18, which increases with \xb.
 No further restrictions are applied as they would have enhanced the strong correlation between the scaling variables \xb and \Q.
\end{enumerate}

\begin{figure}[]
\centering
 \includegraphics[width=0.7\textwidth,keepaspectratio]{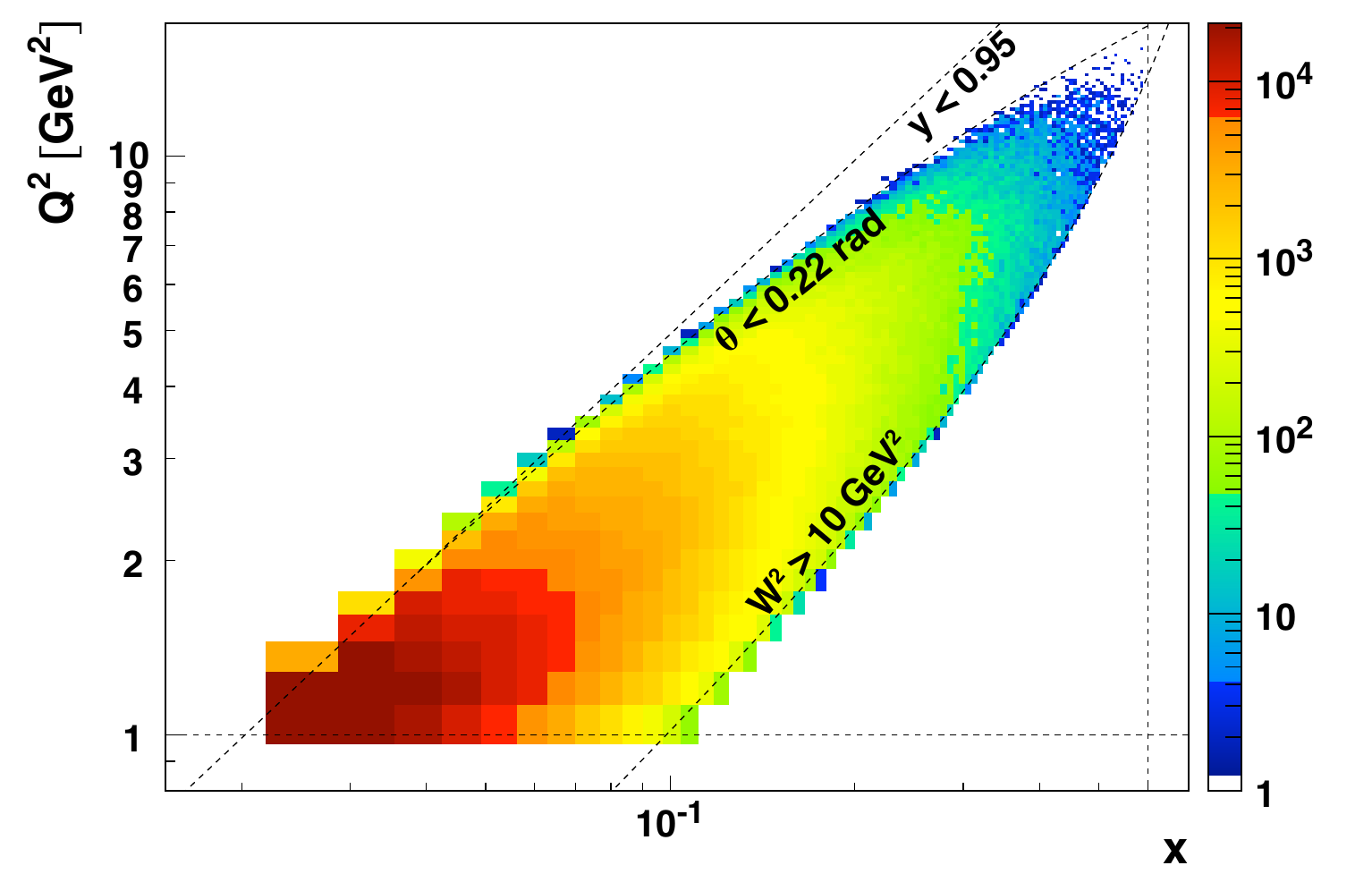}
 \caption{Event distribution in the kinematic space in (\xb, \Q), including the various boundaries arising from constraints on \x, \y, \Q, \W, and the upper reach in the lepton scattering angle \(\theta\).}
  \label{fig:xQ2plane}
\end{figure}

The resulting kinematic phase-space in the \xb--\Q plane is shown in figure \ref{fig:xQ2plane}, where also the constraints on \xb, \y, \Q, \W, and the upper reach in the lepton scattering angle are indicated. 
The strong correlation between \xb and \Q is apparent.

\begin{figure}[]
\centering
 \includegraphics[width=0.5\textwidth,keepaspectratio]{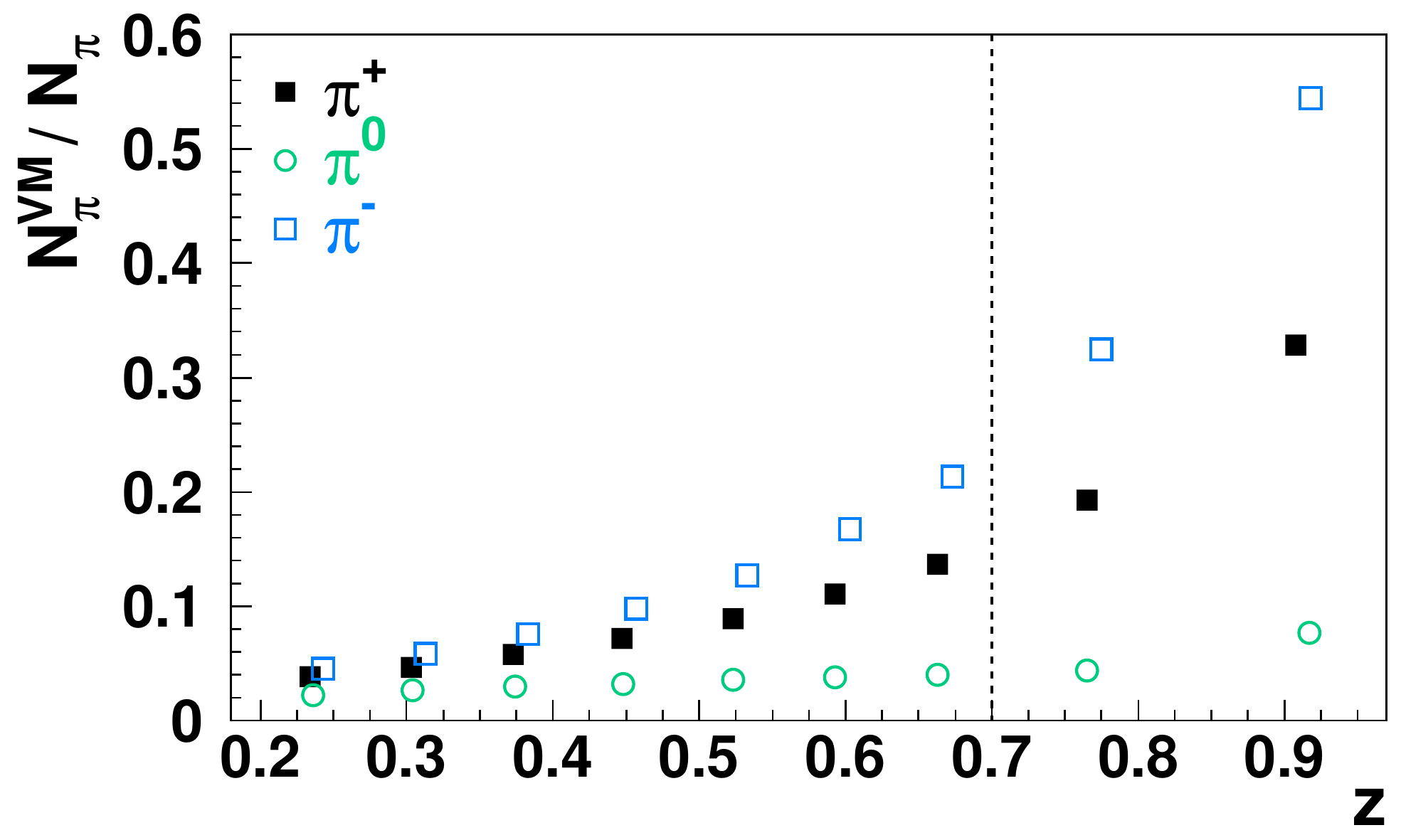}
 \caption{The simulated fraction of pions
   originating from diffractive vector-meson production and decay is 
   shown as a function of \z. (The open squares indicating \piminus 
   are slightly shifted horizontally). 
   The contributions are simulated by a version of \pythia~\cite{Sjostrand:2000wi,Sjostrand:2001yu}
   tuned for \hermes kinematics. By limiting \z
   to \(\z<0.7\), a kinematic region is probed where the vector-meson
   contribution to the electroproduction of pions 
  is suppressed, in particular for charged pions.
  For charged kaons, the contribution from \(\phi\) decay is at maximum 10\%~\cite{Airapetian:2012ki}.}
  \label{measurement-exclusivechannels}
\end{figure}

\begin{figure}[t]
\centering
 \includegraphics[width=0.49\textwidth,keepaspectratio]{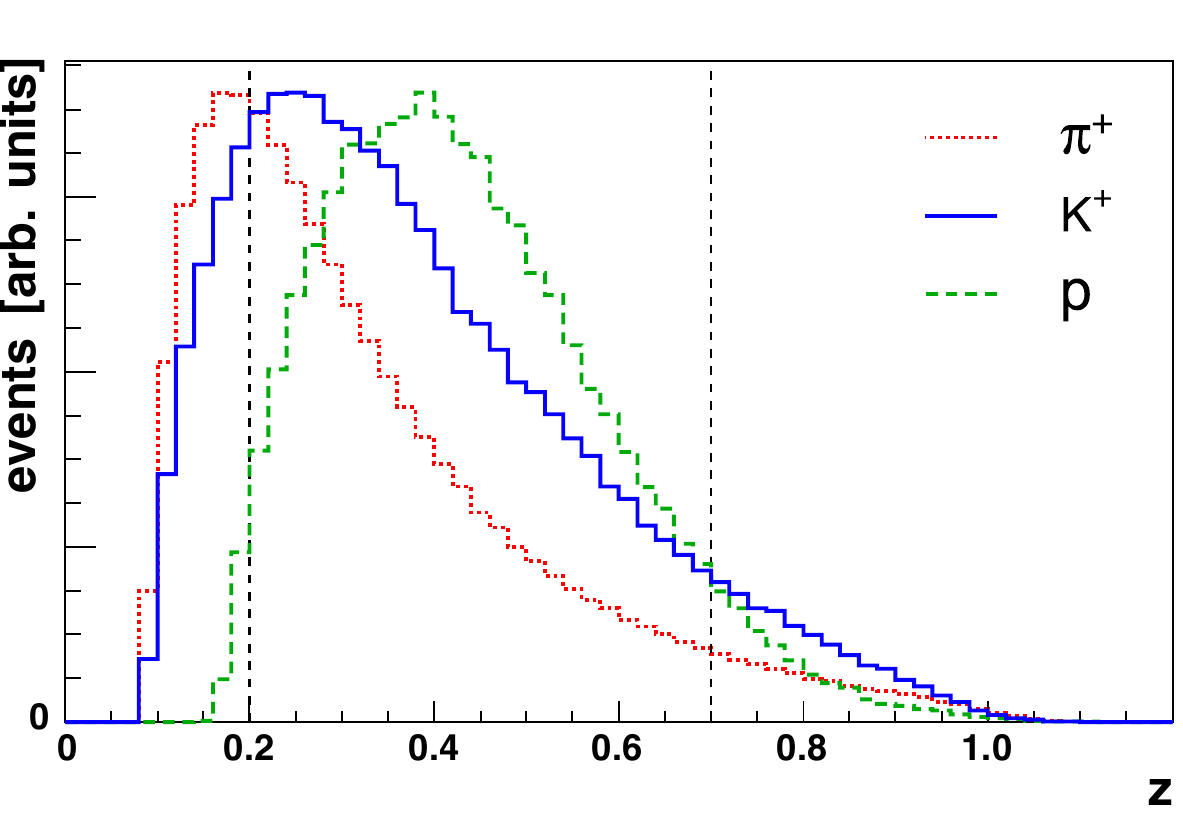}~
 \includegraphics[width=0.49\textwidth,keepaspectratio]{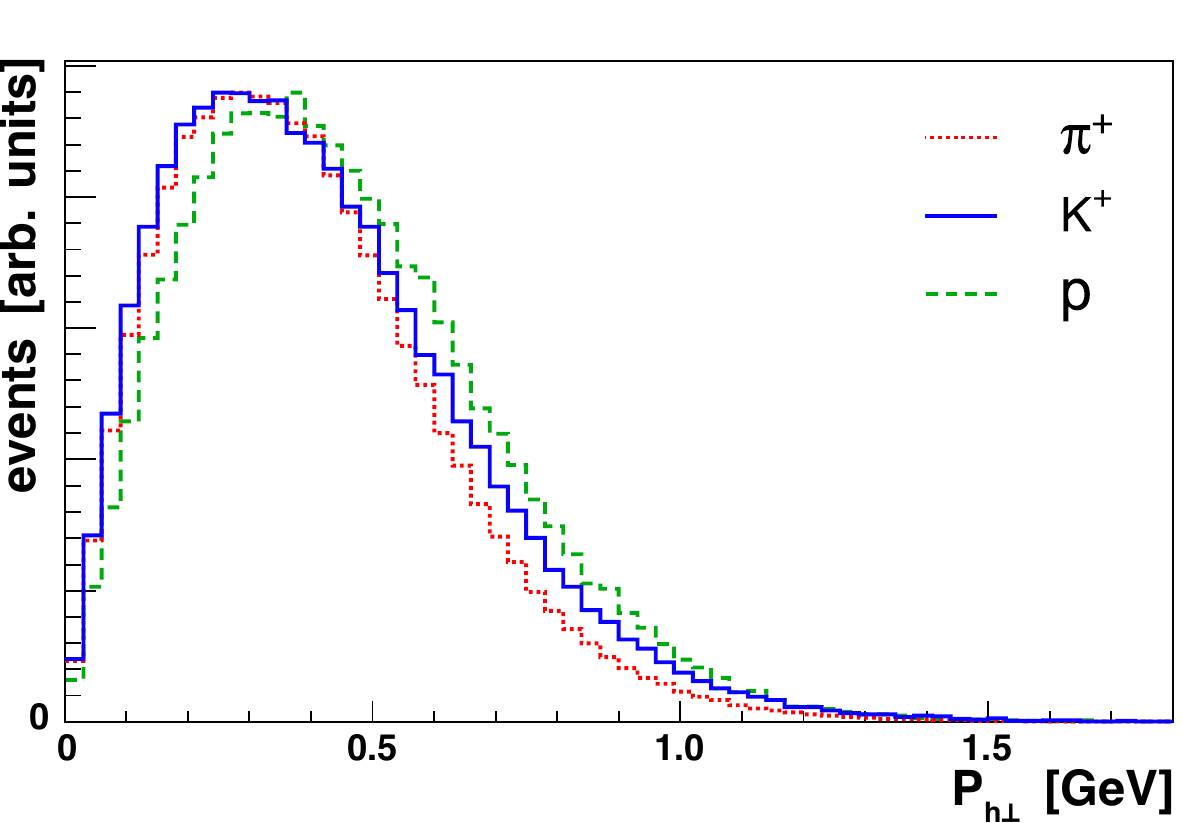}
 \caption{
  	Shape comparison of arbitrarily normalized \piplus (red dotted line), \kplus (blue line), and proton (green dashed line) yield distributions in the hadron variables \z (left) and \Phperpabs (right). 
	The region between the two vertical dashed lines indicates the range in \z used for semi-inclusive DIS sample, while events in the extended range \(0.7<\z<1.2\) are 
	analyzed only in the one-dimensional \z binning. 
	}
 \label{fig:hadron-variable-dist}
\end{figure}

\begin{figure}[t]
 \centering
 \includegraphics[width=0.7\textwidth,keepaspectratio]{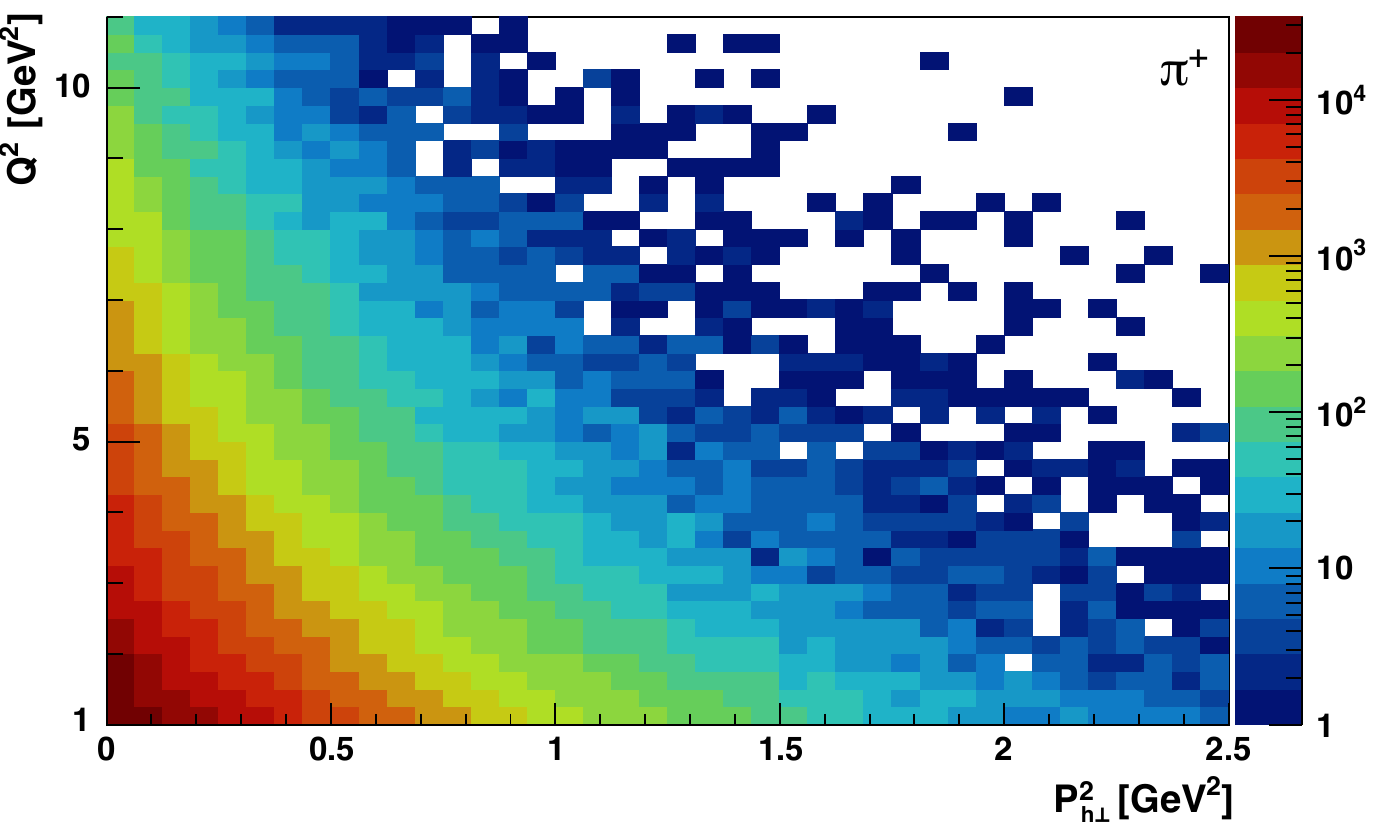}
 \caption{Distribution in \Q versus \Phperpsqr of the semi-inclusive \piplus yield.}
 \label{measurement-factorisation}
\end{figure}

The ``semi-inclusive \dis'' sample fulfills in addition the following criteria:
\begin{enumerate}[label=(\roman*)]
\item All identified hadrons are selected (and not only the leading
  hadron, i.e., the one with the highest momentum in the
  event).
\item A lower limit \(\z > 0.2\) is applied to suppress contributions from the target fragmentation region.
\item An upper limit  \(\z < 0.7\) is generally applied to suppress 
contributions from hadrons originating from the decay of diffractively produced 
vector-mesons.  
As shown in figure~\ref{measurement-exclusivechannels},
contributions due to exclusive channels (in particular for charged pions) become sizable at large \z. 
However, when looking at only the one-dimensional \z dependence of the azimuthal asymmetries, 
this requirement is lifted and instead an upper limit of 1.2 (driven by the detector resolution) is imposed, 
in order to probe this ``semi-exclusive'' transition region. 
The resulting yield distributions for the positively charged hadrons are shown in figure~\ref{fig:hadron-variable-dist} (left).
The shift towards higher \z in the distribution of protons mainly results from the larger hadron mass 
and the 4~\GeV\ minimum-momentum requirement (compared to 2~\GeV\ for charged mesons).
\item  The formalism of TMD factorization involves one hard scale, \Q, and 
  transverse momenta that are small in comparison.
  While no lower limit on \Phperpabs is imposed, an upper limit of \( \Phperpabs < 2\)~\GeV\ is applied in this analysis 
  (cf.~figure \ref{fig:hadron-variable-dist}, right).
  On average, the constraint \(\Phperpabs^2\ll\Q\) is
  fulfilled for most \dis events (cf.~figure~ \ref{measurement-factorisation}),
  while the stricter constraint \( \Phperpabs^2  \ll z^{2} \Q\) is often violated at large \Phperpabs in the kinematic region of low \xb (which corresponds to low \Q) and low \z.%
  \footnote{A more detailed discussion is presented in appendix~\ref{app:TMDfactorization}, 
  including further distributions, e.g., for the more critical region of low \z and \Q.} 
\end{enumerate}

\begin{figure}
 \centering
 \includegraphics[width=0.7\textwidth]{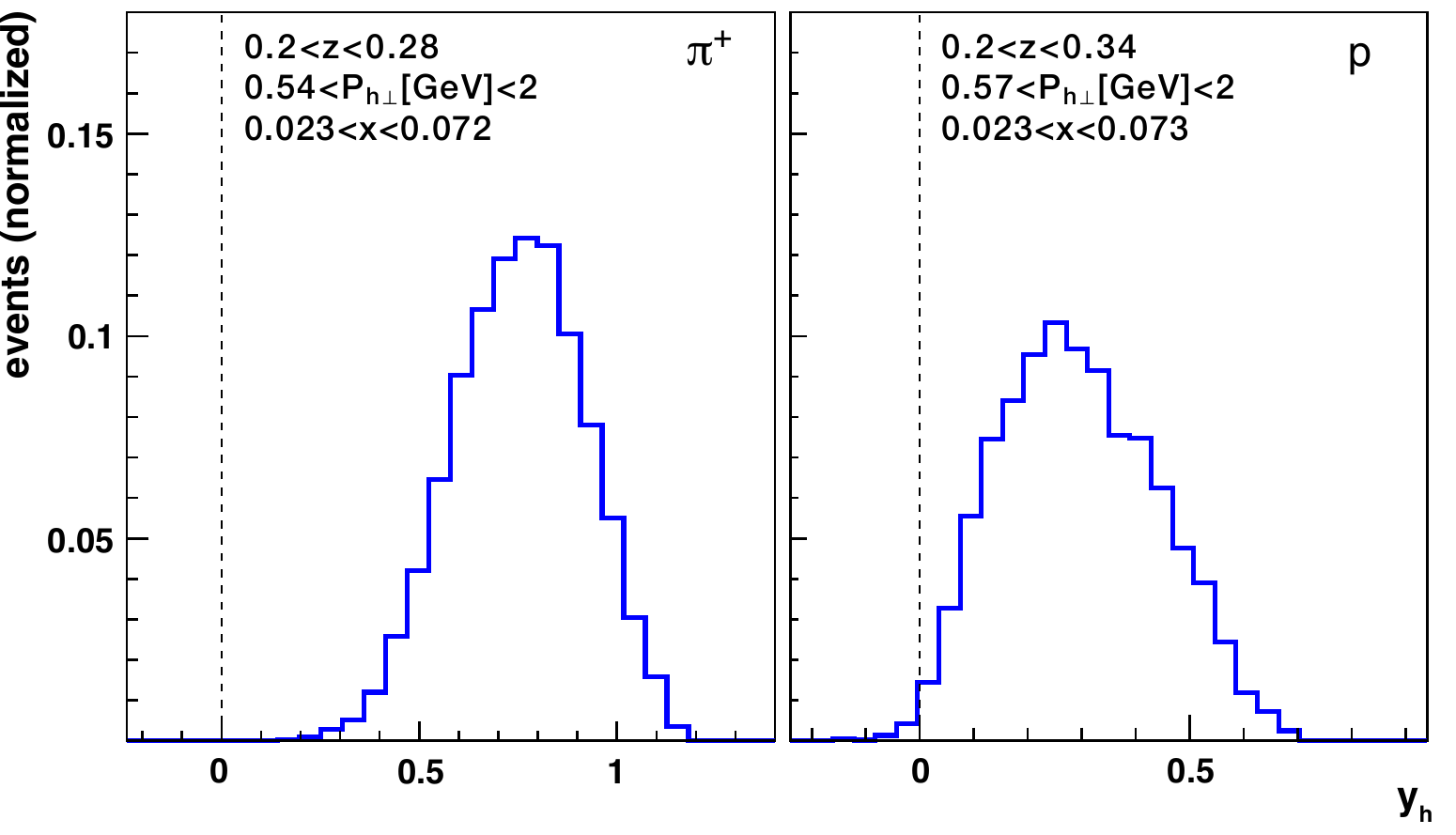}
 \caption{Rapidity distributions for \piplus (left) and protons (right) in the kinematic region indicated. (Distributions are normalized to unity.)}
 \label{fig:rapidity}
\end{figure}

Recently, separation of current and target fragmentation has been revisited for semi-inclusive \dis involving transverse momentum~\cite{Boglione:2016bph}.
In particular, low-\z hadrons with large transverse momentum might originate from the remnants of the target 
and not from the fragmentation of the struck quark~\cite{Berger:1987zu,Mulders:2000jt}, the region that is described here in terms of \tmd distribution and fragmentation functions. 
While no general recipe, e.g., a quantitative limit on kinematic variables, is available,
it appears appropriate to provide additional information about the kinematic distributions in this measurement.
For this it is useful to introduce both Feynman-\x, \xf, the ratio of the longitudinal hadron momentum \( P_{h\parallel}^{\text{CM}} \) along the 
virtual-photon direction to its maximum possible value in the virtual-photon--nucleon center-of-mass system (CM), and
rapidity,
\begin{equation}
y_{h} \equiv \frac{1}{2} \ln  \frac{ P^{+}_h }{ P^{-}_h }   \, ,
\end{equation}
where \( P^{\pm}_h \) are the \(\pm\) light-cone momenta, i.e., \( E_h^{\text{CM}} \pm P_{h\parallel}^{\text{CM}} \), of the hadron in the virtual-photon--nucleon center-of-mass system.
Both are measures of the ``forwardness'' of the hadron in that system. Positive values of \xf and \(y_{h}\) are more likely associated with hadrons produced from the struck quark,
while negative values point at target fragmentation.
As an example, the rapidity distributions for \piplus and protons are shown in figure~\ref{fig:rapidity} for a specific kinematic bin of small \z and large \Phperpabs.
Even though proton production is more susceptible to contributions from target fragmentation, the proton's rapidity remains, like that of pions, mainly positive.
Further discussion including more distributions can be found in appendix~\ref{app:TMDfactorization}.

\tableenv{t}{
\begin{tabular}{|lrcl|}
\hline
\textbf{Scattered lepton:} &    	&\Q&  	\(> 1\GeV^2\) 	\\
	& 			 		&\W&       \(> 10\GeV^2\)  	\\
	& ~~~\(0.023<\)			&\xb&   	\(<0.6\)    		\\
	& \(0.1 <\)				&\y&	  	\(<0.95\)   		\\
\textbf{Detected hadrons:} ~~~& \(2 \GeV<\)	&\Phabs&  \(<15\GeV\)  ~~charged mesons \\
	& \(4 \GeV<\)			&\Phabs&	\(<15\GeV\)  ~~(anti)protons \\
	& 					&\Phabs&  \(> 2 \GeV\, \)	~~~neutral pions \\
	& 					&\Phperpabs& \(<2\GeV\) 	\\
	& \(0.2<\)				&\z&		\(<0.7\) (1.2 for the ``semi-exclusive'' region)    \\
\hline
\end{tabular}}
{Restrictions on selected kinematics variables. The upper limit on \z of 1.2 applies only to the analysis of the \z dependence.}{table:measurement-cutssummary}

\tableenv{}{
\begin{tabular}{|lccccccc|}
\hline
				& \hspace*{0.4cm}\piplus\hspace*{0.4cm}	& \hspace*{0.4cm}\pizero\hspace*{0.4cm}	& \hspace*{0.4cm}\piminus\hspace*{0.4cm}& \hspace*{0.4cm}\kplus\hspace*{0.4cm}	& \hspace*{0.4cm} \kminus\hspace*{0.4cm}& \hspace*{0.4cm}\( p \)\hspace*{0.4cm}	& \hspace*{0.4cm}\( \bar{p} \)\hspace*{0.4cm}	\\
				\hline
\( 0.2 < \z< 0.7 \) 	& 755k		& 158k		& 543k		&  136k		& 57k		& 94k	&  14k		\\ 
\( 0.7< \z < 1.2 \)	& 68k		& 10k		& 40k		&    14k		&  1k			& 6k		&  \(<\)1k		\\	
\hline
\end{tabular}}
{Hadron yields for the semi-inclusive DIS range and the high-\z region.}{table:hadronyields}

The criteria for the selection of scattered leptons and of hadrons detected in coincidence  
are summarized in table~\ref{table:measurement-cutssummary}. 
They have been chosen to ensure a good semi-inclusive \dis measurement, e.g., adequate detector resolutions and minimal backgrounds, 
but have not  be tuned to the requirements of current \tmd factorization only. 
The data are thus sensitive to kinematic regions in semi-inclusive \dis, including various transition regions that are under theoretical investigation.
The final number of hadrons after the application of all selection criteria is provided in table~\ref{table:hadronyields}
for both the semi-inclusive range of \( 0.2<z<0.7 \) and the extended range of \( 0.7<z<1.2 \).

\subsection{The extraction of the asymmetry amplitudes}

Signals for {\tmd}s are extracted using an unbinned maximum-likelihood fit to
their distinctive signatures in the azimuthal angles \phih and
\phis. The extracted Fourier components are studied as a function of
the kinematic variables \xb, \z, and \Phperpabs. 
As the three-dimensional dependence of the
asymmetry amplitudes does not factorize {\em a priori},
the primary results of this analysis are provided in a three-dimensional
binning in those kinematic variables. Binning the data also in \Q (or alternatively \y) is 
not applicable by lack of statistical precision.
The bin sizes and boundaries are optimized for 
the various hadrons in order to have results in all bins. This results in two sets of \(4\times4\times4\) grids 
with a total of 64 bins each for charged mesons and for protons (see tables~\ref{tab:3d-binning-mesons} and \ref{tab:3d-binning-baryons}, respectively).
The yields for neutral pions and for antiprotons
are insufficient for using such three-dimensional binning. 
 
\tableenv{t}{
\begin{tabular}{|ccc|}
\hline
 \hspace*{1cm}\xb\textbf{bins}\hspace*{1cm} & 
	\hspace*{1cm}\z\textbf{bins}\hspace*{1cm} & \hspace*{1cm}\Phperpabs\textbf{bins}\hspace*{1cm}\\
\hline
  \rangeho{0.023}{0.072} 	& \rangeho{0.20}{0.28} & \rangeho{0.00\GeV}{0.23\GeV} \\
 \rangeho{0.072}{0.098} 	& \rangeho{0.28}{0.37} & \rangeho{0.23\GeV}{0.36\GeV} \\
 \rangeho{0.098}{0.138} 	& \rangeho{0.37}{0.49} & \rangeho{0.36\GeV}{0.54\GeV} \\
 \rangeho{0.138}{0.600} 	& \rangeho{0.49}{0.70} & \rangeho{0.54\GeV}{2.00\GeV} \\
 \hline
\end{tabular}}
{Definition of the three-dimensional binning for charged mesons: the first, second, and third columns list the limits in the kinematic variables \xb, \z, and \Phperpabs, respectively.}{tab:3d-binning-mesons}

\tableenv{t}{
\begin{tabular}{|ccc|}
\hline
 \hspace*{1cm}\xb\textbf{bins}\hspace*{1cm} & 
	\hspace*{1cm}\z\textbf{bins}\hspace*{1cm} & \hspace*{1cm}\Phperpabs\textbf{bins}\hspace*{1cm}\\
\hline
 \rangeho{0.023}{0.073} 	& \rangeho{0.20}{0.34} & \rangeho{0.00\GeV}{0.24\GeV} \\
 \rangeho{0.073}{0.107} 	& \rangeho{0.34}{0.43} & \rangeho{0.24\GeV}{0.40\GeV} \\
 \rangeho{0.107}{0.157} 	& \rangeho{0.43}{0.52} & \rangeho{0.40\GeV}{0.57\GeV} \\
 \rangeho{0.157}{0.600} 	& \rangeho{0.52}{0.70} & \rangeho{0.57\GeV}{2.00\GeV} \\
\hline
\end{tabular}}
{Definition of the three-dimensional binning for protons: the first, second, and third columns list the limits in the kinematic variables \xb, \z, and \Phperpabs, respectively.}{tab:3d-binning-baryons}

In addition to the full information given in the three-dimensional representations, 
results for one-dimensional projections are provided, 
for which the data are subdivided into seven bins in either \xb, \z, or \Phperpabs.
This allows presenting results also for neutral pions and antiprotons, but also a much faster 
evaluation of key characteristics of the results. Furthermore, the range in \z is extended by
further three bins to include also the high-\z ``semi-exclusive'' region. 
As before, the binning differs slightly for
mesons and (anti)protons due to the different kinematic requirements and underlying distributions.
The resulting bin boundaries are given for mesons in table~\ref{tab:1d-binning-mesons} 
and for (anti)protons in table~\ref{tab:1d-binning-baryons}.

\tableenv{t}{
\begin{tabular}{|cccc|}
\hline
\textbf{Bin} & \hspace*{0.4cm}\xb\textbf{dependence}\hspace*{0.4cm} &
\hspace*{0.4cm}\z\textbf{dependence}\hspace*{0.4cm} & \hspace*{0.4cm}\Phperpabs\textbf{dependence}\hspace*{0.4cm} \\
\hline
 1 	& \rangeho{0.023}{0.046} 	& \rangeho{0.20}{0.26} & \rangeho{0.00\GeV}{0.17\GeV} \\
 2 	& \rangeho{0.046}{0.067} 	& \rangeho{0.26}{0.32} & \rangeho{0.17\GeV}{0.25\GeV} \\
 3 	& \rangeho{0.067}{0.082} 	& \rangeho{0.32}{0.38} & \rangeho{0.25\GeV}{0.31\GeV} \\
 4 	& \rangeho{0.082}{0.105} 	& \rangeho{0.38}{0.45} & \rangeho{0.31\GeV}{0.38\GeV} \\
 5 	& \rangeho{0.105}{0.134} 	& \rangeho{0.45}{0.52} & \rangeho{0.38\GeV}{0.52\GeV} \\
 6	& \rangeho{0.134}{0.186} 	& \rangeho{0.52}{0.60} & \rangeho{0.52\GeV}{0.69\GeV} \\
 7 	& \rangeho{0.186}{0.600} 	& \rangeho{0.60}{0.70} & \rangeho{0.69\GeV}{2.00\GeV} \\
 8  	& 					& \rangeho{0.70}{0.76} &  \\ 
 9  	& 					& \rangeho{0.76}{0.84} &  \\
10 	& 					& \rangeho{0.84}{1.20} &  \\
\hline
\end{tabular}}
{Definition of the one-dimensional binning for mesons: the first column lists the bin number; the second, third, and fourth columns give the corresponding
  limits in the kinematic variables \xb, \z, and \Phperpabs, respectively.}{tab:1d-binning-mesons}

\tableenv{t}{
\begin{tabular}{|cccc|}
\hline
\textbf{Bin} & \hspace*{0.4cm}\xb\textbf{dependence}\hspace*{0.4cm} &
\hspace*{0.4cm}\z\textbf{dependence}\hspace*{0.4cm} & \hspace*{0.4cm}\Phperpabs\textbf{dependence}\hspace*{0.4cm} \\
\hline
 1 	& \rangeho{0.023}{0.040} 	& \rangeho{0.20}{0.27} & \rangeho{0.00\GeV}{0.23\GeV} \\
 2 	& \rangeho{0.040}{0.057} 	& \rangeho{0.27}{0.34} & \rangeho{0.23\GeV}{0.34\GeV} \\
 3 	& \rangeho{0.057}{0.075} 	& \rangeho{0.34}{0.41} & \rangeho{0.34\GeV}{0.43\GeV} \\
 4 	& \rangeho{0.075}{0.098} 	& \rangeho{0.41}{0.47} & \rangeho{0.43\GeV}{0.52\GeV} \\
 5 	& \rangeho{0.098}{0.136} 	& \rangeho{0.47}{0.53} & \rangeho{0.52\GeV}{0.62\GeV} \\
 6	& \rangeho{0.136}{0.185} 	& \rangeho{0.53}{0.61} & \rangeho{0.62\GeV}{0.74\GeV} \\
 7 	& \rangeho{0.185}{0.600} 	& \rangeho{0.61}{0.70} & \rangeho{0.74\GeV}{2.00\GeV} \\
 8  	& 					& \rangeho{0.70}{0.78} &  \\ 
 9  	& 					& \rangeho{0.78}{0.88} &  \\
10 	& 					& \rangeho{0.88}{1.20} &  \\
\hline
\end{tabular}}
{Definition of the one-dimensional binning for (anti)protons: the first column lists the bin number; the second, third, and fourth columns give the corresponding limits in the kinematic variables \xb, \z, and \Phperpabs, respectively.}{tab:1d-binning-baryons}

\subsubsection{The choice of the probability-density function}\label{measurement-pdfselection}

Ideally, the various structure functions of the semi-inclusive cross section~\eqref{theory-fourier}  
are extracted directly. However, experimentally such an extraction would require 
precision knowledge of the luminosity and all the instrumental effects, e.g., efficiencies and geometrical acceptance. 
Instead, in the measurement reported here the rapid spin reversal of the target protons is exploited to effectively extract spin asymmetries.
While avoiding many of the experimental uncertainties, theoretical uncertainties arise in the interpretation
of the results as they constitute {\em relative} quantities by normalizing the polarization-dependent 
structure functions to the polarization-averaged and \(\phih\)-integrated cross section, which is proportional to \( \structureh{UU,T} + \epsilon \structureh{UU,L} \). 
The detailed knowledge of the latter is still limited, in particular the transverse-momentum dependence, 
but also the contribution from longitudinal photons. 
In the case of inclusive \dis at \hermes kinematics, the contribution from longitudinal photons can reach values of up to 30\% 
compared to the one from transverse photons  (used to interpret the structure functions 
in the parton model at leading order in \(\alpha_{S}\)).

An experimental limitation is the inability to polarize the target on an event-by-event basis with respect to the virtual-photon direction. 
The latter is used in eq.~\eqref{theory-fourier} as a reference axis because it is a more convenient and natural choice for the decomposition.
In contrast, in an actual experiment, target-polarization states are chosen with respect to the incident-lepton direction. 
The coordinate transformation from the lepton-beam system to the virtual-photon system and its effects are worked out in ref.~\cite{Diehl:2005pc}. 
It involves the usually small polar angle \(\theta_{\gamma^{*}}\) between the incident-lepton and virtual-photon three-momenta.
As discussed in more detail in appendix~\ref{sec-app-longpol}, the observable azimuthal modulations, labeled henceforth by \(\perp\) (\(\parallel\)) instead of T (L) for the transverse (longitudinal) target-polarization component, are in general a mixture of contributions from the target-polarization terms labeled with T and L in eq.~\eqref{theory-fourier}. Moreover, the Fourier decomposition of the azimuthal distribution for the \(\perp\) (\(\parallel\)) configuration includes additional terms not present in eq.~\eqref{theory-fourier}. In particular, for \(\perp\) target polarization an additional \(\sin(2\phih+\phis)\) [\(\cos(\phih+\phis)\)] modulation is possible when the lepton beam is unpolarized [longitudinally polarized].
The number of azimuthal modulations for hadron leptoproduction on a target polarized perpendicular to the direction of the incident lepton are thus ten: six single-spin and four double-spin asymmetries. 
Of those, three [\(\sin\phis\), \(\sin(3\phih-\phis)\), and \(\cos(2\phih-\phis)\)] arise genuinely from transverse target polarization, 
five [\(\sin(\phih+\phis)\), \(\sin(\phih-\phis)\), \(\sin(2\phih-\phis)\), \(\cos\phis\), and \(\cos(\phih-\phis)\)] are dominantly transverse-polarization effects with a small admixture from longitudinal target polarization, and the remaining two are genuine contributions from the small but non-vanishing longitudinal target-polarization component.

In this measurement, a maximum-likelihood fit is employed
that incorporates the reversal of both the beam and target polarization in the probability density function.
The probability density for the combined Fourier analysis of
single-spin
and double-spin azimuthal asymmetries
is modeled according to the cross-section contributions
\sigmaut{h}{\perp} and \sigmalt{h}{\perp}. As such it includes a total of ten modulations:  
the six sine modulations of the cross section \sigmaut{h}{\perp}
and, when including the longitudinal lepton-beam polarization, four cosine modulations.

Another choice has to be made concerning which kinematic terms of the cross section to include as part of the parameters to be fit. 
Two possibilities are presented here: The {\em cross-section asymmetries} (CSA),
which involve --- up to prefactors common to all cross-section terms --- the entire Fourier amplitude of each cross-section modulation, 
e.g., also the \(\epsilon\)-dependent kinematic prefactors. 
In contrast, the {\em structure-function asymmetries} (SFA) are to first approximation ratios of only the structure functions
discussed in more detail in section~\ref{sec:qpm}, obtained by including explicitly the \(\epsilon\)-dependent kinematic prefactors in the likelihood function separated from the fit parameters. 
A compelling advantage of the latter asymmetries is their simple interpretation. The strongly experiment-dependent kinematic prefactors contain
little additional information and cloud direct comparisons to results from different experiments.
The advantages of the former include the possibility to correct in a straightforward way for the contributions 
from the longitudinal target-polarization component (cf.~appendix~\ref{sec-app-longpol}),\footnote{For example, 
the contributions from the transverse and longitudinal components of the target polarization may have different kinematic prefactors, which cannot be taken into account in the fit of structure-function asymmetries.} 
and the independence of the analysis from the particular assumptions made in the 
expansion of the modulations in terms of structure functions. 
The primary results presented here will be the {\em structure-function asymmetries}.

In the case of {\em perfect} acceptance in the azimuthal angles, each Fourier amplitude could be 
extracted separately due to orthogonality of the azimuthal modulations. 
However, under realistic experimental conditions cross-contamination may occur. 
Hence, both the single-spin and double-spin Fourier amplitudes are extracted
simultaneously. 
The corresponding probability-density function for the
Fourier decomposition of the cross section (\csa decomposition) is then defined as
\begin{align}
\mathbb{P}\Big(\xb,\z,  \Phperpabs, & \phih,\phis , P_{l}\, , S_\perp :   \siversexp{h},\dots \, \cosphilexp{h}\Big) \nonumber \\
= \Big[1+S_\perp\Big(&
\siversexp{h}\sinemodulation{\phih-\phis} \, + 
\collinsexp{h}\sinemodulation{\phih+\phis} \, + \nonumber \\
&\sinthreephiexp{h}\sinemodulation{3\phih-\phis} \,+
\sinphisexp{h} \sinemodulation{\phis} \, + \nonumber \\ 
& \sintwophiexp{h}\sinemodulation{2\phih-\phis} \, + 
\sintwophilexp{h}\sinemodulation{2\phih+\phis} \Big)   \nonumber \\
+P_{l}\,S_\perp\Big(&
\cosphiexp{h}\cosinemodulation{\phih-\phis}  \, + 
\cosphisexp{h}\cosinemodulation{\phis} \, + \nonumber \\
&\costwophiexp{h} \cosinemodulation{2\phih-\phis} \, +
\cosphilexp{h} \cosinemodulation{\phih+\phis}  \Big)
\Big]^w   , 
\label{eq:CSA-pdf}
\end{align}
where \(P_{l}\) and \(S_\perp\) represent the degree of longitudinal beam polarization and target
polarization perpendicular to the lepton beam, respectively, 
and \(w\) is an event weight further detailed below. The Fourier amplitudes \( \siversexp{h},\dots \, \cosphilexp{h}\) appearing as parameters in eq.~\eqref{eq:CSA-pdf} are the cross-section asymmetry amplitudes to be fit to the data.
Likewise, the probability-density function for the \sfa amplitudes reads
%
\begin{align}
\mathbb{P}\Big(\xb,\z, \epsilon, \Phperpabs,  & \phih,\phis , P_{l}\, , S_\perp : \siversexpSFA{h},\dots \cosphilexpSFA{h}\Big) \quad \quad\quad\quad  \nonumber \\
= \Big[1+ S_\perp\Big(&
			\siversexpSFA{h} \sinemodulation{\phih-\phis} + 
   \epsilon \; 	\collinsexpSFA{h} \sinemodulation{\phih+\phis} \, +                 \nonumber \\
& \epsilon \;  	\sinthreephiexpSFA{h} \sinemodulation{3\phih-\phis}  \, +       \nonumber \\
& \sqrt{2\epsilon(1+\epsilon)} \; 	\sinphisexpSFA{h}    \sinemodulation{\phis} \, + \nonumber \\ 
& \sqrt{2\epsilon(1+\epsilon)} \; 	\sintwophiexpSFA{h}\sinemodulation{2\phih-\phis}  \, + \nonumber \\ 
& \epsilon \; 	\sintwophilexpSFA{h} \sinemodulation{2\phih+\phis} \Big)          \nonumber \\
+P_{l}\,S_\perp\Big(&
   \sqrt{1-\epsilon^{2}} \; 		\cosphiexpSFA{h}   \cosinemodulation{\phih-\phis} \, + \nonumber \\  
& \sqrt{2\epsilon(1-\epsilon)} \; 	\cosphisexpSFA{h} \cosinemodulation{\phis} \, 		+ \nonumber \\
& \sqrt{2\epsilon(1-\epsilon)} \; 	\costwophiexpSFA{h} \cosinemodulation{2\phih-\phis} \, + \nonumber \\ 
& \sqrt{2\epsilon(1-\epsilon)} \; 	\cosphilexpSFA{h}   \cosinemodulation{\phih+\phis}   \Big)
\Big]^w \, .
\label{eq:SFA-pdf}
\end{align}

Charged-hadron [pion, kaon, and (anti-)proton] weights are assigned to each hadron track selected
to account for the efficiency of the \rich detector and the contamination
of the pion, kaon, and proton identification. When the charge of the
scattered lepton does not correspond to the charge of the incoming
beam leptons, the weights are multiplied by \(-1\) in order to subtract
the background arising from the pair-production process. 
In a similar way, combinatorial background in the \pizero signal region of the
two-photon invariant-mass spectrum is subtracted using events from the sidebands (cf.~figure~\ref{fig:pi0spectrum}) 
and assigning a negative weight equal to \(-R\), where the ratio \(R\) is the {\em relative} population of 
combinatorial background in the signal region and the sidebands, as given by the invariant-mass fit.

As the sum over all weights does not
coincide with the number of hadrons detected, i.e., 
\begin{equation}
\sum\limits_{i=1}^{N^h} w_i \ne N^h ,
\end{equation}  
the statistical
uncertainties of the asymmetry amplitudes extracted have to be
corrected for the event weighting. 
The covariance matrix \(C\), obtained in the maximum-likelihood fit, 
is corrected by the covariance matrix \(K\)
that is obtained in a maximum-likelihood fit to the same data 
but weighting the events with \(w_i^2\) instead of \(w_i\). 
The statistical uncertainties are then evaluated from the corrected covariance matrix~\cite{Solmitz:1964xw}
\begin{align}
C' = C K^{-1}C\, .\label{measuremente-solmitzcorrection}
\end{align}

In the likelihood formalism, not only the target polarization but also
the beam polarization is applied on event level, i.e., 
for each identified hadron of a given semi-inclusive \dis event, 
the actual beam and target polarization values of that event are used in the likelihood function.

The normalization of the probability density function is not required
as in the data set selected the net target polarization is found to be
negligible. Thus, the normalization integral is independent of the
asymmetry amplitudes extracted and cannot influence the shape of the
likelihood dependence on the azimuthal amplitudes. 

The \csa amplitudes  are then extracted from the semi-inclusive \dis events by minimizing
\begin{equation}
-\ln \mathbb{L}=-\sum_{i=1}^{N_{h}} w_i \ln \mathbb{P} \left( \xb_{i}, \z_{i}, {\Phperpabs}_{,i} ,  \phih_{i},\phi_{S,i}, P_{l,i}, S_{\perp,i} : \siversexp{h},\ldots  \right) 
\label{eq:loglikelihood}
\end{equation}
using  eq.~\eqref{eq:CSA-pdf}  for \(\mathbb{P}(\ldots)\). In a similar fashion, eq.~\eqref{eq:SFA-pdf} is used to extract the \sfa amplitudes, including now also the event-wise value of the photon-polarization parameter \(\epsilon\).

Comparing eqs.~\eqref{eq:CSA-pdf} and \eqref{eq:SFA-pdf} with eq.~\eqref{theory-fourier}, 
it becomes clear that in the probability density the azimuthally uniform contribution to the
cross section, \( \structureh{UU,T} + \epsilon \structureh{UU,L} \), has been factored out, which corresponds 
to normalizing all the Fourier amplitudes to \( \structureh{UU,T} + \epsilon \structureh{UU,L} \). 
Setting \structureh{UU,L} equal to zero, as valid up to subleading twist and leading order in \(\alpha_{S}\),
results in Fourier amplitudes normalized to
\begin{equation}
 \structureh{UU,T} = \convolution{f_1^{q} D_1^{q\rightarrow h}} \, ,
 \label{eq:f1D1}
\end{equation}
e.g., to SFA amplitudes of the form \( \structuresinh{UT}{\phih-\phis} / \structureh{UU,T} \).
\subsubsection{Systematic uncertainties}\label{analysis-systematics}

Systematic uncertainties in the asymmetry amplitudes arise from
\begin{enumerate}[label=(\roman*)]
\item the accuracy of the beam and target polarization measurements, 
\item the choice of the probability-density function,
\item acceptance effects caused by limitations in the geometric
  acceptance and kinematic requirements in the event selection,
\item higher-order \QED processes and kinematic smearing effects due
  to finite spectrometer resolution,
\item the hadron identification using the \rich detector,
\item the stability of the detector over the course of data taking.
\end{enumerate}
In addition, further sources of systematic effects are studied for neutral pions:
\begin{enumerate}[label=(\roman*)]
\item variation of the parameterization for the background shape of the two-photon invariant-mass spectrum: third-order Chebyshev polynomial versus Weibull distribution~\cite{Weibull:1951aa},
\item variation of the sideband positions with respect to the signal range,
\item variation of both sidebands and signal ranges.
\end{enumerate}

The accuracy of the polarization measurements is taken into
account as a scale uncertainty on the amplitudes extracted. 
They amount to 7.3\% and 8.0\% for the single- and double-spin asymmetries, respectively.
As they affect both the central values and all other uncertainties as a multiplicative factor, 
they are provided separately.

Inclusion of additional cosine modulations related to the polarization-averaged cross section, e.g., 
arising from the Boer--Mulders or Cahn~\cite{Cahn:1978se} effects, has negligible effects on the
single- and double-spin Fourier amplitudes extracted. 
For that study, an empirical model of those cosine modulations, fit to \hermes data~\cite{Airapetian:2012yg},
was added to the probability density functions. 
Furthermore, the results for the single-spin asymmetries extracted using either the full probability density function, 
e.g., eq.~\eqref{eq:CSA-pdf}, or one containing only the single-spin asymmetry terms (as done, e.g., in 
the previous publications~\cite{Airapetian:2009ae,Airapetian:2010ds}) are fully consistent.

Systematic uncertainties due to experimental acceptance, kinematic smearing, and the hadron identification are estimated
simultaneously. Results presented here involve integration over finite bin sizes 
and partially larger ranges in kinematic variables not explicitly binned in, e.g., in the one-dimensional 
projections. They are cross-section asymmetries folded with the experimental acceptance, which 
in general depends on the same set of kinematic variables. They thus represent 
averages of not only the kinematic dependences of the underlying physics modulations 
but also include often unaccounted instrumental effects~\cite{Schnell:2015gaa}. 
In particular, these {\em average} asymmetries, in general, do not coincide with the 
cross-section asymmetries at the average kinematics provided with each data point.
The size of such deviations is estimated using a full Monte Carlo simulation of the experiment 
based on a version of \pythia~\cite{Sjostrand:2000wi,Sjostrand:2001yu} tuned for \hermes kinematics  
and extended with \radgen~\cite{Akushevich:1998ft} to account for \QED radiative
effects. The simulation also uses a \geant\cite{Brun:1978fy,Brun:1987ma} description of the \hermes 
apparatus, including the beam trajectory and particle deflection in the holding field of the target magnet 
as well as the efficiency and the cross contamination of the hadron identification using the \rich detector.

The \pythia event generator does not simulate polarization effects such as those studied here. 
For this reason, empirical parameterizations (based on a Taylor expansion) of the single- and double-spin asymmetries 
as a function of \xb, \Q, \z, and \Phperpabs are used to assign a beam-helicity and target-spin state to each 
semi-inclusive DIS event of the simulation, as detailed in appendix~\ref{app:systematics}, to effectively ``polarize'' the \pythia simulation.
The set of parametric models is obtained from fits to the experimental data, separately for each hadron type, using the method described
above (section \ref{measurement-pdfselection}) but now unbinned in {\em all} kinematic variables.
These models, representing the four-dimensional kinematic dependence (\xb, \Q, \z, \Phperpabs) of the ten asymmetry amplitudes of interest, 
are virtually unaffected by acceptance and instrumental effects, 
though somewhat biased by the finite number of terms included in the fit (compared to the {\em a priori} infinite Taylor expansion).

The systematic uncertainties for the combined effect of limited acceptance, higher-order \QED effects, and the hadron identification using the \rich[short] detector 
are then estimated from the difference of the
asymmetry amplitudes extracted from the simulated data and their model
evaluated at their average kinematic values (further details are provided in appendix~\ref{app:systematics}). 
These systematic uncertainties thus correspond to the case of interpreting 
the data as asymmetry values for the given average kinematics in each bin, in contrast to ratio of
cross sections integrated over all the kinematics in the ranges applicable to each particular bin. 

The stability of the results was tested in various ways: 
comparing the results extracted for positron and electron beam separately, 
and comparing double-spin asymmetries for periods of different beam-helicity states. 
The studies found in general full consistency of the data for different beam charges as well as of the data for different beam helicities. 
The only notable exception are the \pizero results. 
Various statistical tests
result in a statistical incompatibility at 90\% confidence level\footnote{The results of these tests are, however, limited by the small number of data points.} 
for the one-dimensional extraction of the Sivers case, 
with hints of statistical incompatibility for some of the other modulations.
As a consequence, conservatively, half the difference between constant fits to the \pizero results from electron and positron data are assigned as additional systematic uncertainties.
They are added in quadrature to those related to other instrumental effects and kinematic smearing.

The remaining sources of systematic uncertainties considered are found to have a negligible effect on the results.

\subsection{Differences with previous analyses}\label{sec:analysis-differences}

Though the general framework has much in common with that in the prior
\hermes publications on the Sivers and Collins effects~\cite{Airapetian:2009ae,Airapetian:2010ds}, there are 
several obvious differences and some minor improvements in the data analysis:
\begin{enumerate}[label=(\roman*)]
\item The analysis is based on a later data production, which among others included updated tracking and alignment information, as well as corrections for minimal beam-energy variations.
\item The first such measurement of asymmetries for protons and antiprotons is presented.
\item The extraction of asymmetries for neutral pions is improved in various aspects, among others a different treatment of photons that start showering already in the pre-shower detector and adjusted ranges in the two-photon invariant mass used for the signal and the background subtraction. Also, only photon pairs that are detected in the same detector half are kept in the analysis.
\item The analysis is performed in a three-dimensional kinematic binning; the \xb range is extended to an upper limit of 0.6.
\item The one-dimensional binning has been adapted to permit extraction of asymmetry amplitudes for also the low-statistic hadrons;
 in addition, the binning in \z is extended to include the high-\z region of \(0.7<\z<1.2\).
\item The extraction of all the various \ssa and \dsa is performed in one combined fit to minimize potential cross talk between moments.
\item The standard set of results comprises the structure-function asymmetries and thus includes corrections for the \(\epsilon\)-dependent kinematic prefactors.
\end{enumerate}

%% file: sections/interpretation.tex
\section{Results and interpretation}\label{section-interpretation}

The \ssa and \dsa amplitudes are extracted in a three-dimensional kinematic binning in 
\x, \z, and \Phperpabs to allow the exploration of correlated dependences. In comparison to earlier 
measurements, e.g., in refs.~\cite{Airapetian:2009ae, Airapetian:2010ds}, this provides measurements in
kinematic corners that are suppressed when integrating over all but one variable. Three further principal advancements 
are worth mentioning: 
(i) the total number of  data points per particle species increases to 64, 
(ii) none of those 64 data points is statistically correlated with any of the other,\footnote{While data points for one particular azimuthal moment are uncorrelated, results for the different azimuthal moments in one kinematic bin may still be correlated. That degree of correlation is provided in the Supplemental Material \cite{supplemental}.} and 
(iii) the multi-dimensional binning avoids integration over large regions of the kinematic space and results in a much reduced systematic uncertainty. 
In particular the latter two should significantly increase the reliability of uncertainties resulting from phenomenological fits to combined data of one-dimensional projections as the latter have an unspecified degree of statistical and systematic correlation.

Due to the more limited precision of the antiproton and neutral-pion data, such three-dimensional kinematic binning was not feasible. 
They were thus analyzed as functions of \x, \z, and \Phperpabs individually (cf.~tables~\ref{tab:1d-binning-mesons} and \ref{tab:1d-binning-baryons}), integrating over the corresponding remaining kinematic variables.

\tableenv{t}{%
\begin{tabular}{|cl|ccccc|cc|}
\hline
\multicolumn{2}{|c|}{ \hspace*{0.3cm} \textbf{Azimuthal modulation} \hspace*{0.3cm} } 		& \multicolumn{7}{c|}{ \textbf{Significant non-vanishing Fourier amplitude} }\\
\multicolumn{2}{|c|}{  } 					& \piplus		& \piminus	 	& \kplus		& \kminus		& \(p\) 		& \pizero 		& \( \bar{p} \) \\ 	
\hline
\sinemodulation{\phih+\phis} 	& ~[Collins]		& \checkmark 	& \checkmark	& \checkmark 	&		   	& \checkmark	&			& 			\\
\sinemodulation{\phih-\phis}  	& ~[Sivers] 		& \checkmark 	& 			& \checkmark 	& \checkmark	& \checkmark	& (\checkmark) & \checkmark	\\
\sinemodulation{3\phih-\phis} 	& ~[Pretzelosity]	&	& 	& 	&	&	&	&															\\
\sinemodulation{\phis}          	& 			& (\checkmark)	& \checkmark 	& 			& \checkmark 	&			&			& 			\\
\sinemodulation{2\phih-\phis} 	&			&			&		 	& 			&			&			&			& (\checkmark)	\\
\sinemodulation{2\phih+\phis} 	&			&			&		 	& \checkmark 	&			&			&			&			\\
\cosinemodulation{\phih-\phis}	& ~[Worm-gear]	& \checkmark	& (\checkmark)	& (\checkmark)	&			&			&			& 			\\
\cosinemodulation{\phih+\phis}	&			&	& 	& 	&	&	&	&															\\
\cosinemodulation{\phis}		&			& 			& 			& \checkmark	&			&			&			&			\\
\cosinemodulation{2\phih-\phis}	&	\hspace*{1.cm}	&\hspace*{0.9cm}	& \hspace*{0.9cm}	& \hspace*{0.9cm}	&\hspace*{1.cm}	&\hspace*{0.9cm}	&\hspace*{0.9cm}	&	\hspace*{0.9cm}	\\
\hline					
\end{tabular}}{The various azimuthal modulations of the semi-inclusive cross section and those hadron species whose corresponding Fourier amplitudes are incompatible with the NULL hypothesis at 95\% (90\%) confidence. Antiprotons and \pizero are given separated in the last two columns to indicate that the statistical test of those is based on the one-dimensional projections and hence restricted to using only seven data points.}
{tab:significant-modulations}

Asymmetries in one overall kinematic bin are not presented as their extraction suffers from the largest acceptance effects.
They are also of limited value for phenomenology. 
Instead, the results for all asymmetries were tested against the NULL hypothesis using the two-sided Student's t-test. 
The asymmetry results binned in three dimensions were used, where available, to increase the robustness of the Student's t-test 
by using 64 data points and avoiding cancelation effects from integrating over kinematic dependences.
In the case of \pizero and antiprotons, where results in only the one-dimensional binning are available, 
they are considered to be inconsistent with zero if the Student's t-test established this for at least one of the three projections (versus \x, \z, or \Phperpabs).%
\footnote{It has to be kept in mind that the Student's t-test becomes less reliable when using a small number of data points as, e.g., the case for the one-dimensional binning.}
It is found that most asymmetry amplitudes are consistent with zero in the semi-inclusive region \(0.2<\z<0.7\) used here.
Those asymmetry amplitudes that are found to be inconsistent with zero at 95\% (90\%) confidence level
are listed in table~\ref{tab:significant-modulations}.
Significantly non-zero results were neither found for the pretzelosity \sinthreephiexp{h} Fourier amplitudes nor for the \(M/Q\)-suppressed  
\cosphilexp{h} and \costwophiexp{h} Fourier amplitudes. For the \sintwophiexp{h} Fourier amplitude, only antiprotons were found to be inconsistent with the NULL hypothesis and this only at the 90\% but not at the 95\% confidence level.

\tableenv{t}{%
\begin{tabular}{|lccccc|}
\hline
\textbf{Hadron}\hspace*{0.9cm} & \multicolumn{5}{c|}{\textbf{Mean values of kinematic variables}}\\
         & \hspace*{0.6cm}\mean{\Q}\hspace*{0.6cm} & \hspace*{0.6cm}\mean{\x}\hspace*{0.6cm} & \hspace*{0.6cm}\mean{\y}\hspace*{0.6cm} & \hspace*{0.6cm}\mean{\z}\hspace*{0.6cm} &         \hspace*{0.6cm}\mean{\Phperpabs}\hspace*{0.6cm} \\
\hline
\piplus  	& \(2.445\GeV^2\) & \(0.095\)  & \(0.544\)  & \(0.362\) & \(0.394\GeV\) \\
\pizero  	& \(2.506\GeV^2\) & \(0.089\)  & \(0.588\)  & \(0.357\) & \(0.396\GeV\) \\
\piminus 	& \(2.366\GeV^2\) & \(0.092\)  & \(0.548\)  & \(0.354\) & \(0.393\GeV\) \\
\kplus   	& \(2.524\GeV^2\) & \(0.097\)  & \(0.548\)  & \(0.391\) & \(0.417\GeV\) \\
\kminus 	& \(2.381\GeV^2\) & \(0.089\)  & \(0.569\)  & \(0.356\) & \(0.412\GeV\) \\
\proton 	& \(2.595\GeV^2\) & \(0.095\)  & \(0.574\)  & \(0.421\) & \(0.452\GeV\) \\
\pbar 	& \(2.393\GeV^2\) & \(0.076\)  & \(0.655\)  & \(0.364\) & \(0.477\GeV\) \\
\hline
\end{tabular}}{Mean kinematic values for pions, charged kaons, as well as for protons and antiprotons in the standard semi-inclusive range $0.2<\z<0.7$.}
{interpretation-meankinematics}

In the following, the most important observations and features of the data are discussed.%
\footnote{The complete set of figures are provided as Supplemental Material~\cite{supplemental}, including tables of all the asymmetry results.}
The corresponding mean kinematics for the kinematic region covered within the standard semi-inclusive selection are listed in table~\ref{interpretation-meankinematics}.

The error bars in the following figures indicate the statistical uncertainties of the \ssa and \dsa Fourier amplitudes. 
The uncertainty bands represent the systematic uncertainties of the results arising from acceptance, 
finite detector resolution, higher-order \QED effects, possible misidentification of hadrons, 
and detector instabilities (the latter only for \pizero, while negligible for all other hadrons).
In addition, the uncertainties arising from the measurement precision of beam and target polarization
are provided separately as an overall scale uncertainty: 7.3\% in the case of \ssa amplitudes and 8.0\% for the \dsa amplitudes.

\subsection{Signals for transversity and the Collins fragmentation function}\label{interpretation-collins}

Non-vanishing \sinemodulation{\phih+\phis} modulations (``Collins asymmetries'')
are evidence for two chiral-odd TMDs: the transversity distribution 
and the \Todd Collins fragmentation function. Both have been subject to intense 
experimental and theoretical studies, also at \hermes, which first reported evidence for those~\cite{Airapetian:2004tw}. 
Results for pions, charged kaons, and the pion charge-difference \csa were reported in ref.~\cite{Airapetian:2010ds}
for one-dimensional projections in \x, \z, and \Phperpabs. The most striking feature of those results is a large negative asymmetry 
for negative pions, opposite in sign and even larger in magnitude in comparison to the asymmetry for positive pions.
These results were explained~\cite{Airapetian:2004tw} by a large disfavored Collins function, describing, e.g., 
the fragmentation of up quarks into negative pions, that is opposite in sign to the favored Collins function. 
This explanation was later confirmed by phenomenological fits~\cite{Anselmino:2007fs, Anselmino:2013vqa,Anselmino:2015sxa,Lin:2017stx,Barone:2019yvn,DAlesio:2020vtw,Cammarota:2020qcw} 
to various data sets on semi-inclusive \dis~\cite{Airapetian:2004tw,Airapetian:2010ds,Alexakhin:2005iw,Ageev:2006da,Alekseev:2010rw, Adolph:2012sn,Alekseev:2008aa,Adolph:2014zba,Qian:2011py} 
and on \sia into hadrons~\cite{Seidl:2006,Seidl:2008xc,TheBABAR:2013yha,Ablikim:2015pta},
as well as on hadron collisions in the case of ref.~\cite{Cammarota:2020qcw}. 
While earlier work employed simplified approaches for the Dokshitzer--Gribov--Lipatov--Altarelli--Parisi
evolution in the fits to data at various scales, the focus has moved to employ TMD evolution in more recent works, 
especially in view of the \(B\)-factory data at  \(\Q \sim 100~\GeV^2\).

The results for the transversity distributions from global fits are of the same sign\footnote{Note that the {\em absolute} 
sign can not be determined unambiguously due to the chiral-odd nature of both transversity and the Collins fragmentation function.}
as results for the helicity distribution, but somewhat smaller in magnitude, by as much as a factor of two for 
the $d$-quark distribution.
Flavor decompositions of the collinear transversity distribution, based on analysis of dihadron production
in semi-inclusive \dis~\cite{Airapetian:2008sk,Adolph:2012nw,Adolph:2014fjw}, \sia~\cite{Vossen:2011fk}, and more recently in \(p^{\uparrow}p\) collision~\cite{Adamczyk:2015hri}, confirm this general behavior~\cite{Bacchetta:2012ty,Radici:2015mwa,Radici:2018iag,Benel:2019mcq}. 
In general, the $d$-quark transversity distribution is much less constrained, given the $u$-quark dominance in many of the processes employed in the extractions.
It is interesting to remark that all phenomenological
extractions of the transversity
distribution present some discrepancies with respect to lattice predictions,
especially for what concerns the $u$-quark contribution to the nucleon tensor
charge (see, e.g., refs.~\cite{Yamanaka:2018uud,Gupta:2018qil,Alexandrou:2018eet}). 

\begin{figure}[t]
 \centering
 \includegraphics[bb = 15 60 500 470, clip, width=0.49\textwidth,keepaspectratio]{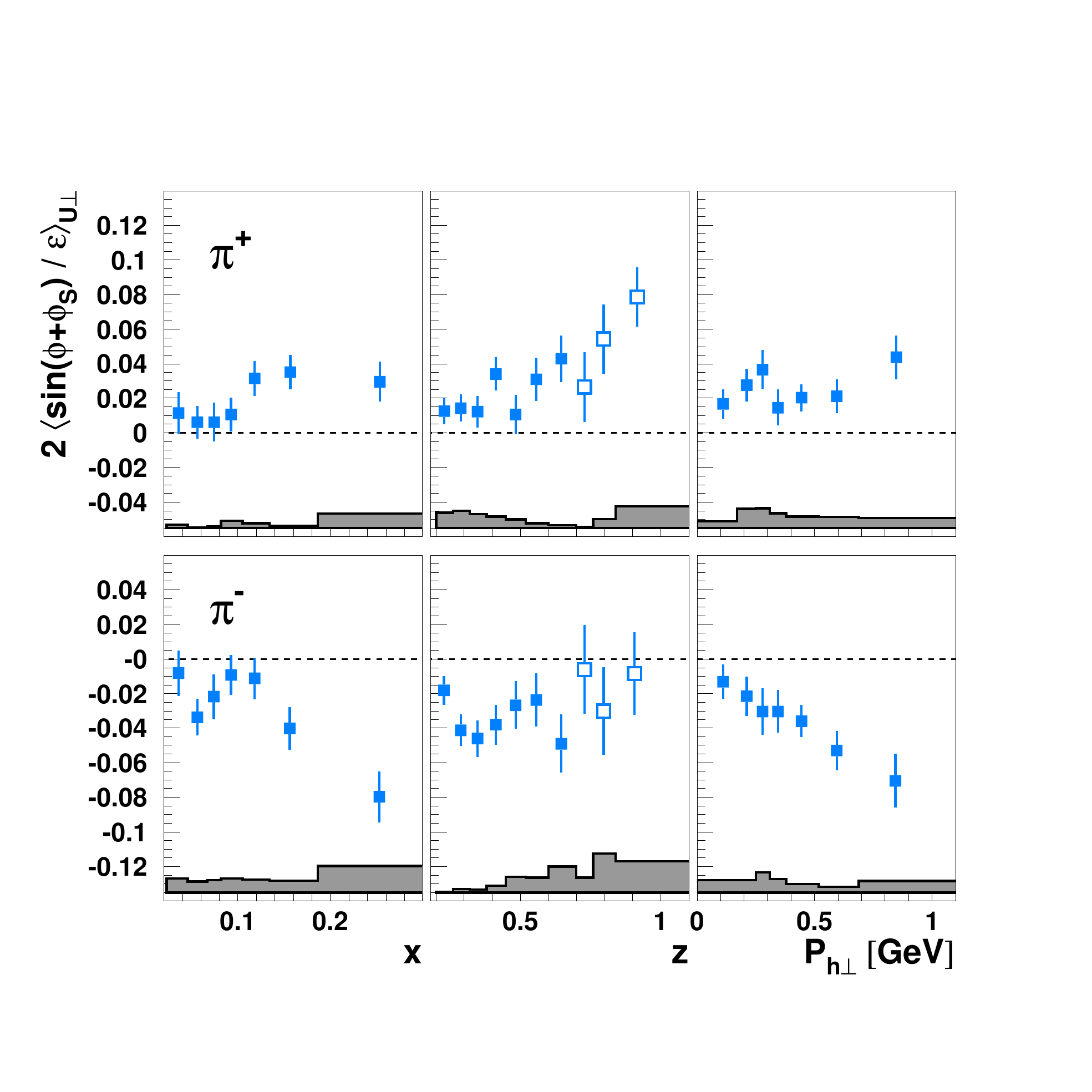}~~\includegraphics[bb = 15 60 500 470, clip, width=0.49\textwidth,keepaspectratio]{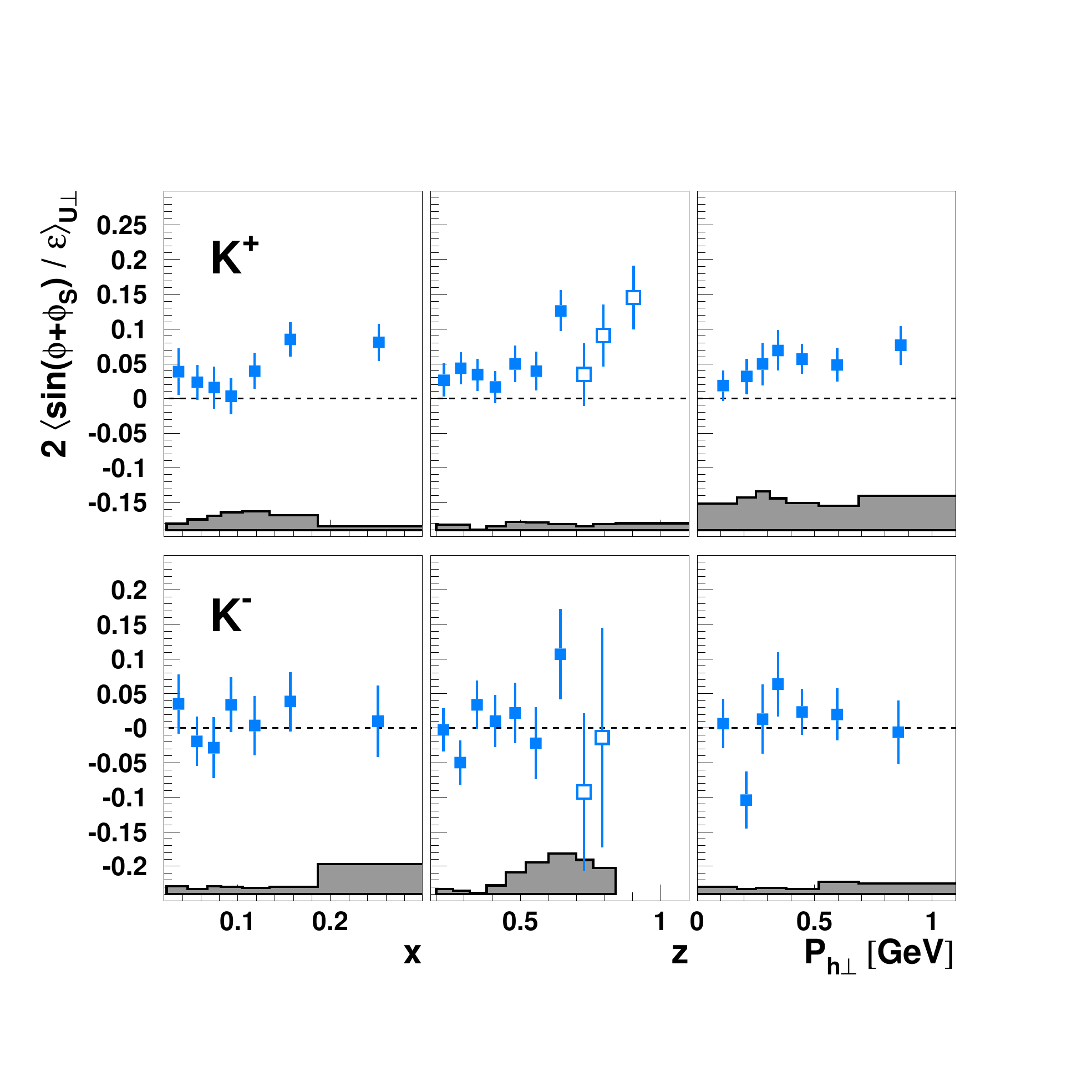}
 \caption{Collins \sfa for charged mesons (left: pions; right: kaons) presented either in bins of \xb, \z, or \Phperpabs. Data at large values of \z, marked by open points in the \z projection, are not included in the other projections. Systematic uncertainties are given as bands, not including the additional scale uncertainty of 7.3\% due to the precision of the target-polarization determination.}
 \label{fig:mesons-1d-collins}
\end{figure}

The Collins asymmetries extracted here for mesons in one-dimensional projections resemble to a high degree those published previously~\cite{Airapetian:2010ds}.
This is expected as based on the same data set, though involving a number of analysis improvements (cf.~section~\ref{sec:analysis-differences}). 
The most significant advancement in the measurement of the \sfa shown in figure~\ref{fig:mesons-1d-collins} 
is the inclusion of the \(\epsilon\)-dependent kinematic prefactors in the probability density function~\eqref{eq:SFA-pdf} of the maximum-likelihood fit.
This leads on average to an amplification of the asymmetry magnitude as, in the case of the Collins asymmetry, this prefactor is smaller than unity and thus diminishes the transversity/Collins-induced modulation.  

The Collins asymmetries for charged pions are opposite in sign and increasing with \xb, 
which can be attributed to transversity predominantly being a valence-quark effect.
The dependence on \z in the semi-inclusive range is a clear increase with \z for \piplus, while first clearly increasing but then leveling out for \piminus. 
As expected, the asymmetries increase with \Phperpabs at low values of \Phperpabs. 
This rise continues in the case of \piminus up to the highest \Phperpabs values probed here. 
In contrast, for \piplus there is a hint of a plateau after the initial rise with \Phperpabs.

In the case of strange mesons, positive kaons exhibit larger though in shape similar Collins asymmetries when compared to those for positive pions. 
In contrast, the Collins asymmetries for negative kaons are found to be consistent with zero. 
Assuming that the nucleon's sea-quark transversity distributions are vanishing (or small), 
only disfavored fragmentation of up and down quarks can contribute to the \kminus moments. 
Being disfavored fragmentation, the contribution is expected to be suppressed. 
Furthermore, being of opposite sign, the up and down contributions cancel to a large extend. 
Recently, data from \sia into kaons and pions~\cite{Aubert:2015hha} were analyzed and the Collins fragmentations functions extracted were then used for the estimate of the Collins asymmetries in semi-inclusive \dis.
Indeed, a largely vanishing \kminus Collins asymmetry, as observed here, was found considering only valence transversity as non-vanishing~\cite{Anselmino:2015fty}.
The data for kaons are interesting in the context of chiral symmetry breaking in QCD, where pions and kaons are considered to be the Goldstone bosons. In the chiral limit, fragmentation into pions and kaons should be the same, in particular, \(\ffcollins{q\to\pi} = \ffcollins{q\to K} \)~\cite{Efremov:2001ia}. In reality, this is already violated in the case of unpolarized fragmentation, e.g., \( \ffd{u\to\piplus} > \ffd{u\to\kplus} \). 
Extractions of the Collins fragmentation function for both pions and kaons will shed light on the (better) validity of the chiral limit for the case of the Collins fragmentation function.

The one-dimensional dependences of the Collins asymmetries measured by the COMPASS Collaboration~\cite{Adolph:2014zba} are consistent with the ones reported here, 
apart from the \kminus asymmetries, which are non-vanishing and negative\footnote{Note that COMPASS uses a different sign convention for the transversity-induced asymmetries.} at COMPASS.
The kaon Collins asymmetries from Jefferson Lab for transversely polarized \(^{3}\)He, effectively a target of transversely polarized neutrons, 
are consistent with zero within large uncertainties, with a hint of a sizable negative asymmetry for \kminus~\cite{Zhao:2014qvx}.

\begin{figure}
 \centering
 \includegraphics[bb = 30 23 515 543, clip, width=0.7\textwidth,keepaspectratio]{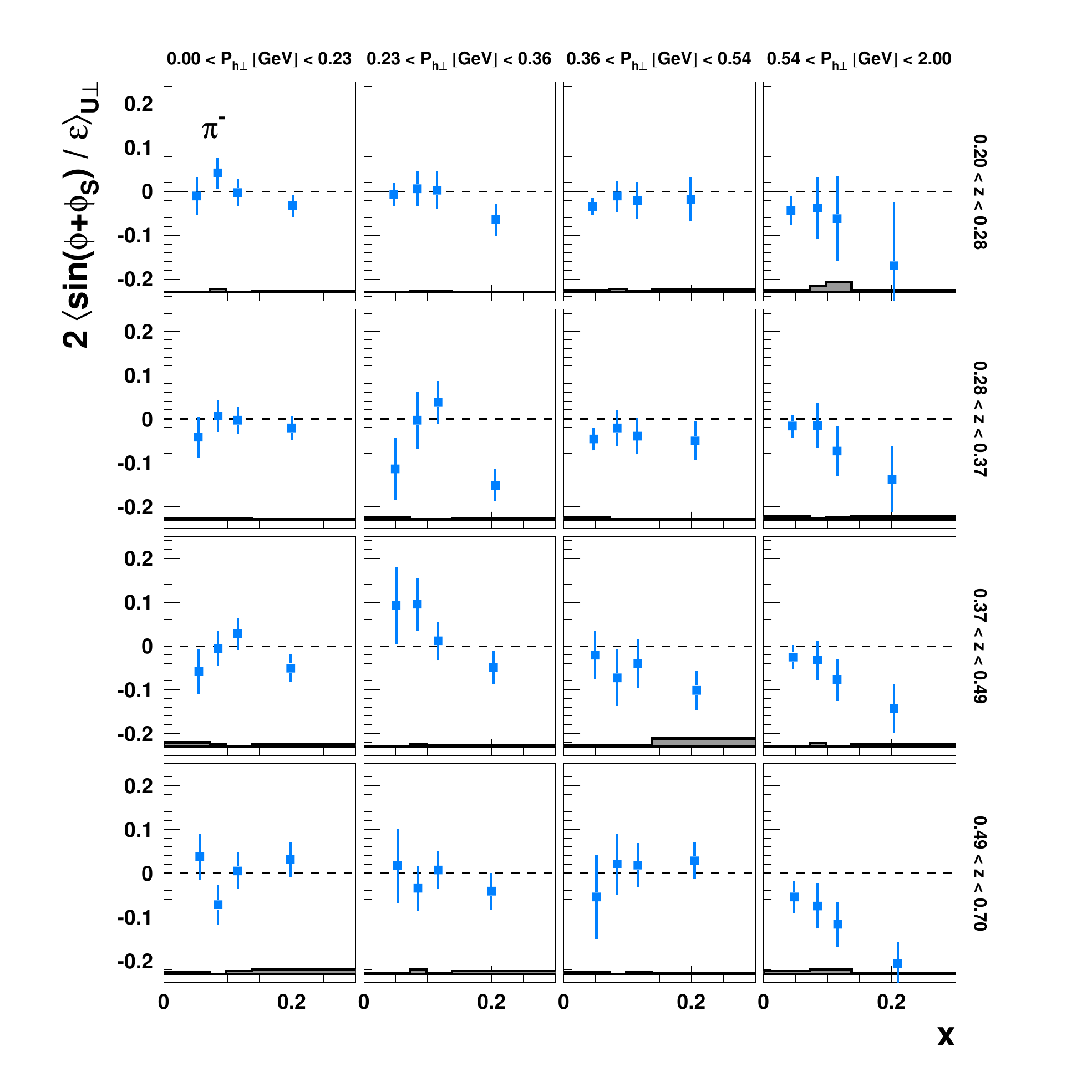}
 \caption{Collins \sfa for \piminus extracted simultaneously in bins of \xb, \z, and \Phperpabs, presented as a function of \xb. Systematic uncertainties are given as bands, not
 		including the additional scale uncertainty of 7.3\% due to the precision of the target-polarization determination.}
 \label{fig:3d-collins-x}
\end{figure}

\begin{figure}
 \centering
 \includegraphics[bb = 30 23 515 543, clip, width=0.7\textwidth,keepaspectratio]{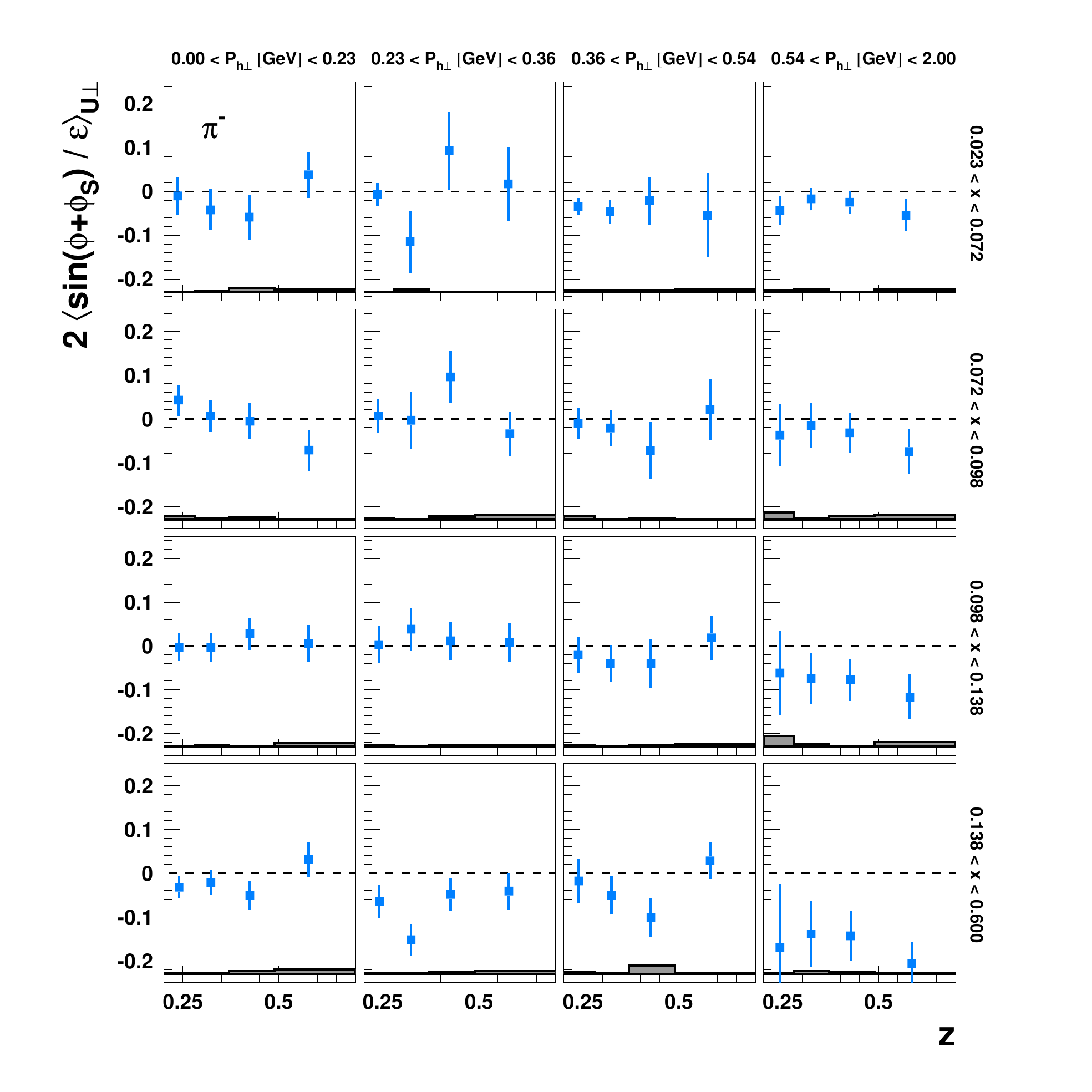}
 \caption{Collins \sfa for \piminus extracted simultaneously in bins of \xb, \z, and \Phperpabs, presented as a function of \z. Systematic uncertainties are given as bands, not including the additional scale uncertainty of 7.3\% due to the precision of the target-polarization determination.}
 \label{fig:3d-collins-z}
\end{figure}

Two examples for the three-dimensionally binned data are presented in figures~\ref{fig:3d-collins-x} and \ref{fig:3d-collins-z}. The \piminus Collins asymmetries are plotted either versus \xb (figure~\ref{fig:3d-collins-x}) or versus \z (figure~\ref{fig:3d-collins-z}), revealing a merely weak dependence on \z but an \xb dependence that is pronounced, mainly at large \Phperpabs (and \z).

\begin{figure}
 \centering
\includegraphics[bb = 15 60 500 470, clip, width=0.49\textwidth,keepaspectratio]{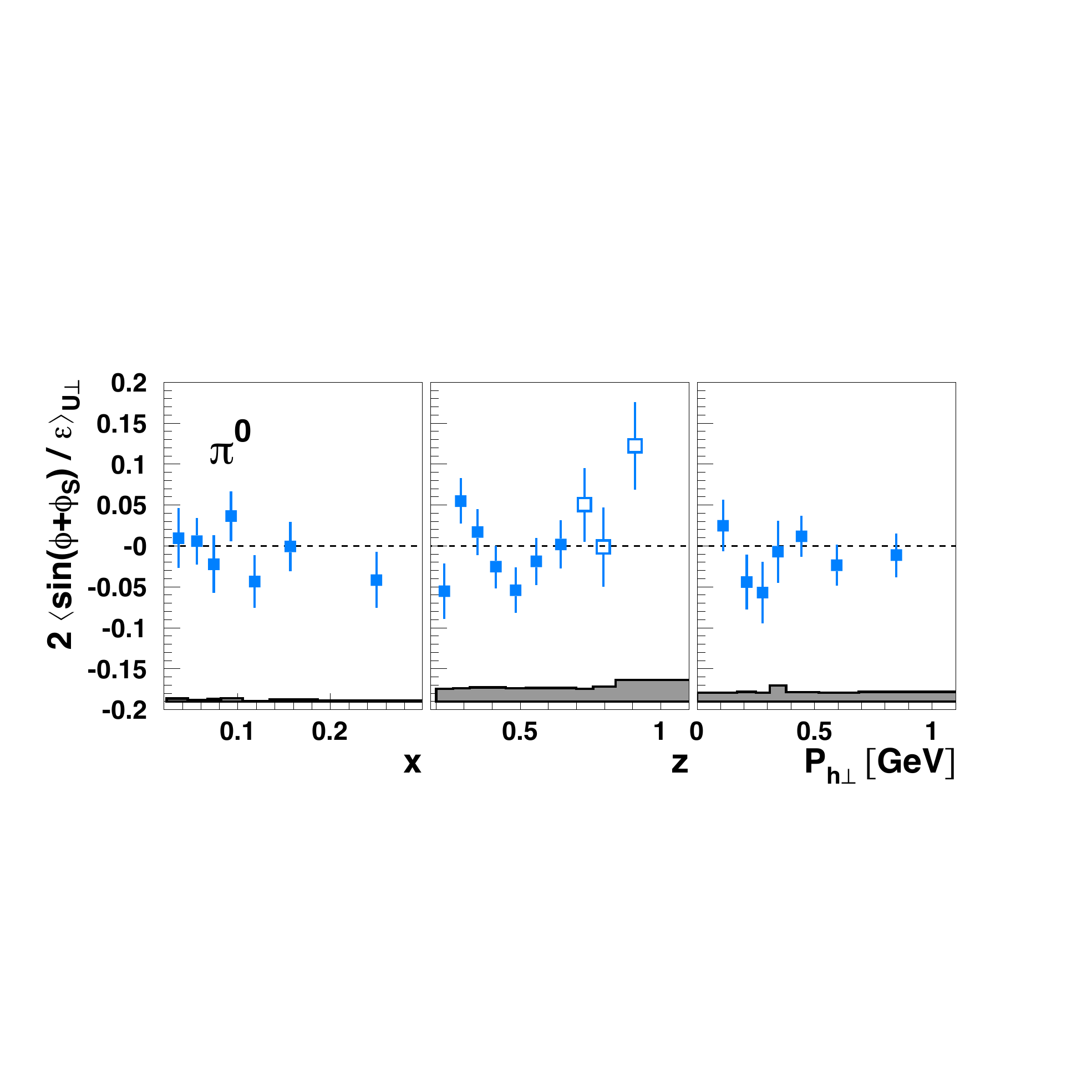}~~%
\includegraphics[bb = 15 60 500 470, clip, width=0.49\textwidth,keepaspectratio]{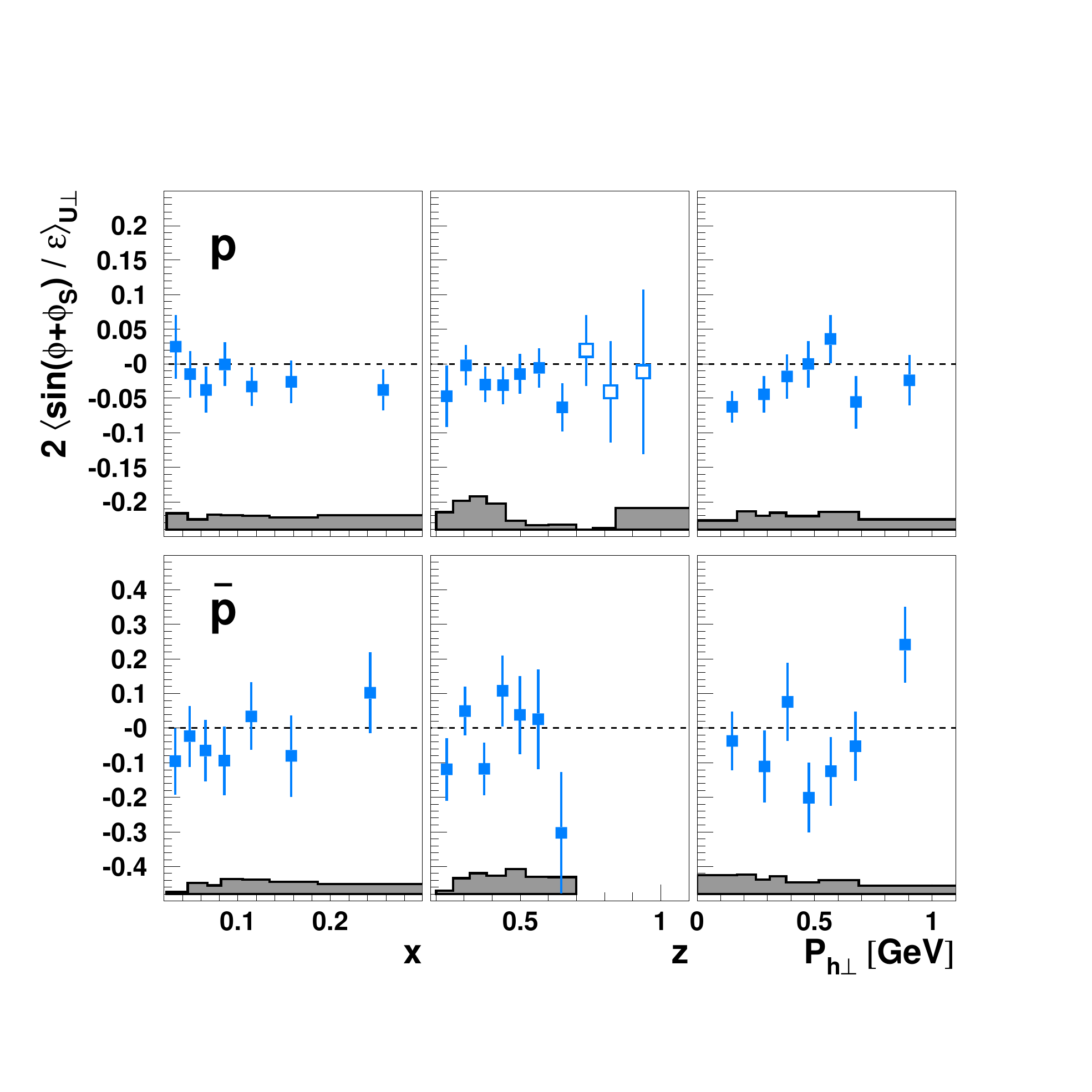}
\caption{Collins \sfa for \pizero (left), protons, and antiprotons (right) presented either in bins of \xb, \z, or \Phperpabs. Data at large values of \z, marked by open points in the \z projection, are not included in the other projections (no such high-\z points are available for antiprotons due to a lack of precision). Systematic uncertainties are given as bands, not including the additional scale uncertainty of 7.3\% due to the precision of the target-polarization determination.}
 \label{fig:protons-1d-collins}
\end{figure}

As discussed above, the Collins fragmentation functions extracted in phenomenological fits are opposite in sign and similar in magnitude for \piplus and \piminus. 
The \pizero Collins fragmentation function can be related through isospin symmetry to the ones of charged pions. 
In particular, it is the average of the latter two and thus approximately vanishes.
The \pizero Collins asymmetries, available only as one-dimensional projections, are shown in figure~\ref{fig:protons-1d-collins}.
They are indeed consistent with zero as expected.

The proton and antiproton Collins asymmetries, measured here for the first time, are depicted in figure~\ref{fig:protons-1d-collins} as one-dimensional projections.
They are mostly negative in case of protons, while the antiproton results are consistent with zero.
The Collins effect is a fragmentation effect, it might be suppressed for spin-\(\frac{1}{2}\) hadron production as compared to meson production. 
Models provide little guidance, and were already severely challenged by the large disfavored Collins fragmentation function for pions.
In the {\em Artru} approach~\cite{Artru:1995bh}, the transverse momentum of pions arises through an interplay of the meson and quark spins, as well as the vacuum structure: 
\( q\bar{q} \) pairs produced in the string-fragmentation model~\cite{Andersson:1983ia} are produced with vacuum quantum numbers, i.e., their spins are aligned 
and possess one unit of angular momentum opposite to their spin orientation. This orbital angular momentum is partially transformed into transverse momentum 
of the produced pion when pairing one of those quarks with the fragmenting quark, with the quark spins anti-aligned to form a spin-zero pion. 
If a favored pion forms in the first string break, a disfavored pion from the next break will inherit transverse momentum from the first break in the direction opposite to that acquired by the first pion, 
leading to a disfavored Collins function that is opposite in sign to that of the favored Collins function, consistent with the data.
The Collins function for baryons, however, is more difficult to predict in this approach as, 
e.g., the role of diquark  production in the fragmentation process or diquark fragmentation is far less understood.
The more complex production might thus easily wash out any transverse-polarization dependence of the fragmentation process.
More recently, a calculation in a diquark spectator model resulted in sizable Collins functions for up and down quarks into \(\Lambda\) hyperons~\cite{Wang:2018wqo}. 
While no such calculation is presented for the case of protons, it is not unplausible that it would result in a non-vanishing Collins effect, as hinted at by the data.

Lastly, looking at the ``semi-exclusive'' large-\z region (figures~\ref{fig:mesons-1d-collins} and \ref{fig:protons-1d-collins}), 
the asymmetries for positive mesons follow the trend of increasing with \z all the way to the highest \z, while such behavior is not visible for the other hadrons.%
\footnote{Due to insufficient yields, results for only two high-\z bins are available for \kminus and none for antiprotons.}
With increasing \z disfavored fragmentation decreases in importance. As a result the sensitivity to the struck quark --- mainly up quarks --- increases, 
leading to a further enhancement of the \piplus and \kplus asymmetries.

\subsection{Evidence for the Sivers function}\label{interpretation-sivers}

The naive-T-odd Sivers effect, first observed in semi-inclusive \dis by \hermes for positive pions~\cite{Airapetian:2004tw}, has been discussed 
already in detail in ref.~\cite{Airapetian:2009ae}, where one-dimensional projections versus \x, \z, and \Phperpabs 
of the \(\sin(\phi-\phis)\) Fourier amplitudes
were presented for pions, charged kaons, and the pion charge difference. 
Significantly positive asymmetries
were observed for positive pions and kaons, again larger for kaons than for pions. 
Significant positive values were also seen for \pizero as well as the pion charge-difference asymmetry, 
while results for negative pions and kaons were found to be consistent with zero.
These findings were interpreted as originating from up and down valence-quark Sivers distributions that are opposite in sign,
in accordance with the prediction~\cite{Burkardt:2002ks} based on the quark contributions to the proton's anomalous magnetic moment.
Phenomenological fits~\cite{Efremov:2004tp,Vogelsang:2005cs,Anselmino:2005ea,Anselmino:2005nn,Collins:2005ie,Anselmino:2008sga,Bacchetta:2011gx,Anselmino:2012aa,Sun:2013dya,Gamberg:2013kla,Echevarria:2014xaa,Anselmino:2016uie,Boglione:2018dqd,Cammarota:2020qcw}
to the HERMES and other semi-inclusive \dis data~\cite{Ageev:2006da,Alekseev:2008aa,Alekseev:2010rw,Qian:2011py,Adolph:2012sp,Adolph:2014zba,Adolph:2016dvl} 
(as well as to hadron-collision data in the case of ref.~\cite{Cammarota:2020qcw})
mainly result in Sivers distributions that are indeed significant only for valence quarks.\footnote{In ref.~\cite{Boglione:2018dqd} only the $u$-quark Sivers function is unambiguously found non-zero and the experimental data can be described with assigning the still required contributions either to \(d\) quarks or to the other remaining parton flavors, with further data needed for a more conclusive evaluation of the situation.} 
Those fit results suggest that valence quarks are sufficient to saturate the Burkardt sum rule~\cite{Burkardt:2003yg,Burkardt:2004ur}, 
which states that the net transverse momentum carried by partons inside a transversely polarized nucleon (which is related to the Sivers function) vanishes when summing over all partons (quarks and gluons).

\begin{figure}
\centering
\includegraphics[bb = 15 60 500 470, clip, width=0.49\textwidth,keepaspectratio]{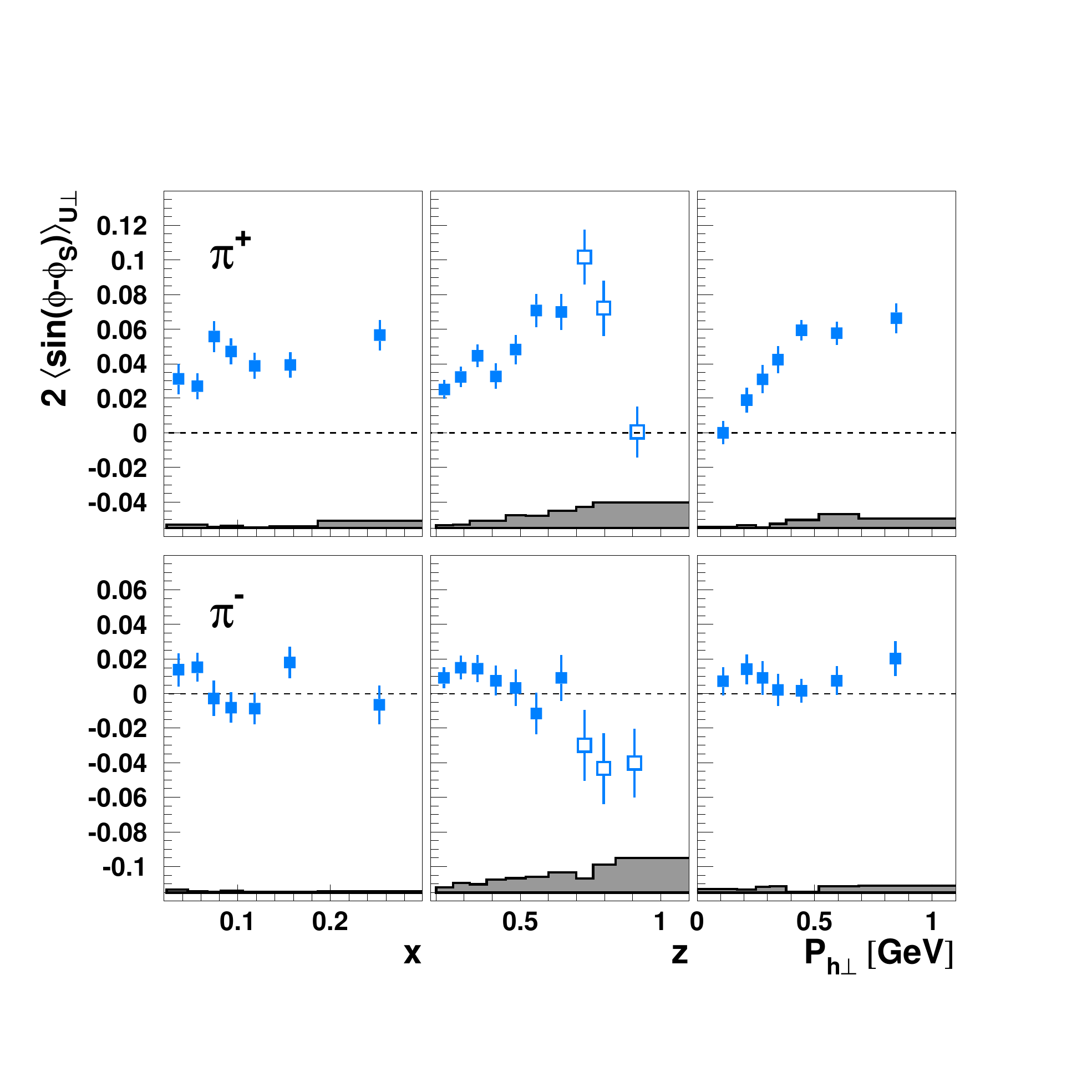}~~%
\includegraphics[bb = 15 60 500 470, clip, width=0.49\textwidth,keepaspectratio]{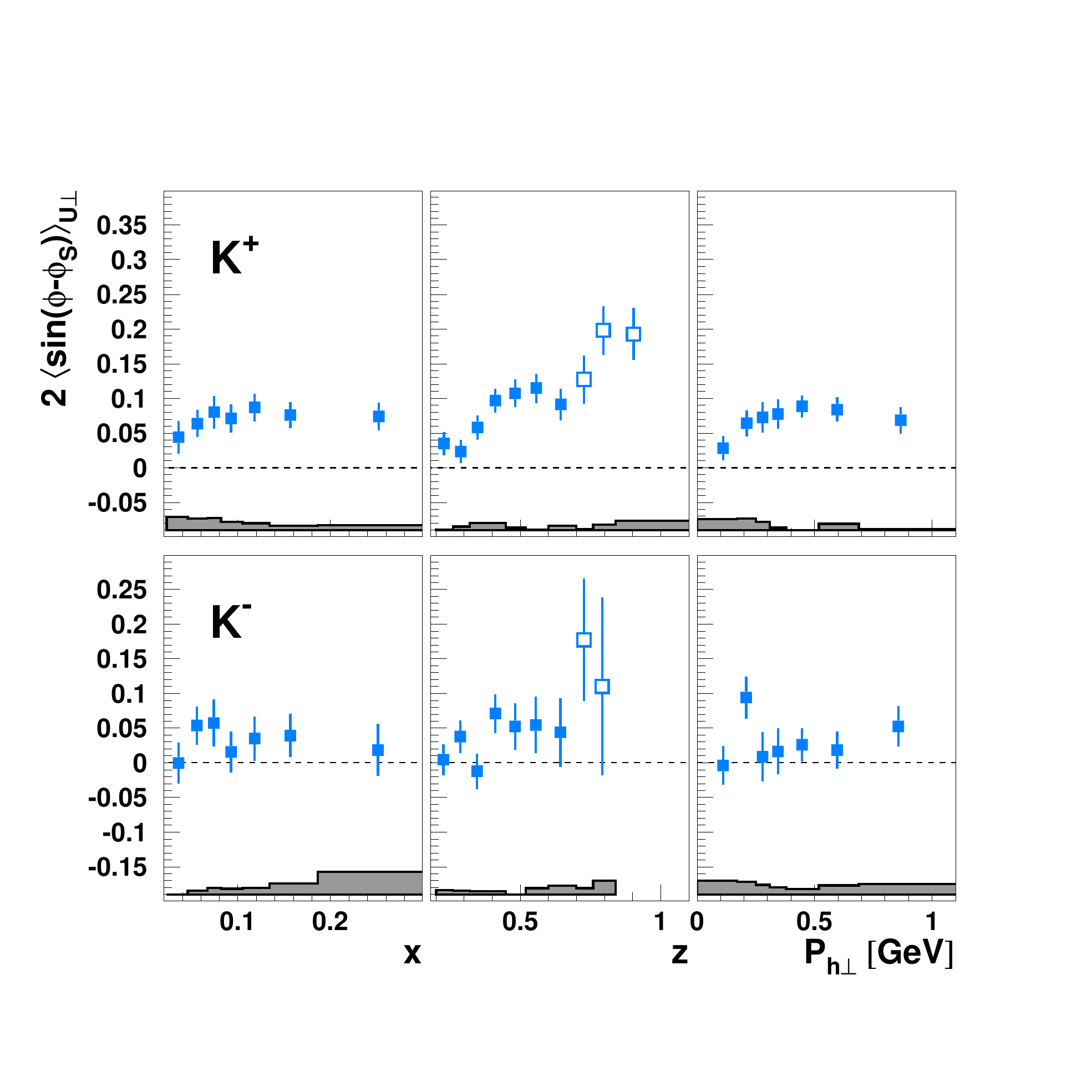}
 \caption{Sivers \sfa for charged mesons (left: pions; right: kaons) presented either in bins of \xb, \z, or \Phperpabs. Data at large values of \z, marked by open points in the \z projection, are not included in the other projections. Systematic uncertainties are given as bands, not including the additional scale uncertainty of 7.3\% due to the precision of the target-polarization determination.}
 \label{fig:mesons-1d-sivers}
\end{figure}

The Sivers asymmetries extracted here for charged pions and kaons in one-dimensional projections are presented in figure~\ref{fig:mesons-1d-sivers}. 
The Sivers modulation is the only one analyzed here that does not involve an \(\epsilon\)-dependent kinematic prefactor, i.e., \sfa and \csa should coincide. 
This is indeed found up to negligible variations introduced through correlations with other modulations in the fit.
Hence, even though the previously published results~\cite{Airapetian:2009ae} were obtained from a fit of the \csa to the data, 
while the ones shown in this section are the outcome of the \sfa fit, 
the slight differences between them --- though consistent --- stem solely from the updated analysis (changes in binning, newer calibrations of the data, etc.).

As in the previous publication~\cite{Airapetian:2009ae}, significantly positive Sivers amplitudes are observed for positive pions.
The asymmetries rise slightly with \xb, though remain significantly non-zero even at the lowest \xb values probed in this experiment. 
The rise with \z and \Phperpabs is much more pronounced. However, while the rise continues throughout the semi-inclusive \z range, it is leveling off at larger values of \Phperpabs.

The \piminus Sivers asymmetry in the one-dimensional \xb projection is consistent with zero. 
While \piplus electroproduction off protons is dominated by up-quark scattering, 
\piminus receives large contributions from down quarks. The vanishing Sivers asymmetry for negative pions can thus be understood as a cancelation
of a Sivers effect that is opposite in sign for up and down quarks. This may also explain the peculiar behavior of the \z dependence: at low values of \z
disfavored fragmentation plays a significant role and thus contributions from up quarks can push the asymmetry towards positive values. 
At large values of \z, however, disfavored fragmentation dies out and the favored production off down quarks prevails leading to a negative asymmetry. 
Some caution with this argumentation is deserved as
at large values of \z, the contribution from the decay of exclusive \( \rho^{0} \) electroproduction to both the \piplus and \piminus samples becomes sizable, 
as can be concluded from a \pythia Monte Carlo simulation (cf.~figure~\ref{measurement-exclusivechannels}), even more so for \piminus than for \piplus. 
Charge-conjugation dictates that the decay pions from the  \( \rho^{0} \) exhibit the same asymmetry regardless of their charge.\footnote{This is also one motivation for looking at the charge-difference asymmetry in ref.~\cite{Airapetian:2009ae} in which such contributions cancel.}
Examining the large-\z behavior of the charged-pion asymmetries, indeed a clear change of trend can be observed for positive pions. 
Still, the significant difference between the charged-pion asymmetries over most of the kinematic range suggests that the non-vanishing asymmetries observed are not driven merely by exclusive \( \rho^{0} \) electroproduction.

\begin{figure}
\centering
\includegraphics[bb = 22 246 499 473, clip, width=0.6\textwidth,keepaspectratio]{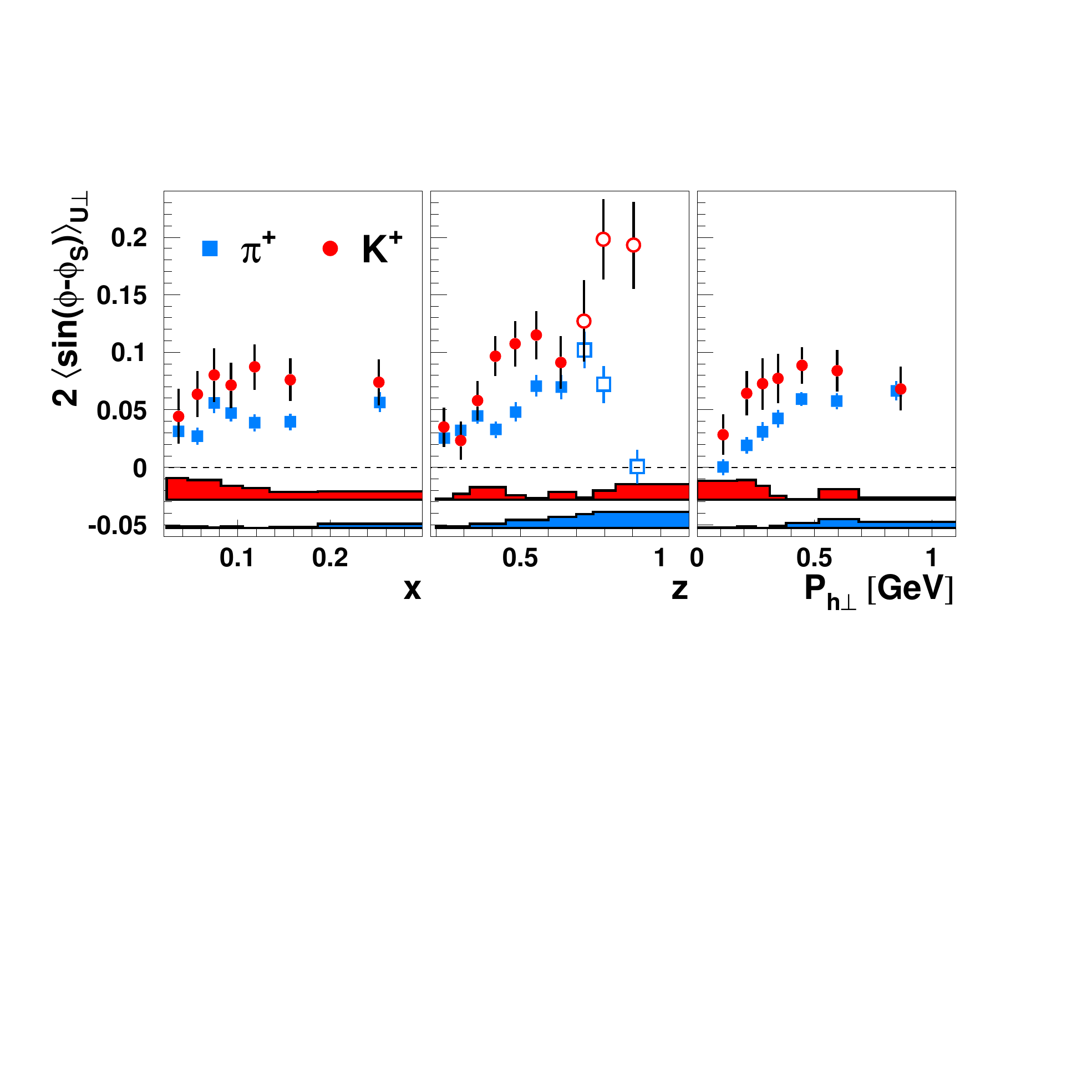}
  \caption{Comparison of Sivers \sfa for positive pions (squares) and kaons (circles) presented either in bins of \xb, \z, or \Phperpabs. Data at large values of \z, marked by open points in the \z projection, are not included in the other projections. Systematic uncertainties are given as bands, not including the additional scale uncertainty of 7.3\% due to the precision of the target-polarization determination.}
 \label{fig:pion-kaon-1d-sivers}
\end{figure}

The \kplus Sivers asymmetry follows a similar kinematic behavior as the one for \piplus, but is larger in magnitude, as can be seen in figure~\ref{fig:pion-kaon-1d-sivers}. 
While \(u\)-quark scattering should dominate production off protons of both positive pions and kaons, various differences between pion and kaon production might point to the origin for the larger \kplus asymmetry: 
(i) differences in the relative strengths of the disfavored \(d\)-quark fragmentation compared to the favored \(u\)-quark fragmentation for positive pions and kaons might lead to a reduced canceling contribution from the \(d\)-quark Sivers function; 
(ii) in general, differences in the role of sea quarks; 
(iii) differences --- as hinted in a phenomenological analysis~\cite{Signori:2013mda} of \hermes multiplicity data~\cite{Airapetian:2012ki} --- in the transverse-momentum dependence of hadronization for different quark flavors that enters the convolution over transverse momentum in eq.~\eqref{eq:QPM-sivers};
(iv) and also higher-twist effects as it was observed in ref.~\cite{Airapetian:2009ae} that the \piplus--\kplus difference was more pronounced at lower values of \Q. 
Notwithstanding those differences, acknowledging \(u\)-quark dominance in both \piplus and \kplus production and relating their positive Sivers asymmetries to eq.~\eqref{eq:QPM-sivers} leads immediately to the conclusion that the \(u\)-quark Sivers function, \tmdsivers{u}, must be negative. Adding the \piminus data, as argued before, results in a positive \tmdsivers{d}.

Looking at the newly explored large-\z region, the similarity of \piplus and \kplus Sivers asymmetries disappears: in contrast to the drop at large \z of the asymmetry values in the case of positive pions, the \kplus Sivers asymmetry continues its trend to increase with \z, which is indeed the expected behavior. This divergence of behavior for positive pions and kaons can also be seen in the corresponding data of the \compass Collaboration~\cite{Adolph:2014zba}, in particular in the \x region overlapping with \hermes. As decay products from exclusively produced vector-mesons contribute significantly less to \kplus production, this might be another indication of a non-negligible role of those in the case of the pion data.

While the data on negative kaons is more limited in precision, also here a positive asymmetry is clearly visible in the right plot of figure~\ref{fig:mesons-1d-sivers}. Negative kaons and the target proton have no valence quarks in common. While sensitive to the nucleon's sea-quark, \(u\)-quark scattering will still be a dominant contribution, as can be concluded from the \kminus purity in ref.~\cite{Airapetian:2004zf}. 
However, in contrast to \kplus, the \(u\)-quark contribution is suppressed and {\em diluted} \footnote{``Diluted'' in the literal sense or through competing/canceling contributions from other quark flavors, e.g., \(d\)-quarks.} in the case of the \kminus asymmetry.

\begin{figure}
\centering
\includegraphics[bb = 21 157 500 376, clip, width=0.6\textwidth,keepaspectratio]{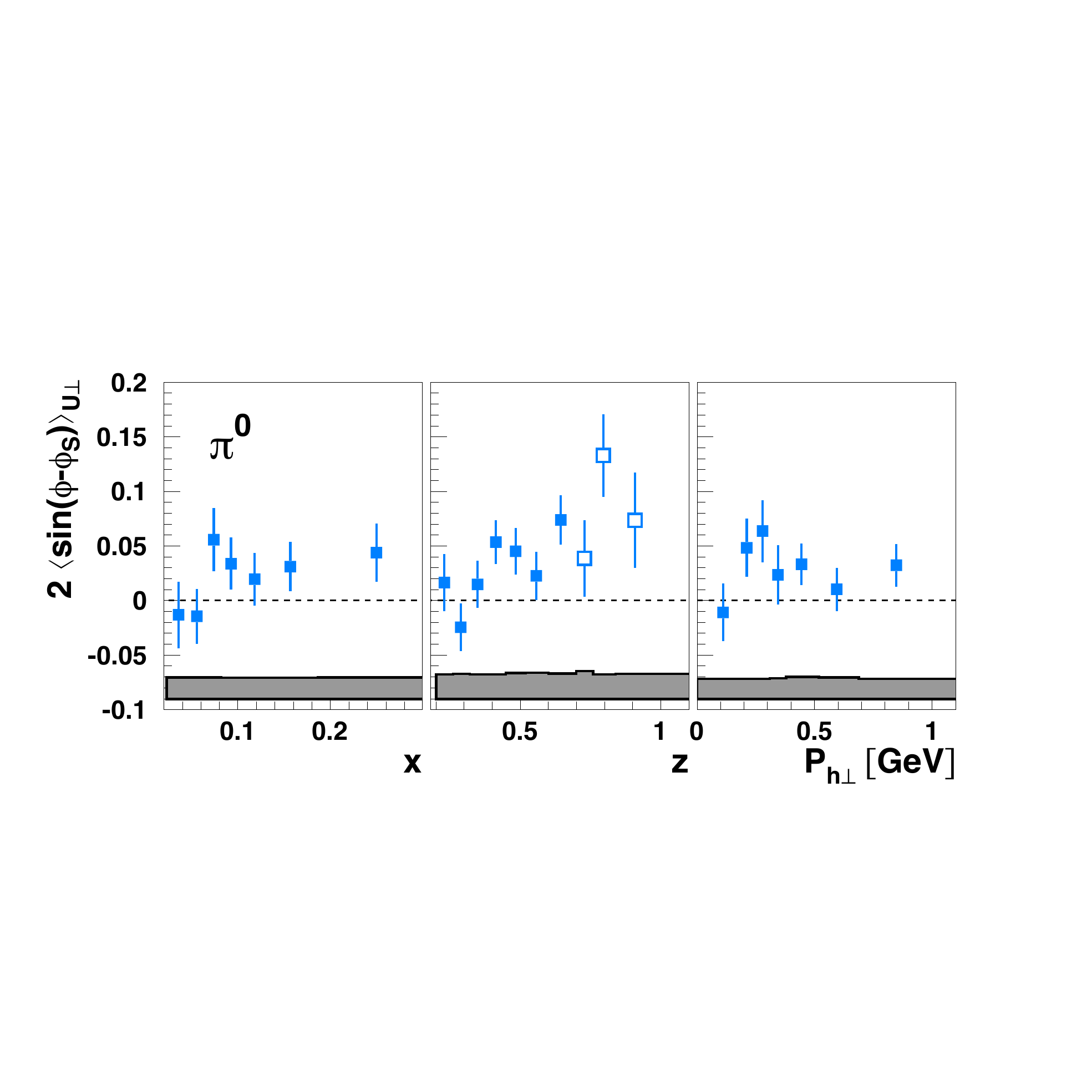}
  \caption{Sivers \sfa for \pizero presented either in bins of \xb, \z, or \Phperpabs. Data at large values of \z, marked by open points in the \z projection, are not included in the other projections. Systematic uncertainties are given as bands, not including the additional scale uncertainty of 7.3\% due to the precision of the target-polarization determination.}
 \label{fig:pizero-1d-sivers}
\end{figure}

As is the case for \kminus, the \pizero results, presented in figure~\ref{fig:pizero-1d-sivers}, have poor statistical precision but still indicate a positive asymmetry.
This can be expected from the results for charged pions due to isospin symmetry in semi-inclusive \dis. 
In the high-\z range, the \pizero asymmetries remain positive around 5--10\%, thus not following the strongly falling trend of the \piplus asymmetries. 
Also here the contribution from exclusive vector-meson production is much smaller than for \piplus (cf.~figure~\ref{measurement-exclusivechannels}); 
thus, an interpretation in terms of ordinary fragmentation is likely much more applicable, leading to a positive asymmetry due to \(u\)-quark dominance.

\begin{figure}
\centering
\includegraphics[bb = 30 23 515 543, clip, width=0.7\textwidth,keepaspectratio]{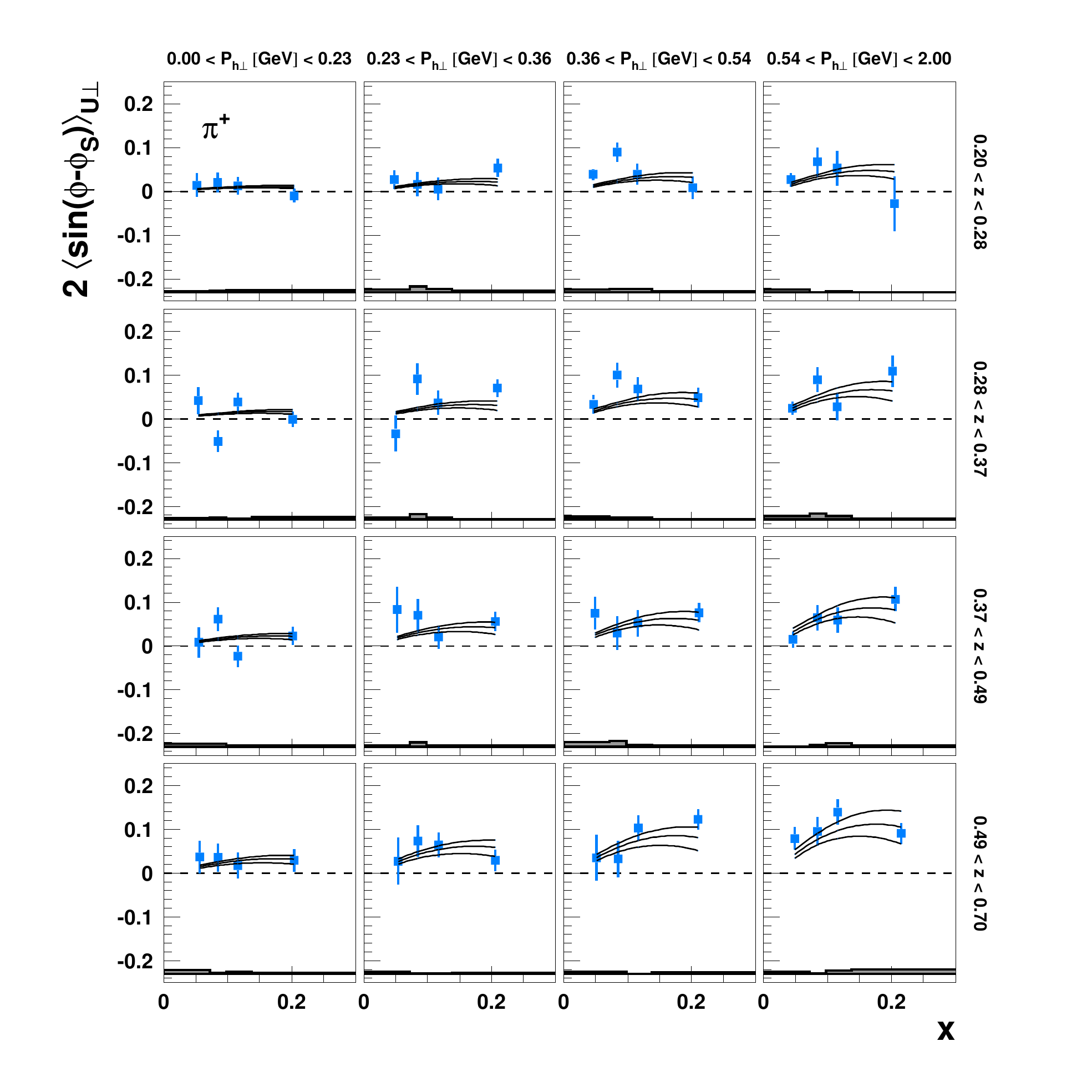}
  \caption{Sivers \sfa for \piplus extracted simultaneously in bins of \xb, \z, and \Phperpabs, presented as a function of \xb. Systematic uncertainties are given as bands, not including the additional scale uncertainty of 7.3\% due to the precision of the target-polarization determination. Overlaid is a phenomenological fit~\cite{Anselmino:2012aa} to previously available data, with the three lines corresponding to the central value of the fit and the fit uncertainty.}
 \label{fig:sivers-torino-comparison}
\end{figure}

Figure \ref{fig:sivers-torino-comparison} shows, as an illustrative example, the Sivers asymmetry for \piplus mesons in the three-dimensional binning, compared to a phenomenological fit~\cite{Anselmino:2012aa}. 
The latter, being based on previous versions of these data (as well as data from \compass),
describes the overall behavior well. 
The multi-dimensional binning as well as the much reduced systematics of the data presented here should help to better constrain future phenomenological analyses.

\begin{figure}
\centering
\includegraphics[bb = 15 60 500 470, clip, width=0.6\textwidth,keepaspectratio]{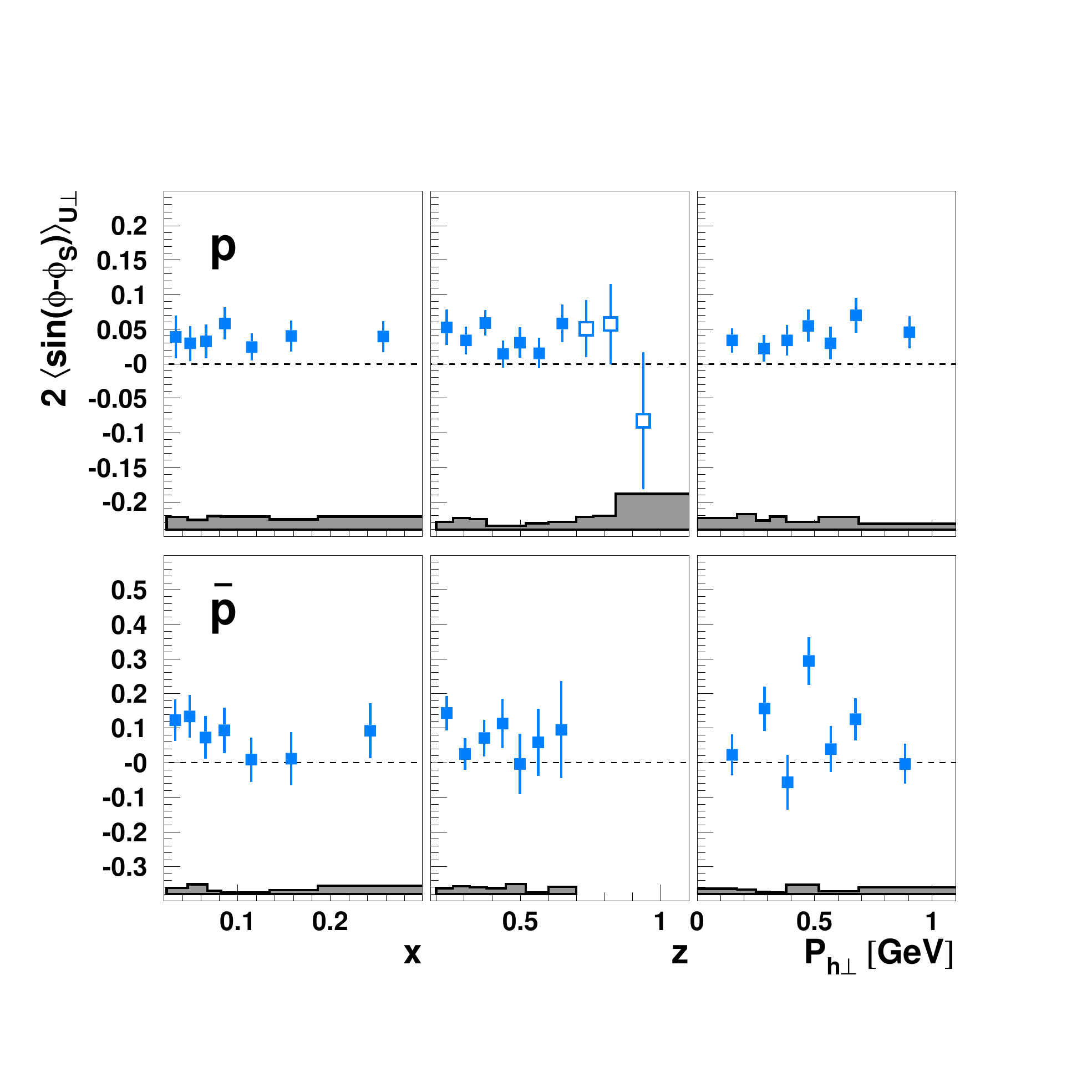}
  \caption{Sivers \sfa for protons (upper row) and antiprotons (lower row) presented either in bins of \xb, \z, or \Phperpabs. Data at large values of \z, marked by open points in the \z projection, are not included in the other projections (no such high-\z points are available for antiprotons due to a lack of precision). Systematic uncertainties are given as bands, not including the additional scale uncertainty of 7.3\% due to the precision of the target-polarization determination.}
 \label{fig:protons-1d-sivers}
\end{figure}

\begin{figure}
\centering
\includegraphics[bb = 15 60 500 470, clip, width=0.6\textwidth,keepaspectratio]{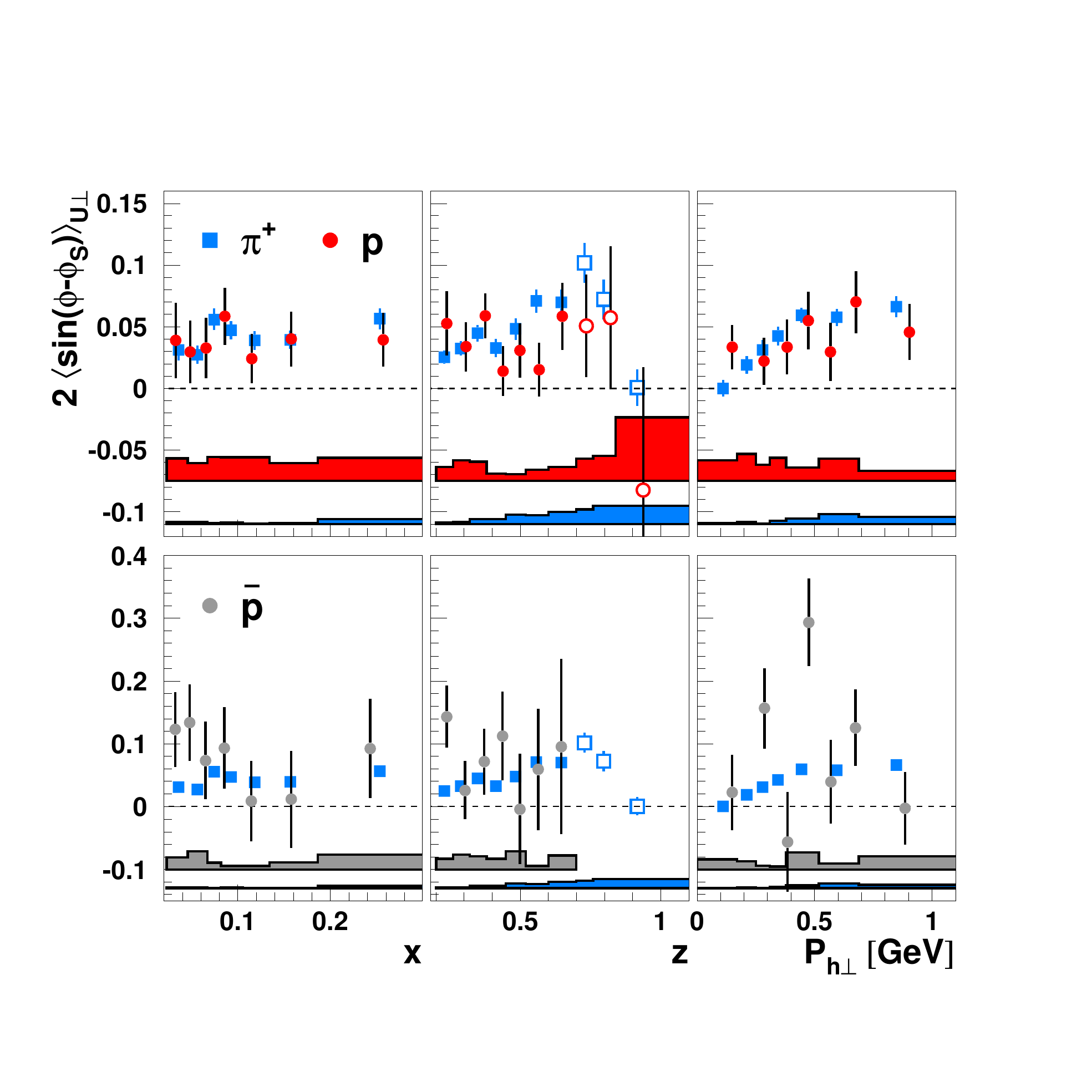}
 \caption{Comparison of Sivers \sfa for positive pions and protons (upper plot) or antiprotons (lower plot) presented either in bins of \xb, \z, or \Phperpabs. Data at large values of \z, marked by open points in the \z projection, are not included in the other projections (no such high-\z points are available for antiprotons due to a lack of precision). Systematic uncertainties are given as bands, not including the additional scale uncertainty of 7.3\% due to the precision of the target-polarization determination.}
 \label{fig:pions-protons-pbar-1d-sivers}
\end{figure}

In figure~\ref{fig:protons-1d-sivers}, the first measurement of Sivers asymmetries for proton and antiprotons is presented. A clearly positive Sivers asymmetry is observed for protons. Also the less precise antiproton data favor a positive Sivers asymmetry. Baryon production is a less understood process at lower center-of-mass energies. Therefore, care must be taken when interpreting those in the usual factorized way. Leaving this warning aside and assuming quark fragmentation as the dominant process here, \(u\)-quark fragmentation prevails proton production, and --- having no valence quark in common with the target proton --- antiprotons as well are likely to originate from \(u\)-quarks, in particular at these values of \x, where sea quarks are still scarce in the target proton.  
Dominance of  \(u\)-quarks in proton and antiproton leptoproduction is supported by results from global fits of fragmentation functions~\cite{deFlorian:2007hc}. 
The Sivers effect is sometimes referred to as a ``quark-jet effect'', e.g., already before forming the final hadron, the transverse-momentum distribution of the fragmenting quark exhibits the Sivers signature of a left-right asymmetry with respect to the direction of the target polarization. It is thus natural to expect similar asymmetries for ``current-fragmentation'' protons and antiprotons as those for the other  hadrons whose electroproduction off the proton is dominated by \(u\)-quark scattering~\cite{Echevarria2018}.
Figure~\ref{fig:pions-protons-pbar-1d-sivers} compares the Sivers asymmetries for both protons and antiprotons with those for positive pions. Within the available precision an almost surprising agreement of proton and \piplus asymmetries is visible. Also the asymmetries for antiprotons are very similar, however, the present measurement is plagued by large uncertainties.

\begin{figure}
\centering
\includegraphics[width=0.4\textwidth,keepaspectratio]{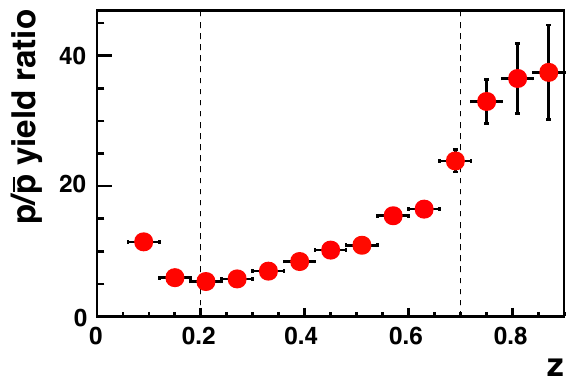}
  \caption{Ratio of raw proton to antiproton yields at HERMES as a function of \z. The bin boundaries for the semi-inclusive DIS range are marked by dashed lines. The ratio exhibits a clear rise towards very low \z, which might indicate the onset of significant target-fragmentation contributions, excluded in the data sample used by the minimum-\z requirement of 0.2.}
 \label{fig:protons-pbar-rates}
\end{figure}

In order to investigate slightly more the nature of proton and antiproton production at \hermes, figure~\ref{fig:protons-pbar-rates} 
depicts the ratio of their raw production rates, e.g., yields not corrected for instrumental effects. 
The sudden increase of the proton-over-antiproton ratio towards very low $z$ might indicate the onset of target fragmentation, 
while in most of the \z range studied here the ratio exhibits a behavior consistent with {\em current fragmentation}. 
In particular, with increasing \z the production of antiprotons, which have no valence quarks in common with the target nucleons, 
is increasingly suppressed compared to protons. 
A second qualitative argument supporting the hypothesis of dominance of current fragmentation is the sign of the Sivers asymmetry for protons. The current jet is dominated by \(u\)-quark scattering, which exhibits a positive Sivers asymmetry. The recoiling target fragments are thus expected to exhibit a Sivers asymmetry of opposite sign. As the proton Sivers asymmetry is positive, it appears less likely that those protons came from the fragmenting target. 
All these features are, however, also not sufficient to establish that the protons and antiprotons are dominantly produced in the hadronization of the current-quark jet, which needs to be kept in mind when interpreting the results in such framework.

\subsection{The vanishing signals for the pretzelosity function}\label{interpretation-pretzelosity}

The chiral-odd pretzelosity distribution, \tmdhtpt{q}, provides information about the non-spherical shape of transversely polarized protons in momentum space
caused by significant contributions from orbital angular momentum to a quadrupole modulation of the parton distributions~\cite{Miller:2007ae}. 
It can be accessed coupled to the chiral-odd Collins fragmentation function in semi-inclusive \dis through the 
\sinemodulation{3\phih-\phis} modulation of the cross section. 
So far, only the measurement of this amplitude using a transversely polarized \(^{3}\)He target by the Jefferson Lab Hall A Collaboration has been published~\cite{Zhang:2013dow}.
In a combination with preliminary data from both the \compass and \hermes collaborations as well as the Collins fragmentation function from a phenomenological analysis~\cite{Anselmino:2013vqa}, 
\tmdhtpt{q} was extracted both for up and down quarks and found to be consistent with zero albeit within large uncertainties~\cite{Lefky:2014eia}.

The underlying transverse-momentum convolution in eq.~\eqref{eq:sinthreephi} 
involves a weight that is expected to scale with \(\Phperpabs^3\). 
As relatively low transverse momenta are observed, \(\mean{\Phperpabs} < 1 \GeV\), 
the amplitude of the \sinemodulation{3\phih-\phis} modulation is suppressed with respect to, e.g., the Collins
amplitude, which also involves a convolution of a chiral-odd parton distribution with the Collins fragmentation function, 
but which scales with \(\Phperpabs\).

\begin{figure}
\centering
\includegraphics[bb = 15 60 500 470, clip, width=0.49\textwidth,keepaspectratio]{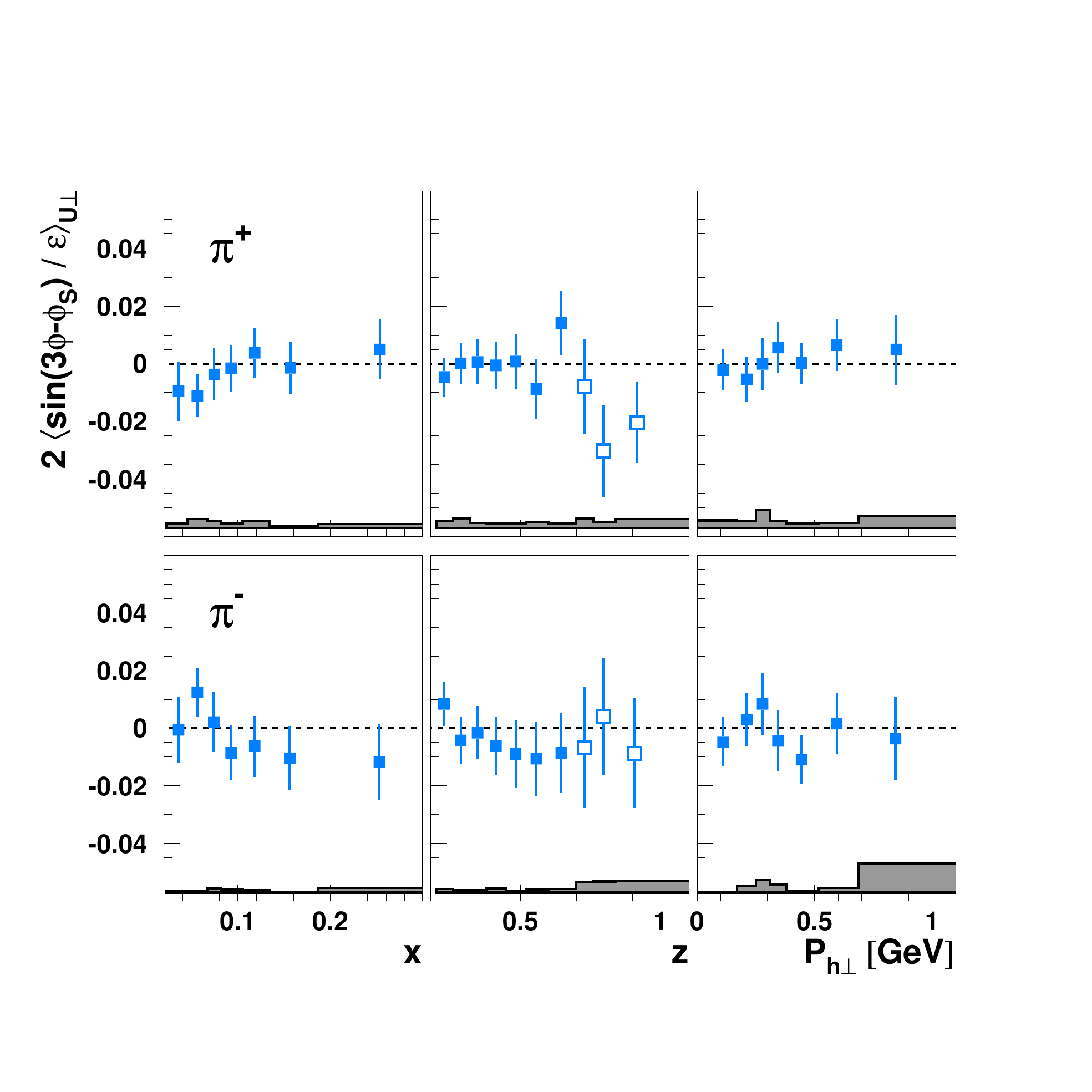}~~%
\includegraphics[bb = 15 60 500 470, clip, width=0.49\textwidth,keepaspectratio]{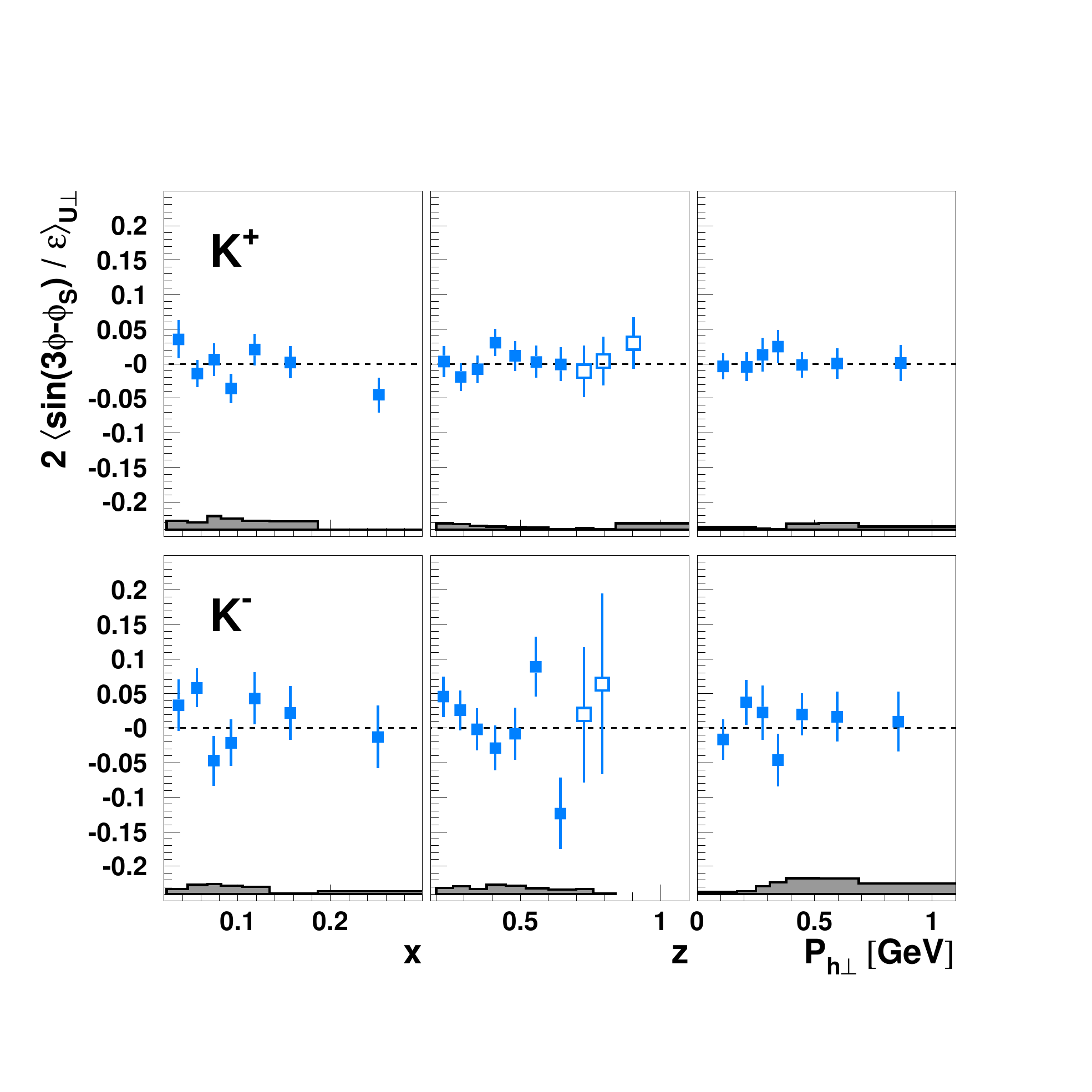}
  \caption{Pretzelosity \sfa for charged mesons (left: pions; right: kaons) presented either in bins of \xb, \z, or \Phperpabs. Data at large values of \z, marked by open points in the \z projection, are not included in the other projections. Systematic uncertainties are given as bands, not including the additional scale uncertainty of 7.3\% due to the precision of the target-polarization determination.}
 \label{fig:mesons-pretzelosity}
\end{figure}

\begin{figure}
\centering
\includegraphics[bb = 15 60 500 490, clip, width=0.49\textwidth,keepaspectratio]{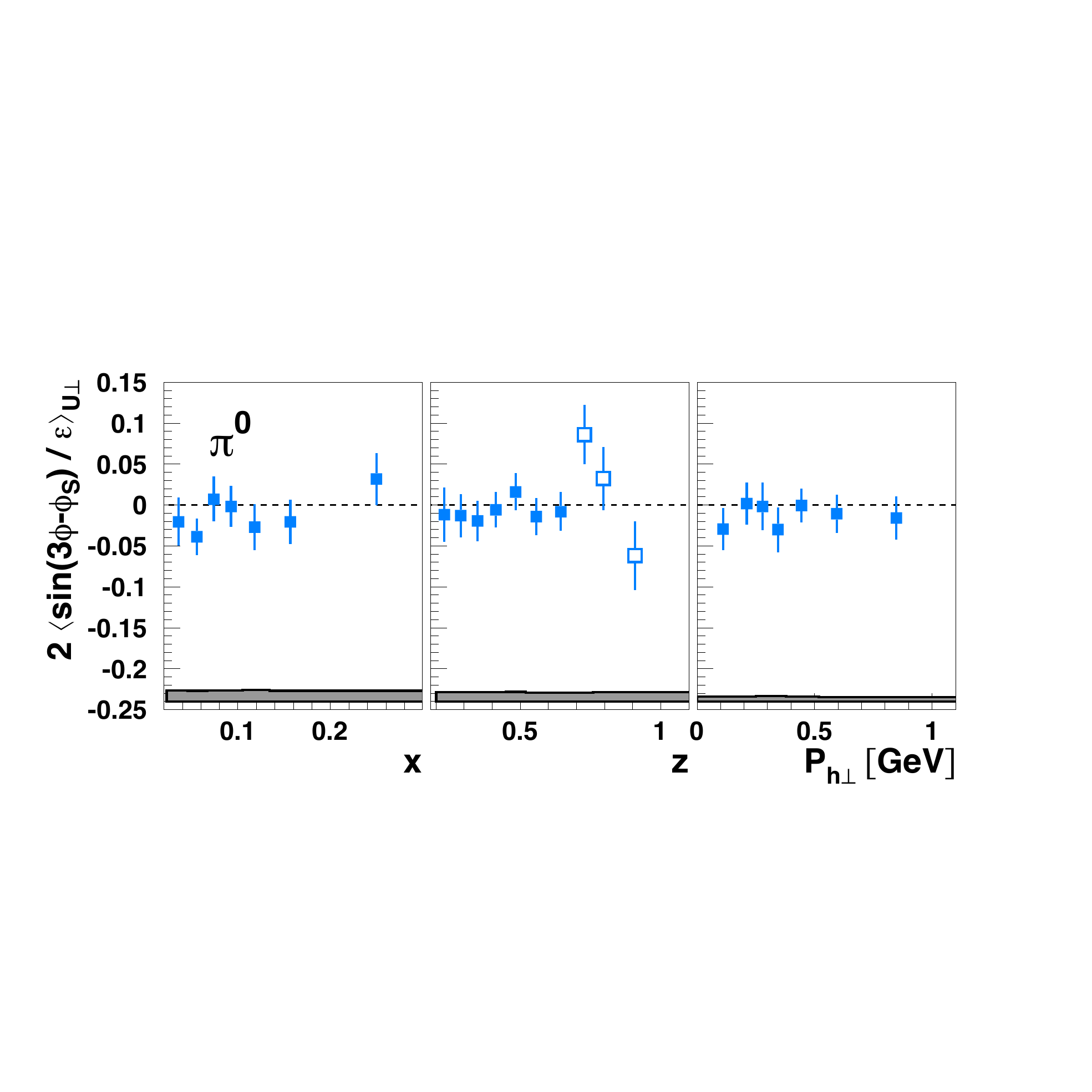}~~%
\includegraphics[bb = 15 60 500 490, clip, width=0.49\textwidth,keepaspectratio]{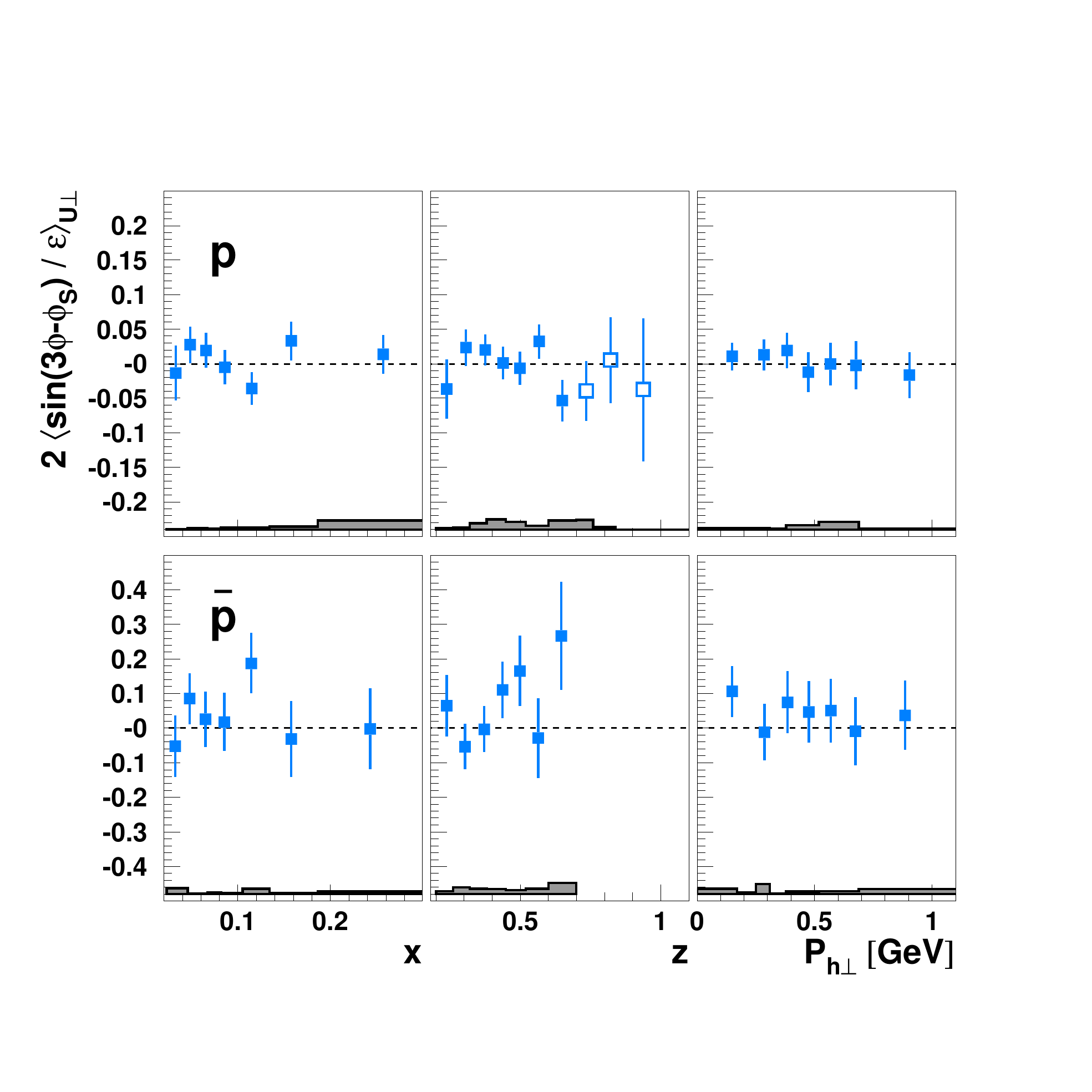}
  \caption{Pretzelosity  \sfa for \pizero (left), protons, and antiprotons (right) presented either in bins of \xb, \z, or \Phperpabs. Data at large values of \z, marked by open points in the \z projection, are not included in the other projections (no such high-\z points are available for antiprotons due to a lack of precision). Systematic uncertainties are given as bands, not including the additional scale uncertainty of 7.3\% due to the precision of the target-polarization determination.}
 \label{fig:pizero-protons-pretzelosity}
\end{figure}

In this analysis, the \sinthreephiexpSFA{h} amplitudes, shown in figure~\ref{fig:mesons-pretzelosity} for charged mesons 
and in figure~\ref{fig:pizero-protons-pretzelosity} for neutral pions as well as for (anti)protons, are found to be consistent with zero. 
There is a hint of a small negative amplitude for negative pions
that is, however, statistically not sufficiently significant to claim a non-vanishing pretzelosity.

As noted before, the pretzelosity amplitudes are expected to be suppressed.
Cancelations, e.g., from the Collins function that changes sign for favored and disfavored fragmentation,
might also contribute to the vanishing signal.
Model calculations thus predict in general small asymmetries below 0.01 (see, e.g., ref.~\cite{Boffi:2009sh}),
beyond the precision of this measurement.

\subsection{Signals for the worm-gear (II) distribution \tmdgtpt{q}}\label{interpretation-wormgear}

The \Teven and chiral-even worm-gear (II) distribution \tmdgtpt{q} is unique in the sense
that it is the only \tmd that vanishes when integrating over \pt but
neither entails nor is affected by final-state interactions. At
leading twist, this \tmd cannot contribute to \Todd effects that cause
\ssa[long]. Its spin-orbit correlation, \(\lambda S_T^ip_T^i\),
involves a common product of the helicity of the struck quark and the
transverse spin direction of the nucleon. In combination with the
selection of quarks with a certain helicity by a longitudinally
polarized lepton beam, the worm-gear (II) distribution \tmdgtpt{q} can be
related to the \cosinemodulation{\phih-\phis} modulation of the
double-spin asymmetry in the scattering of longitudinally polarized leptons by transversely polarized nucleons.

\begin{figure}
\centering
\includegraphics[bb = 15 60 500 470, clip, width=0.49\textwidth,keepaspectratio]{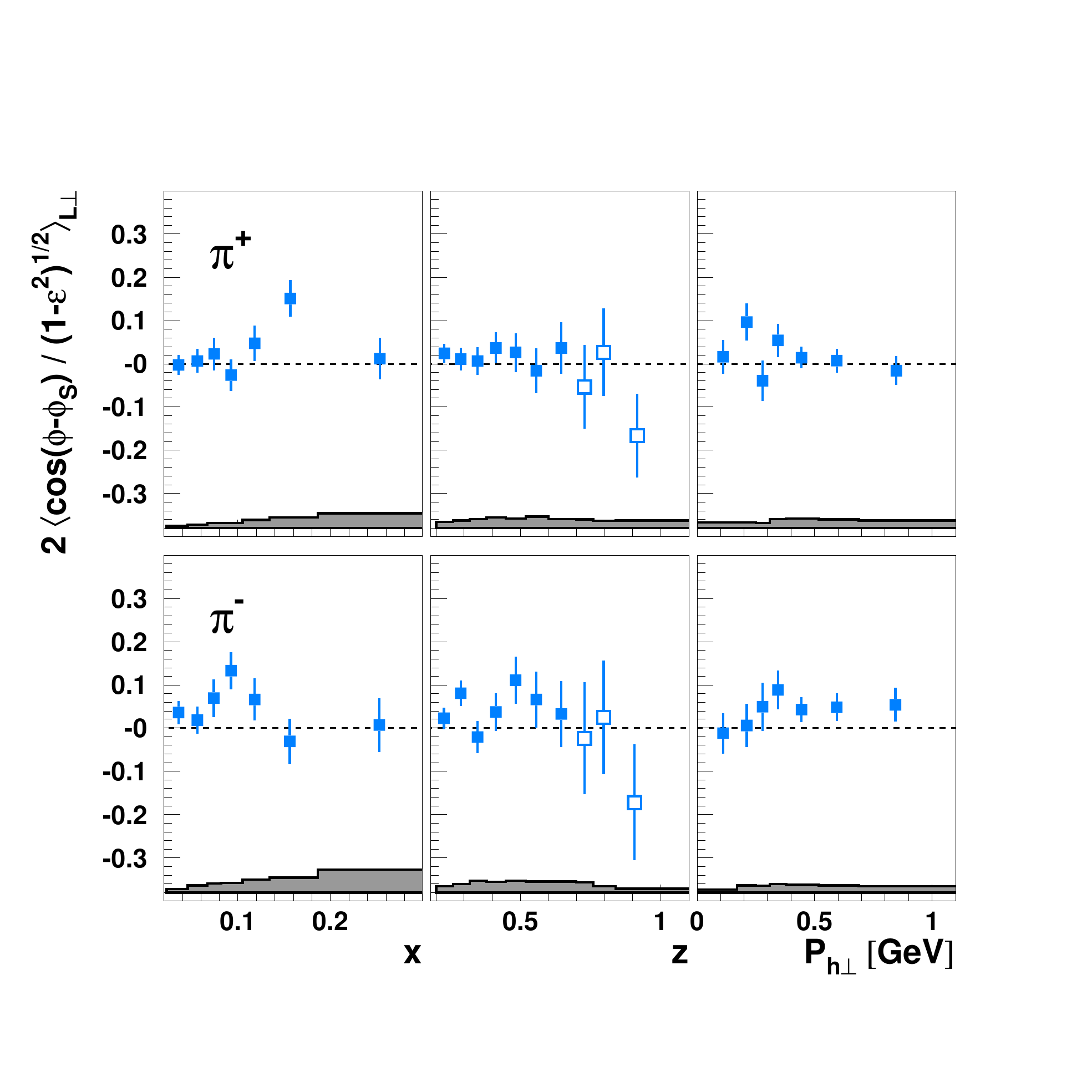}~~%
\includegraphics[bb = 15 60 500 470, clip, width=0.49\textwidth,keepaspectratio]{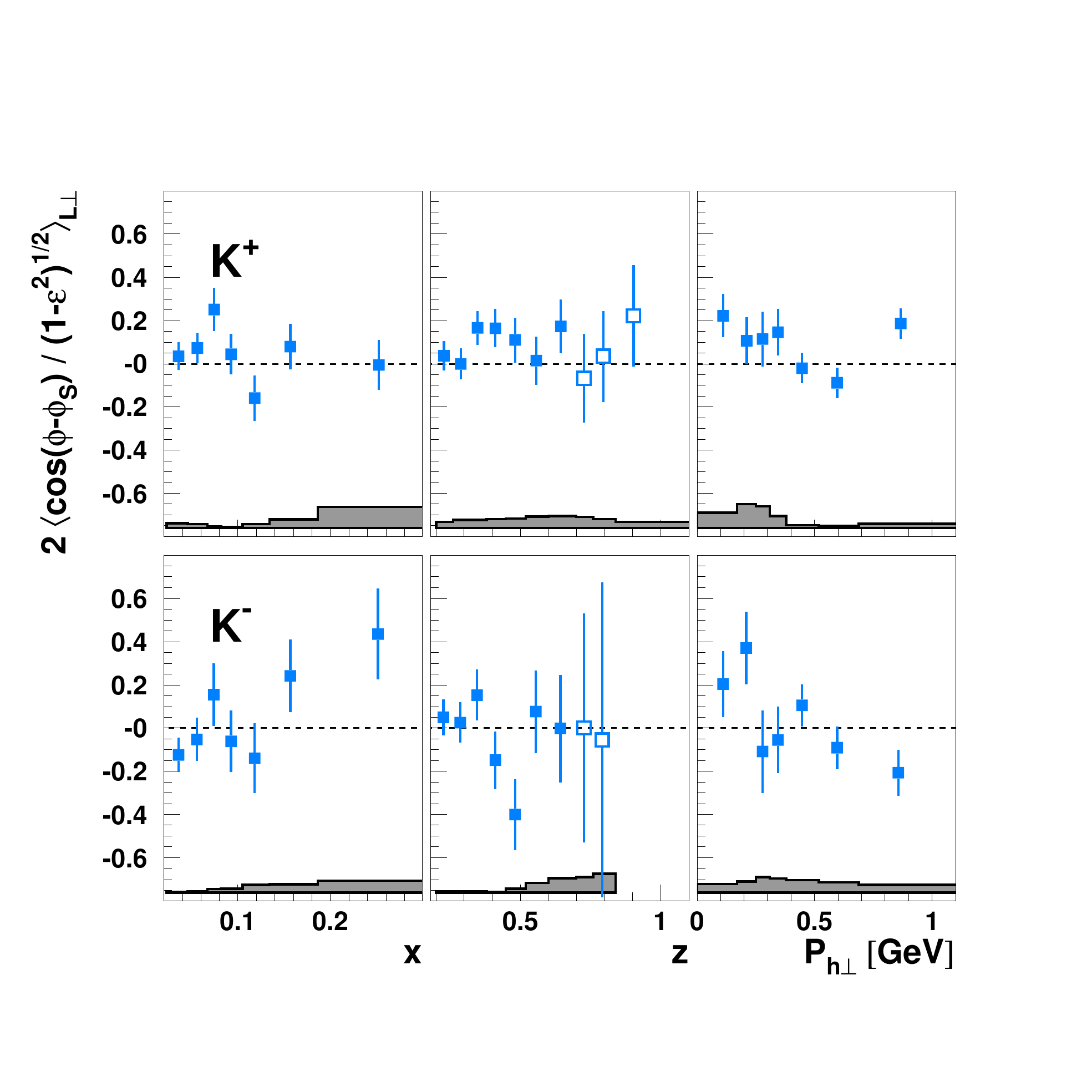}
  \caption{The \cosphiexpSFA{h}  amplitudes  for charged mesons (left: pions; right: kaons) presented either in bins of \xb, \z, or \Phperpabs. Data at large values of \z, marked by open points in the \z projection, are not included in the other projections. Systematic uncertainties are given as bands, not including the additional scale uncertainty of 8.0\% due to the precision in the determination of the target and beam polarizations.}
 \label{fig:mesons-g1Tperp}
\end{figure}
\begin{figure}
\centering
\includegraphics[bb = 15 60 500 490, clip, width=0.49\textwidth,keepaspectratio]{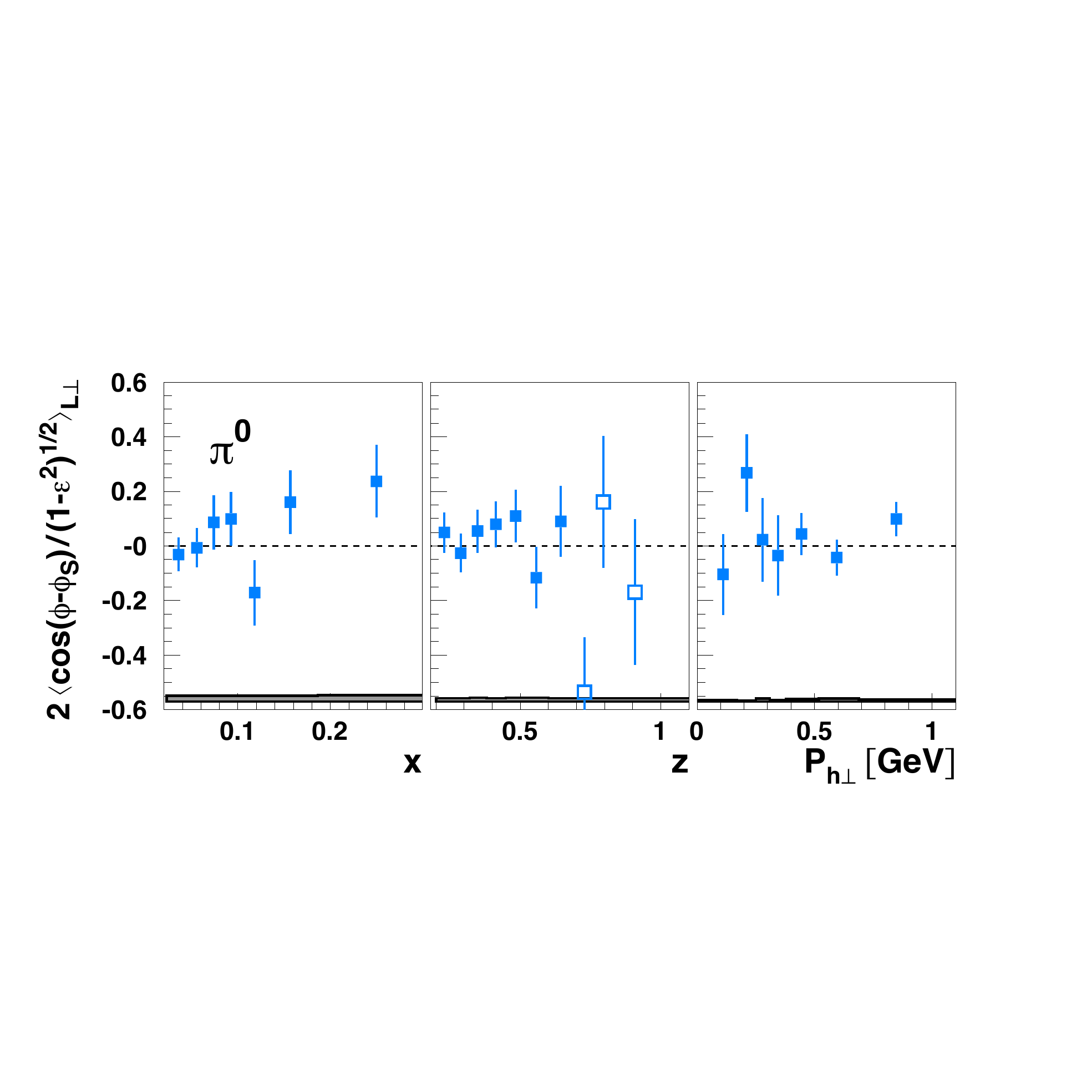}~~%
\includegraphics[bb = 15 60 500 490, clip, width=0.49\textwidth,keepaspectratio]{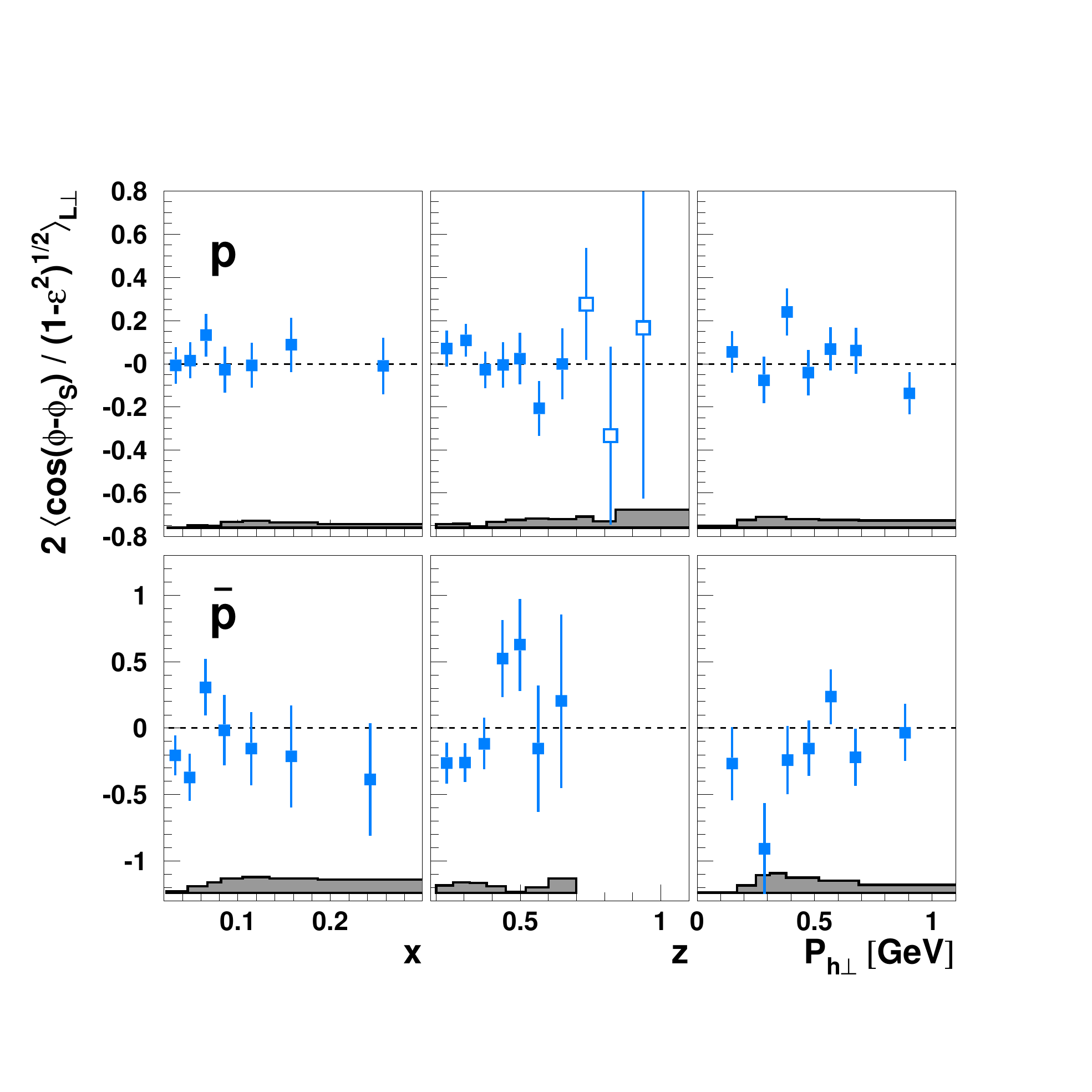}
  \caption{The \cosphiexpSFA{h} amplitudes for \pizero (left), protons, and antiprotons (right) presented either in bins of \xb, \z, or \Phperpabs. Data at large values of \z, marked by open points in the \z projection, are not included in the other projections (no such high-\z points are available for antiprotons due to a lack of precision). Systematic uncertainties are given as bands, not including the additional scale uncertainty of 8.0\% due to the precision in the determination of the target and beam polarizations.}
 \label{fig:pizero-protons-g1Tperp}
\end{figure}

This \cosinemodulation{\phih-\phis} modulation provides a leading-twist signal for
the worm-gear (II) distribution \tmdgtpt{q} in
combination with the spin-independent fragmentation function \ffdkt{q\to h} [c.f.~eq.~\eqref{eq:cosphi}].
As such it is not additionally suppressed in the asymmetry amplitude by the relative magnitude of
\ffcollinskt{q\to h} compared to \ffdkt{q\to h}.

In figures~\ref{fig:mesons-g1Tperp} and \ref{fig:pizero-protons-g1Tperp}, the 
\cosphiexpSFA{h} Fourier amplitudes of the double-spin asymmetry  \alp{h} are presented 
for pions, charged kaons, as well as for (anti)protons. 
As a consequence of the relatively small degree of polarization of the
\hera lepton beam during the years 2002--2005, the statistical
uncertainties are generally larger than those for the Fourier amplitudes of the transverse single-spin asymmetry \aup{h}.

For positively charged pions, non-vanishing \cosphiexpSFA{h} Fourier amplitudes
are extracted, providing an indication for a non-vanishing worm-gear (II) distribution \tmdgtpt{q}. 
Results for \piminus and \kplus are inconsistent with zero at 90\% but not at 95\% confidence level.

When comparing the meson results to the Sivers asymmetries, which also involve only the ordinary 
\ffdkt{q} fragmentation function and are thus easier to interpret in terms of separate quark-flavor contributions,
a similar picture becomes apparent: mainly the positively charged mesons exhibits a 
(positive) \cosphiexpSFA{} amplitude.
In analogy to the Sivers discussion, taking into account the additional minus sign in the 
Sivers convolution~\eqref{eq:QPM-sivers} compared to \eqref{eq:cosphi}, 
the data suggest that  \tmdgtpt{u} is positive.

However, all of the above discussion is merely qualitative in view of the large uncertainties of this measurement. 
In that respect, it should be emphasized that tremendous progress has been made predicting \tmdgtpt{q} based on models and by now also lattice-QCD calculations~\cite{Musch:2010ka,Yoon:2017qzo}. 
A common thread among the calculations is a positive \tmdgtpt{u} and a negative \tmdgtpt{d}, not at variance with the above discussion. 
For example, the calculation in ref.~\cite{Boffi:2009sh} --- based on the light-cone constituent quark model --- predicts positive \cosphi{} Fourier amplitudes for charged pions of the order of 2--3\%, larger for \piplus than for \piminus, which qualitatively agrees with the results presented here. 
The results by the Jefferson Lab Hall A Collaboration~\cite{Huang:2011bc} using a transversely polarized \(^{3}\)He target, which essentially can be regarded as a neutron target, show a large positive asymmetry for \piminus while the \piplus asymmetry is consistent with zero, also consistent with the model predictions.

\subsection{The subleading-twist \ssa and \dsa amplitudes}\label{interpretation-subleadingtwist}

Four modulations contributing to the cross sections \eqref{theory-fourier} involving transverse target polarization (two of which require in addition longitudinal beam polarization) vanish at twist-2 level and thus involve either twist-3 distribution or fragmentation functions, as detailed in section~\ref{theory-subleadingtwist}. 
As such they offer a way to constrain multi-parton correlations, while on the other hand being expected to be small as formally suppressed by  \(M/Q\).
Interpretation of those modulations is hampered by the multitude of twist-3 functions contributing, often lacking clear guidance from phenomenology.  Wandzura--Wilczek-type approximations~\cite{Bastami:2018xqd} help to reduce the number of terms, but have their own limitations. 
For example, the clearly non-vanishing beam-helicity asymmetry in, e.g., ref.~\cite{Airapetian:2019mov} challenges the Wandzura--Wilczek-type approximation, the latter predicting asymmetries identical to zero.

The results presented below constitute the first measurement of those subleading-twist Fourier amplitudes in semi-inclusive \dis by transversely polarized protons.

\begin{figure}
\centering
\includegraphics[bb = 15 60 500 470, clip, width=0.49\textwidth,keepaspectratio]{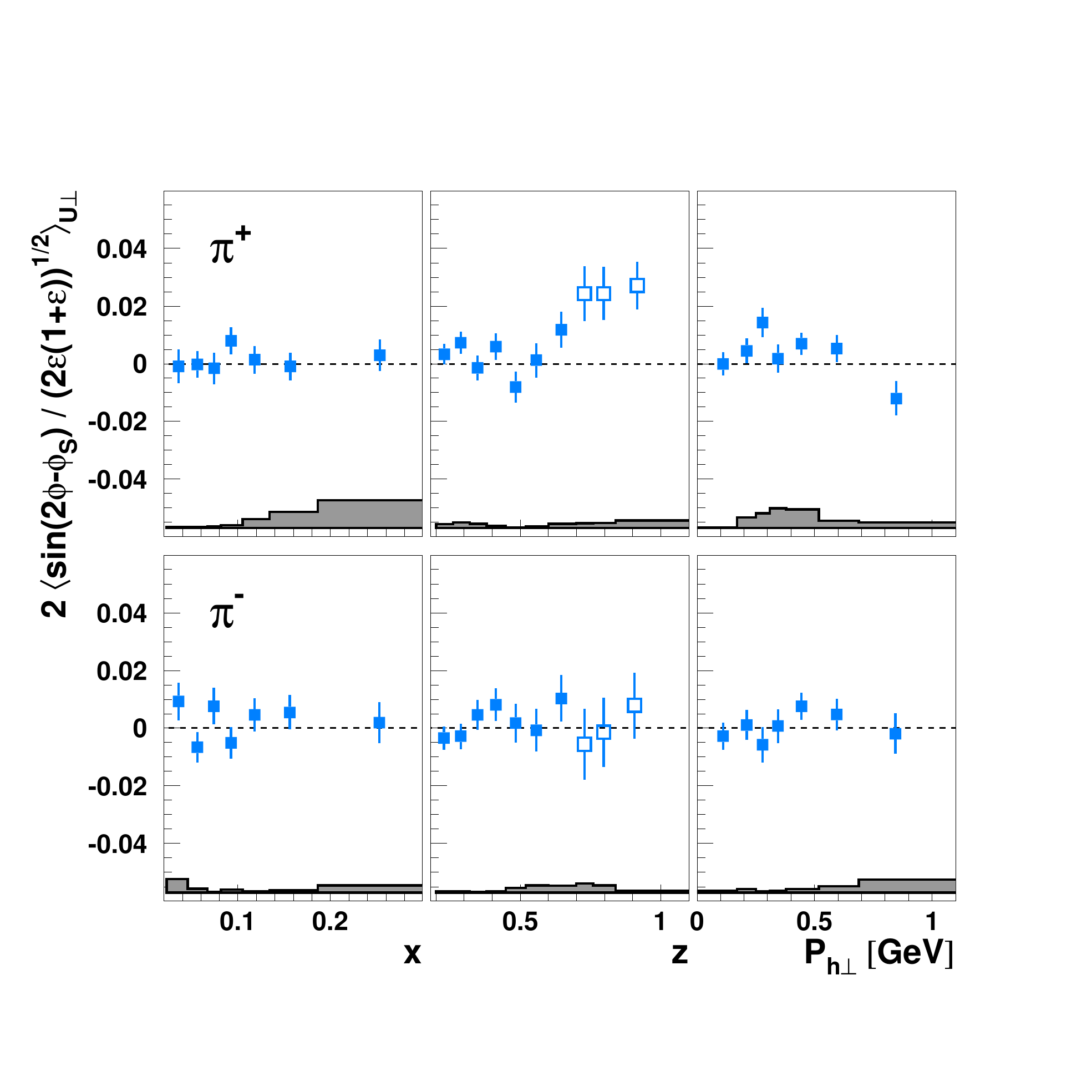}~~%
\includegraphics[bb = 15 60 500 470, clip, width=0.49\textwidth,keepaspectratio]{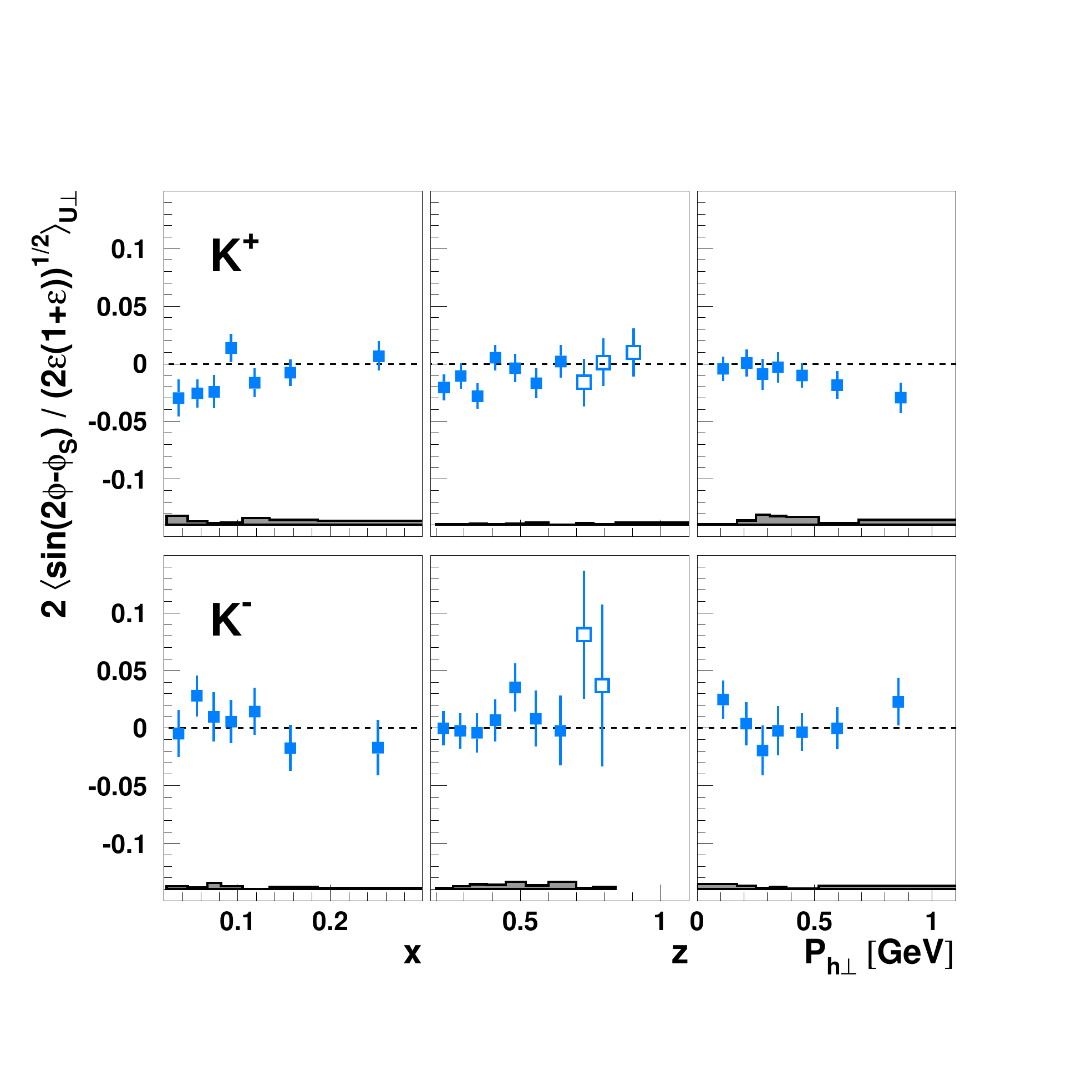}
  \caption{The \sintwophiexpSFA{h}  amplitudes for charged mesons (left: pions; right: kaons) presented either in bins of \xb, \z, or \Phperpabs. Data at large values of \z, marked by open points in the \z projection, are not included in the other projections. Systematic uncertainties are given as bands, not including the additional scale uncertainty of 7.3\% due to the precision of the target-polarization determination.}
 \label{fig:mesons-sin2phi}
\end{figure}
\begin{figure}
\centering
\includegraphics[bb = 15 60 500 490, clip, width=0.49\textwidth,keepaspectratio]{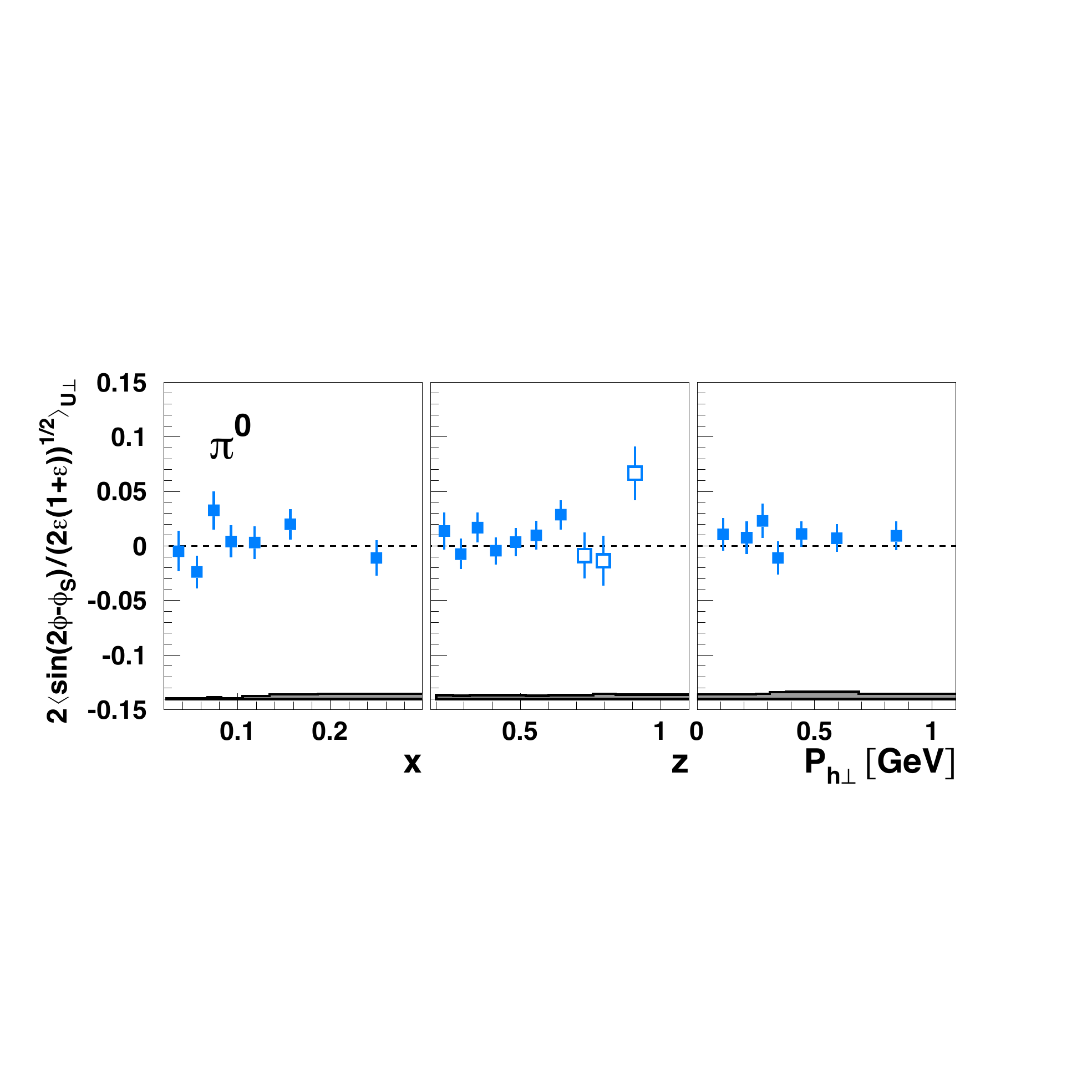}~~%
\includegraphics[bb = 15 60 500 490, clip, width=0.49\textwidth,keepaspectratio]{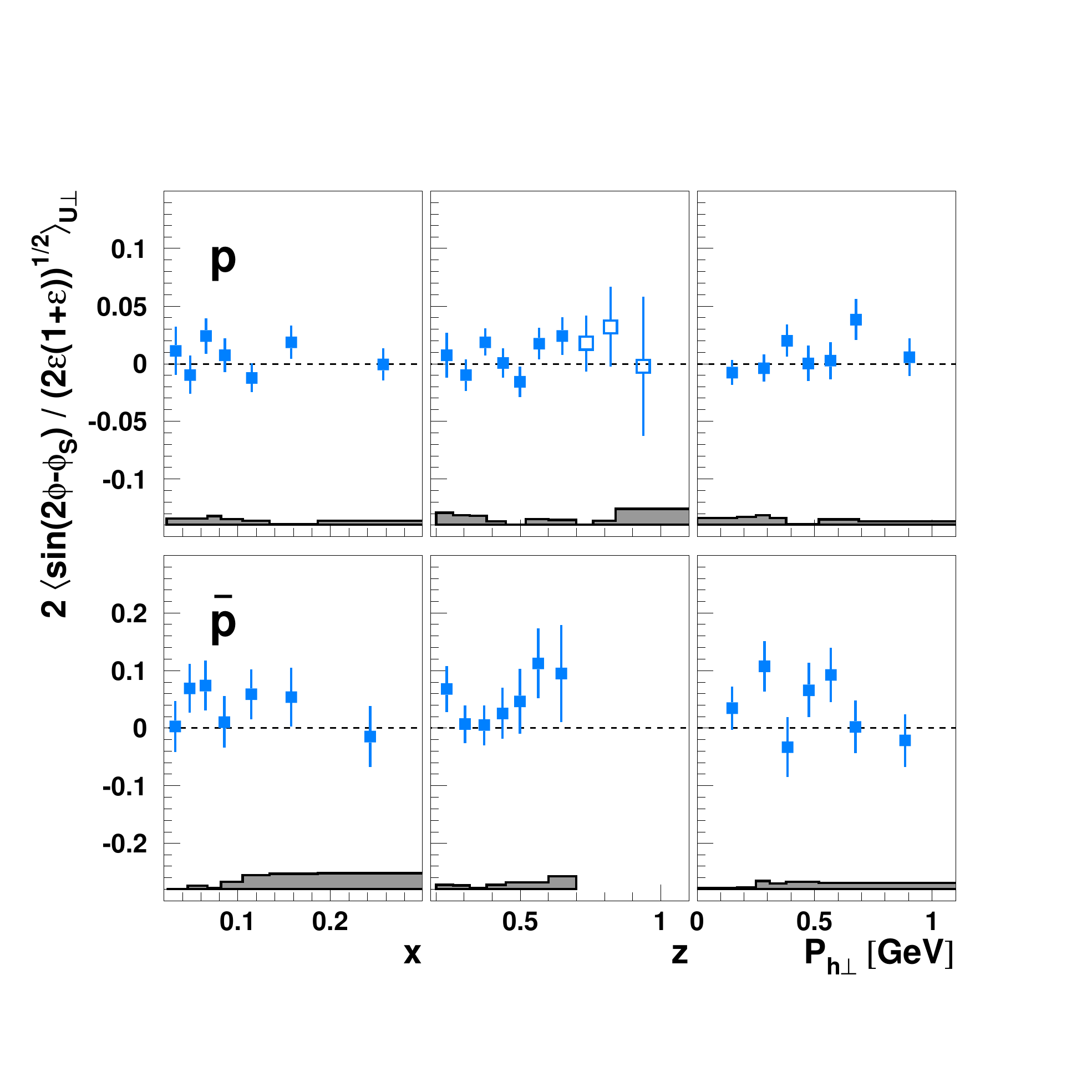}
  \caption{The \sintwophiexpSFA{h}  amplitudes for \pizero (left), protons, and antiprotons (right) presented either in bins of \xb, \z, or \Phperpabs. Data at large values of \z, marked by open points in the \z projection, are not included in the other projections (no such high-\z points are available for antiprotons due to a lack of precision). Systematic uncertainties are given as bands, not including the additional scale uncertainty of 7.3\% due to the precision of the target-polarization determination.}
 \label{fig:pizero-protons-sin2phi}
\end{figure}

The \sintwophiexpSFA{h} Fourier amplitudes are found to be mostly consistent with zero as shown in
figures \ref{fig:mesons-sin2phi} and \ref{fig:pizero-protons-sin2phi}. 
Within the semi-inclusive DIS kinematic range of the measurement, 
they are consistent with zero at 95\% confidence level for all hadrons and only at 90\% confidence level inconsistent with zero for antiprotons (cf.~table~\ref{tab:significant-modulations}).

Besides the suppression from being a twist-3 observable, the Fourier amplitude of the \sinemodulation{2\phih-\phis} modulation 
is subject to a \Phperpabs suppression arising through the transverse-momentum convolution. 
This is similar to what was discussed for pretzelosity in section~\ref{interpretation-pretzelosity}. 
However, in comparison to, e.g., the Collins and Sivers modulations, it is only one additional power of \Phperpabs and not two.
Looking at the \kplus results, which hint a slightly negative modulation at low \x, non-vanishing asymmetries are indeed only visible at large \Phperpabs, where such \Phperpabs suppression should die out.

Small asymmetries on the sub-percent level consistent with these data are predicted for pions in Wandzura--Wilczek-type approximations~\cite{Bastami:2018xqd}, 
in which only the terms involving the twist-3 TMDs \twistftperp{q}, \twistht{q}, and \twisthtperp{q} give contributions. 
Similarly, a calculation based on a spectator-diquark model for those three TMDs results again in only a small \sinemodulation{2\phih-\phis} modulation consistent with the measurement presented here~\cite{Mao:2014aoa}.

In the semi-exclusive region of \(\z>0.7\), a positive \sintwophiexpSFA{} Fourier amplitude on the order of 0.02 is extracted for positive pions. In general, the interpretation of asymmetries in this region in terms of TMDs is to be taken with caution; still, an attempt is provided below. 
From the various terms contributing to the related structure function 
in eq.~\eqref{eq:sintwophi}, three are increasingly suppressed with increasing \z.
The very first term reduces in the Wandzura--Wilczek-type approximation~\cite{Bastami:2018xqd} to the Sivers effect, albeit with the opposite sign compared to the leading-twist Sivers asymmetry. 
The measured Sivers asymmetries are indeed large at high \z. However, as they exhibit the same sign as the \sintwophiexpSFA{} Fourier amplitudes,
either the Wandzura--Wilczek-type approximation predicts the wrong sign (and thus appears to fail) or the positive \sintwophiexpSFA{} Fourier amplitudes in the high-\z region need to be attributed to other contributions.
A possibility could be the second contribution in eq.~\eqref{eq:sintwophi} that is not formally suppressed for large values of \z: the combined contribution of \( \twistht{q} + \twisthtperp{q} \) coupled to the Collins fragmentation function. In the Wandzura--Wilczek-type approximation it is related to pretzelosity, but generally found to be very small~\cite{Bastami:2018xqd,Mao:2014aoa}.
There is some similarity of the large-\z behavior of the \sintwophiexpSFA{\piplus} to that of the \sintwophilexpSFA{\piplus} Fourier amplitude discussed further below (cf.~section \ref{sec:ALU_ALL}).
As both modulations receive the same cross-section contribution from the longitudinal target-polarization component, the source 
for the non-vanishing asymmetries at large \z might indeed stem from a \sintwophiul{\piplus} Fourier amplitude of the longitudinal \ssa.
Unfortunately, not much is known about the latter amplitude in the kinematic regime of this measurement. \hermes data for the related \sintwophiulexp{h} Fourier amplitude for charged pions~\cite{Airapetian:1999tv} are consistent with zero when integrated over the semi-inclusive \z range of \(0.2<\z<0.7\), without presenting data binned in \z or for \(\z>0.7\). 
Likewise, preliminary \compass data, both for the semi-inclusive \z region and for large \z, do not exhibit a sizable \sintwophiulexp{h} asymmetry~\cite{Parsamyan:2018ovx}.
Only the \clas collaboration reported non-vanishing \sintwophiulexp{h} asymmetry amplitudes for charged pions~\cite{Avakian:2010ae}, 
however, not for the \(\z>0.7\) range considered here. 
In contrast to the earlier \hermes measurement of \sintwophiulexp{h}, 
the \clas data are on average at larger \z since they are integrated over the range \(0.4<\z<0.7\).
Thus, the non-zero \clas data might be a hint of an increase in magnitude of these asymmetry amplitudes with increasing \z. 
On the other hand, the negative values of these asymmetry amplitudes are not compatible with the positive \sintwophiexpSFA{\piplus} amplitudes presented here.
Last but not least, positive \sinemodulation{2\phih-\phis} modulations have been observed in {\em exclusive} \piplus electroproduction off transversely polarized protons~\cite{Airapetian:2009ac}, 
which suggests a smooth transition from the semi-exclusive high-\z region studied here to exclusive \piplus production.

\begin{figure}
\centering
\includegraphics[bb = 15 60 500 470, clip, width=0.49\textwidth,keepaspectratio]{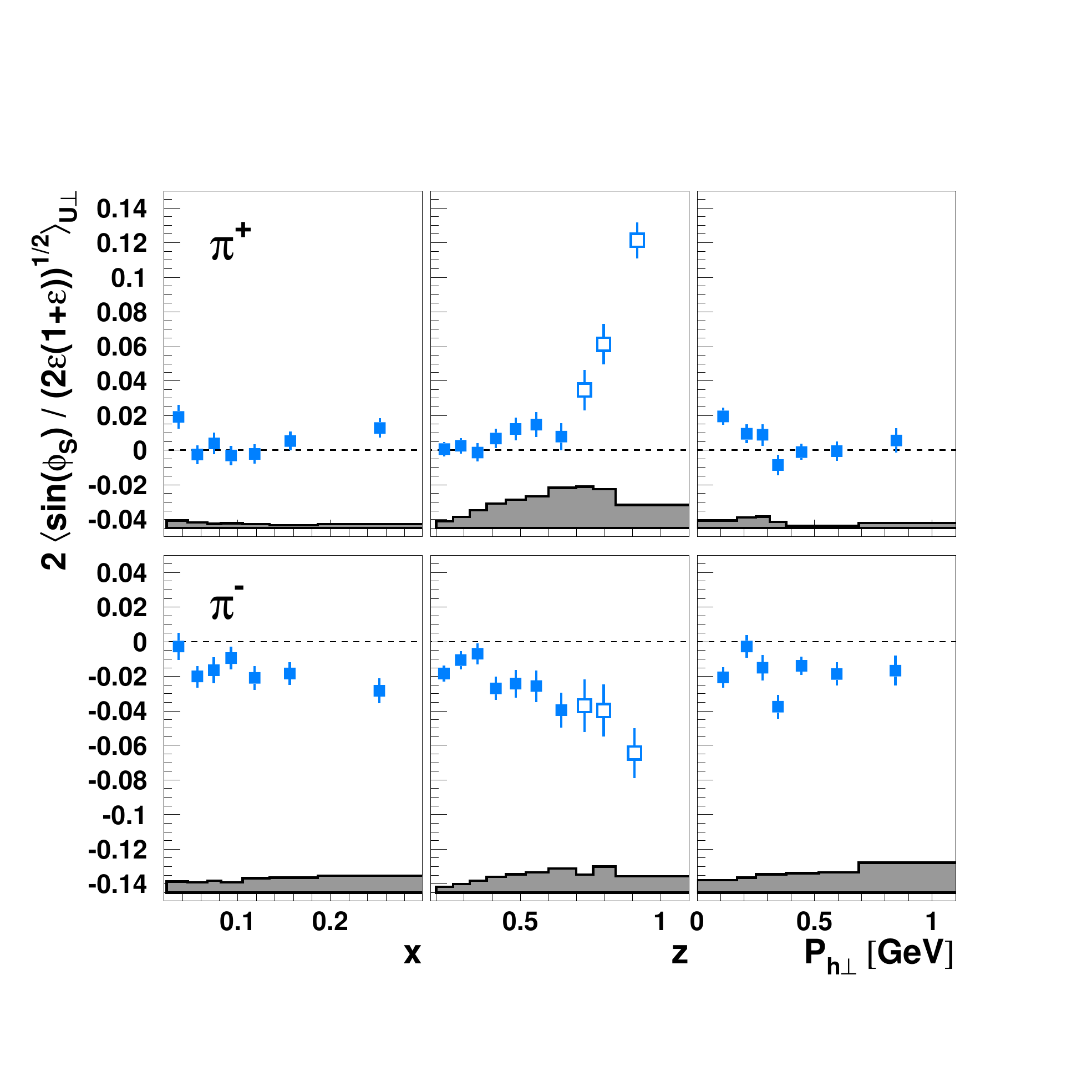}~~%
\includegraphics[bb = 15 60 500 470, clip, width=0.49\textwidth,keepaspectratio]{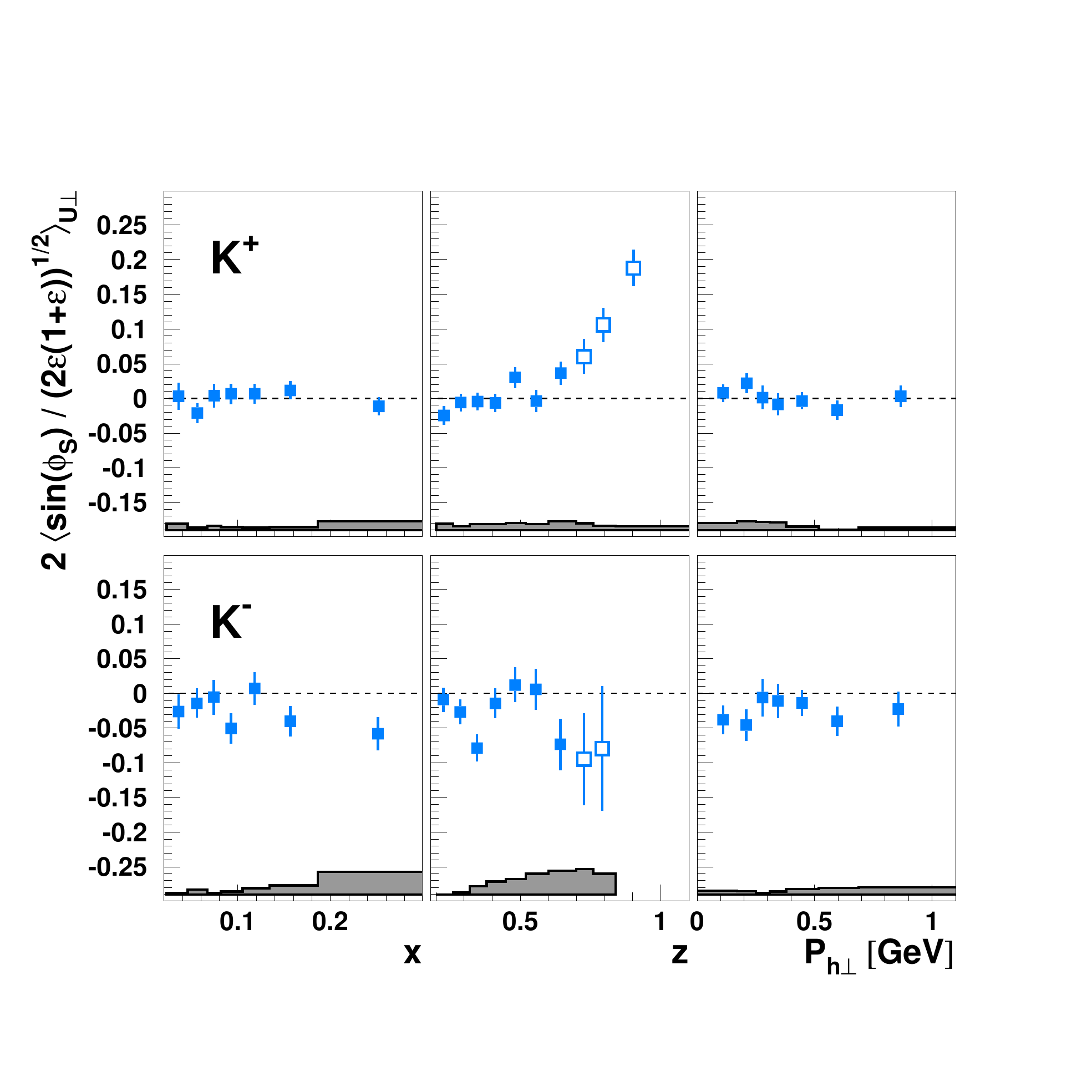}
  \caption{The \sinphisexpSFA{h}  amplitudes for charged mesons (left: pions; right: kaons) presented either in bins of \xb, \z, or \Phperpabs. Data at large values of \z, marked by open points in the \z projection, are not included in the other projections. Systematic uncertainties are given as bands, not including the additional scale uncertainty of 7.3\% due to the precision of the target-polarization determination.}
 \label{fig:mesons-sinphis}
\end{figure}

\begin{figure}
\centering
\includegraphics[bb = 27 18 515 543, clip, width=0.7\textwidth,keepaspectratio]{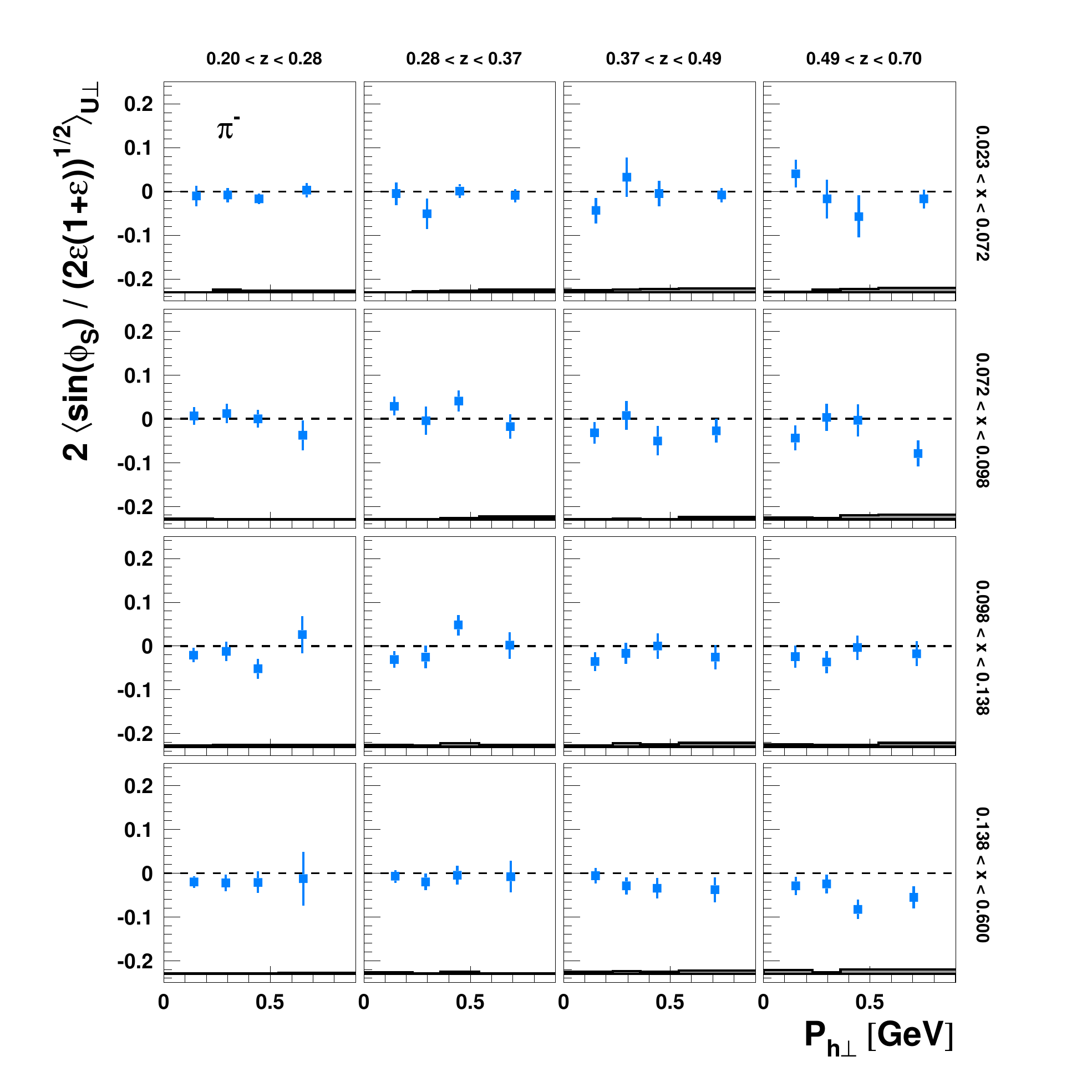}
  \caption{The \sinphisexpSFA{} Fourier amplitudes for \piminus extracted simultaneously in bins of \xb, \z, and \Phperpabs, presented as a function of \Phperpabs. Systematic uncertainties are given as bands, not including the additional scale uncertainty of 7.3\% due to the precision of the target-polarization determination.}
 \label{fig:pim-sinphis3d}
\end{figure}

One of the more striking results of this analysis is the observation of large subleading-twist \sinphisexpSFA{h} Fourier amplitudes.
In particular, they provide the largest twist-3 signal in this measurement.
They surprise also with a large kinematic dependence as visible in figure~\ref{fig:mesons-sinphis}, where they are shown for charged mesons.
In the semi-inclusive \dis region, mainly the Fourier amplitudes for negative mesons are significantly different from zero, being of order -0.02.
The three-dimensional binning, depicted in figure~\ref{fig:pim-sinphis3d} for the \piminus, 
reveals that those non-vanishing asymmetries stem predominantly from the large-\x and large-\z region, where they reach even larger magnitudes.
The amplitudes clearly rise with \z for charged pions and positive kaons.
The precision for \kminus and neutral pions in that region is insufficient for drawing a 
strong conclusion, though also here an increase in magnitude with \z is hinted.
A noteworthy characteristic of the results is the clearly opposite sign for the \piminus results compared to both \piplus and \kplus, 
reminiscent of what is observed for the Collins asymmetries.

\begin{figure}
\centering
\includegraphics[width=0.5\textwidth,keepaspectratio]{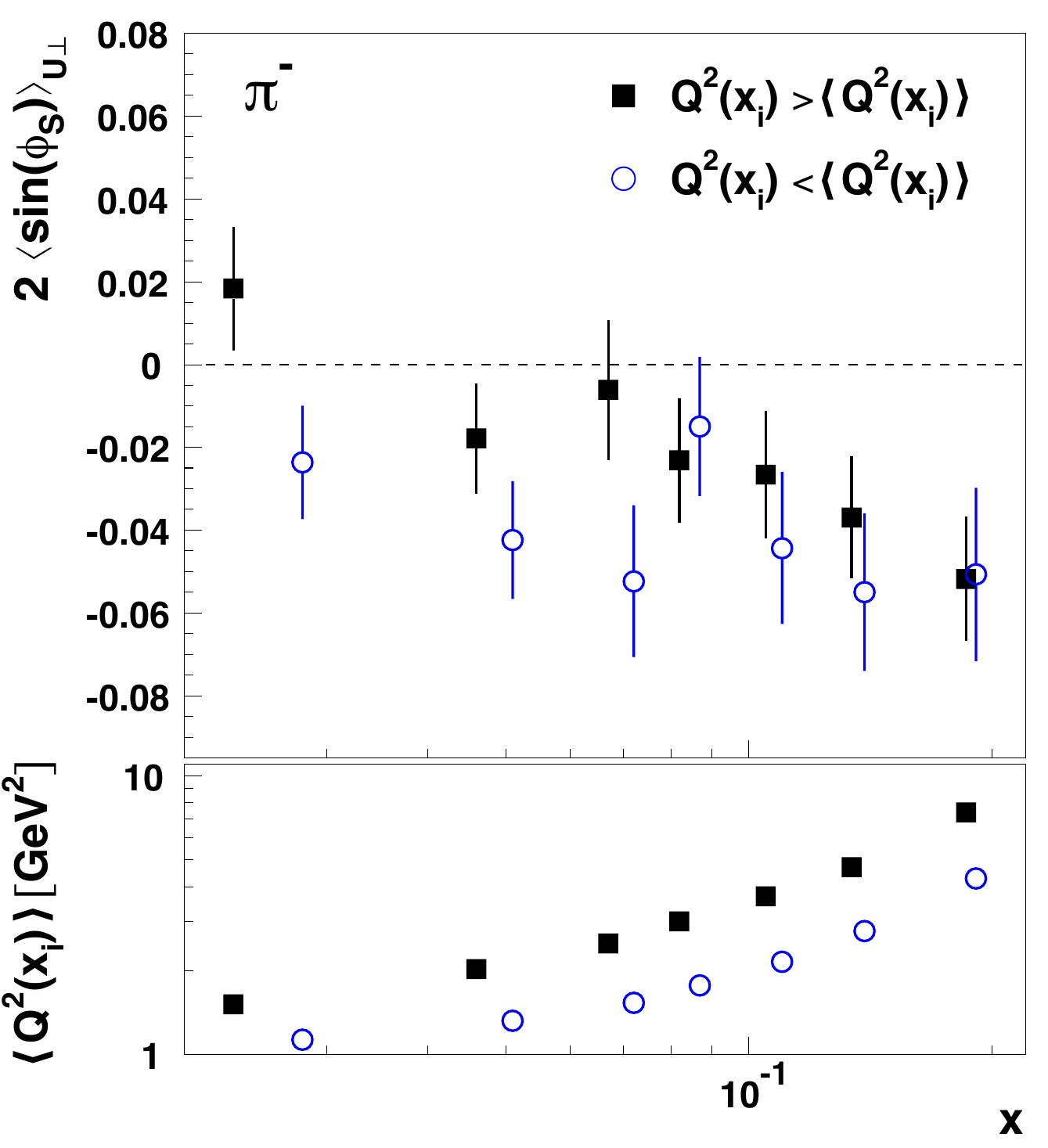}
  \caption{The \sinphisexp{h}  \csa amplitudes for \piminus as a function of \xb. The \Q region for each bin was divided into the two regions above (squares) and below (circles) the average \Q of that bin. The average \Q is given in the bottom for all bins separately for the two \Q regions. The error bars represent statistical uncertainties only.}
 \label{fig:pim-sinphislba-q2}
\end{figure}

The Fourier amplitudes of the \sinemodulation{\phis} modulations 
are related to subleading-twist cross-section contributions (cf.~eq.~\eqref{theory-sinphis}).
As such it is interesting to explore the \Q dependence of this azimuthal asymmetry. Because \xb and \Q are highly correlated, a one-dimensional binning in \Q mixes effects from the twist-3 suppression with the inherent \xb dependence of the asymmetry. Therefore, an approach employed already in the previous \hermes Collins and Sivers publications~\cite{Airapetian:2010ds,Airapetian:2009ae} has been adopted here that splits each \xb bin into the two regions of \Q: below and above the average \Q of each \xb bin.
The resulting \piminus \csa amplitudes are shown in figure~\ref{fig:pim-sinphislba-q2}. A hint of a suppression is visible for the regions of larger \Q, though not very pronounced, which might be a consequence of the relatively small lever arm in \Q as apparent from the difference in average \Q for the two regions, plotted in the bottom panel of the figure.

The structure function \structuresin{UT}{\phis} is of particular
interest as it is the only contribution to the cross section
\sigmaut{h}{T} that survives integration over transverse hadron
momentum:
\begin{equation}
 \structuresin{UT}{\phis}\left(\x,\Q,\z\right) = 
 \int \! d^2 \Phperp \, \structuresin{UT}{\phis}\left(\x,\Q,\z,\Phperpabs\right) = 
 -x \frac{2M_h}{Q} \sum\limits_q e_q^2 \, \pdfh{q} \frac{\tilde{H}^{\,q}\left(z\right)}{z}.
\label{eq:sinphisint}
\end{equation}
It thus provides, in principle, sensitivity to the transversity distribution without
involving a convolution over intrinsic transverse momenta~\cite{Mulders:1995dh}. 
In addition, the modulation does not necessarily have to vanish in the limit of \Phperpabs going to zero. 
Another rather interesting aspect of the \sinemodulation{\phis} modulation --- as pointed out already in section~\ref{theory-subleadingtwist} --- is the fact 
that the inclusive analogue, i.e., summing over all final-state hadrons and integrating over their four-momenta, must vanish in the one-photon-exchange approximation, which was tested at \hermes to the \(10^{-3}\) level~\cite{Airapetian:2009ab}.

A serious experimental drawback in using the relation \eqref{eq:sinphisint} to extract transversity could be the
systematic effect arising from the usually incomplete integration
over \Phperp due to limitations in the geometric acceptance or kinematic requirements in experiments.
Furthermore, a current drawback of such measurement is the lack of knowledge about the
interaction-dependent fragmentation function \(\tilde{H}^{\,q}\left(z\right)\). 
However, it has been shown that the latter, the Collins fragmentation function, as well as the collinear twist-3 fragmentation function 
that is suspected to cause the transverse-spin asymmetries in inclusive pion production in single-polarized proton-proton collisions
are related~\cite{Gamberg:2017gle}. This may explain the similar qualitative behavior of the Collins asymmetries and of
the \sinphisexp{\pi} Fourier amplitudes.

\begin{figure}
\centering
\includegraphics[bb = 15 60 500 470, clip, width=0.49\textwidth,keepaspectratio]{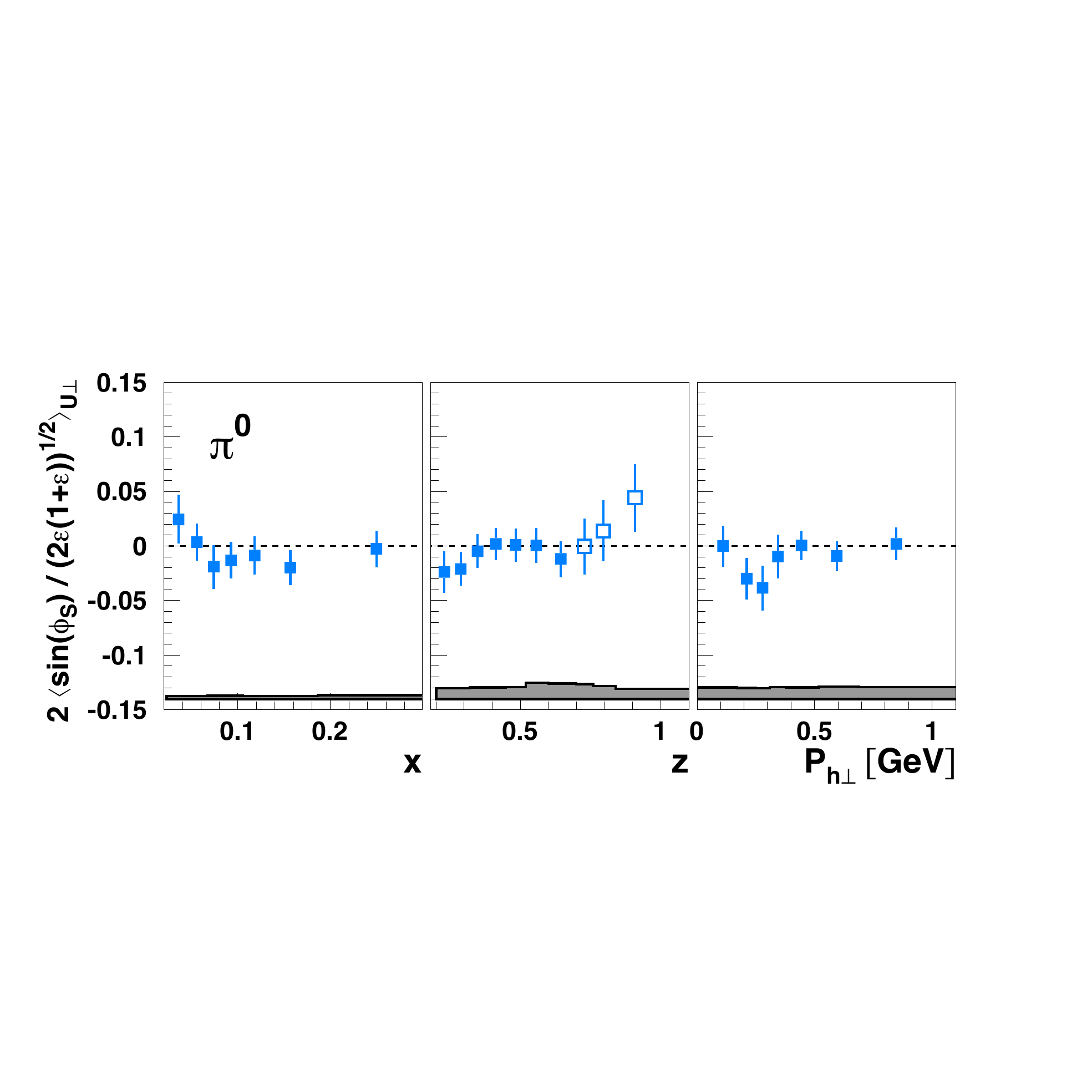}~~%
\includegraphics[bb = 15 60 500 470, clip, width=0.49\textwidth,keepaspectratio]{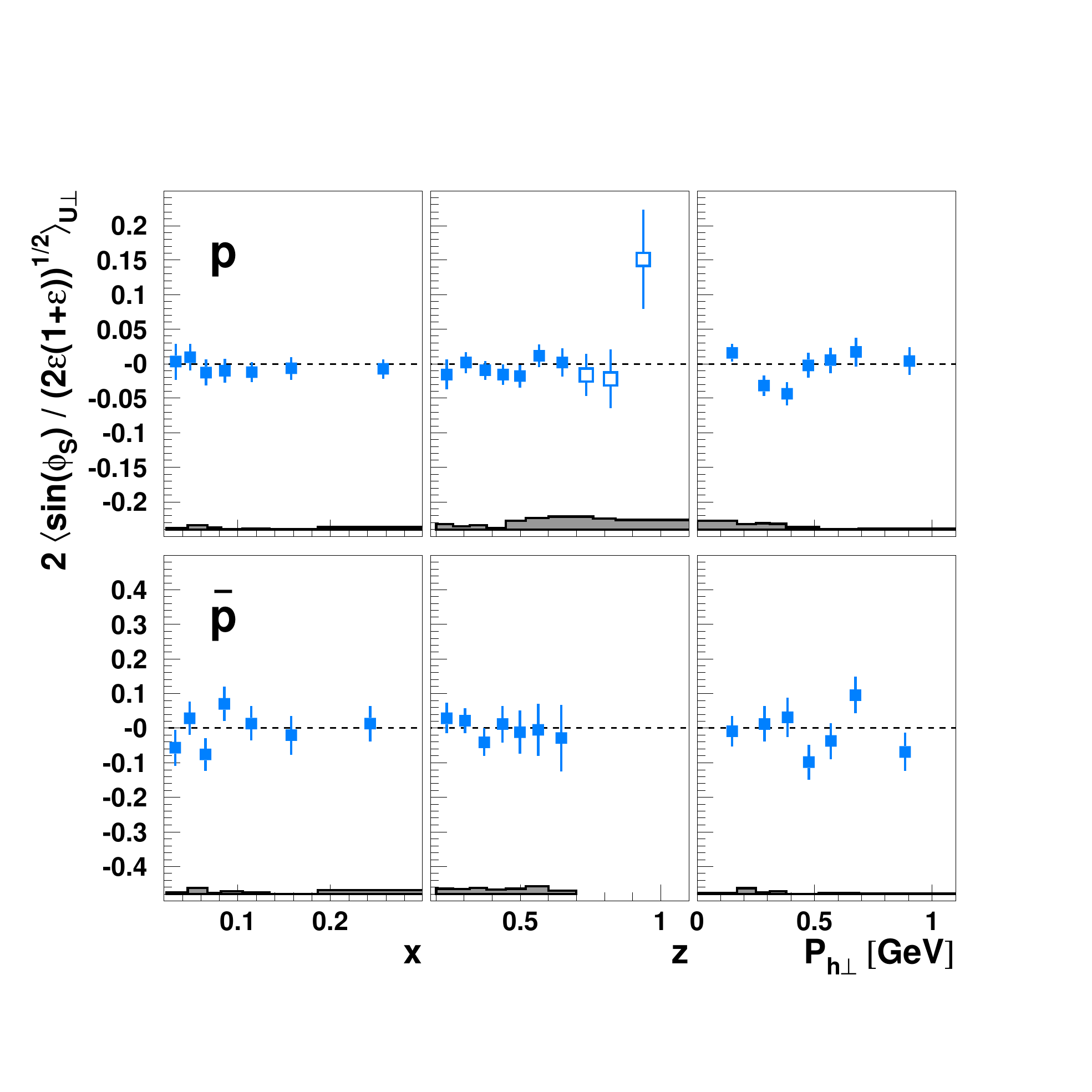}
  \caption{The \sinphisexpSFA{h}  amplitudes for \pizero (left), protons, and antiprotons (right) presented either in bins of \xb, \z, or \Phperpabs. Data at large values of \z, marked by open points in the \z projection, are not included in the other projections (no such high-\z points are available for antiprotons due to a lack of precision). Systematic uncertainties are given as bands, not including the additional scale uncertainty of 7.3\% due to the precision of the target-polarization determination.}
 \label{fig:pizero-protons-sinphis}
\end{figure}

The relation to the Collins effect might also explain why the results for protons and antiprotons are consistent with zero, as shown in figure~\ref{fig:pizero-protons-sinphis} (where also the vanishing signal for \pizero is presented). 
As novel spin-dependent fragmentation is involved, it is reasonable to expect a fundamental difference for production of spin-0 versus spin-\(\frac{1}{2}\) hadrons.

The vanishing effect for protons and the negative asymmetry for \piminus also disfavor a sizable contribution of  \( \twistft{q} \ffdmod{q\to h} \) in eq.~\eqref{theory-sinphis} --- which can be related to the Sivers effect in the Wandzura--Wilczek-type approximation --- being in conflict with the behavior of the Sivers asymmetry for those hadrons. Furthermore, \twistft{q} has to fulfill the sum rule \( \int  \text{d}^2 \pt \, \twistft{q} (\xb, \pt^{2}) = 0 \), which poses a problem when using currently available parameterization for the Sivers function in the Wandzura--Wilczek-type approximation for \twistft{q} because they are violating the sum rule. 
For that reason, it is not further considered here. 

Staying within the Wandzura--Wilczek-type approximation, from the remaining terms in eq.~\eqref{theory-sinphis} contributing to the \sinphis{\pi} Fourier amplitude only the ones involving the Collins fragmentation function survive. The combined contribution involves \(-\xb (\twistht{q} - \twisthtperp{q}) \stackrel{\text{WW}}{=}  \pdfh{q} \) and thus the product of transversity and the Collins fragmentation function. As in the above discussion of the \(\tilde{H}^{\,q}\left(\z\right)\) contribution, this might explain the qualitative similarity of the charged-pion Collins and  \sinphisexp{\pi} Fourier amplitudes.

In contrast to the \sintwophiexp{h} Fourier amplitude, there is no additional term contributing through the longitudinal target-polarization component. As a consequence, the \sinphis{h} and \sinphisexp{h} Fourier amplitudes differ only by the factor \(\cos\theta_{\gamma^*}\simeq 1\) in the kinematic region of this measurement (cf.~appendix~\ref{sec-app-longpol}).

While disentangling all the different contributions to the \sinemodulation{\phis} modulation will clearly require further detailed studies, the marked increase in magnitude of those modulations for charged pions and \kplus with \z in the semi-inclusive region is especially intriguing. In that respect, it appears worthwhile to point out that very sizable \sinemodulation{\phis} modulations were observed in exclusive \piplus electroproduction off transversely polarized protons~\cite{Airapetian:2009ac}.

The remaining two twist-3 Fourier amplitudes, the \cosinemodulation{2\phih-\phis} and  \cosinemodulation{\phis} modulations, require longitudinally polarized leptons in addition to transverse target polarization. As such, their statistical precision suffers from the relatively small lepton-beam polarization in these data.
Again, several (and partially similar) terms contribute to those Fourier amplitudes as can be seen from eqs.~\eqref{theory-cos2phi-phis} and \eqref{eq:theory-cosphis}, making {\em a priori} the interpretation in terms of specific TMDs difficult. Also in this case, Wandzura--Wilczek-type approximations might help to focus on only a few of the terms.

\begin{figure}
\centering
\includegraphics[bb = 15 60 500 470, clip, width=0.49\textwidth,keepaspectratio]{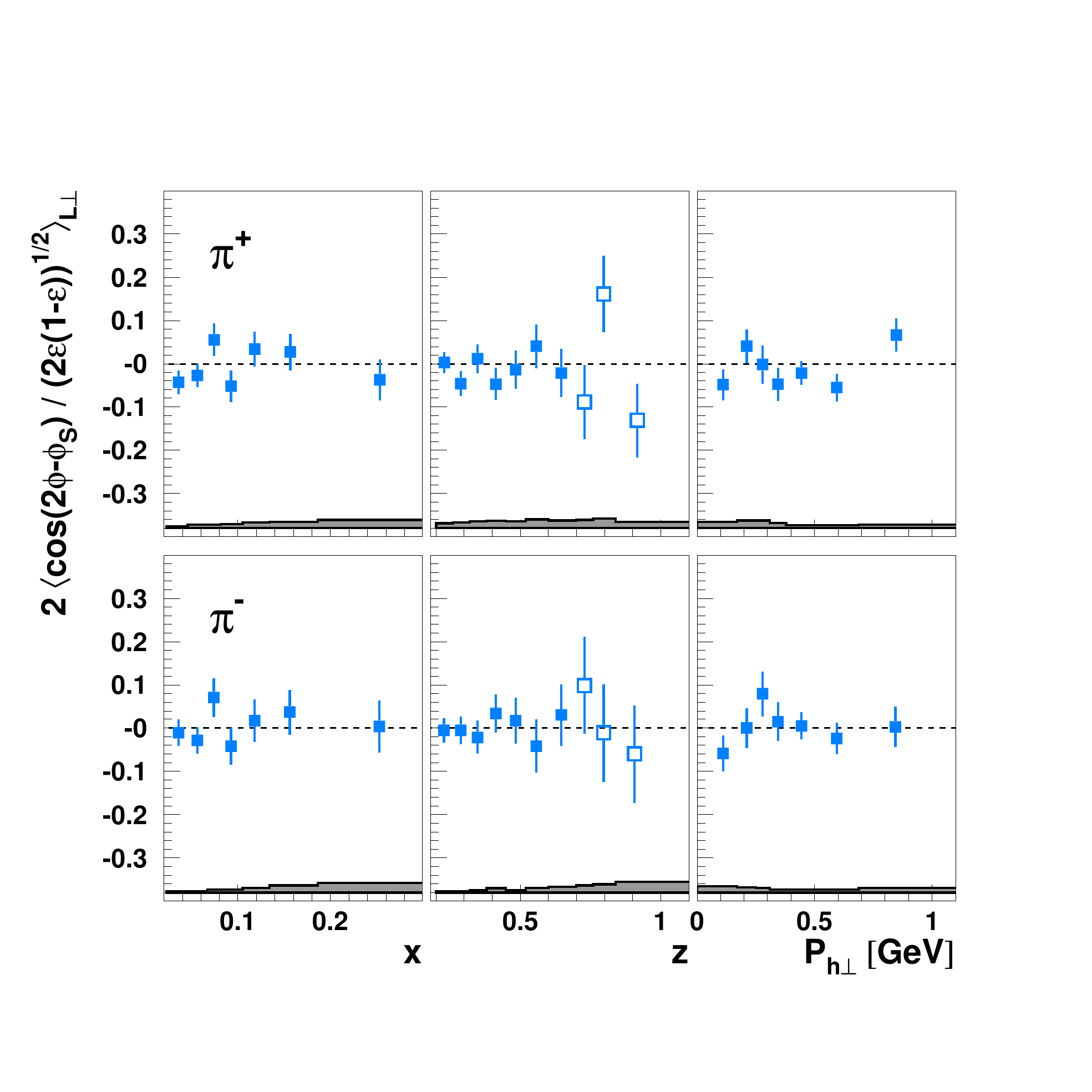}~~%
\includegraphics[bb = 15 60 500 470, clip, width=0.49\textwidth,keepaspectratio]{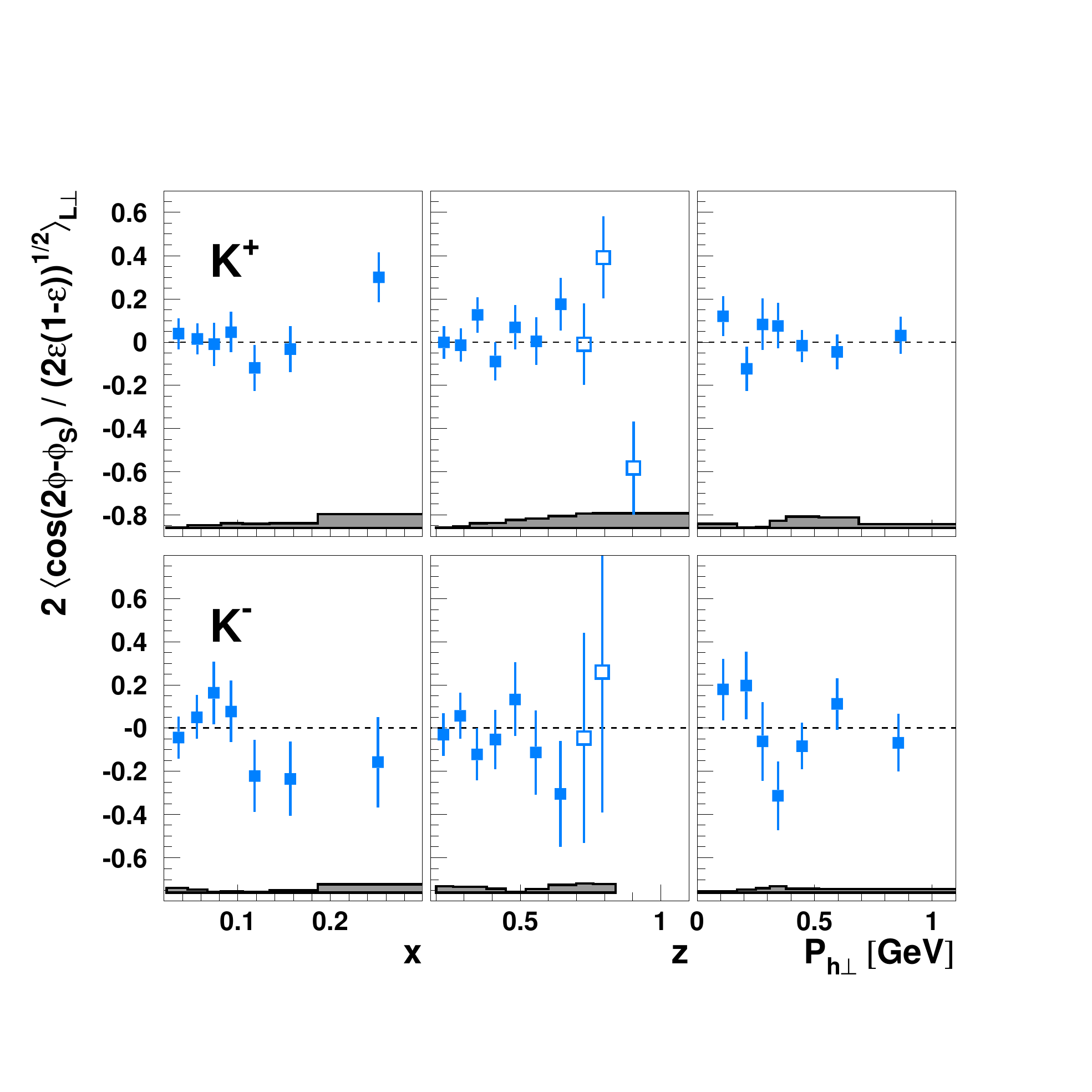}
  \caption{The \costwophiexpSFA{h}  amplitudes for charged mesons (left: pions; right: kaons) presented either in bins of \xb, \z, or \Phperpabs. Data at large values of \z, marked by open points in the \z projection, are not included in the other projections. Systematic uncertainties are given as bands, not including the additional scale uncertainty of 8.0\% due to the precision of the target-polarization determination.}
 \label{fig:mesons-cos2phimphis}
\end{figure}
\begin{figure}
\centering
\includegraphics[bb = 15 60 500 490, clip, width=0.49\textwidth,keepaspectratio]{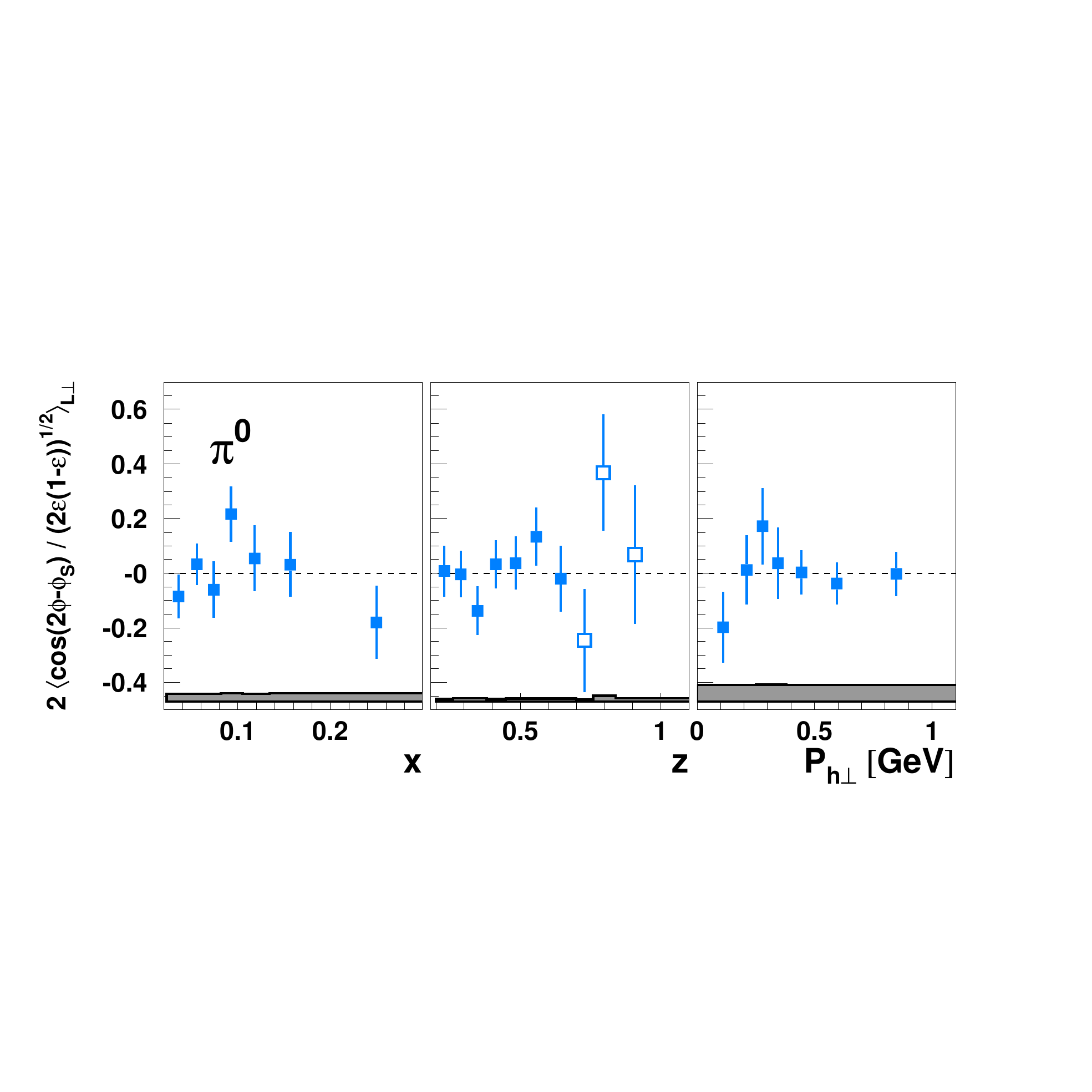}~~%
\includegraphics[bb = 15 60 500 490, clip, width=0.49\textwidth,keepaspectratio]{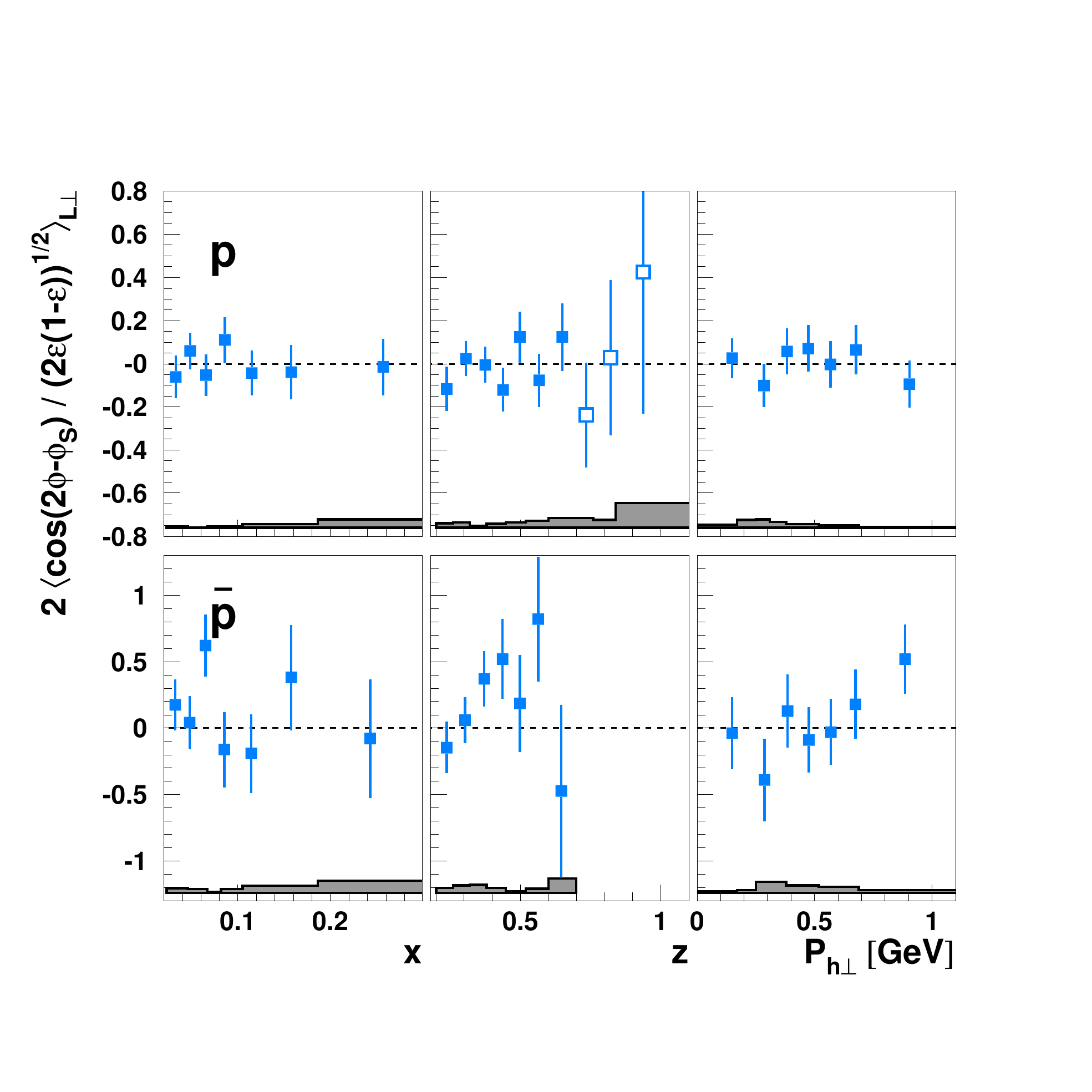}
  \caption{The \costwophiexpSFA{h}  amplitudes for \pizero (left), protons, and antiprotons (right) presented either in bins of \xb, \z, or \Phperpabs. Data at large values of \z, marked by open points in the \z projection, are not included in the other projections (no such high-\z points are available for antiprotons due to a lack of precision). Systematic uncertainties are given as bands, not including the additional scale uncertainty of 8.0\% due to the precision of the target-polarization determination.}
 \label{fig:pizero-protons-cos2phimphis}
\end{figure}

The \costwophiexpSFA{h} Fourier amplitudes for pions, charged kaons, and (anti)protons are presented in figures~\ref{fig:mesons-cos2phimphis} and \ref{fig:pizero-protons-cos2phimphis}. None of those are found to be significantly different from zero. This is consistent with expectations~\cite{Bastami:2018xqd} using Wandzura--Wilczek-type approximations of sub-percent level asymmetries. In such an approximation, only a term proportional to the worm-gear (II) \tmdgt{q} and the ordinary \ffdmod{q\rightarrow h}  fragmentation function survives. 

As in the case of the \sinphisexp{h} Fourier amplitude, there is no contribution to the \costwophiexp{h} Fourier amplitude from the longitudinal target-polarization component. 
Therefore, the \costwophiexp{h}  and \costwophi{h} Fourier amplitudes differ only by the factor \(\cos\theta_{\gamma^*}\simeq 1\) in the kinematic region of this measurement (cf.~appendix~\ref{sec-app-longpol}).

\begin{figure}
\centering
\includegraphics[bb = 15 60 500 470, clip, width=0.49\textwidth,keepaspectratio]{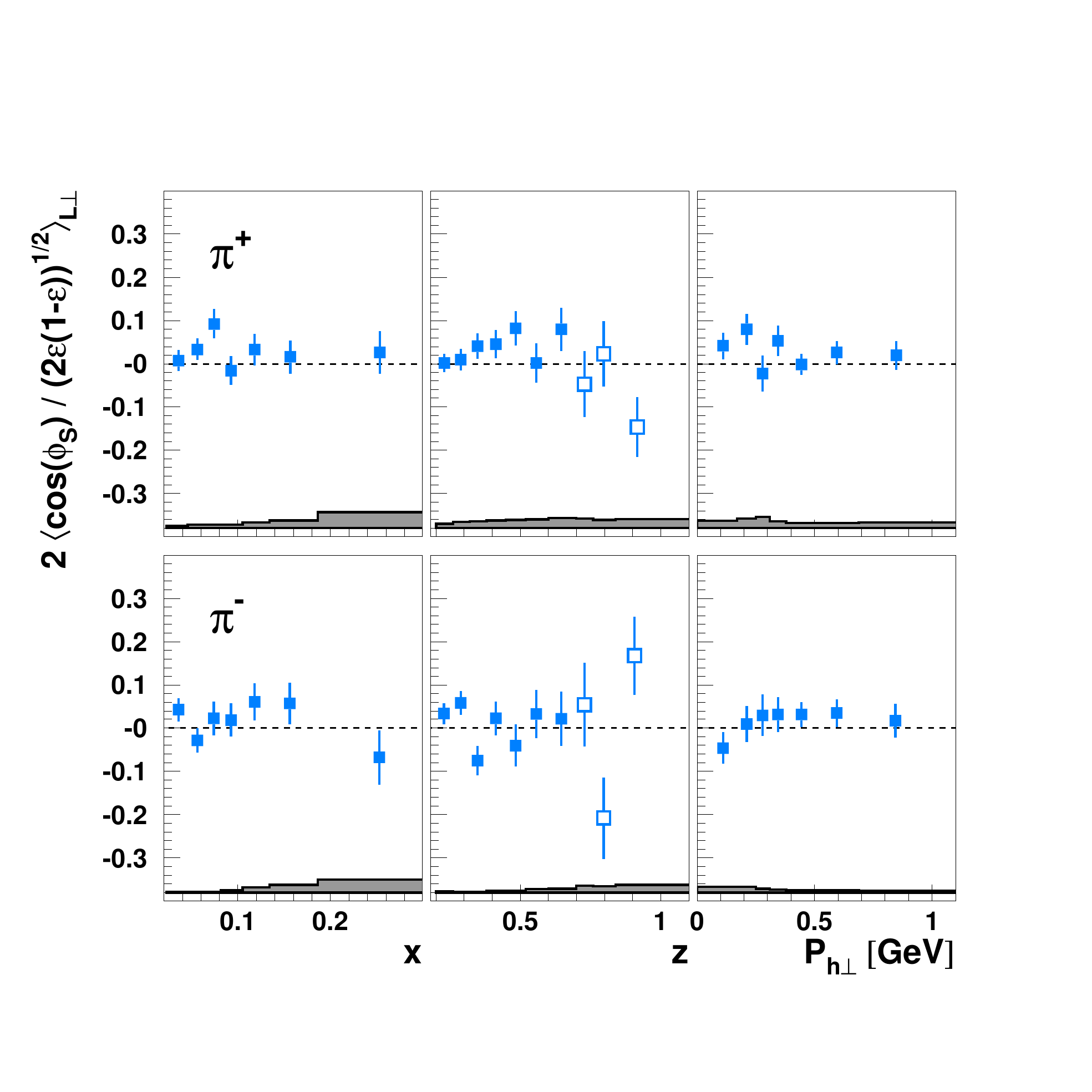}~~%
\includegraphics[bb = 15 60 500 470, clip, width=0.49\textwidth,keepaspectratio]{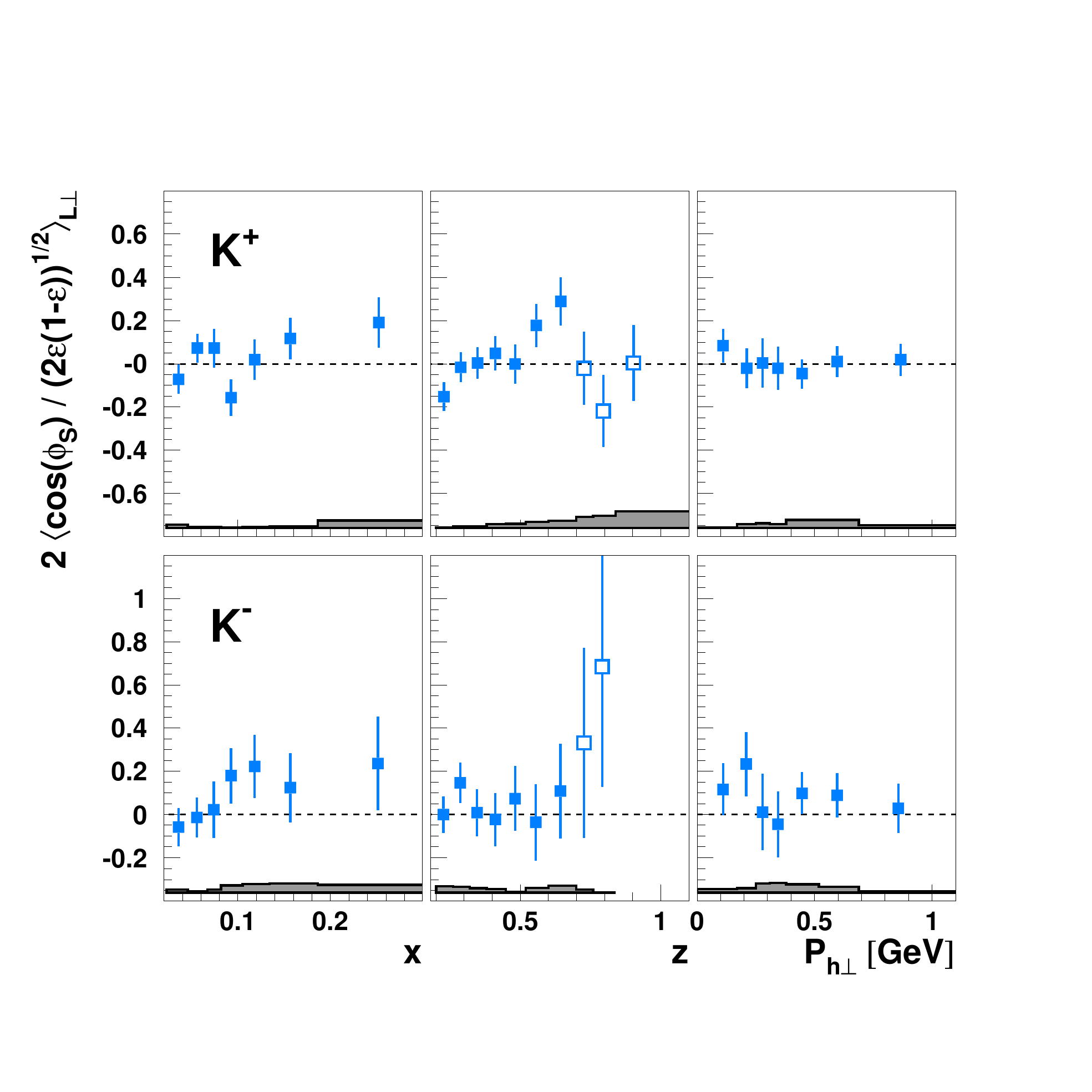}
  \caption{The \cosphisexpSFA{h}  amplitudes for charged mesons (left: pions; right: kaons) presented either in bins of \xb, \z, or \Phperpabs. Data at large values of \z, marked by open points in the \z projection, are not included in the other projections. Systematic uncertainties are given as bands, not including the additional scale uncertainty of 8.0\% due to the precision of the target-polarization determination.}
 \label{fig:mesons-cosphis}
\end{figure}

\begin{figure}
\centering
\includegraphics[bb = 15 60 500 490, clip, width=0.49\textwidth,keepaspectratio]{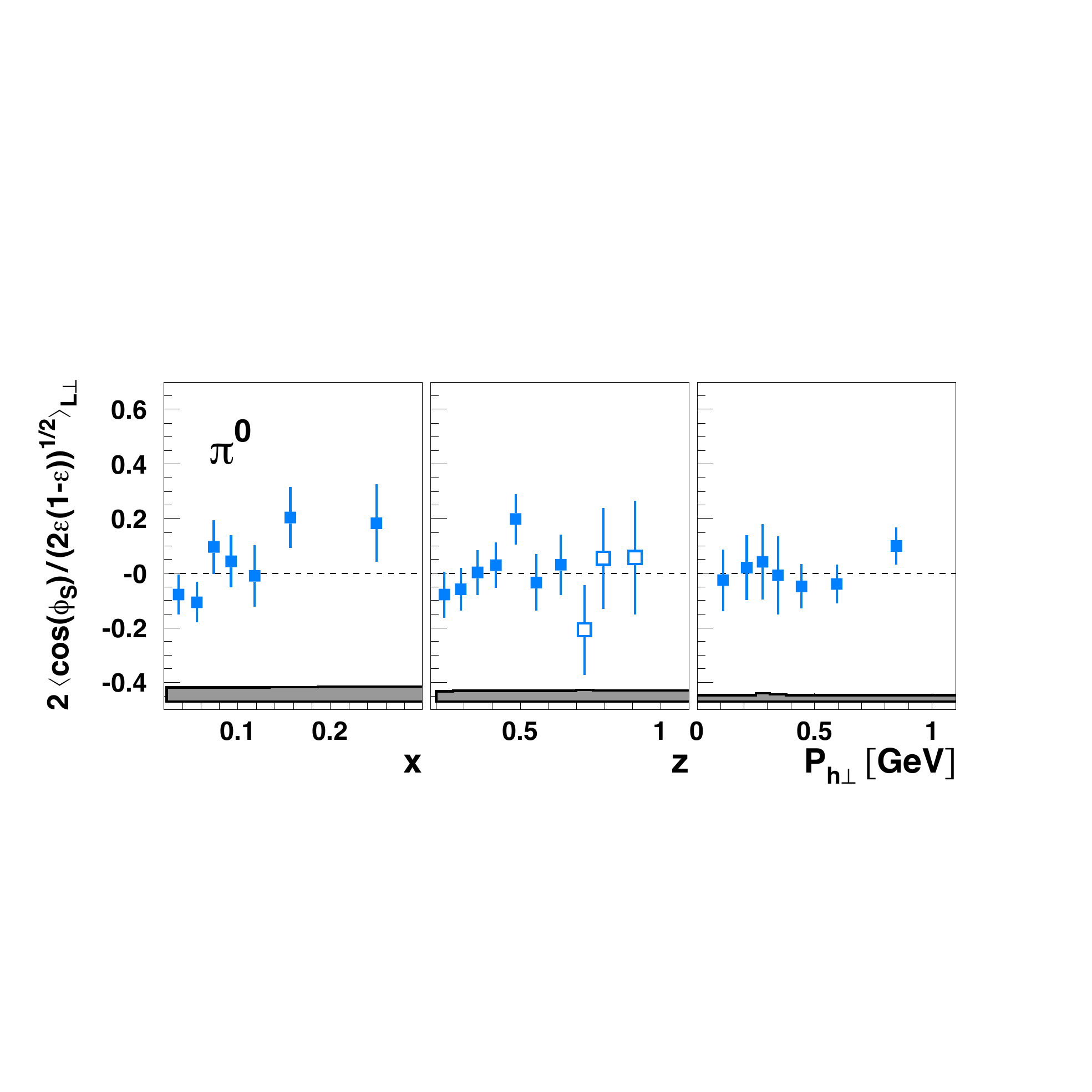}~~%
\includegraphics[bb = 15 60 500 490, clip, width=0.49\textwidth,keepaspectratio]{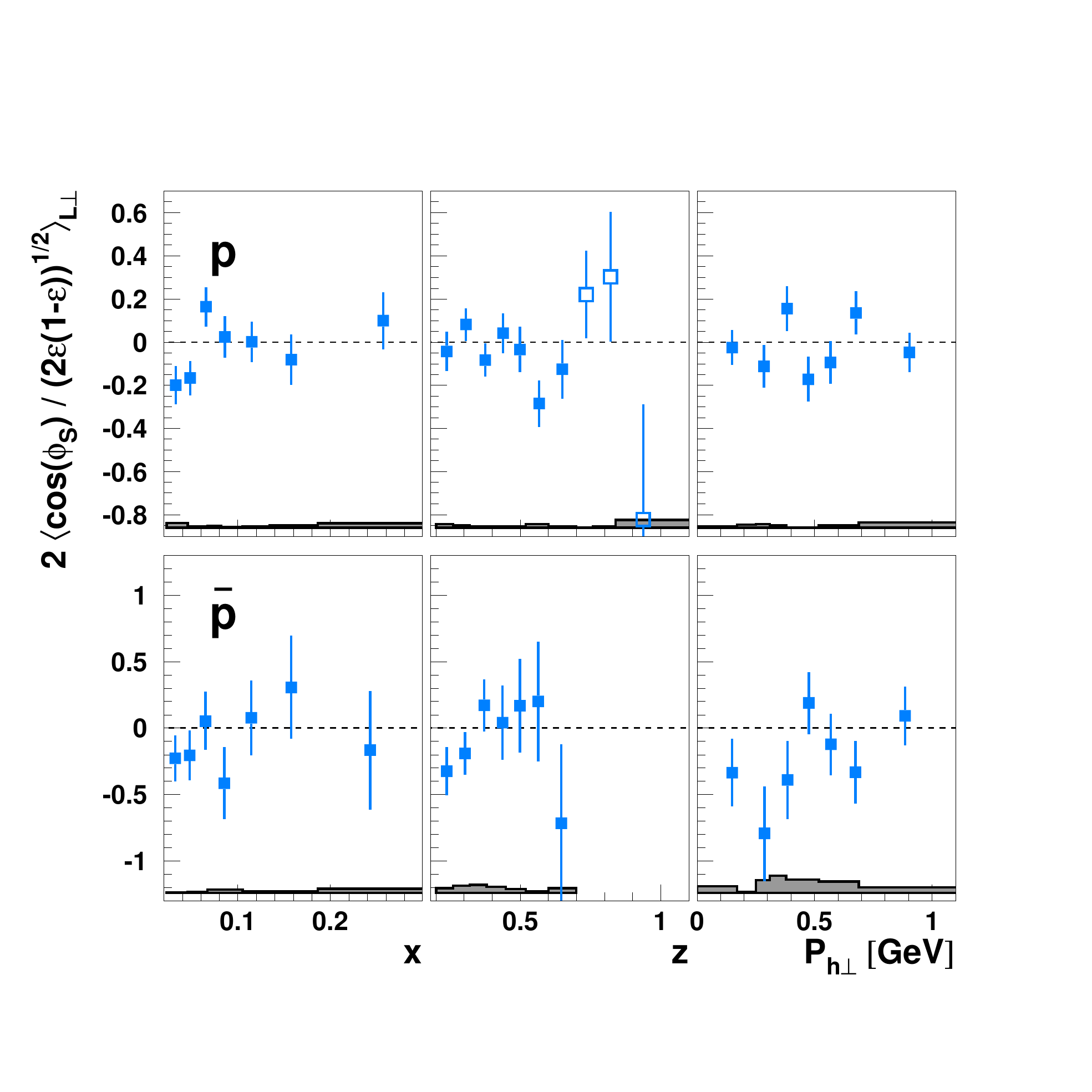}
  \caption{The \cosphisexpSFA{h}  amplitudes for \pizero (left), protons, and antiprotons (right) presented either in bins of \xb, \z, or \Phperpabs. Data at large values of \z, marked by open points in the \z projection, are not included in the other projections (no such high-\z points are available for antiprotons due to a lack of precision). Systematic uncertainties are given as bands, not including the additional scale uncertainty of 8.0\% due to the precision of the target-polarization determination.}
 \label{fig:pizero-protons-cosphisp}
\end{figure}

Finally, the subleading-twist \cosphisexpSFA{h} Fourier amplitudes are depicted in figures~\ref{fig:mesons-cosphis} and \ref{fig:pizero-protons-cosphisp}.
They are mostly consistent with zero, except for \kminus, whose Fourier amplitudes are found to be incompatible with the NULL hypothesis at 95\% confidence level.

In the Wandzura--Wilczek-type approximation, only the term proportional to \twistgt{q} times  \ffdmod{q\rightarrow h} survives. The former quantifies the quark-flavor contribution to the inclusive-DIS structure function \pdfgtwo{} via eq.~\eqref{eq:g2gT}. In this approximation, a small negative \cosphis{h} Fourier amplitude of the order of 1--2\% is predicted. While not necessarily favored by the data, such small negative asymmetries are not excluded in view of the overall precision of the data.
 
Without resorting to a Wandzura--Wilczek-type approximation, one can still reduce the number of contributing terms to \cosphis{h} by looking at the \cosinemodulation{\phis} modulation integrated over transverse momentum because --- like in the case of the \sinemodulation{\phis} modulation --- the \cosphis{h} Fourier amplitude is not required to vanish upon integration over transverse hadron momentum. But in contrast to the \sinphis{h} Fourier amplitude, two terms survive: the one discussed above involving \twistgt{q} and the product of transversity and the twist-3 collinear \(\widetilde{E}^{\,q\to h}\left(\z\right)\)~\cite{Jaffe:1993xb}, as can be seen from eq.~\eqref{eq:cosphisintegrated}. This allows for a collinear extraction of transversity, at least in principle as the contribution of the \twistgt{q} term needs to be subtracted.
Furthermore, there exist similar considerations as for the \sinemodulation{\phis} modulation, namely the usually incomplete integration
over \Phperp due to limitations in the geometric acceptance or kinematic requirements in experiments and 
the presently rather limited knowledge of the twist-3 fragmentation function \(\widetilde{E}^{\,q\to h}\left(\z\right)\).
 

Unlike the case of \sinphisexp{h} and \costwophiexp{h}, in the experimental measurement of \cosphisexp{} amplitudes, relatively large contributions from the longitudinal target-polarization component can be expected due to the mixing discussed in appendix~\ref{sec-app-longpol}. The double-spin asymmetry associated with the longitudinal polarization component is the typically sizable \( A_{\parallel}^{h} \) related to the \( A_{1}^{h} \) helicity asymmetry. 
It reaches values of 0.5 and higher~\cite{Airapetian:2004zf,Airapetian:2018rlq}, and thus values that are in general much larger than those measured for azimuthal asymmetries. While suppressed because of the small value of \( \theta_{\gamma^{*}} \), this contribution could still be substantial in comparison to the subleading-twist contributions from eq.~\eqref{eq:theory-cosphis} (cf.~discussion in appendix~\ref{sec-app-longpol}).

\subsection{Fourier moments arising solely from the longitudinal component of the target polarization}\label{sec:ALU_ALL}

In total ten Fourier components dependent on the transverse target polarization are extracted here. 
Of those, two arise solely because of a small longitudinal component of the proton polarization along the virtual-photon direction (cf.~appendix~\ref{sec-app-longpol}).
They are the \sintwophilexp{h} Fourier amplitude of the transverse \ssa 
and the \cosphiexp{h} Fourier amplitude of the \dsa, which are related to 
the \sintwophiul{h} Fourier amplitude of the longitudinal \ssa 
and the \cosphill{h} Fourier amplitude of the longitudinal \dsa, respectively.
While \sintwophiul{h} receives contributions at leading twist, \cosphill{h} is of subleading twist.

The \sintwophiul{h} Fourier amplitude provides access to the  
chiral-odd worm-gear (I) distribution \tmdhlpt{q}, 
which describes the distribution of
transversely polarized quarks in a longitudinally polarized
nucleon. As the final state involves unpolarized hadrons only, this
chiral-odd \tmd must couple to the chiral-odd Collins
fragmentation function.

Vanishing \sintwophiul{h} amplitudes for pions have been reported
by the \hermes Collaboration in an analysis of  \ssa[long] using longitudinally polarized hydrogen~\cite{Airapetian:1999tv,
Airapetian:2001eg} and deuterium~\cite{Airapetian:2002mf} targets.
The latter included also a measurement for \kplus mesons, which was found to be consistent with zero as well.
The only non-vanishing signal so far has been reported by the \clas Collaboration using a longitudinally polarized ammonia (\(^{15}\)NH\(_{3}\)) target 
(providing longitudinally polarized protons)~\cite{Avakian:2010ae}. The \sintwophiul{\pi} Fourier amplitudes for charged pions are negative and of the order of 5\% in magnitude.

\begin{figure}
\centering
\includegraphics[bb = 15 60 500 470, clip, width=0.49\textwidth,keepaspectratio]{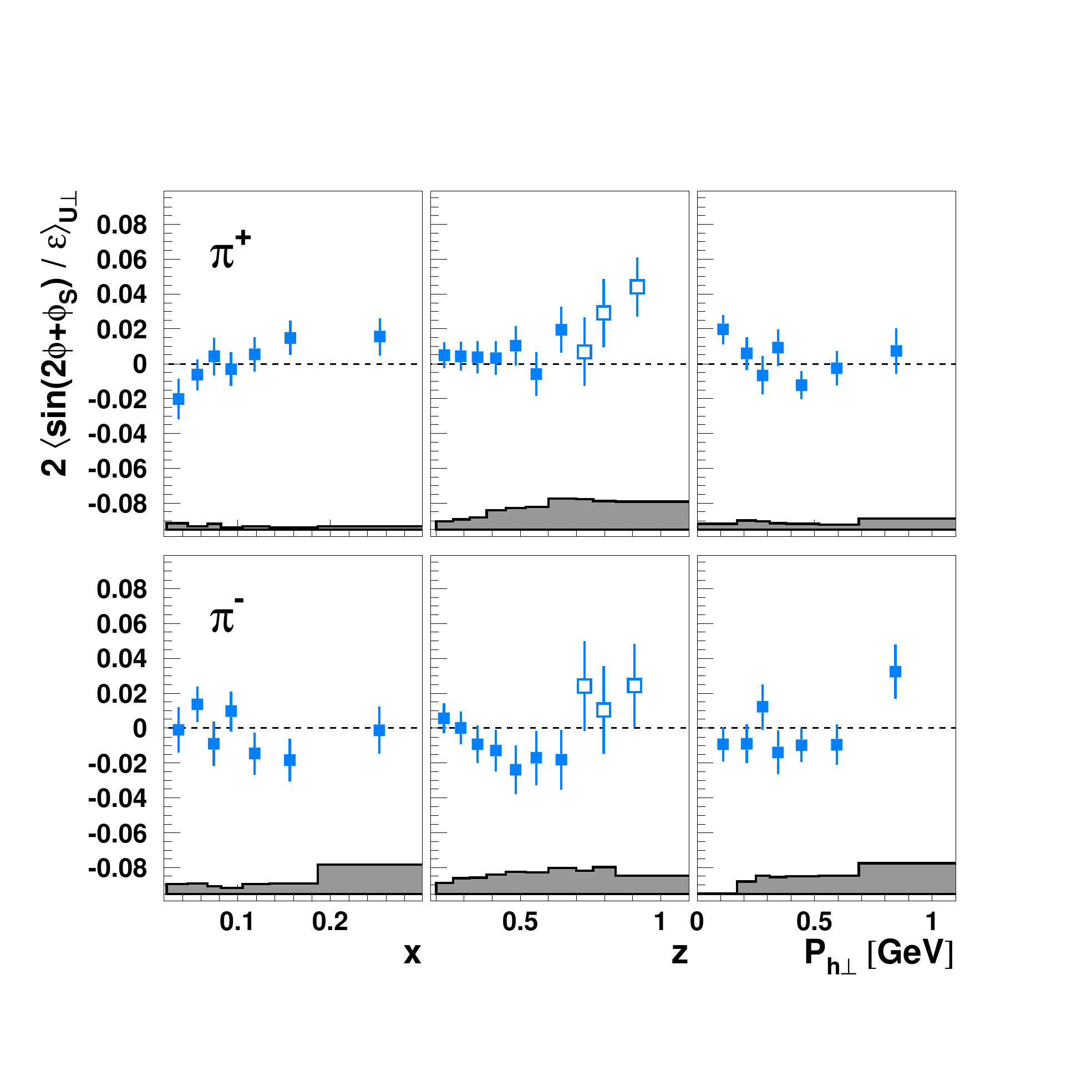}~~%
\includegraphics[bb = 15 60 500 470, clip, width=0.49\textwidth,keepaspectratio]{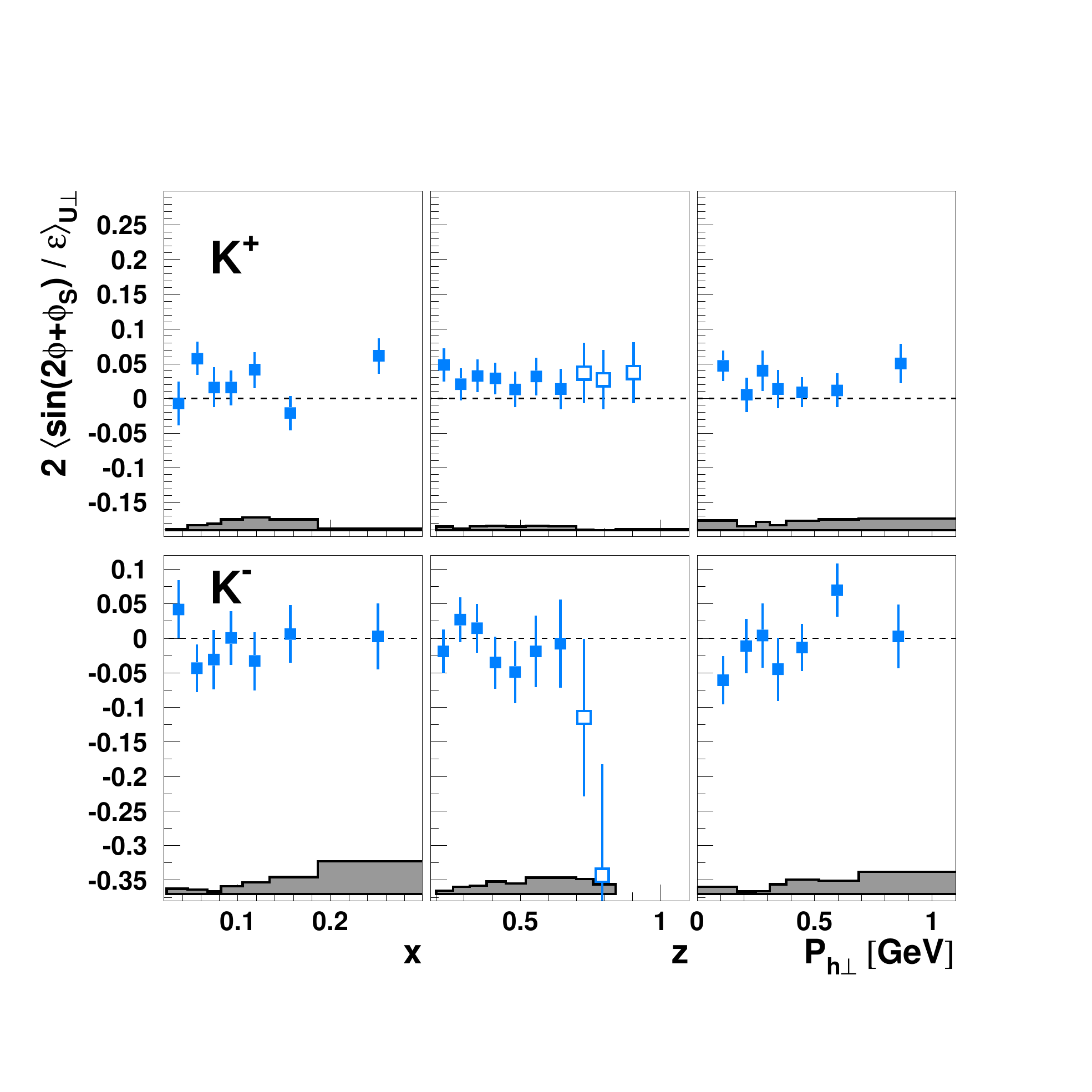}
  \caption{The \sintwophilexpSFA{h}  amplitudes for charged mesons (left: pions; right: kaons) presented either in bins of \xb, \z, or \Phperpabs. Data at large values of \z, marked by open points in the \z projection, are not included in the other projections. Systematic uncertainties are given as bands, not including the additional scale uncertainty of 7.3\% due to the precision of the target-polarization determination.}
 \label{fig:mesons-h1Lperp}
\end{figure}
\begin{figure}
\centering
\includegraphics[bb = 15 60 500 490, clip, width=0.49\textwidth,keepaspectratio]{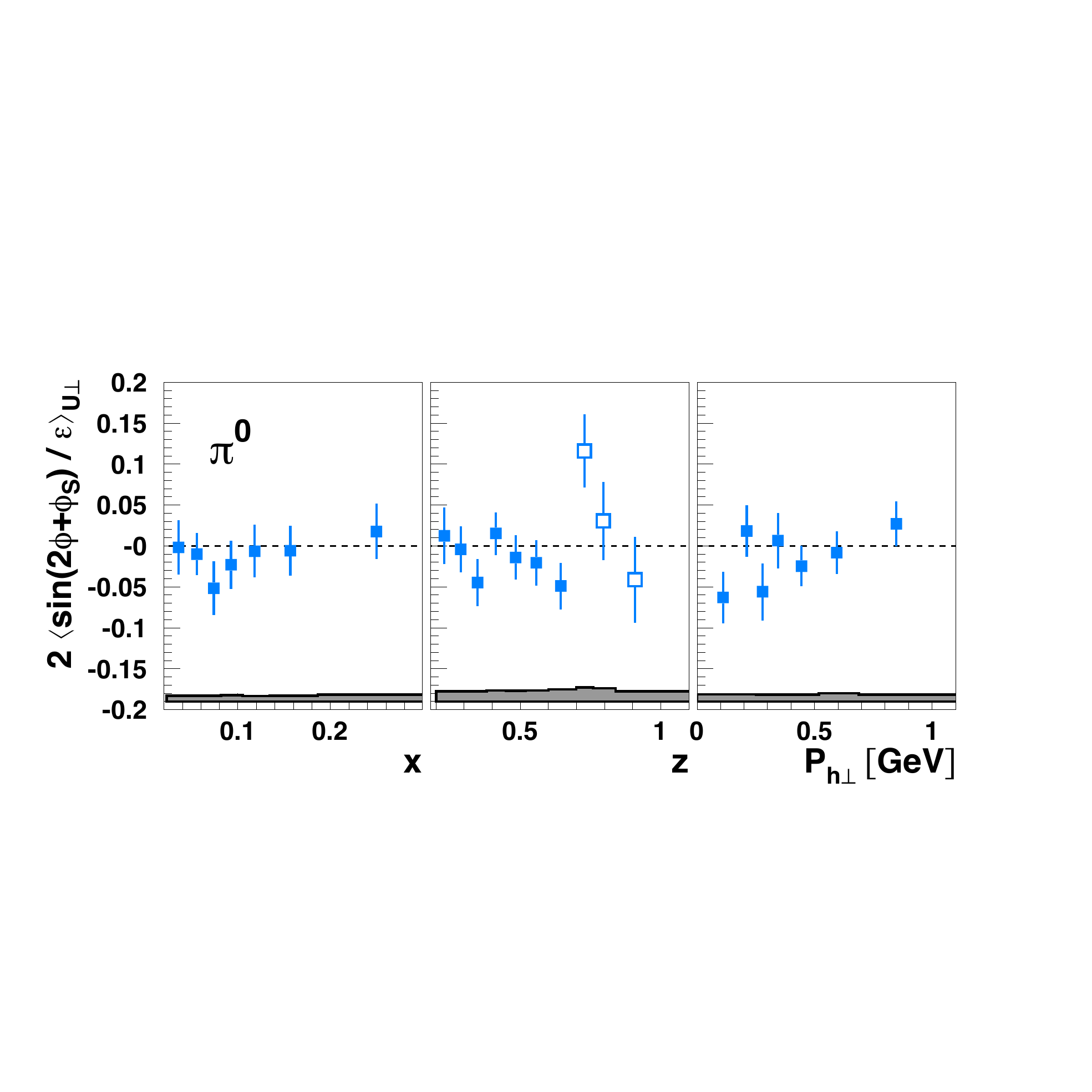}~~%
\includegraphics[bb = 15 60 500 490, clip, width=0.49\textwidth,keepaspectratio]{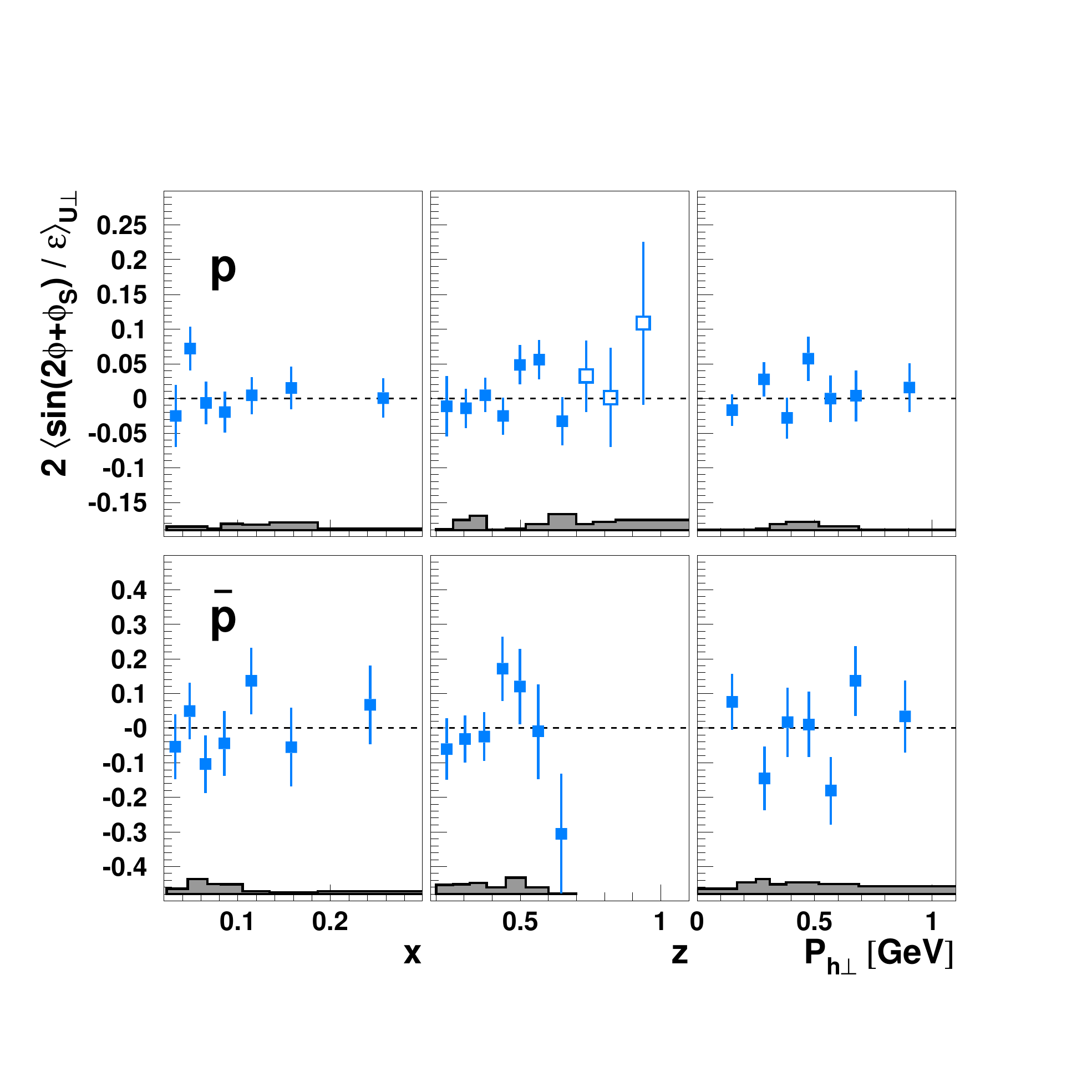}
  \caption{The \sintwophilexpSFA{h}  amplitudes for \pizero (left), protons, and antiprotons (right) presented either in bins of \xb, \z, or \Phperpabs. Data at large values of \z, marked by open points in the \z projection, are not included in the other projections (no such high-\z points are available for antiprotons due to a lack of precision). Systematic uncertainties are given as bands, not including the additional scale uncertainty of 7.3\% due to the precision of the target-polarization determination.}
 \label{fig:pizero-protons-h1Lperp}
\end{figure}

The \sintwophilexp{h} amplitudes in the Fourier decomposition of the
experimentally measured cross section, which are not sensitive to a
structure function of the transverse target spin-dependent
cross-section contribution \sigmaut{h}{T} in eq.~\eqref{theory-fourier}, 
are related to the \sintwophiul{h} amplitudes through 
\begin{equation} 
 \sintwophilexp{h} \simeq \frac{1}{2} \sin\theta_{\gamma^{*}} \,\, \sintwophiul{h}
 \label{eq:sintwophilexp}
\end{equation}
(cf.~appendix~\ref{sec-app-longpol}).\footnote{Through the same longitudinal target-polarization component, \tmdhlpt{q} contributes with equal magnitude also to 
\sintwophiexp{h} as discussed before. However, in that case it has to compete with the genuine transverse-polarization effects introduced in section~\ref{theory-subleadingtwist}.}
Therefore, a potential signal for \tmdhlpt{q} is additionally suppressed by at least an order of magnitude 
compared to corresponding measurements using longitudinally polarized targets.

The \sintwophilexpSFA{h} Fourier amplitudes for pions, charged kaons, as well as for (anti)protons extracted
in the presented analysis are shown in figures.~\ref{fig:mesons-h1Lperp} and \ref{fig:pizero-protons-h1Lperp}.
They are primarily consistent with zero and thus in agreement with the
previous \sintwophiul{h} related measurements, where data are available.
There is a tendency for a non-vanishing signal for positive pions at very large \z, e.g., when approaching the exclusive region, 
similar to what has been discussed for the \sintwophiexp{h} Fourier amplitude in section~\ref{interpretation-subleadingtwist}.
An analogous Fourier decomposition of the transverse SSA in exclusive \piplus electroproduction~\cite{Airapetian:2009ac} 
does result in \sinemodulation{2\phih+\phis} modulations not dissimilar to the behavior observed here in the large-\z region,
hinting at a non-vanishing \sintwophiul{\piplus} Fourier amplitude in the exclusive regime and possibly in the semi-exclusive region probed here. 
On the other hand, the direct measurement of the \sintwophiulexp{\piplus} Fourier amplitude in exclusive \piplus electroproduction
gives only \(0.05\pm0.05\)~\cite{Airapetian:2001iy}, likely too small to produce a sizable \sintwophilexp{\piplus} 
Fourier amplitude through the mixing of longitudinal and transverse target-polarization components.\footnote{A conclusive evaluation is hampered by the precision of the data and the possibility that the \sintwophiulexp{\piplus} result for exclusive \piplus electo-production received in turn contributions from a \sintwophi{\piplus} term in exclusive \piplus electroproduction, requiring a combined analysis of all three modulations, \sintwophiulexp{\piplus}, \sintwophiexp{\piplus}, and \sintwophilexp{\piplus}, along the lines of what was done in ref.~\cite{Airapetian:2005jc}.}

The \kplus \sintwophilexpSFA{} Fourier amplitude presented here might be the notable exception as --- somewhat unexpectedly --- it is positive over essentially the entire \z range. It is incompatible with the NULL hypothesis at 95\% confidence as already presented in table~\ref{tab:significant-modulations}. This points to a possibly sizable \sintwophiul{\kplus} asymmetry and thus indirectly to a sizable worm-gear (I) \tmdhlpt{q}. In particular, taking into account the factor \(\frac{1}{2}\sin\theta_{\gamma^*}\) that relates the two Fourier amplitudes and that amounts on average to 0.04 (cf.~eq.~\eqref{eq:sintwophilexp} and appendix~\ref{sec-app-longpol}), \sintwophiul{\kplus} Fourier amplitudes of the order 30\% can be expected.
No measurement of \sintwophiul{\kplus} for a proton target is presently available. A direct comparison of \sintwophiul{\kplus} to the \sintwophilexp{\kplus} presented here is thus not possible. Results for the \sintwophiulexp{\kplus} Fourier amplitude for a deuteron target are consistent with zero and within the achieved precision incompatible with magnitudes of tens of percent. On the other hand, there have been various instances where the \kplus result exceeds significantly the magnitudes for pions, prominent cases being the Sivers as well as the Collins asymmetries presented here. Recalling that \tmdhlpt{q} couples to the Collins fragmentation function in \sintwophiul{h} and that for \piplus there are significant cancelation effects due to the opposite signs for the favored and disfavored Collins fragmentation functions of pions, larger \kplus modulations can be expected if, for example, only \ffcollinsmod{u\rightarrow \kplus} is sizable as found in ref.~\cite{Anselmino:2015fty}.

\begin{figure}
\centering
\includegraphics[bb = 15 60 500 470, clip, width=0.49\textwidth,keepaspectratio]{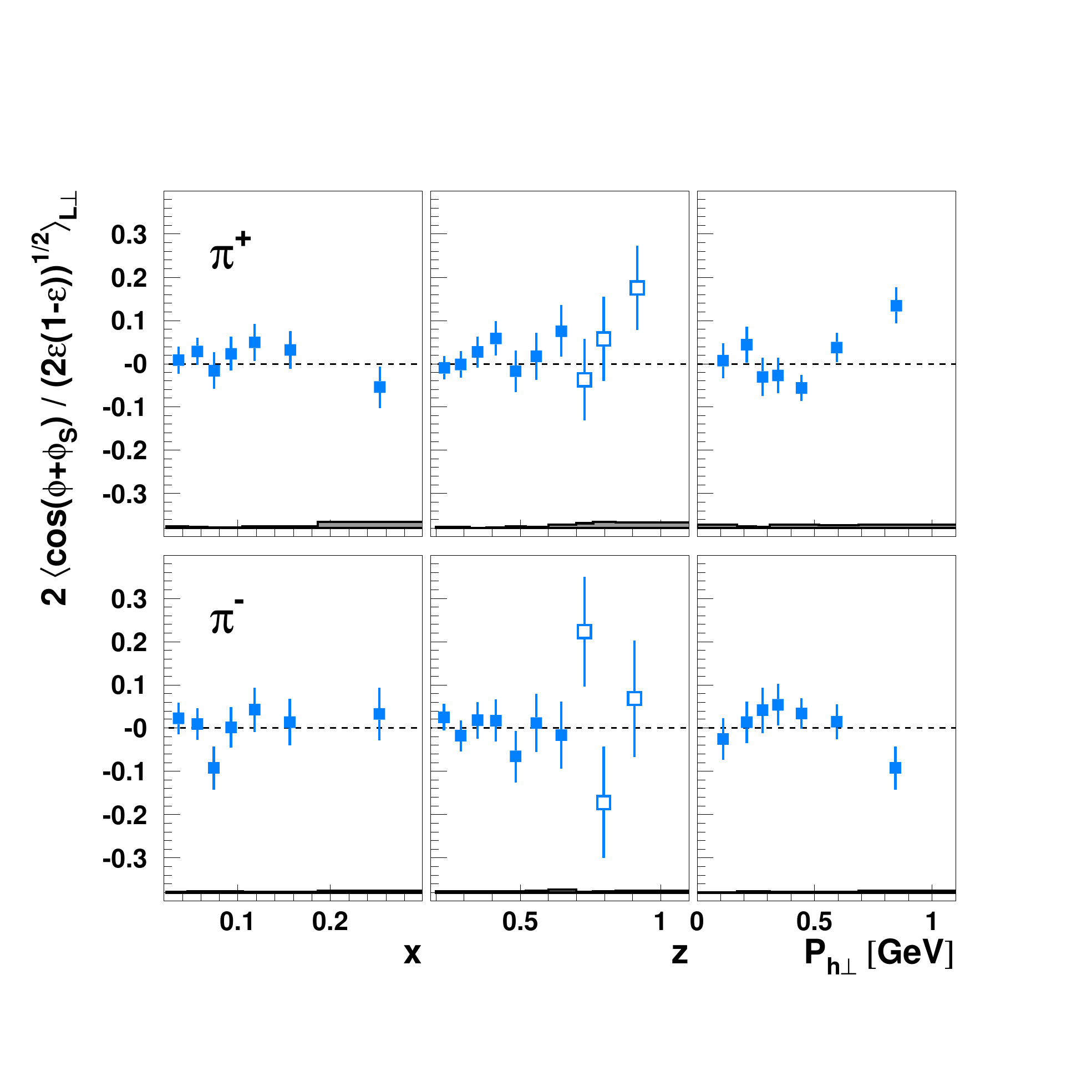}~~%
\includegraphics[bb = 15 60 500 470, clip, width=0.49\textwidth,keepaspectratio]{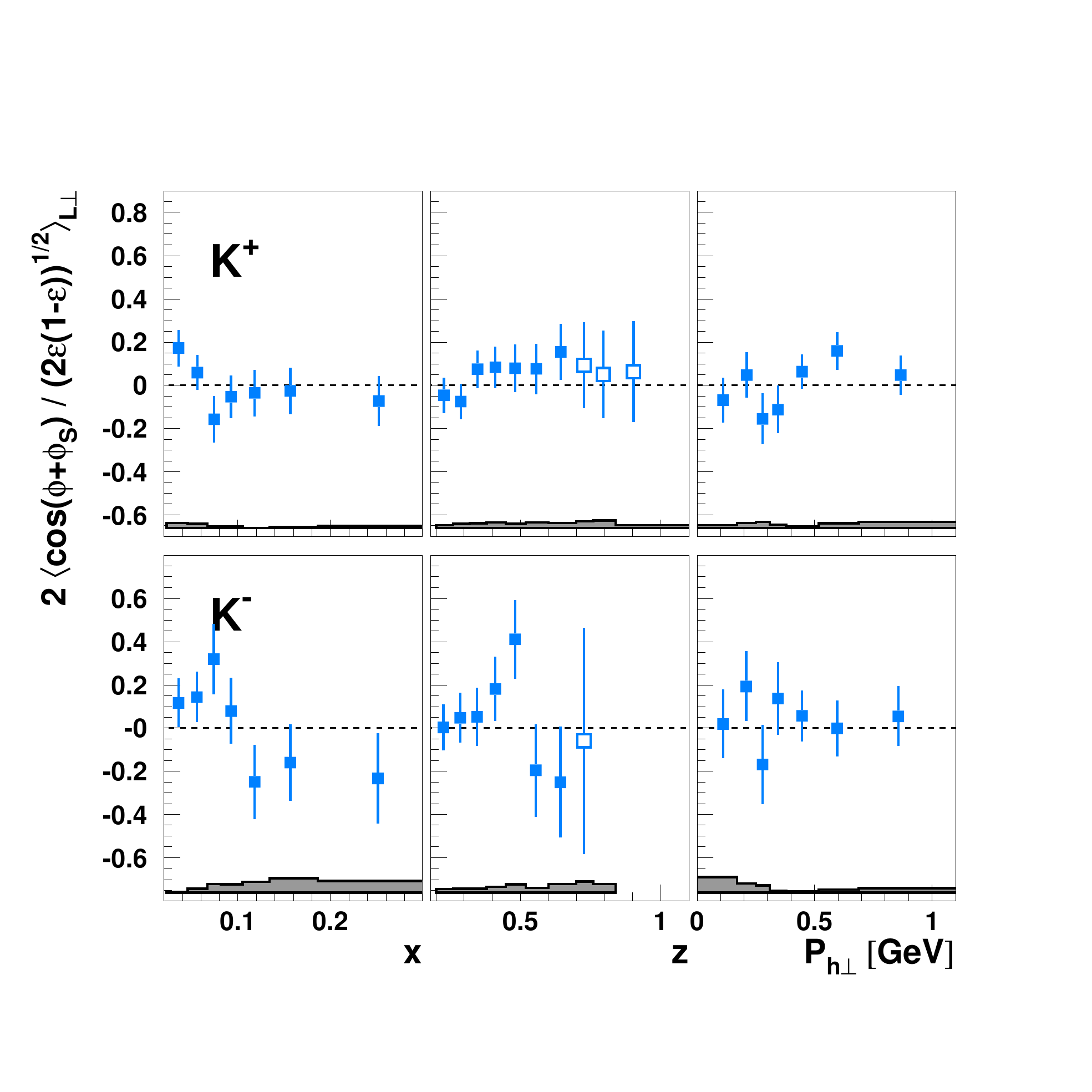}
  \caption{The \cosphilexpSFA{h}  amplitudes for charged mesons (left: pions; right: kaons) presented either in bins of \xb, \z, or \Phperpabs. Data at large values of \z, marked by open points in the \z projection, are not included in the other projections. Systematic uncertainties are given as bands, not including the additional scale uncertainty of 8.0\% due to the precision of the target-polarization determination.}
 \label{fig:mesons-cosppp}
\end{figure}

\begin{figure}
\centering
\includegraphics[bb = 15 60 500 490, clip, width=0.49\textwidth,keepaspectratio]{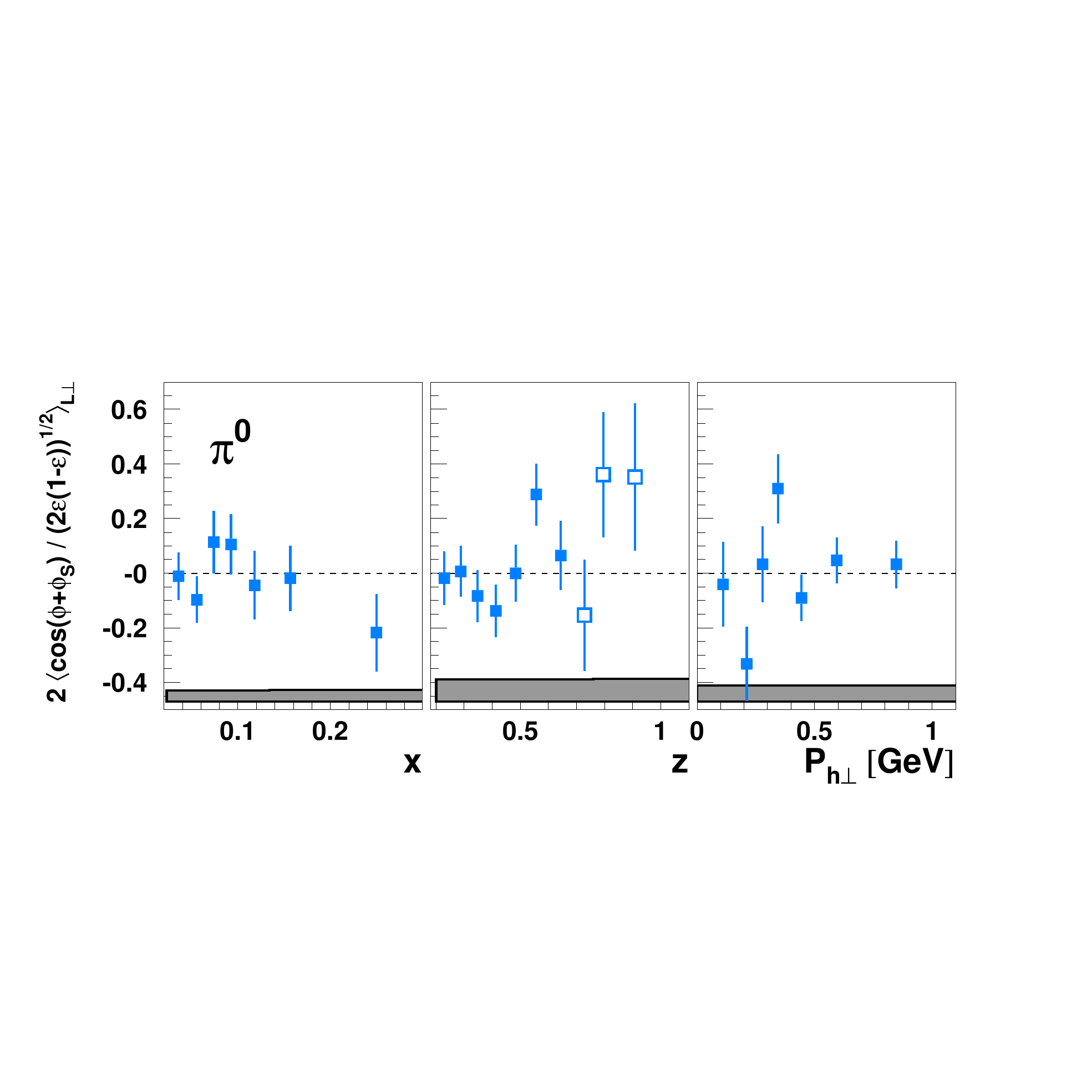}~~%
\includegraphics[bb = 15 60 500 490, clip, width=0.49\textwidth,keepaspectratio]{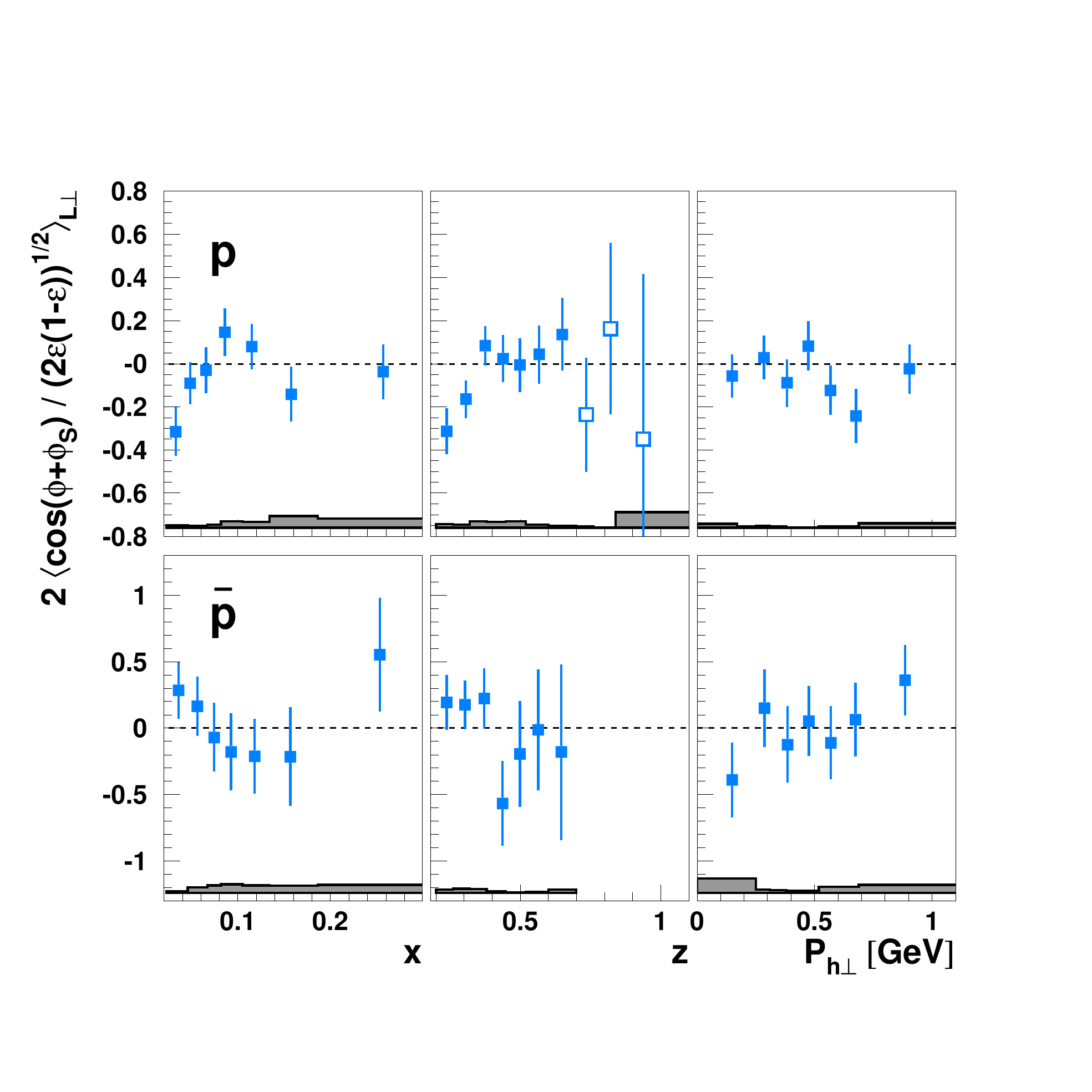}
  \caption{The \cosphilexpSFA{h}  amplitudes for \pizero (left), protons, and antiprotons (right) presented either in bins of \xb, \z, or \Phperpabs. Data at large values of \z, marked by open points in the \z projection, are not included in the other projections (no such high-\z points are available for antiprotons due to a lack of precision). Systematic uncertainties are given as bands, not including the additional scale uncertainty of 8.0\% due to the precision of the target-polarization determination.}
 \label{fig:pizero-protons-cosppp}
\end{figure}

The \cosphilexpSFA{h} Fourier amplitudes for pions, for charged kaons, as well as for (anti-)protons extracted
in this analysis are shown in figures~\ref{fig:mesons-cosppp} and \ref{fig:pizero-protons-cosppp}.
They arise through the small longitudinal target-polarization component from the subleading-twist 
\(\cos(\phi)\) azimuthal modulation of the longitudinal \dsa. The latter may arise through the ``polarized Cahn effect''~\cite{Cahn:1978se,Oganessyan:2002pc,Anselmino:2006yc}, 
which combines transverse momenta of longitudinally polarized quarks in a longitudinally polarized nucleon with 
the transverse momentum acquired in the fragmentation process and as such is sensitive to the transverse-momentum
dependence of the helicity distribution, \pdfgpt{}.
No significant signal for such modulation has been reported so far, 
neither for unidentified hadrons from a longitudinally polarized \(^{6}\)LiD (effectively a deuteron) target
at the \compass experiment~\cite{Alekseev:2010dm} nor for identified pions (and kaons) from a longitudinally polarized hydrogen 
(deuterium) target by the \hermes Collaboration~\cite{Airapetian:2018rlq}. This is consistent with the vanishing signal for \cosphilexp{h} reported here.

%% file: sections/conclusion.tex
\section{Conclusion}

A comprehensive discussion of azimuthal single- and double-spin asymmetries in semi-inclusive leptoproduction of pions, charged kaons, protons, and antiprotons 
from transversely polarized protons at \hermes has been presented. These asymmetries include the previously published \hermes results on Collins and Sivers asymmetries~\cite{Airapetian:2010ds,Airapetian:2009ae}, which have been extended to include protons and antiprotons and also to an extraction in a three-dimensional binning in \xb, \z, and \Phperpabs. In addition, the large-\z region of \(\z>0.7\) is explored to study the transition from the semi-inclusive to the exclusive regime. 

Furthermore, the set of azimuthal asymmetries measured include those arising from the leading-twist pretzelosity and worm-gear (II) distributions, four subleading-twist modulations, and two modulations that contribute to the \(e^{\pm}p\) cross-section through the small but non-vanishing longitudinal target-polarization component in experiments where the target is polarized perpendicular to the beam direction. No sign of a non-zero pretzelosity is found, while the non-vanishing \cosinemodulation{\phih-\phis} modulations for pions provide evidence for a sizable worm-gear (II) distribution, \tmdgtpt{q}. The subleading-twist contributions and the ones from the longitudinal target-spin component are mostly consistent with zero. A rather notable exception are the large \sinemodulation{\phis} modulations for charged pions and  \kplus.

All modulations were studied as functions of \xb, \z, and \Phperpabs individually as well as simultaneously binned in all the three kinematic variables, except for \pizero and antiprotons, in which case the corresponding yields were not sufficient to allow such three-dimensional binning. Fourier amplitudes were extracted including or excluding the kinematic prefactor arising from the photon spin-density matrix accompanying each specific cross-section contribution. This allows for a simpler comparison with other experiments or theoretical calculations as experiment-specific integration over kinematic variables is minimized.
The results for the azimuthal modulations are supplemented by information on the unpolarized cross section, in particular, distributions in rapidity as well as of transverse momentum vs.~the hard scale \Q. Those are expected to facilitate the interpretation of the modulation in global analyses within the TMD framework.

\acknowledgments

This paper is dedicated to our recently deceased colleagues Robert Avakian, Stanislav Belostotski, Harold E. Jackson, and Bohdan Marianski. 

This work would not have been possible without the continuous input by our theory colleagues;
especially appreciated is the input by M. Boglione for providing calculations of the Sivers asymmetry and by M. Diehl in many very valuable discussions.
We gratefully acknowledge the \desy\ management for its support and the staff
at \desy, in particular, the data-preservation group, as well as the collaborating institutions for their significant effort. 
This work was supported by 
the State Committee of Science of the Republic of Armenia, Grant No. 18T-1C180;
the FWO-Flanders and IWT, Belgium;
the Natural Sciences and Engineering Research Council of Canada;
the National Natural Science Foundation of China;
the Alexander von Humboldt Stiftung,
the German Bundesministerium f\"ur Bildung und Forschung (BMBF), and
the Deutsche Forschungsgemeinschaft (DFG);
the Italian Istituto Nazionale di Fisica Nucleare (INFN);
the MEXT, JSPS, and G-COE of Japan;
the Dutch Foundation for Fundamenteel Onderzoek der Materie (FOM);
the Russian Academy of Science and the Russian Federal Agency for 
Science and Innovations;
the Basque Foundation for Science (IKERBASQUE), Spain;
the U.K.~Engineering and Physical Sciences Research Council, 
the Science and Technology Facilities Council,
and the Scottish Universities Physics Alliance;
as well as the U.S.~Department of Energy (DOE) and the National Science Foundation (NSF).

%% file: sections/appendices.tex
\appendix
\section{Contribution from longitudinal target polarization}\label{sec-app-longpol}

\begin{figure}
\centering
\includegraphics[width=0.6\textwidth,keepaspectratio]{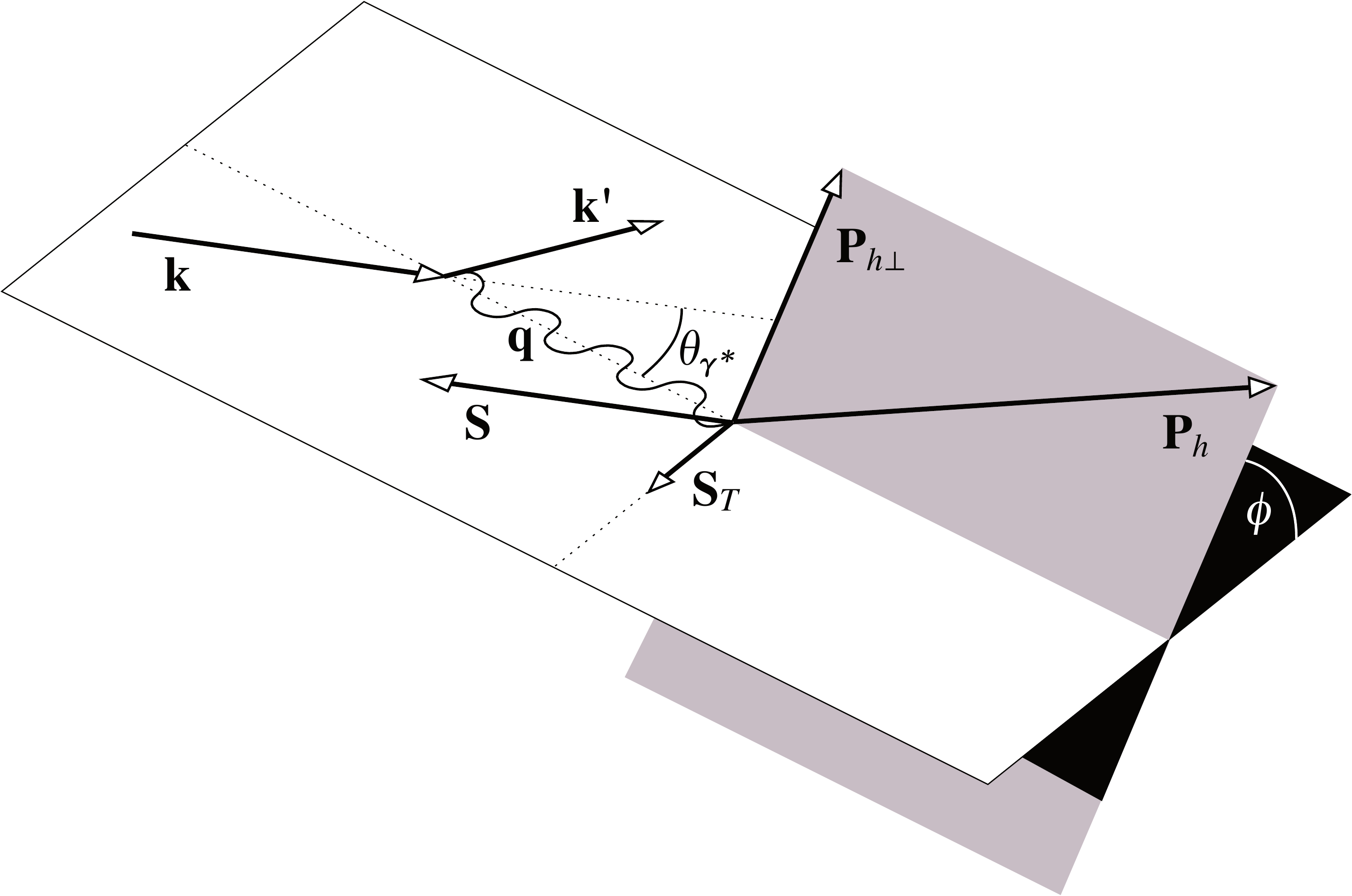}
  \caption{Illustration of the mixing of longitudinal and transverse target polarizations w.r.t.~the virtual-photon direction when polarizing the target along the lepton-beam direction. In this particular case of longitudinal target polarization, a small component transverse to the virtual-photon direction arises that is proportional to \(\sin\theta_{\gamma^{\ast}}\).
  }
 \label{fig:phi}
\end{figure}

The interest in leptoproduction data on transversely polarized protons lies in the various semi-inclusive structure functions discussed in section~\ref{sec:qpm}. As the target-polarization direction in an actual experiment refers to the lepton-beam direction for the reference axis and not to the virtual-photon direction used in theory, most of the azimuthal modulations measured here receive contributions from the resulting non-vanishing longitudinal component of the target polarization with respect to the virtual-photon direction (see figure~\ref{fig:phi}).
This leads to additional moments as compared to, e.g., eq.~\eqref{theory-fourier}, 
resulting in~\cite{Diehl:2005pc}
\begin{align}
\lefteqn{\sigma^h(\phih,\phis)  =
 \sigmauu{h} \Bigg\lbrace 
    1 + \unpolcosphi{h}\cosinemodulation{\phih}
        + \unpolcostwophi{h}\cosinemodulation{2\phih} }  		\nonumber \\ 
 & \qquad \quad  + S_L \left[ 
	\sinphiexp{h} \sinemodulation{\phih} + 
  	\sintwophiulexp{h} \sinemodulation{2\phih}  + 
	\sinthreephiulexp{h} \sinemodulation{3\phih}    \right. 		\nonumber \\ 
 & \qquad \qquad 	\left. + \lambda_l \left( \coszerophillexp{h} \cosinemodulation{0\phih}  + 
				\cosphillexp{h} \cosinemodulation{\phih}    + 
				\costwophillexp{h} \cosinemodulation{2\phih}  \right)
	\right]  										\nonumber \\
 & \qquad \quad + S_T \left[ 
	\siversexp{h}\sinemodulation{\phih-\phis}  +
       	\collinsexp{h}\sinemodulation{\phih+\phis} 
\right. 											\nonumber \\
 & \qquad \qquad	+ \sinthreephiexp{h}\sinemodulation{3\phih-\phis} + 
	\sinphisexp{h}\sinemodulation{\phis}     				\nonumber \\
 & \qquad \qquad	+ \sintwophiexp{h}\sinemodulation{2\phih-\phis} +
	\sintwophilexp{h}\sinemodulation{2\phih+\phis} 
	  											\nonumber \\
  & \qquad \qquad + \lambda_l \left( 
	\cosphiexp{h}\cosinemodulation{\phih-\phis} 	+
	\cosphilexp{h}\cosinemodulation{\phih+\phis} 	
	\right.  
	 											\nonumber \\
 & \qquad \qquad \quad \quad \left.
\left.
	+ \cosphisexp{h}\cosinemodulation{\phis} +
	\costwophiexp{h}\cosinemodulation{2\phih-\phis} 
	\right)
\right]  \Bigg\rbrace   \, ,
\label{DS-fourier}
\end{align}
where the cross section averaged over the polarization states and integrated over \phih and \phis is represented by \sigmauu{h} and has been factored out.

\begin{figure}
 \centering
 \includegraphics[width=0.75\textwidth,keepaspectratio]{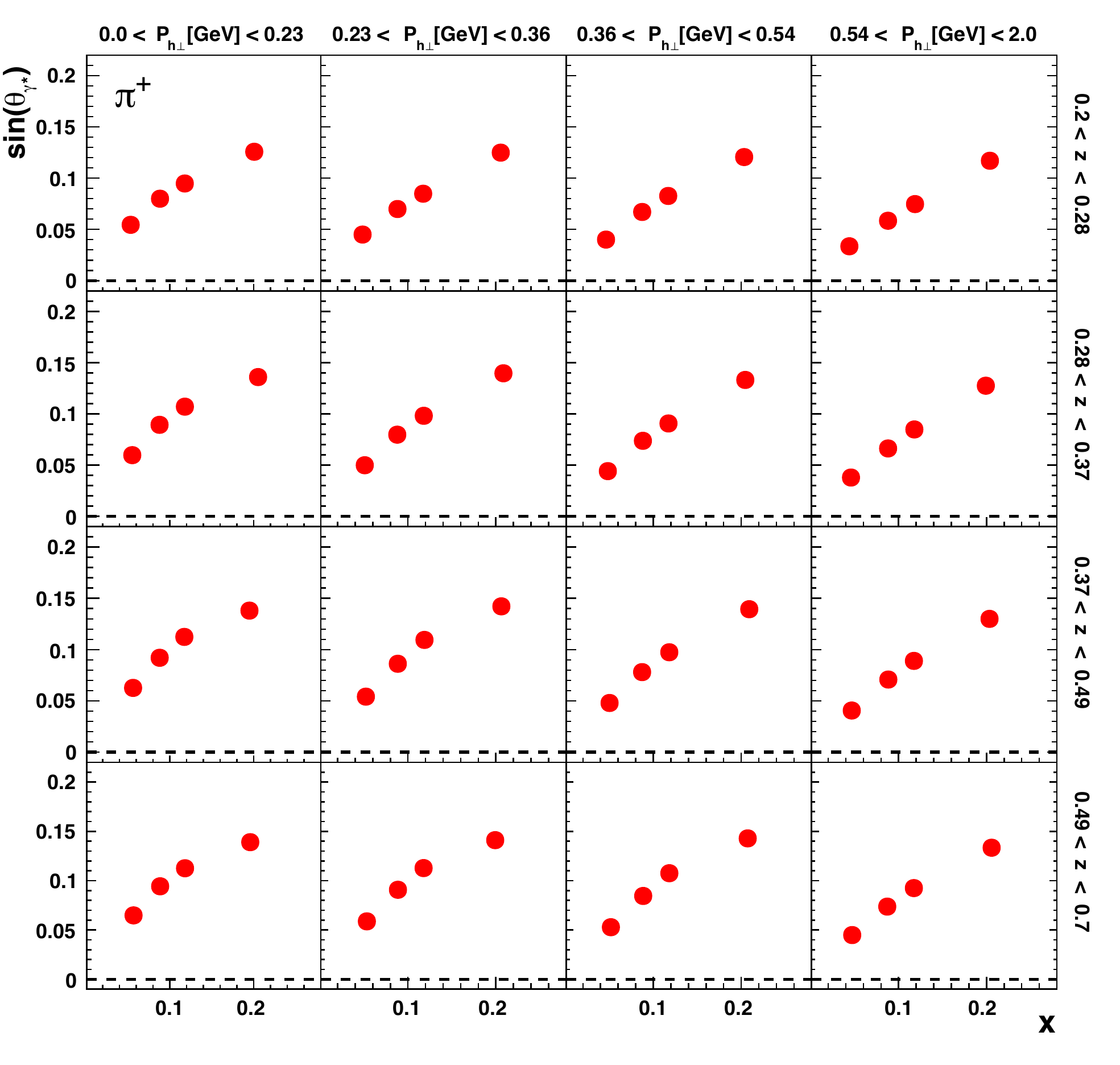}
 \caption{Average \(\sin\theta_{\gamma^*}\) as a function of \xb in bins of \z and  \Phperpabs in the same three-dimensional binning 
                used for the analysis of azimuthal asymmetries of charged mesons (using here \piplus data as an example).
  }
 \label{fig:sintheta-3d-mesons}
\end{figure}

\begin{figure}
 \centering
 \includegraphics[width=0.68\textwidth,keepaspectratio]{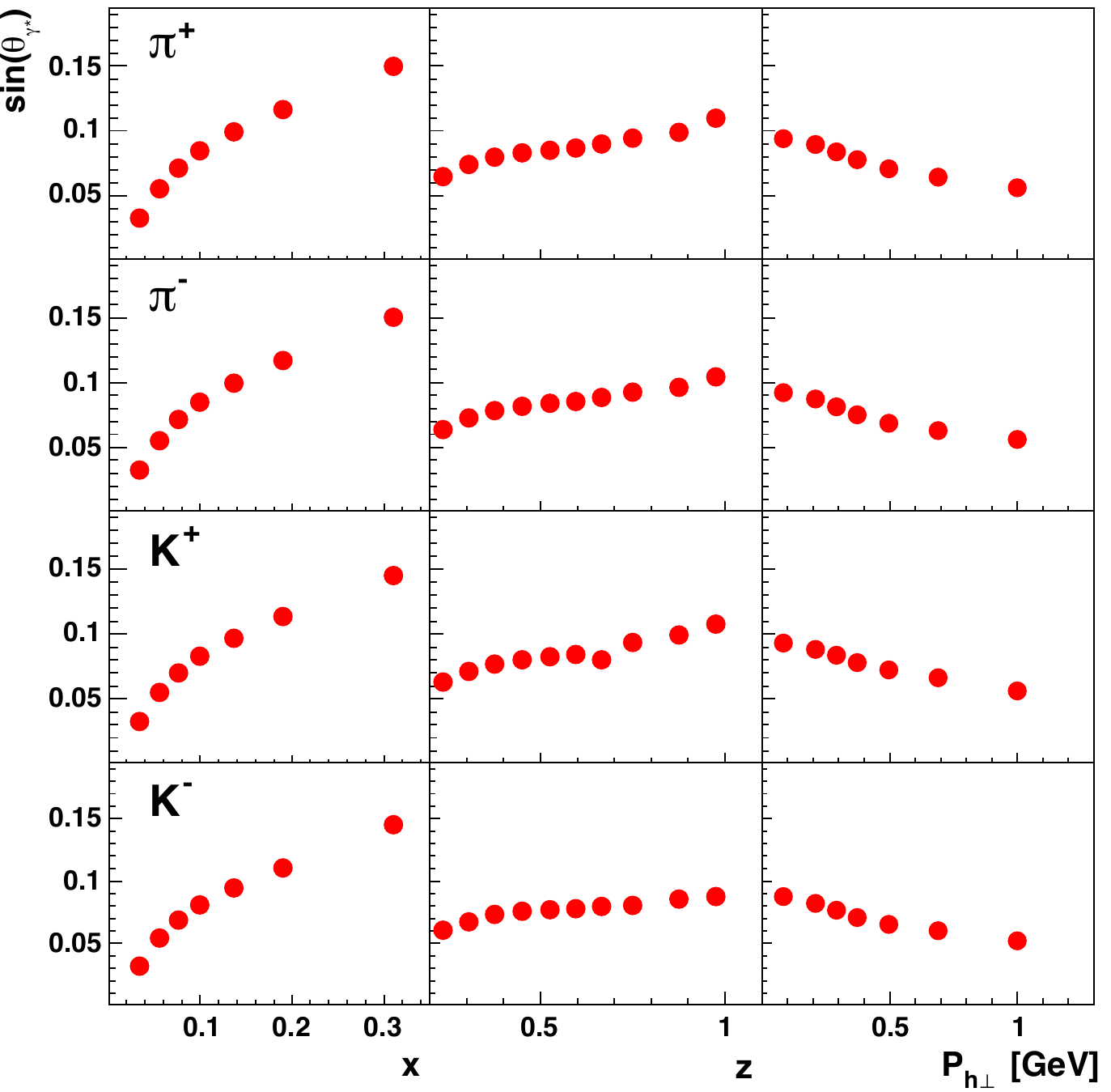}
 \caption{Average \(\sin\theta_{\gamma^*}\) as a function of \xb,\z, or  \Phperpabs for charged mesons as labeled.}
 \label{fig:sintheta-1d-mesons}
\end{figure}

The size of the component of the nucleon-spin vector that is longitudinal to the virtual-photon direction 
depends on  \(\theta_{\gamma^*}\), the polar angle between the incoming-beam and the virtual-photon directions.
Hence it strongly depends on the event kinematics.
At \hermes \ kinematics, \(\sin\theta_{\gamma^*}\) is of the order of 0.1 but can be as large as 0.2 for events at very large \xb.
Here, \(\sin \theta_{\gamma^*}\) is evaluated from the lepton kinematics as 
\begin{equation}\label{eq:sinthetastar}
 \sin\theta_{\gamma^*} =  \frac{2 \xb M}{Q}  \sqrt{\frac{1 - \y - \y^2 \xb^2 M^2/\Q}{1 +  4\xb^2 M^2/\Q}}. 
\end{equation}
Its average values are presented in figure~\ref{fig:sintheta-3d-mesons} for \piplus (similar for the other hadrons)
in the same three-dimensional kinematic binning used also for the asymmetry measurement. 
Likewise, they are presented in figure~\ref{fig:sintheta-1d-mesons} for the one-dimensional binning of mesons. 
The longitudinal polarization components are also provided as tabulated values for all particle types \cite{supplemental}.

The contributions from the transverse and longitudinal components can only be disentangled using data from targets with both polarization orientations. Such analysis was presented in ref.~\cite{Airapetian:2005jc} by the \hermes Collaboration, using both data for the Sivers and Collins type modulations for transverse target polarization~\cite{Airapetian:2004tw} and the \(\sin\phi\) modulation for longitudinal polarization~\cite{Airapetian:1999tv}. It is based on the inversion of the {\em mixing matrix} 
\begin{equation}\label{eq:matrix}
\left( \begin{array}{c}
\sinphiexp{}\\[1.5mm]
\siversexp{}\\[1.5mm]
\collinsexp{}
\end{array} \right)
=
\left( \begin{array}{ccc}
\cos\theta_{\gamma^*}            & -\sin\theta_{\gamma^*}&-\sin\theta_{\gamma^*}  \\[1.5mm]
\frac{1}{2}\sin\theta_{\gamma^*} & \cos\theta_{\gamma^*} &      0        \\[1.5mm]
\frac{1}{2}\sin\theta_{\gamma^*} &       0               & \cos\theta_{\gamma^*} 
\end{array} \right)
\left( \begin{array}{c}
\sinphi{}\\[1.5mm]
\sivers{}\\[1.5mm]
\collins{}
\end{array} \right),
\end{equation}
which is valid up to corrections of order \(\sin^2\theta_{\gamma^*}\)~\cite{Diehl:2005pc}.
Similar expressions are obtained for the other modulations studied here by interchanging in eq.~\eqref{eq:matrix}
\begin{enumerate}[label=(\roman*)]
 \item \(\sin(\phi\pm\phis) \leftrightarrow \sin(n\phi\pm\phis)\) and \(\sin(\phi)\leftrightarrow\sin(n\phi)\) for \( n > 0 \),
 \item \(\sin \leftrightarrow \cos\)  in case of longitudinal beam polarization.
\end{enumerate}
Note that some of the elements of the moments vectors might then be identical to zero (cf.~eqs.~\eqref{theory-fourier} and \eqref{DS-fourier}), e.g., \(\langle \sin(3\phi) \rangle_{\text{LL}} \), 
at least in the one-photon-exchange approximation. 

The \phih-independent \ssa relates to its theory counterpart via
\begin{equation}
\sinphisexp{}  = \cos\theta_{\gamma^*} \,\, \sinphis{},
\end{equation}
while the mixing of the \phih-independent \dsa[long]  can be expressed as
\begin{equation}\label{eq:matrixLL}
\left( \begin{array}{c}
\langle\cos 0\rangle_{\text{L}\parallel} \\[1.5mm]
\cosphisexp{}
\end{array} \right)
=
\left( \begin{array}{cc}
\cos\theta_{\gamma^*}            &-\sin\theta_{\gamma^*}  \\[1.5mm]
\sin\theta_{\gamma^*} 	      & \cos\theta_{\gamma^*} 
\end{array} \right)
\left( \begin{array}{c}
\langle\cos 0\rangle_{\text{LL}} \\[1.5mm]
\cosphis{}
\end{array} \right) \, .
\end{equation}

The experimental challenge consists in combining measurements using transversely and longitudinally polarized targets under similar conditions, 
which among others requires identical kinematic binning for the two data sets.
For the analysis presented here, such a matching data set is missing. In particular, the use of a threshold Cherenkov counter during data taking with a longitudinally polarized hydrogen target prohibits the measurement of the relevant kaon and also (anti)proton asymmetries. Therefore, no attempt has been made to disentangle the structure functions related to transversely and longitudinally polarized protons. Future data and/or parameterization of the relevant longitudinal-spin asymmetries might be used instead to extract the purely transverse structure functions.

\begin{figure}
 \centering
 \includegraphics[width=0.55\textwidth,keepaspectratio]{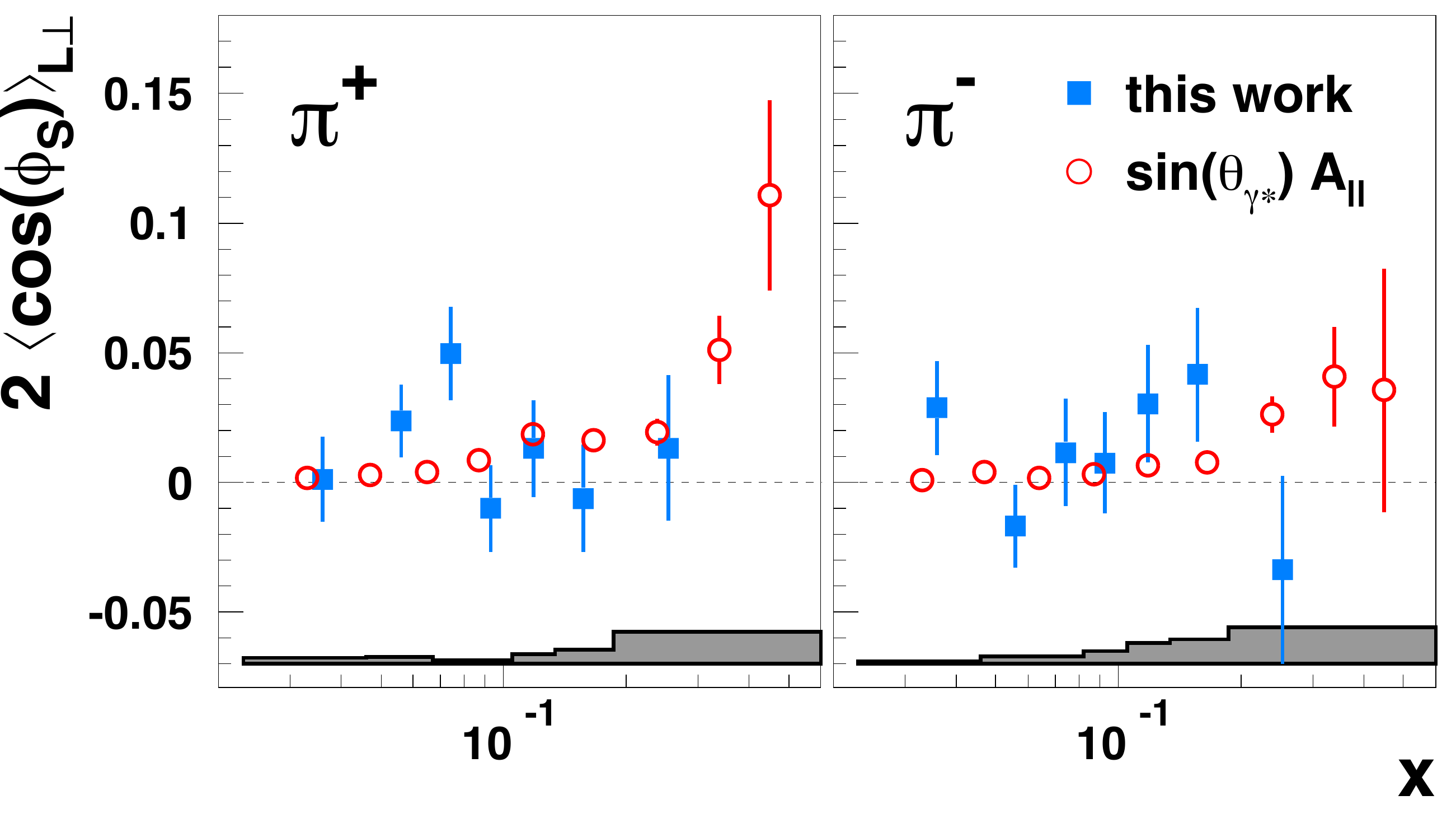}
  \caption{The \xb dependences of the charged-pion \cosphisexp{} asymmetries measured here (blue squares) and the contribution to this observable from \(A_{\parallel}\) (red circles). For the latter, the \hermes measurement~\cite{Airapetian:2018rlq}, which is taken in the range \(0.2<\z<0.8\), is scaled by the average \(\sin\theta_{\gamma^{\ast}}\) in each bin.}
 \label{fig:A_LLinA_LT}
\end{figure}

Nevertheless, while a precise quantitative evaluation of the effect for all the SSAs and DSAs of this measurement is currently out of reach, a few qualitative comments might be in order.
In general, most azimuthal moments presented here and elsewhere for longitudinal target polarization are of similar order of magnitude, e.g., below 0.1 in magnitude.
The corrections are thus small as already noted for the Sivers and Collins asymmetries in ref.~\cite{Airapetian:2005jc}.
This is not quite the case though for  the \phih-independent \dsa, \cosphisexp{},
which receives rather large contributions from the azimuthally uniform structure function arising from the quark-helicity distribution.
Those are up to an order of magnitude larger~\cite{Airapetian:2018rlq} than the typical azimuthal moments and increase with \xb as does the longitudinal target-polarization component, which needs to be considered when interpreting the \cosphisexp{} results. As an example, the contribution from \(A_{\parallel}\) has been evaluated using the HERMES measurement~\cite{Airapetian:2018rlq}  scaled by the corresponding average longitudinal target-spin component of each \xb bin, shown in figure~\ref{fig:A_LLinA_LT}.

There are notable exceptions to this general discussion.
Three azimuthal asymmetries, namely the \sinphisexp{}, the \sinthreephiexp{}, and the \costwophiexp{} Fourier amplitudes,
do not receive contributions from the longitudinal component of the target polarization.
The experimentally measured azimuthal asymmetries are thus only diluted. 
The correction factor \(1/\cos\theta_{\gamma^*}\), however, can be taken as unity under the kinematic conditions here.
The second class of exception concerns the \sintwophiexp{} and \cosphiexp{} Fourier amplitudes.
The contributions from the longitudinal component to those are equal to the contributions to \sintwophilexp{} and \cosphilexp{}.
In contrast to the \sintwophiexp{} and \cosphiexp{} Fourier amplitudes, the \sintwophilexp{} and \cosphilexp{} Fourier amplitudes arise solely because of the contribution from the longitudinal component of the target polarization and are thus a measure for that contribution to also \sintwophiexp{} and \cosphiexp{}.

The mixing of target-spin components occurs on the level of the lepton-proton cross sections. Disentangling the contributions thus works in a straightforward way for the \csa by solving the set of linear equations~\eqref{eq:matrix} as well as \eqref{eq:matrixLL}. By contrast, the extraction of \sfa already includes a compensation for the \(\epsilon\)-dependent prefactors, which are in general not the same for the longitudinal and transverse target-spin contributions. As a consequence, a similar separation of the terms from longitudinal and transverse target polarization requires the inclusion of these prefactors in the matrices of eqs~\eqref{eq:matrix} and \eqref{eq:matrixLL}.

\section{Transverse-momentum factorization and the separation of current and target fragmentation}\label{app:TMDfactorization}

This measurement has been performed in the approach presented by Mulders and Tangerman~\cite{Mulders:1995dh} and subsequent works, assuming that the hard scale given by \Q is sufficiently large compared to the transverse momenta involved, and that hadrons are produced in the commonly denoted {\em current} or {\em quark fragmentation region}~\cite{Berger:1987zu}, i.e., during the hadronization of the quark struck by the virtual photon. 
In the kinematic region of typical fixed-target \dis experiments, the clear separation of the current from the target fragmentation --- where the hadron originates from the target remnants (see, e.g., refs.~\cite{Trentadue:1993ka,Anselmino:2011ss}) --- is not always granted~\cite{Mulders:2000jt}. As outlined already in section~\ref{measurement-kinematicrequirements}, 
the situation is even more vague
when looking at transverse-momentum-dependent observables as in this work, because in that case, a hadron produced with large enough transverse momentum in the target fragmentation may mimic a hadron with large transverse momentum from current fragmentation. This complication has attracted increased attention, e.g., through the works of refs.~\cite{Boglione:2016bph,Boglione:2019nwk}.
There is no unique recipe to ensure complete separation of current and target fragmentation and the applicability of QCD factorization theorems may be questioned in the more extreme kinematic regions of growing overlap of the two, e.g., at very low \z and large transverse momentum and especially at low values of \Q. But where exactly to draw the boundary remains an open issue.

Rather than explicitly applying stringent constraints on the kinematic variables, in this work a large part of the available kinematic phase space is explored within reasonable limits and the azimuthal modulations of interest studied in that kinematic region.
In addition, in order to facilitate interpretation of the results, kinematic distributions are provided for the various choices of kinematic binning and hadron species. In this way, the door is open for phenomenology to explore in more detail whether and where the factorized picture might break down for these spin asymmetries.

The particular choice of kinematic distributions provided here are driven by the two aspects considered in the beginning of this section, namely (i) the separation of current and target fragmentation as studied through rapidity distributions, and (ii) the small transverse-momentum requirement as explored by looking at both \Q versus \(\Phperpabs^{2}\) and \Q versus \(\Phperpabs^{2}/\z^{2}\).

A presentation in this paper of the distributions for all kinematic bins and hadron species is not practical, they will hence be made available elsewhere \cite{supplemental}.
Instead, a selection of those are presented for the more extreme cases.

\subsection{Separation of target and current fragmentation}\label{sec:rapidity}

\begin{figure}
 \centering
 \includegraphics[width=0.48\textwidth,keepaspectratio]{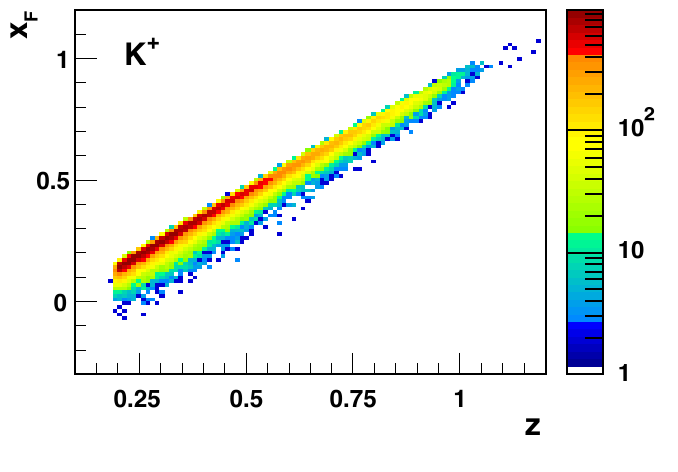}%
 \includegraphics[width=0.48\textwidth,keepaspectratio]{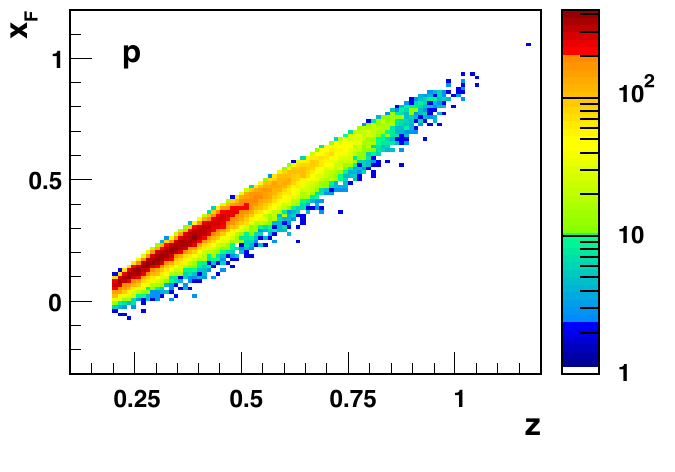}
 \caption{Distributions in \xf vs.~\z  of the \kplus (left) and proton (right) yields.}
 \label{fig:xFdistribrution}
\end{figure}

\begin{figure}
 \centering
 \includegraphics[width=\textwidth,keepaspectratio]{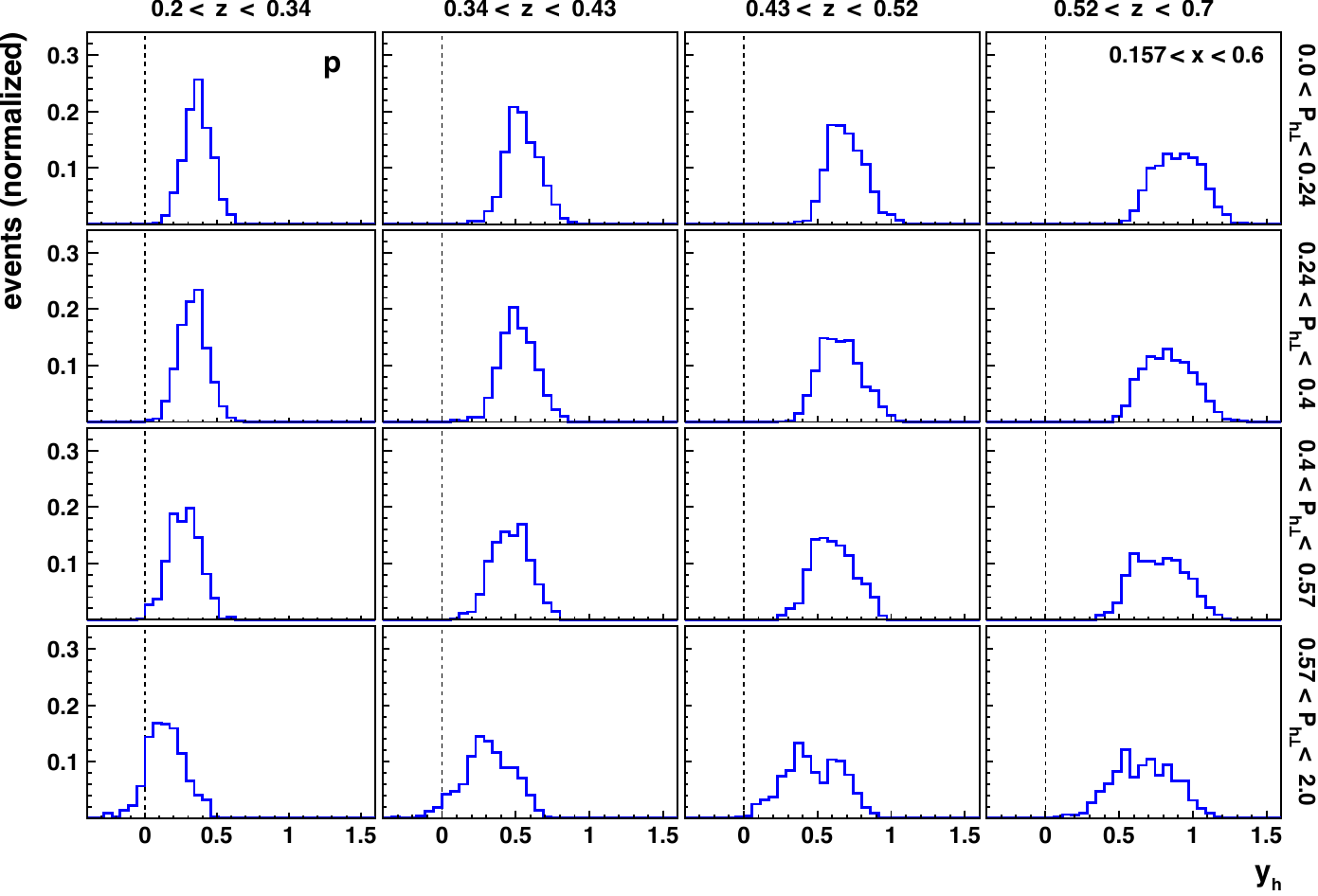}
 \caption{Rapidity distributions for protons in the various (\z, \Phperpabs) bins of the last \xb bin. The dashed lines indicate zero rapidity.}
  \label{fig:rapidity-proton}
\end{figure}

\begin{figure}
 \centering
 \includegraphics[width=\textwidth,keepaspectratio]{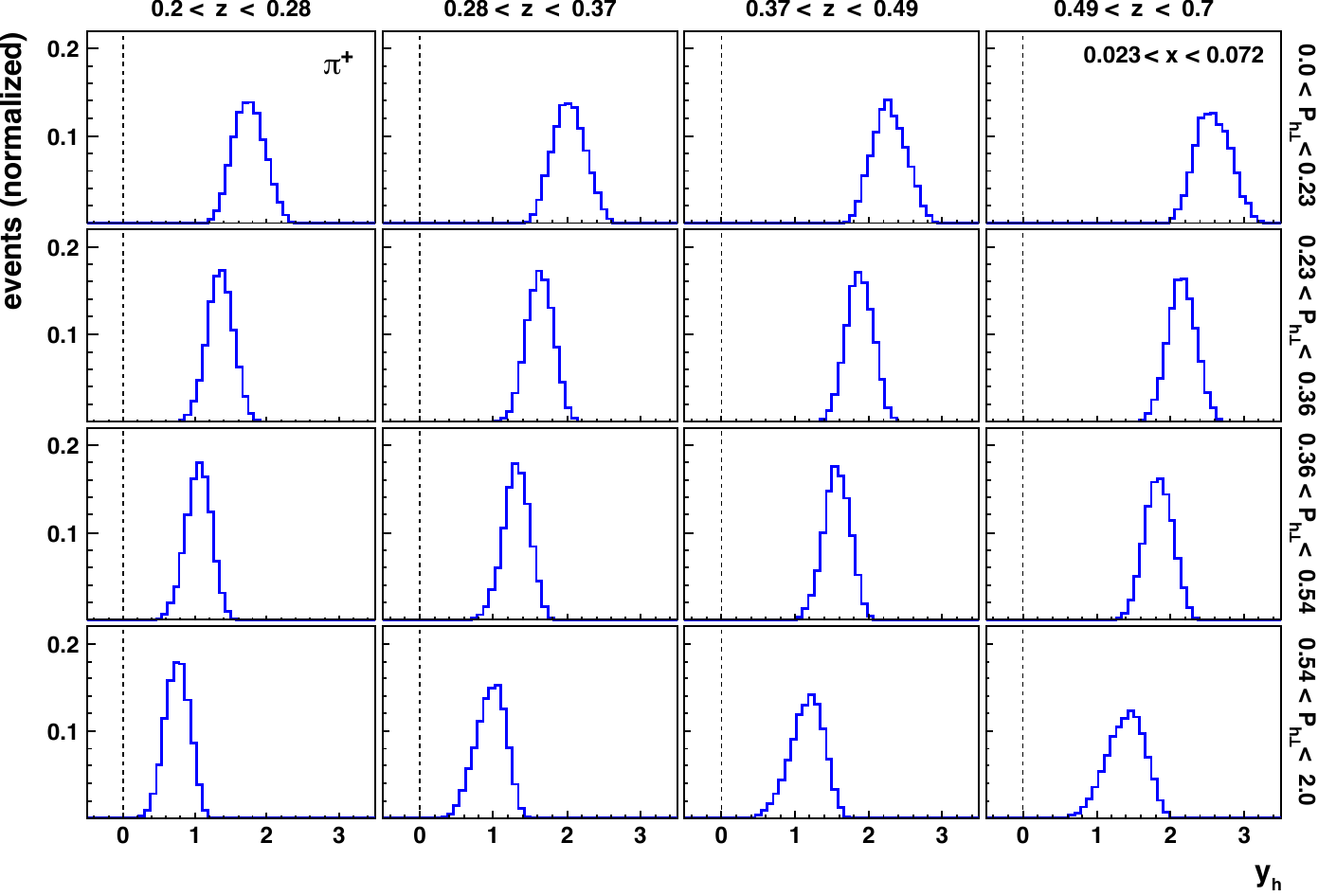}\\[8mm]
 \includegraphics[width=\textwidth,keepaspectratio]{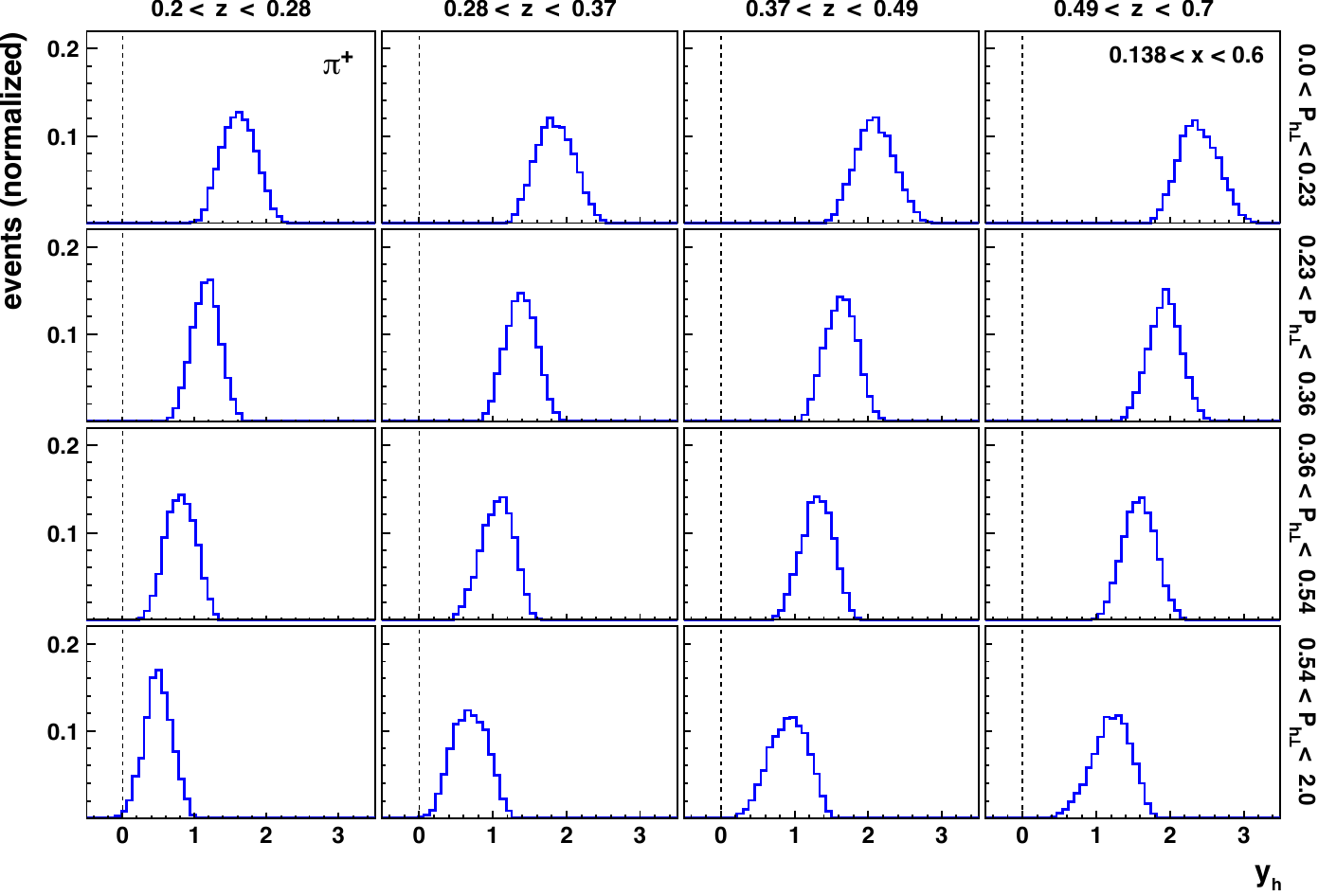}
 \caption{Rapidity distributions for \piplus in the various (\z, \Phperpabs) bins of the first (top) and last (bottom) \xb bins. The dashed lines indicate zero rapidity.}
 \label{fig:rapidity-piplus}
\end{figure}

In this measurement, hadrons were selected that have a high probability to stem from the current fragmentation. 
For that a minimum \z of 0.2 is required, which predominantly selects forward-going hadrons in the virtual-photon--proton center-of-mass system,
forward being the direction of the virtual photon. 
This is visible in figure~\ref{fig:xFdistribrution}, where the correlation between \z and \xf is plotted for both \kplus and protons. 
For kaons (and likewise pions), \(\z>0.2 \) corresponds to positive \xf. 
The situation is slightly less favorable for protons,
where still a notable fraction of the yield in the lowest \z bin falls in the category of negative \xf. 
This can be seen also in the rapidity distributions. 
They are depicted in figure~\ref{fig:rapidity-proton} for the last \xb bin, 
while those for pions are shown for the first and last \x bin in figure~\ref{fig:rapidity-piplus}.
From those distributions it is evident that the majority of events
is at forward rapidity. 
Only a small fraction of events, mainly in the case of protons, populates the region of negative rapidity and do so only for large \Phperpabs and small \z.
Furthermore, clearly visible in the \piplus figure is a general increase of rapidity with increasing  \z as well as when decreasing \Phperpabs and \xb.

\subsection{Transverse-momentum versus hard scale}\label{sec:TMDscale}

\begin{figure}
 \centering
 \includegraphics[width=\textwidth,keepaspectratio]{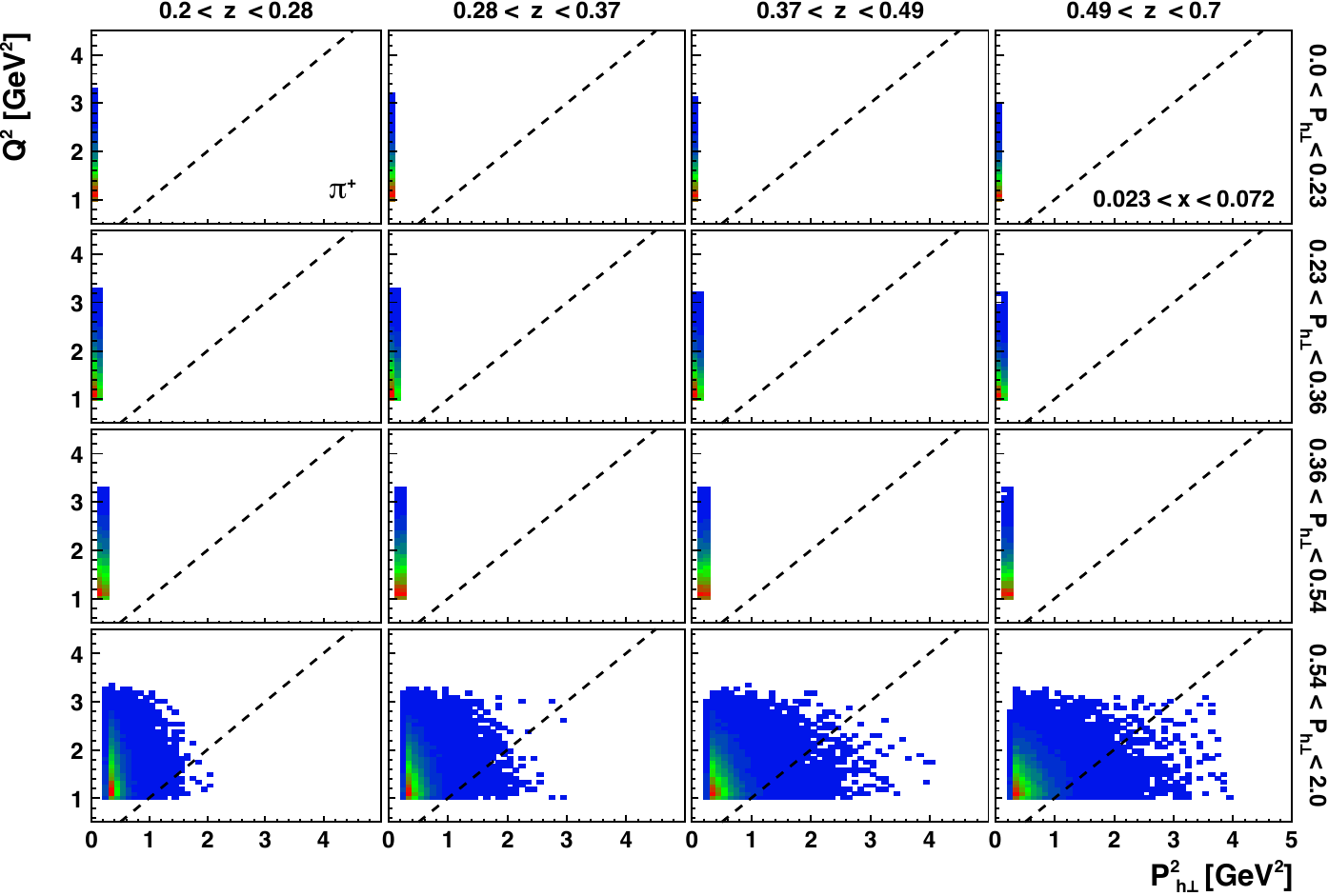}\\[8mm]
 \includegraphics[width=\textwidth,keepaspectratio]{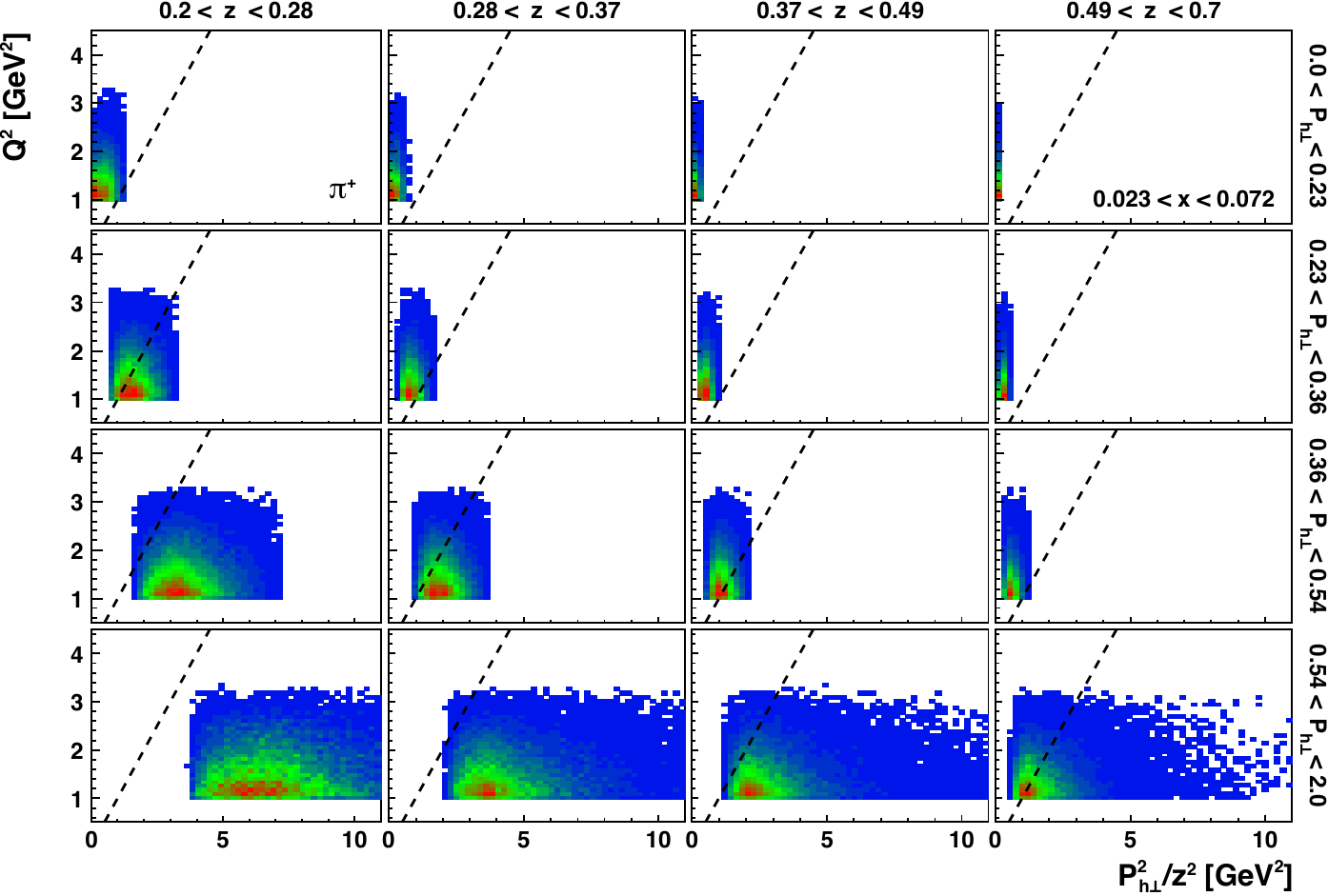}
 \caption{The distribution of \piplus events in the \Q--\Phperpsqr (top) and \Q--\(\Phperpsqr/\z^{2}\) (bottom) planes 
                for the various (\z, \Phperpabs) bins of the lowest \xb bin. The \(\Q = \Phperpsqr\) (top) and 
                \(\Q = \Phperpsqr/\z^{2}\) (bottom) boundaries are indicated by dashed lines.}
 \label{fig:piplus_TMDscale}
\end{figure}

The interpretation of transverse-momentum-dependent azimuthal distributions in terms of TMD PDFs and FFs as discussed in section~\ref{sec-theory} requires the presence of one hard scale (\Q) --- which is much larger than a typical nonperturbative-QCD scale like the proton mass or $\Lambda_\text{QCD} \cong 0.3$ GeV, the QCD-scale parameter --- and transverse momentum that is small in comparison to \Q.
Under these conditions, the transverse momentum of the hadron observed can be interpreted as originating from non-pertubative sources in the initial proton structure and in the fragmentation process (including their calculable variations with the hard scale).
By contrast, in the region of large transverse momentum, perturbative-QCD radiation is the primary source of the observed transverse momentum of the final-state hadron. This is typically accompanied by a 1/\Phperpabs suppression of the observable, which usually can be interpreted in terms of collinear PDFs and FFs. In the intermediate region of relatively large transverse momentum but still larger \Q, these two descriptions are expected to match their behaviors for a number of azimuthal modulations studied here~\cite{Bacchetta:2008xw}.

In this measurement, \Phperpabs is of the order of the QCD scale. However, \Q is neither always very large compared to the proton mass nor compared to the transverse momentum. Under such conditions, subtleties in the definition of the transverse momentum can also become relevant. 
One way of testing the requirement of small transverse momentum is comparing directly \Phperpabs and \Q.
A different choice of transverse momentum, one that is in particular convenient in factorization proofs of transverse-momentum-dependent processes, is that of the virtual boson in the frame where the two hadrons involved (initial and final in case of semi-inclusive \dis) are collinear; this choice is commonly denoted as \qt. For large enough \Q, \(\qtsqr \simeq \Phperpsqr/\z^{2}\), from which follows the requirement of \(\Phperpsqr/\z^{2}\ll\Q\).

In figure~\ref{fig:piplus_TMDscale}, the two different transverse momentum scales, \Phperpsqr and  \qtsqr, are compared to \Q for \piplus in the 16 (\z, \Phperpabs) bins of the lowest \xb bin.
Because \xb and \Q are highly correlated in this measurement, the lowest \xb bin corresponds to the region of lowest \Q and hence the region for which the TMD-factorization requirement of small transverse momentum relative to a single hard scale is the more difficult one to fulfill. As visible in the top plot of the figure, for \(\Phperpabs < 0.54\)~GeV all events are above the \(\Q = \Phperpsqr\) diagonal, i.e., the ``safe'' region. Only in the highest \Phperpabs bin, a small fraction of events are below that diagonal. For larger values of \xb, the situation is even more favorable with a completely negligible fraction of events in the region of  \(\Q < \Phperpsqr\).
Even though only presented here for the \piplus sample, these observations equally hold for the other hadrons considered in this measurement.

\begin{figure}
\centering
\includegraphics[width=\textwidth,keepaspectratio]{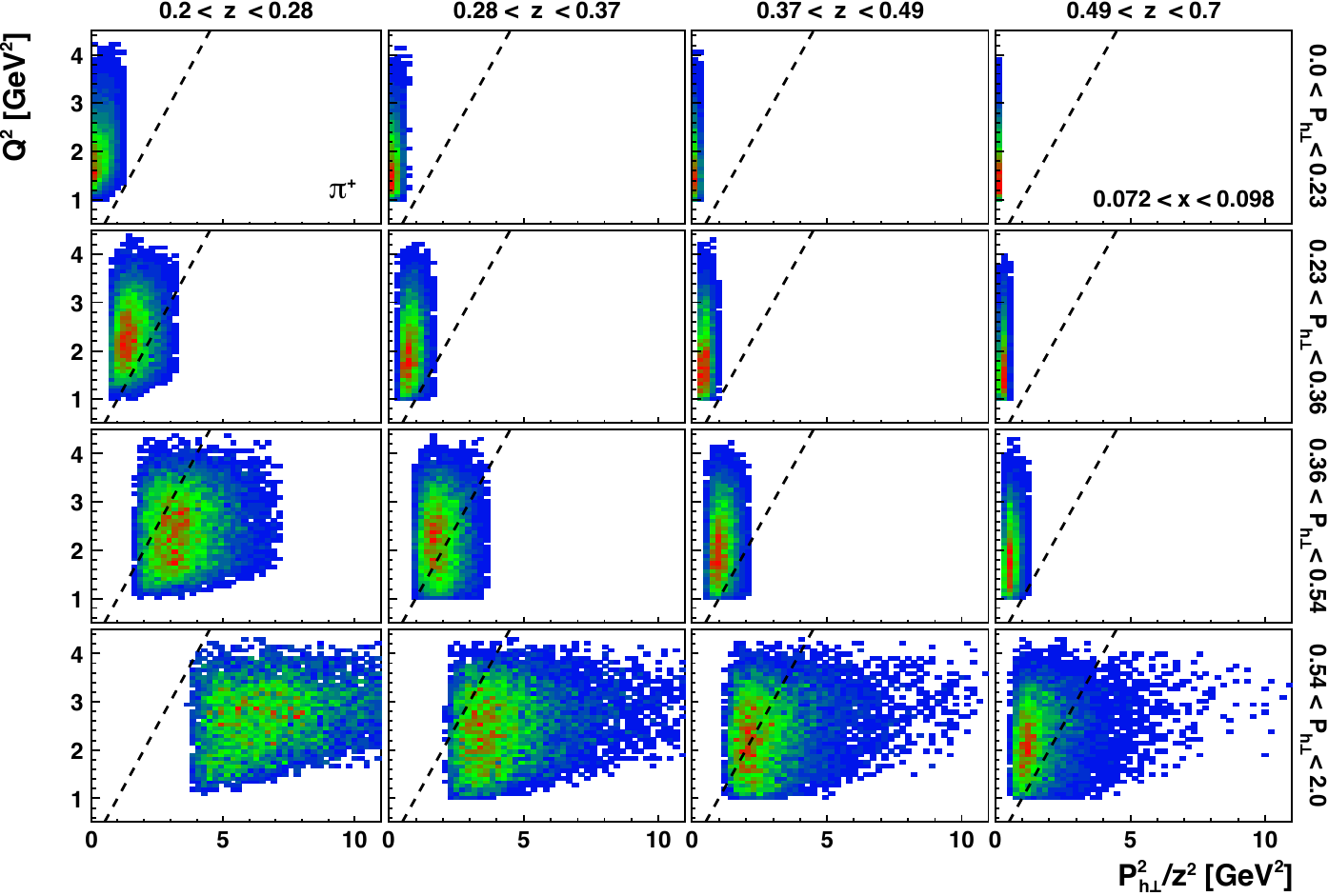}\\[8mm]
\includegraphics[width=\textwidth,keepaspectratio]{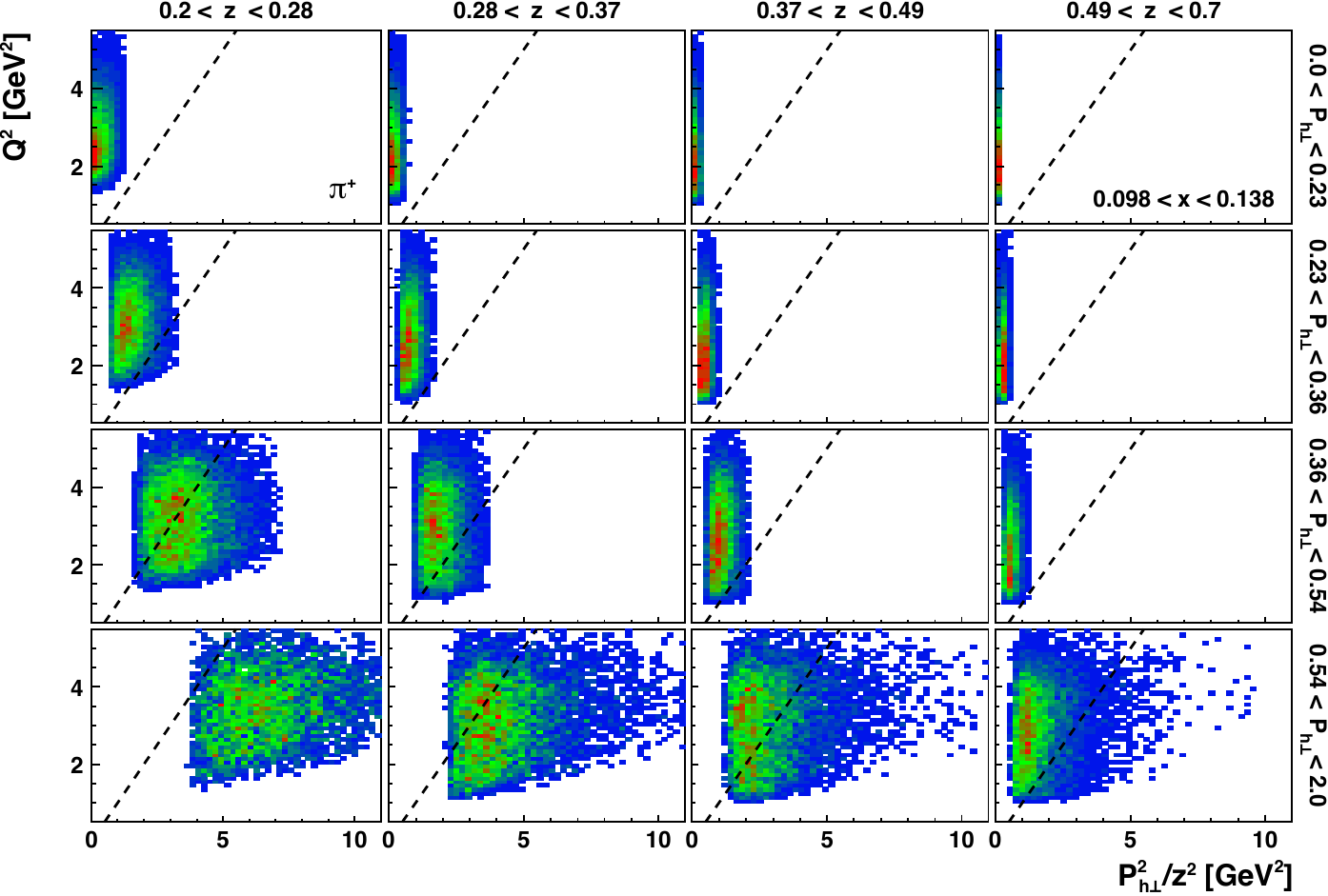}
  \caption{The distribution of \piplus events in the \Q--\(\Phperpsqr/\z^{2}\) plane 
                for the various (\z, \Phperpabs) bins of the second (top) and third (bottom) \xb bin. The  
                \(\Q = \Phperpsqr/\z^{2}\) boundaries are indicated by dashed lines.}
 \label{fig:piplus_TMDscale23}
\end{figure}

\begin{figure}
\centering
\includegraphics[width=\textwidth,keepaspectratio]{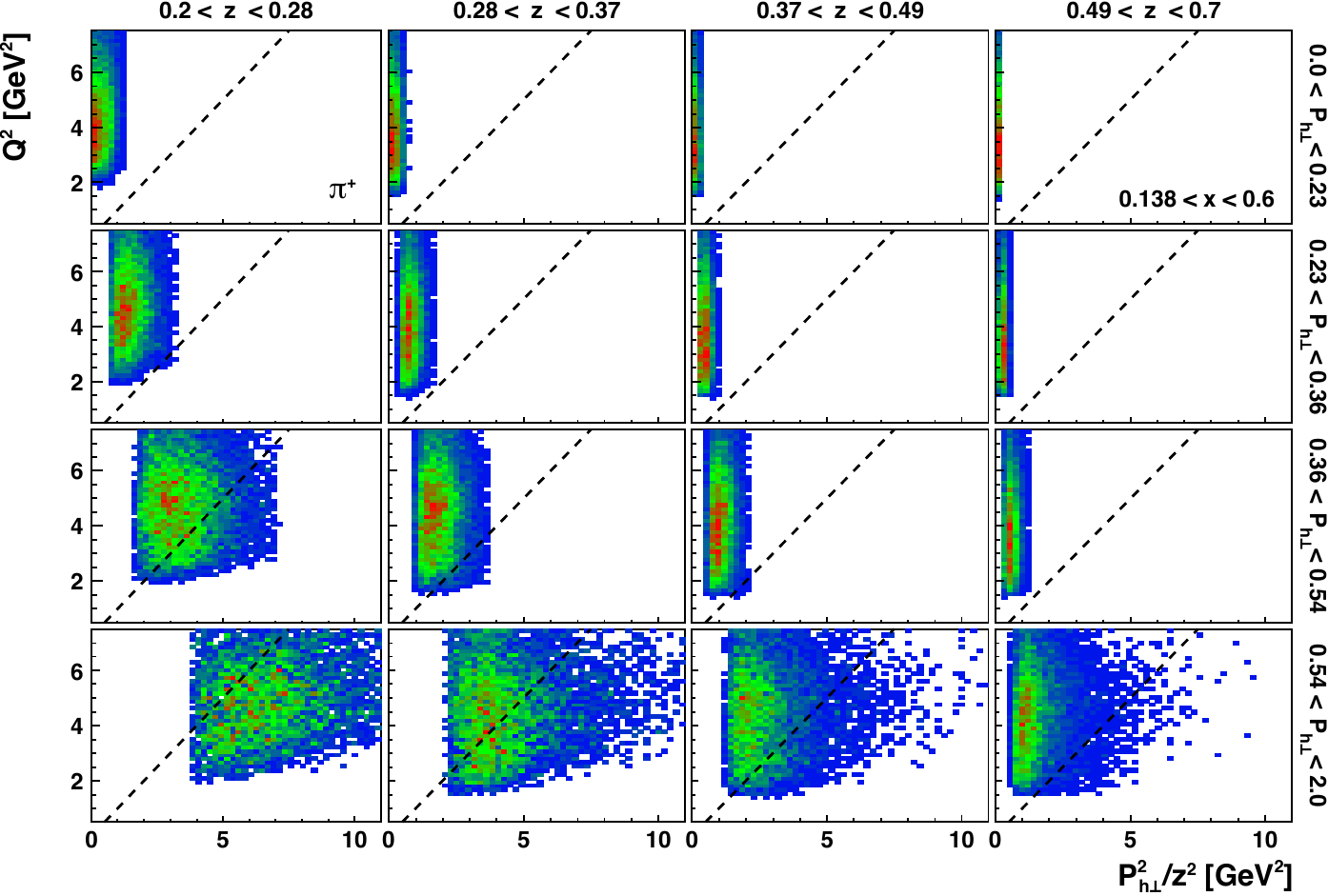}
  \caption{The distribution of \piplus events in the \Q--\(\Phperpsqr/\z^{2}\) plane 
                for the various (\z, \Phperpabs) bins of the highest \xb bin. The  
                \(\Q = \Phperpsqr/\z^{2}\) boundaries are indicated by dashed lines.}
 \label{fig:piplus_TMDscale4}
\end{figure}

The behavior changes significantly when instead the \Q is plotted against \(\Phperpsqr/\z^{2}\), shown in the bottom plot of figure~\ref{fig:piplus_TMDscale}.
The requirement of much larger \Q becomes more stringent due the rescaling of the transverse momentum by \(1/\z^{2}\), which becomes a large factor for the low-\z region. As a consequence, only in the lowest \Phperpabs bin of the lowest \z bin the majority of \piplus events fall in the region above the  \(\Q = \Phperpsqr/\z^{2}\) diagonal. Already in the second \Phperpabs bin the opposite is the case: most of the events populate the region below that diagonal. Going to bins of larger \Phperpabs aggravates this situation, up to a point where the majority of events falls in the ``unsafe''  region for all \z bins of the semi-inclusive region. As before, the \piplus case is exemplary for all the hadrons considered in this measurement. 

The situation improves, as expected from the existing \xb--\Q correlation, when considering larger values of \xb. This is demonstrated in figures~\ref{fig:piplus_TMDscale23} and \ref{fig:piplus_TMDscale4}, where the \Q vs.~\(\Phperpsqr/\z^{2}\) distributions for \piplus are shown for successively increased \xb.

\begin{figure}
\centering
\includegraphics[width=0.94\textwidth,keepaspectratio]{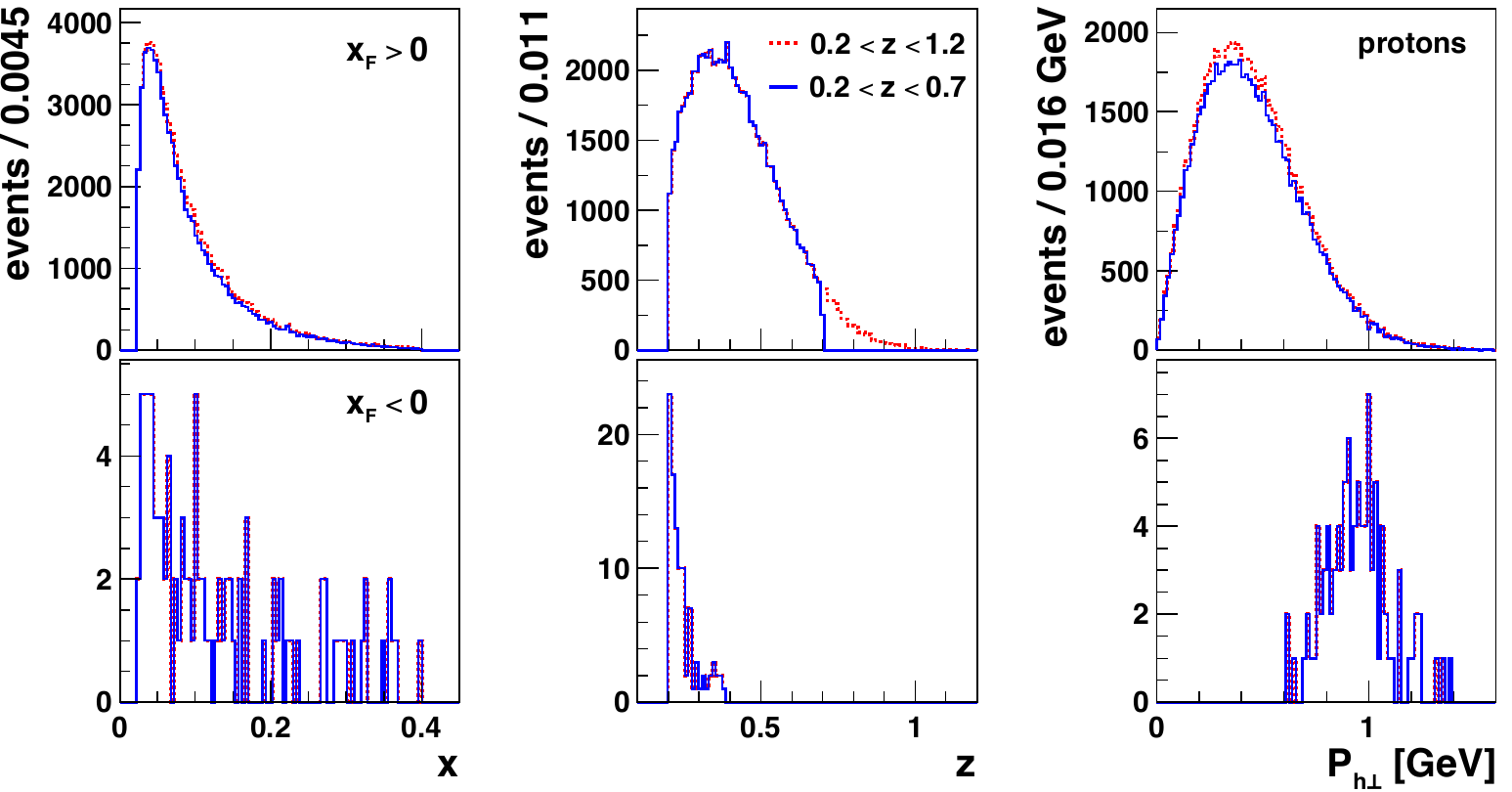}
  \caption{The distribution of proton events versus \xb, \z, and \Phperpabs 
                for both the full range in \z of 0.2--1.2 (red dashed lines) and only the semi-inclusive \z range (blue lines).
               Data with positive \xf are presented in the top row. In the bottom, the corresponding distributions for negative \xf are shown.}
 \label{fig:xzPhperp-xF-protons}
\end{figure}

Figure~\ref{fig:xzPhperp-xF-protons} illustrates both the effect of the upper \z constraint of 0.7 and of a minimum requirement of \(\xf > 0\) (not applied in this measurement) on the proton-yield distributions as functions of \xb, \z, and \Phperpabs. In particular, apart from extending the \z spectrum to larger values, there is no visible qualitative change of the various distributions when including the high-\z range. On the other hand, data for negative \xf are populated in the region of large \Phperpabs as is expected. This feature of the data is observed for all hadrons, albeit even further suppressed in case of mesons.

\section{``Polarizing'' \pythia for the estimate of systematic uncertainties}\label{app:systematics}

One of the major challenges of such semi-inclusive measurements as presented here is the
evaluation of detector effects, in particular the influence of a finite kinematic acceptance on the
Fourier amplitudes extracted. A rigorous analysis procedure involves a fully differential unfolding
as done, e.g., for the \hermes measurement of the cosine modulations in the 
polarization-averaged semi-inclusive \dis cross section~\cite{Airapetian:2012yg}. 
Here, the limited number of events precludes an unfolding in six dimensions. However, being effectively an
asymmetry measurement results in various approximate cancelations of detector effects.
Nevertheless, even though the angular Fourier decomposition uses a maximum-likelihood fit unbinned in the azimuthal angles, 
the limited instrumental acceptance in the remaining kinematic variables can still influence the
measurement~\cite{Schnell:2015gaa}, especially if not performed differential in {\em all} the remaining kinematic variables.

Monte Carlo simulations of both the underlying physics as well as of the detector 
response have become a vital tool for evaluating such systematic effects. 
The basis for those is a reliable modeling of the experimental setup but also
realistic simulations of the physics processes.
The measurements presented here enter a territory for which the latter are scarce, 
mainly due to a lack of knowledge about the various TMDs. 
Several dedicated physics generators have become available, but none that covers all the
TMDs and modulations examined here.

The approach chosen in this analysis makes use of an already very good description of the spin-independent semi-inclusive \dis cross section provided by \pythia~\cite{Sjostrand:2000wi,Sjostrand:2001yu}. 
\pythia events come with event weights equal to unity and are hence easy to reshuffle. 
This is exploited to introduce spin dependence into the otherwise spin-independent event generator~\cite{Pappalardo:2008zza,Diefenthaler:2010zz,Schnell:2015gaa}. 
A polarization state \(\mathcal{P}\) is assigned to each event $i$ based on a model of the spin asymmetry of interest, e.g.,
\begin{align}
 \rho &< \frac{1}{2} \left[ 1 + \mathcal{A}_{\text{U}\perp}^{\sin(\phih-\phis)} (\Omega^{i}) \sin(\phih^{i}-\phis^{i}) \right]  \quad  \Rightarrow \quad \mathcal{P} = +1 \label{eq:plus}\\
 \rho &> \frac{1}{2} \left[ 1 + \mathcal{A}_{\text{U}\perp}^{\sin(\phih-\phis)} (\Omega^{i}) \sin(\phih^{i}-\phis^{i}) \right]  \quad  \Rightarrow \quad \mathcal{P} = -1 \label{eq:minus}
\end{align}
in case of the Sivers Fourier amplitude, by throwing a random variable \(0<\rho<1\). Here, \((\Omega^{i},\phih^{i},\phis^{i})\) are the fully differential {\em true} kinematics for that particular event and \(\mathcal{A}_{\text{U}\perp}^{\sin(\phih-\phis)} \) is a suitable parameterization for the Sivers modulation. In the specific analysis, eqs.~\eqref{eq:plus} and \eqref{eq:minus} are to include all ten azimuthal modulations including the double-spin asymmetries. Virtually any parameterization of the spin dependence can be implemented (as long as fulfilling positivity constraints) without limiting oneself to, e.g., the Gaussian Ansatz for the transverse-momentum dependence. In addition, the full event will remain available, which allows a more thorough study of systematics due to event-topology--dependent detector responses.

\begin{figure*}
 \centering
 \includegraphics[bb = 18 10 500 423, clip, width=.495\textwidth]{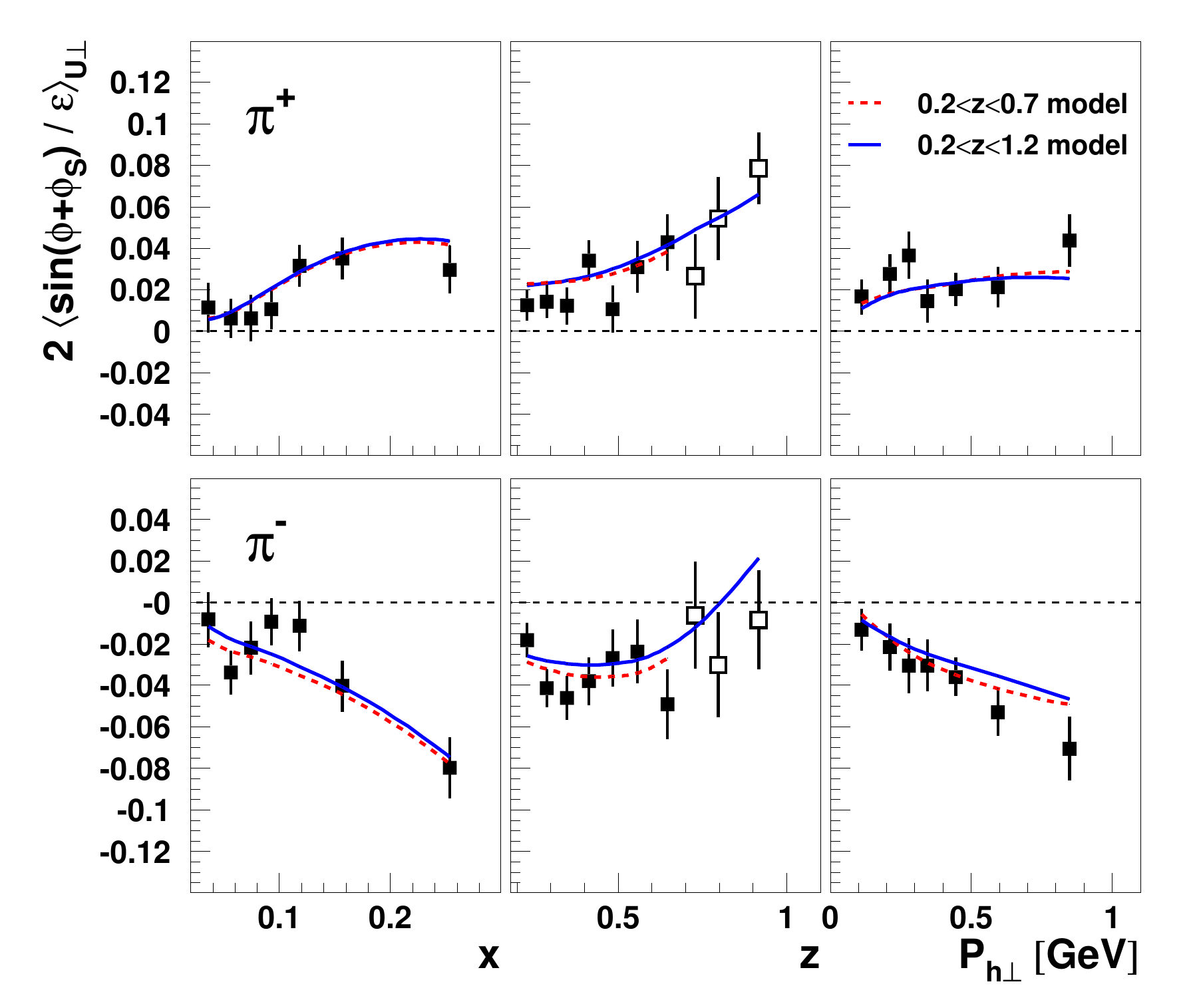}
 \includegraphics[bb = 18 10 500 423, clip, width=.495\textwidth]{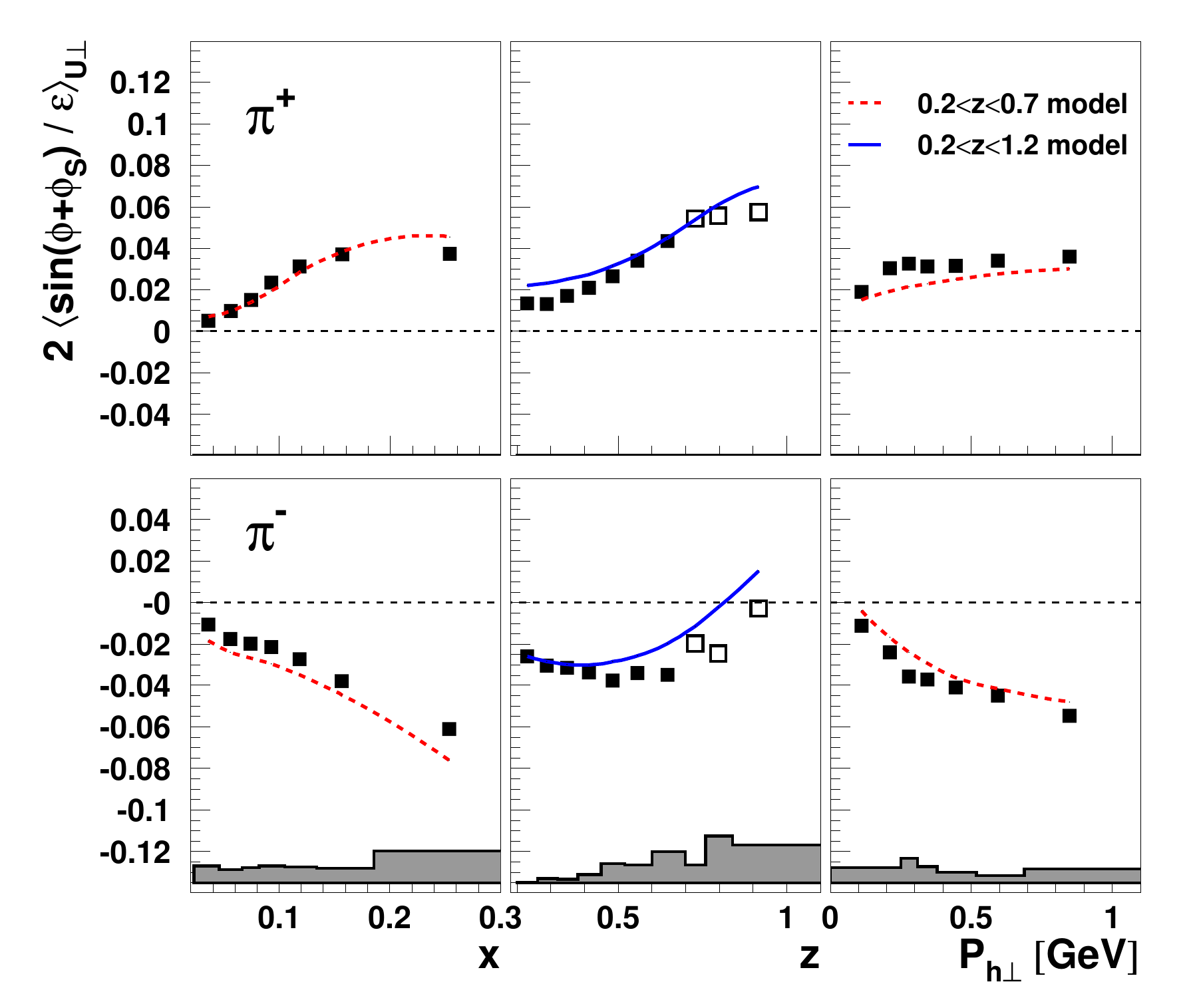}
 \caption{Left: Comparison of the \hermes data for charged-pion Collins \sfa amplitudes 
 		with the fully differential model of those evaluated at the average kinematics of each bin. 
		The dashed curve (red) uses the model based on data in the standard \(0.2 < \z < 0.7\) range, 
		while the solid line (blue) includes also the high-\z data. 
		Right: Comparison of the fully differential models evaluated at the average kinematics of each bin 
		with the fully reconstructed ``polarized \pythia'' simulation (in \hermes acceptance) based on those models. 
		The difference is assigned as systematic uncertainty and shown as uncertainty bands at the bottom of each panel.}
\label{fig:mcdada-comparison}       
\end{figure*}

Given the scant availability of parameterizations for {\em all} modulations studied here, a data-driven approach is employed. An approximate model of reality is obtained by expanding the various Fourier amplitudes measured in a Taylor series in all kinematic variables. A maximum-likelihood fit is employed to extract the coefficients of the fully differential (though truncated) Taylor series for every single azimuthal amplitude appearing in the cross section and for every hadron type. 
These parameterizations are then used to assign spin states to the \pythia Monte Carlo simulation --- augmented with  \radgen~\cite{Akushevich:1998ft} to account for \QED radiative
effects and passed through a \geant\cite{Brun:1978fy} description of the \hermes 
apparatus (including the RICH particle-identification inefficiencies) --- according to eqs.~\eqref{eq:plus} and \eqref{eq:minus}, with the proper inclusion of all the modulations. The resulting asymmetry amplitudes, reconstructed in the same way as those of the actual \hermes data, are  compared to the latter to further tune the truncation of the Taylor series. 
As an example, in figure~\ref{fig:mcdada-comparison} (left) a comparison of the fully differential model extracted with the \hermes data is provided for the Collins \sfa amplitudes of charged pions. 
Limitations stemming from the truncation of the Taylor series might be present. While it is not a principle problem to include additional terms, it turns into a more practical problem, especially when attempting to parameterize all spin-dependent terms in the semi-inclusive \dis cross section, and as a result approaching the usual limit of, e.g., standard \minuit~\cite{James:1975dr}, on how many parameters can be determined simultaneously.

In this work most Fourier amplitudes are found to be consistent with zero.
In order to keep a finite number of parameters, the following choice was made concerning the parameterization of the fully differential model: 
\begin{enumerate}[label=(\roman*)]
 \item For the three Fourier components that exhibit larger asymmetries and non-linear kinematic dependences (Sivers, Collins, and the \sinemodulation{\phis} modulation), 
 all the constant and terms linear in \xb, \z, \Phperpabs, and \Q as well as the 2\(^{\text{nd}}\)-order terms in \xb, \z, \Phperpabs, i.e.,  eleven parameters in total for each modulation, are fit to data.
 \item For all remaining Fourier components, only the constant and terms linear in \xb, \z, \Phperpabs, and \Q, i.e., five parameters for each modulation, are included.
\end{enumerate} 
The model was expanded around the mean kinematics and fit to data either in the default semi-inclusive range of \(0.2 < \z < 0.7\) or in the extended \z range.
The same model was used for the systematics of both the \csa and \sfa and was extracted employing the \sfa probability density \eqref{eq:SFA-pdf} in the maximum-likelihood fit.
Variations of the parameterization of the fully differential model were considered and found to give consistent results for these systematic uncertainty.

Antiprotons and neutral pions were treated slightly different due to a lack of statistical precision. More specifically, 
for the antiproton model, only the standard \(0.2 < \z < 0.7\) range is used as there is not a sufficient number of events at larger values of \z.
Furthermore, only the constant and terms linear in \xb, \z, \Phperpabs, and \Q are kept in the Taylor expansions of all ten Fourier amplitudes.
The neutral-pion model is constructed using the much better constrained charged-pion models under the assumption of isospin symmetry, i.e.,
\begin{equation}
\mathcal{A}^{\pizero} \equiv \mathcal{A}^{\text{isospin}} = \frac{\mathcal{A}^{\piplus}+C \cdot \mathcal{A}^{\piminus}}{(1+C)}
  \label{eq:isospin_relation}
\end{equation}
where \( \mathcal{A}^{\piplus} \) and \( \mathcal{A}^{\piminus} \) are the fully differential models for \piplus and \piminus, respectively, and the coefficient $C$ represents the ratio of the polarization-averaged semi-inclusive DIS cross-sections for negative and positive pion production. In the present analysis the value of $C$ was approximated by
\begin{equation}
 C \equiv \frac{\sigmauu{\piminus}}{\sigmauu{\piplus}} \approx \frac{\langle M^{\piminus} \rangle}{\langle M^{\piplus} \rangle} \approx 0.374 \, 
 \label{eq:C}
\end{equation}
using the average, \( \langle M^{\pi^\pm} \rangle \), of the \( \pi^\pm \) multiplicities~\cite{Airapetian:2012ki}.

\begin{figure*}
\centering
\includegraphics[width=.67\textwidth,bb=29 52 514 545, clip]{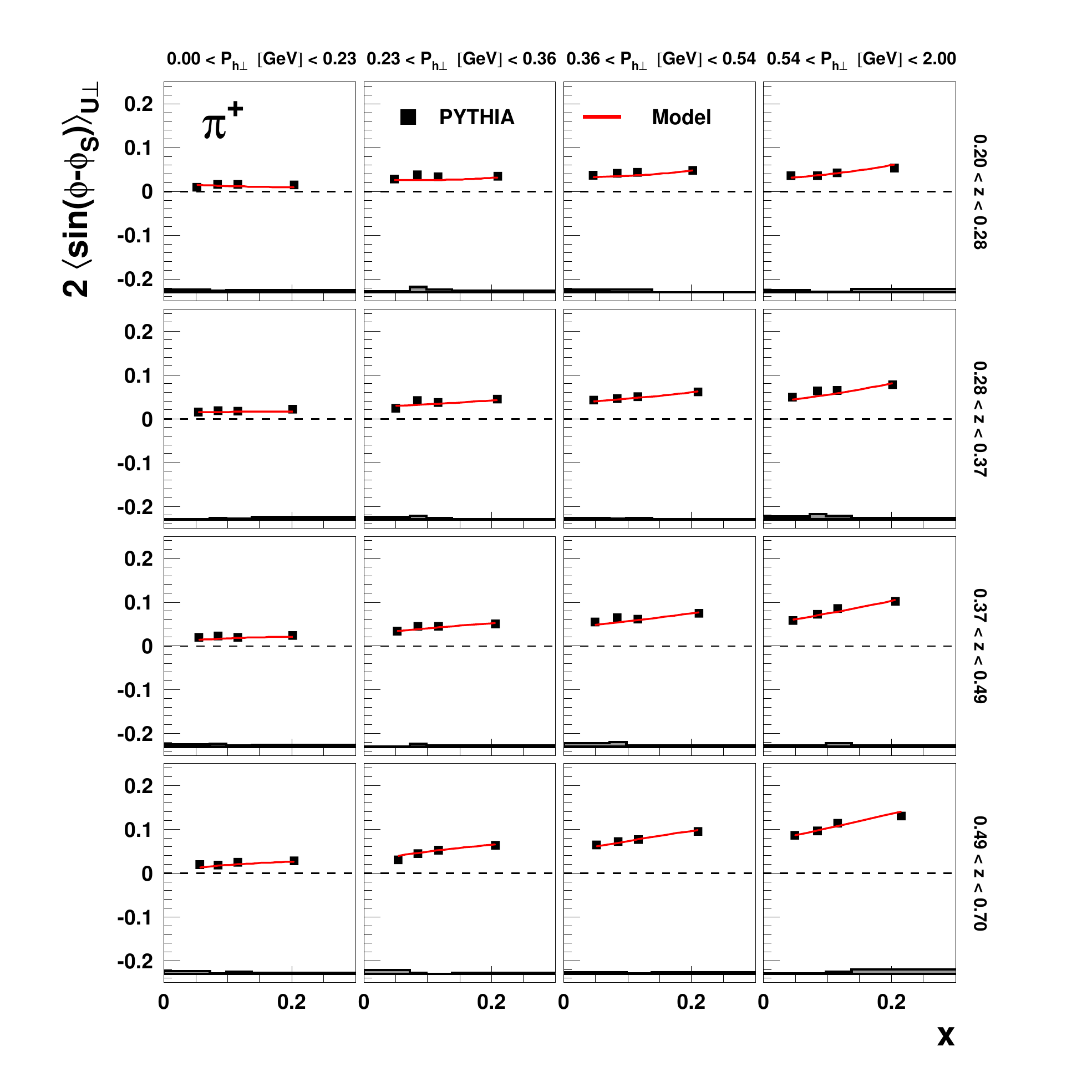}\\[2mm]
\includegraphics[width=.67\textwidth,bb=29 25 514 525, clip]{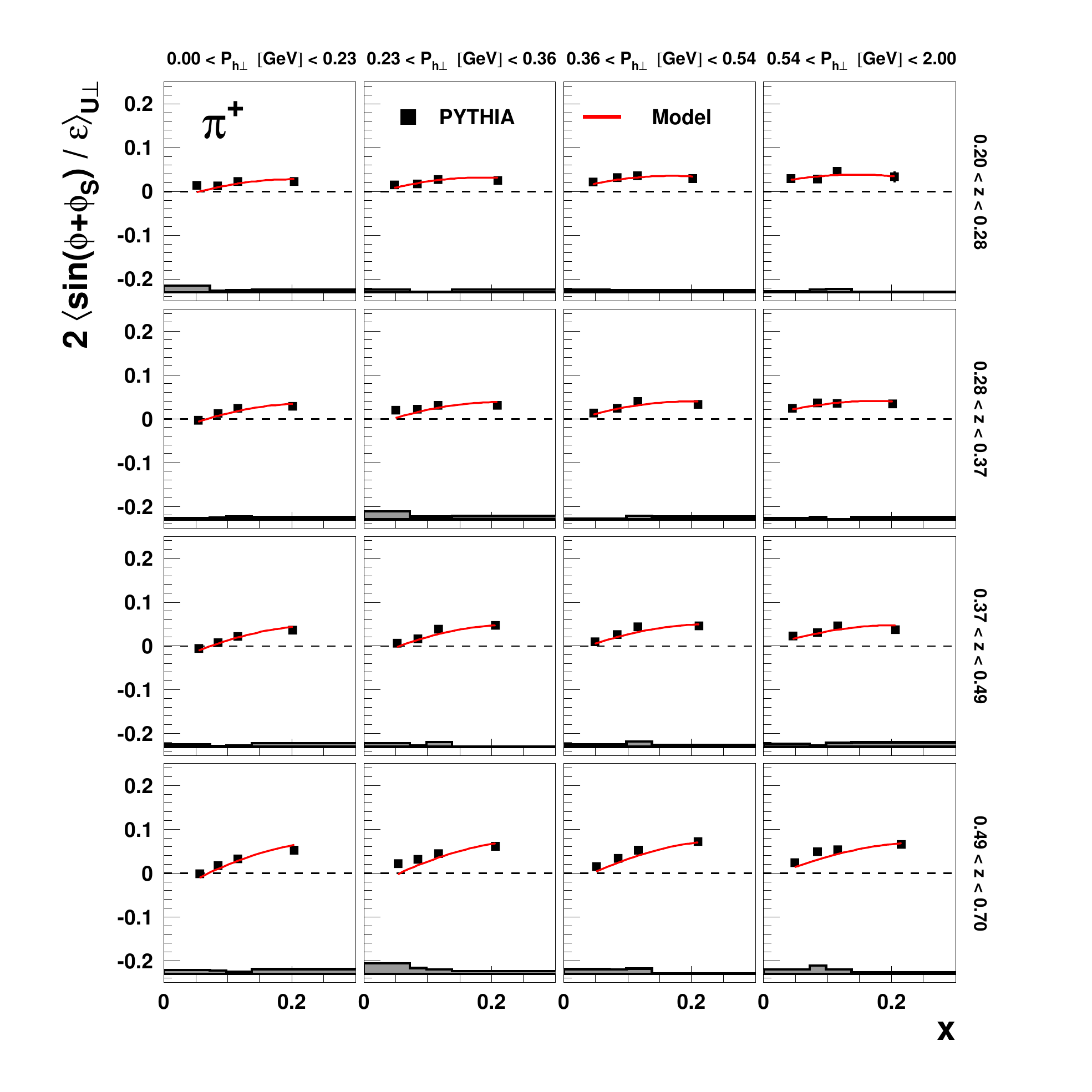}
\caption{Comparison of the Sivers (top) and Collins (bottom) asymmetries extracted from the ``polarized \pythia'' simulation in the \hermes acceptance with the respective input parameterizations evaluated at the average kinematics of each data point.
The difference is assigned as systematic uncertainty and shown as uncertainty bands at the bottom of each panel.}
\label{fig:sys-extraction}       
\end{figure*}

Figures~\ref{fig:mcdada-comparison} (right) and \ref{fig:sys-extraction} illustrate the subsequent extraction of systematic uncertainties.
The ``polarized \pythia'' events were tracked through a realistic simulation of the experiment and analyzed in the same way as normal experimental data. 
The reconstructed asymmetry amplitudes
are compared to the parameterizations evaluated at the mean reconstructed kinematics of each data point, i.e., in each experimental bin.
(This is the same as how the data are usually used in phenomenological fits, e.g., interpreted as the true value of the observable for the average kinematics given alongside.)
In each kinematic bin, the difference of the reconstructed Monte Carlo asymmetries and the parameterization, e.g.,
\begin{equation}
\delta_{\text{sys}} \left( \siversexp{} \right) \; \equiv \; \mid \siversexp{\text{MC}} - \mathcal{A}_{\text{U}\perp}^{\sin(\phih-\phis)} (\langle \Omega \rangle_{\text{bin}} ) \mid
\end{equation}
stems from detector effects including smearing, but more importantly from the integration over kinematic variables, and is assigned as the corresponding systematic uncertainty.

It is worthwhile to highlight that the difference of an average asymmetry in a bin and the asymmetry value at the average kinematics of that bin 
strongly depends on the non-linearity of the asymmetry and the kinematic region integrated over. 
That makes the one-dimensional projections much more susceptible to acceptance effects than the three-dimensional data presented as the main results in this analysis.



%% file: hermesTMDs.bbl
\providecommand{\href}[2]{#2}\begingroup\raggedright\begin{thebibliography}{100}

\bibitem{EPJA52}
M.~Anselmino, M.~Guidal and P.~Rossi, eds., \emph{{The 3-D Structure of the
  Nucleon}}, vol.~52 of \emph{{Eur. Phys. J. A}}, Springer (2016).

\bibitem{Gao:2017yyd}
J.~Gao, L.~Harland-Lang and J.~Rojo, \emph{{The structure of the proton in the
  LHC precision era}},
  \href{https://doi.org/10.1016/j.physrep.2018.03.002}{\emph{Phys. Rept.}
  {\bfseries 742} (2018) 1} [\href{https://arxiv.org/abs/1709.04922}{{\ttfamily
  1709.04922}}].

\bibitem{Collins:2011zzd}
J.~Collins, \emph{{Foundations of perturbative QCD}}, Cambridge University
  Press (2013).

\bibitem{Jaffe:1996zw}
R.~L. Jaffe, \emph{{Spin, Twist and Hadron Structure in Deep Inelastic
  Processes}},  in \emph{Lectures on QCD: Applications}, F.~Lenz,
  H.~Grie{\ss}hammer and D.~Stoll, eds., vol.~496 of \emph{{Lecture Notes in
  Physics}}, (Berlin, New York), pp.~178--249, Springer, 1997.

\bibitem{Engelhardt:2019lyy}
M.~Engelhardt, J.~Green, N.~Hasan, S.~Krieg, S.~Meinel, J.~Negele et~al.,
  \emph{{Quark orbital angular momentum in the proton evaluated using a direct
  derivative method}}, \href{https://doi.org/10.22323/1.334.0115}{\emph{PoS}
  {\bfseries LATTICE2018} (2018) 115}
  [\href{https://arxiv.org/abs/1901.00843}{{\ttfamily 1901.00843}}].

\bibitem{Alexandrou:2020sml}
{\scshape Extended Twisted Mass} collaboration, \emph{{Complete flavor
  decomposition of the spin and momentum fraction of the proton using lattice
  QCD simulations at physical pion mass}},
  \href{https://doi.org/10.1103/PhysRevD.101.094513}{\emph{Phys. Rev. D}
  {\bfseries 101} (2020) 094513}
  [\href{https://arxiv.org/abs/2003.08486}{{\ttfamily 2003.08486}}].

\bibitem{Bacchetta04}
A.~Bacchetta, U.~D'Alesio, M.~Diehl and C.~A. Miller, \emph{{Single-spin
  asymmetries: The Trento conventions}},
  \href{https://doi.org/10.1103/PhysRevD.70.117504}{\emph{Phys. Rev. D}
  {\bfseries 70} (2004) 117504}
  [\href{https://arxiv.org/abs/hep-ph/0410050}{{\ttfamily hep-ph/0410050}}].

\bibitem{Bacchetta:2006tn}
A.~Bacchetta, M.~Diehl, K.~Goeke, A.~Metz, P.~J. Mulders and M.~Schlegel,
  \emph{{Semi-inclusive deep inelastic scattering at small transverse
  momentum}}, \href{https://doi.org/10.1088/1126-6708/2007/02/093}{\emph{JHEP}
  {\bfseries 02} (2007) 093}
  [\href{https://arxiv.org/abs/hep-ph/0611265}{{\ttfamily hep-ph/0611265}}].

\bibitem{Collins:1981uw}
J.~C. Collins and D.~E. Soper, \emph{{Parton distribution and decay
  functions}}, \href{https://doi.org/10.1016/0550-3213(82)90021-9}{\emph{Nucl.
  Phys.} {\bfseries B194} (1982) 445}.

\bibitem{Collins:1984kg}
J.~C. Collins, D.~E. Soper and G.~F. Sterman, \emph{{Transverse momentum
  distribution in Drell-Yan pair and $W$ and $Z$ boson production}},
  \href{https://doi.org/10.1016/0550-3213(85)90479-1}{\emph{Nucl. Phys.}
  {\bfseries B250} (1985) 199}.

\bibitem{Collins:1989gx}
J.~C. Collins, D.~E. Soper and G.~F. Sterman, \emph{{Factorization of Hard
  Processes in QCD}},  in \emph{{Perturbative Quantum Chromodynamics}}, A.~H.
  Mueller, ed., vol.~5 of \emph{{Advanced Series on Directions in High Energy
  Physics}}, pp.~1--91, World Scientific (1989)
  [\href{https://arxiv.org/abs/hep-ph/0409313}{{\ttfamily hep-ph/0409313}}].

\bibitem{Ji:2004wu}
X.~Ji, J.-P. Ma and F.~Yuan, \emph{{QCD factorization for semi-inclusive
  deep-inelastic scattering at low transverse momentum}},
  \href{https://doi.org/10.1103/PhysRevD.71.034005}{\emph{Phys. Rev. D}
  {\bfseries 71} (2005) 034005}
  [\href{https://arxiv.org/abs/hep-ph/0404183}{{\ttfamily hep-ph/0404183}}].

\bibitem{Echevarria:2012js}
M.~G. Echevarr{\'i}a, A.~Idilbi and I.~Scimemi, \emph{{Soft and Collinear
  Factorization and Transverse Momentum Dependent Parton Distribution
  Functions}},
  \href{https://doi.org/10.1016/j.physletb.2013.09.003}{\emph{Phys. Lett. B}
  {\bfseries 726} (2013) 795}
  [\href{https://arxiv.org/abs/1211.1947}{{\ttfamily 1211.1947}}].

\bibitem{Rogers:2015sqa}
T.~C. Rogers, \emph{{An overview of transverse-momentum--dependent
  factorization and evolution}},
  \href{https://doi.org/10.1140/epja/i2016-16153-7}{\emph{Eur. Phys. J. A}
  {\bfseries 52} (2016) 153}
  [\href{https://arxiv.org/abs/1509.04766}{{\ttfamily 1509.04766}}].

\bibitem{Collins:2017oxh}
J.~Collins and T.~C. Rogers, \emph{{Connecting different TMD factorization
  formalisms in QCD}},
  \href{https://doi.org/10.1103/PhysRevD.96.054011}{\emph{Phys. Rev. D}
  {\bfseries 96} (2017) 054011}
  [\href{https://arxiv.org/abs/1705.07167}{{\ttfamily 1705.07167}}].

\bibitem{Boglione:2019nwk}
M.~Boglione, A.~Dotson, L.~Gamberg, S.~Gordon, J.~O. Gonzalez-Hernandez,
  A.~Prokudin et~al., \emph{{Mapping the kinematical regimes of semi-inclusive
  deep inelastic scattering}},
  \href{https://doi.org/10.1007/JHEP10(2019)122}{\emph{JHEP} {\bfseries 10}
  (2019) 122} [\href{https://arxiv.org/abs/1904.12882}{{\ttfamily
  1904.12882}}].

\bibitem{Collins:1981uk}
J.~C. Collins and D.~E. Soper, \emph{{Back-to-back jets in QCD}},
  \href{https://doi.org/10.1016/0550-3213(81)90339-4}{\emph{Nucl. Phys.}
  {\bfseries B193} (1981) 381} [\textit{Erratum ibid:}
  \href{https://doi.org/10.1016/0550-3213(83)90235-3}{\textbf{B213} (1983)
  545}].

\bibitem{Aybat:2011zv}
S.~Aybat and T.~C. Rogers, \emph{{Transverse momentum dependent parton
  distribution and fragmentation functions with QCD evolution}},
  \href{https://doi.org/10.1103/PhysRevD.83.114042}{\emph{Phys. Rev. D}
  {\bfseries 83} (2011) 114042}
  [\href{https://arxiv.org/abs/1101.5057}{{\ttfamily 1101.5057}}].

\bibitem{Echevarria:2012pw}
M.~G. Echevarr{\'i}a, A.~Idilbi, A.~Sch{\"a}fer and I.~Scimemi, \emph{{Model
  independent evolution of transverse momentum dependent distribution functions
  (TMDs) at NNLL}},
  \href{https://doi.org/10.1140/epjc/s10052-013-2636-y}{\emph{Eur. Phys. J. C}
  {\bfseries 73} (2013) 2636}
  [\href{https://arxiv.org/abs/1208.1281}{{\ttfamily 1208.1281}}].

\bibitem{Bacchetta:2017gcc}
A.~Bacchetta, F.~Delcarro, C.~Pisano, M.~Radici and A.~Signori,
  \emph{{Extraction of partonic transverse momentum distributions from
  semi-inclusive deep-inelastic scattering, Drell--Yan and $Z$-boson
  production}}, \href{https://doi.org/10.1007/JHEP06(2017)081}{\emph{JHEP}
  {\bfseries 06} (2017) 081}
  [\href{https://arxiv.org/abs/1703.10157}{{\ttfamily 1703.10157}}],
  [\textit{Erratum ibid:}
  \href{https://doi.org/10.1007/JHEP06(2019)051}{\textbf{06} (2019) 051}].

\bibitem{Bacchetta:2019sam}
A.~Bacchetta, V.~Bertone, C.~Bissolotti, G.~Bozzi, F.~Delcarro, F.~Piacenza
  et~al., \emph{{Transverse-momentum-dependent parton distributions up to
  N$^3$LL from Drell-Yan data}},
  \href{https://arxiv.org/abs/1912.07550}{{\ttfamily 1912.07550}}.

\bibitem{Scimemi:2019cmh}
I.~Scimemi and A.~Vladimirov, \emph{{Non-perturbative structure of
  semi-inclusive deep-inelastic and Drell-Yan scattering at small transverse
  momentum}}, \href{https://doi.org/10.1007/JHEP06(2020)137}{\emph{JHEP}
  {\bfseries 06} (2020) 137}
  [\href{https://arxiv.org/abs/1912.06532}{{\ttfamily 1912.06532}}].

\bibitem{Ji:2019sxk}
X.~Ji, Y.~Liu and Y.-S. Liu, \emph{{TMD soft function from large-momentum
  effective theory}},
  \href{https://doi.org/10.1016/j.nuclphysb.2020.115054}{\emph{Nucl. Phys.}
  {\bfseries B955} (2020) 115054}
  [\href{https://arxiv.org/abs/1910.11415}{{\ttfamily 1910.11415}}].

\bibitem{Vladimirov:2020ofp}
A.~A. Vladimirov and A.~Sch{\"a}fer, \emph{{Transverse-momentum-dependent
  factorization for lattice observables}},
  \href{https://doi.org/10.1103/PhysRevD.101.074517}{\emph{Phys. Rev. D}
  {\bfseries 101} (2020) 074517}
  [\href{https://arxiv.org/abs/2002.07527}{{\ttfamily 2002.07527}}].

\bibitem{Shanahan:2020zxr}
P.~Shanahan, M.~Wagman and Y.~Zhao, \emph{{Collins-Soper Kernel for TMD
  Evolution from Lattice QCD}},
  \href{https://arxiv.org/abs/2003.06063}{{\ttfamily 2003.06063}}.

\bibitem{Jaffe:1991kp}
R.~L. Jaffe and X.~Ji, \emph{{Chiral odd parton distributions and polarized
  Drell--Yan }}, \href{https://doi.org/10.1103/PhysRevLett.67.552}{\emph{Phys.
  Rev. Lett.} {\bfseries 67} (1991) 552}.

\bibitem{Artru:1989zv}
X.~Artru and M.~Mekhfi, \emph{Transversely polarized parton densities, their
  evolution and their measurement},
  \href{https://doi.org/10.1007/BF01556280}{\emph{Z. Phys. C} {\bfseries 45}
  (1990) 669}.

\bibitem{Airapetian:2004tw}
{\scshape \textsc{HERMES}} collaboration, \emph{{Single-spin Asymmetries in
  Semi-Inclusive Deep-Inelastic Scattering on a Transversely Polarized Hydrogen
  Target}}, \href{https://doi.org/10.1103/PhysRevLett.94.012002}{\emph{Phys.
  Rev. Lett.} {\bfseries 94} (2005) 012002}
  [\href{https://arxiv.org/abs/hep-ex/0408013}{{\ttfamily hep-ex/0408013}}].

\bibitem{Airapetian:2010ds}
{\scshape \textsc{HERMES}} collaboration, \emph{{Effects of transversity in
  deep-inelastic scattering by polarized protons}},
  \href{https://doi.org/10.1016/j.physletb.2010.08.012}{\emph{Phys. Lett. B}
  {\bfseries 693} (2010) 11} [\href{https://arxiv.org/abs/1006.4221}{{\ttfamily
  1006.4221}}].

\bibitem{Sivers:1989cc}
D.~W. Sivers, \emph{Single-spin production asymmetries from the hard scattering
  of pointlike constituents},
  \href{https://doi.org/10.1103/PhysRevD.41.83}{\emph{Phys. Rev. D} {\bfseries
  41} (1990) 83}.

\bibitem{Boer:1997nt}
D.~Boer and P.~J. Mulders, \emph{Time-reversal odd distribution functions in
  leptoproduction}, \href{https://doi.org/10.1103/PhysRevD.57.5780}{\emph{Phys.
  Rev. D} {\bfseries 57} (1998) 5780}
  [\href{https://arxiv.org/abs/hep-ph/9711485}{{\ttfamily hep-ph/9711485}}].

\bibitem{Airapetian:2012yg}
{\scshape \textsc{HERMES}} collaboration, \emph{{Azimuthal distributions of
  charged hadrons, pions, and kaons produced in deep-inelastic scattering off
  unpolarized protons and deuterons}},
  \href{https://doi.org/10.1103/PhysRevD.87.012010}{\emph{Phys. Rev. D}
  {\bfseries 87} (2013) 012010}
  [\href{https://arxiv.org/abs/1204.4161}{{\ttfamily 1204.4161}}].

\bibitem{Adolph:2014pwc}
{\scshape COMPASS} collaboration, \emph{{Measurement of azimuthal hadron
  asymmetries in semi-inclusive deep inelastic scattering off unpolarised
  nucleons}},
  \href{https://doi.org/10.1016/j.nuclphysb.2014.07.019}{\emph{Nucl. Phys.}
  {\bfseries B886} (2014) 1046}
  [\href{https://arxiv.org/abs/1401.6284}{{\ttfamily 1401.6284}}].

\bibitem{Antille:1980th}
J.~Antille, L.~Dick, L.~Madansky, D.~Perret-Gallix, M.~Werlen, A.~Gonidec
  et~al., \emph{{Spin dependence of the inclusive reaction $p + p$ (polarized)
  $\to \pi^0 + X$ at 24~GeV/$c$ for high-$p_T~\pi^0$ produced in the central
  region}}, \href{https://doi.org/10.1016/0370-2693(80)90933-8}{\emph{Phys.
  Lett. B} {\bfseries 94} (1980) 523}.

\bibitem{Collins:2002kn}
J.~C. Collins, \emph{{Leading-twist single-transverse-spin asymmetries:
  Drell--Yan and deep-inelastic scattering}},
  \href{https://doi.org/10.1016/S0370-2693(02)01819-1}{\emph{Phys. Lett. B}
  {\bfseries 536} (2002) 43}
  [\href{https://arxiv.org/abs/hep-ph/0204004}{{\ttfamily hep-ph/0204004}}].

\bibitem{Brodsky:2002cx}
S.~J. Brodsky, D.~S. Hwang and I.~Schmidt, \emph{Final-state interactions and
  single-spin asymmetries in semi-inclusive deep inelastic scattering},
  \href{https://doi.org/10.1016/S0370-2693(02)01320-5}{\emph{Phys. Lett. B}
  {\bfseries 530} (2002) 99}
  [\href{https://arxiv.org/abs/hep-ph/0201296}{{\ttfamily hep-ph/0201296}}].

\bibitem{Ji:2002aa}
X.~Ji and F.~Yuan, \emph{Parton distributions in light-cone gauge: where are
  the final-state interactions?},
  \href{https://doi.org/10.1016/S0370-2693(02)02384-5}{\emph{Phys. Lett. B}
  {\bfseries 543} (2002) 66}
  [\href{https://arxiv.org/abs/hep-ph/0206057}{{\ttfamily hep-ph/0206057}}].

\bibitem{Belitsky:2002sm}
A.~V. Belitsky, X.~Ji and F.~Yuan, \emph{Final state interactions and gauge
  invariant parton distributions},
  \href{https://doi.org/10.1016/S0550-3213(03)00121-4}{\emph{Nucl. Phys.}
  {\bfseries B656} (2003) 165}
  [\href{https://arxiv.org/abs/hep-ph/0208038}{{\ttfamily hep-ph/0208038}}].

\bibitem{Burkardt:2003uw}
M.~Burkardt, \emph{{Chromodynamic lensing and transverse single spin
  asymmetries}},
  \href{https://doi.org/10.1016/j.nuclphysa.2004.02.008}{\emph{Nucl. Phys.}
  {\bfseries A735} (2004) 185}
  [\href{https://arxiv.org/abs/hep-ph/0302144}{{\ttfamily hep-ph/0302144}}].

\bibitem{Airapetian:2009ae}
{\scshape \textsc{HERMES}} collaboration, \emph{{Observation of the
  Naive-\(T\)-odd Sivers Effect in Deep-Inelastic Scattering}},
  \href{https://doi.org/10.1103/PhysRevLett.103.152002}{\emph{Phys. Rev. Lett.}
  {\bfseries 103} (2009) 152002}
  [\href{https://arxiv.org/abs/0906.3918}{{\ttfamily 0906.3918}}].

\bibitem{Alekseev:2010rw}
{\scshape \textsc{COMPASS}} collaboration, \emph{{Measurement of the Collins
  and Sivers asymmetries on transversely polarised protons}},
  \href{https://doi.org/10.1016/j.physletb.2010.08.001}{\emph{Phys. Lett. B}
  {\bfseries 692} (2010) 240}
  [\href{https://arxiv.org/abs/1005.5609}{{\ttfamily 1005.5609}}].

\bibitem{Adamczyk:2015gyk}
{\scshape STAR} collaboration, \emph{{Measurement of the Transverse Single-Spin
  Asymmetry in $p^\uparrow+p \to W^{\pm}/Z^0$ at RHIC}},
  \href{https://doi.org/10.1103/PhysRevLett.116.132301}{\emph{Phys. Rev. Lett.}
  {\bfseries 116} (2016) 132301}
  [\href{https://arxiv.org/abs/1511.06003}{{\ttfamily 1511.06003}}].

\bibitem{Aghasyan:2017jop}
{\scshape COMPASS} collaboration, \emph{{First Measurement of
  Transverse-Spin-Dependent Azimuthal Asymmetries in the Drell--Yan Process}},
  \href{https://doi.org/10.1103/PhysRevLett.119.112002}{\emph{Phys. Rev. Lett.}
  {\bfseries 119} (2017) 112002}
  [\href{https://arxiv.org/abs/1704.00488}{{\ttfamily 1704.00488}}].

\bibitem{Burkardt:2005km}
M.~Burkardt and G.~Schnell, \emph{Anomalous magnetic moments and quark orbital
  angular momentum},
  \href{https://doi.org/10.1103/PhysRevD.74.013002}{\emph{Phys. Rev. D}
  {\bfseries 74} (2006) 013002}
  [\href{https://arxiv.org/abs/hep-ph/0510249}{{\ttfamily hep-ph/0510249}}].

\bibitem{Burkardt:2002ks}
M.~Burkardt, \emph{{Impact parameter dependent parton distributions and
  transverse single spin asymmetries}},
  \href{https://doi.org/10.1103/PhysRevD.66.114005}{\emph{Phys. Rev. D}
  {\bfseries 66} (2002) 114005}
  [\href{https://arxiv.org/abs/hep-ph/0209179}{{\ttfamily hep-ph/0209179}}].

\bibitem{Bacchetta:2011gx}
A.~Bacchetta and M.~Radici, \emph{{Constraining Quark Angular Momentum Through
  Semi-Inclusive Measurements}},
  \href{https://doi.org/10.1103/PhysRevLett.107.212001}{\emph{Phys. Rev. Lett.}
  {\bfseries 107} (2011) 212001}
  [\href{https://arxiv.org/abs/1107.5755}{{\ttfamily 1107.5755}}].

\bibitem{Pasquini:2019evu}
B.~Pasquini, S.~Rodini and A.~Bacchetta, \emph{{Revisiting model relations
  between T-odd transverse-momentum-dependent parton distributions and
  generalized parton distributions}},
  \href{https://doi.org/10.1103/PhysRevD.100.054039}{\emph{Phys. Rev. D}
  {\bfseries 100} (2019) 054039}
  [\href{https://arxiv.org/abs/1907.06960}{{\ttfamily 1907.06960}}].

\bibitem{Tangerman:1995hw}
R.~D. Tangerman and P.~J. Mulders, \emph{{Probing transverse quark polarization
  in deep-inelastic leptoproduction}},
  \href{https://doi.org/10.1016/0370-2693(95)00485-4}{\emph{Phys. Lett. B}
  {\bfseries 352} (1995) 129}
  [\href{https://arxiv.org/abs/hep-ph/9501202}{{\ttfamily hep-ph/9501202}}].

\bibitem{Avakian:2008dz}
H.~Avakian, A.~V. Efremov, P.~Schweitzer and F.~Yuan, \emph{{Transverse
  momentum dependent distribution function \(h_{1T}^\perp\) and the single spin
  asymmetry \(A_{UT}^{\sin(3\phi-\phi_S)}\)}},
  \href{https://doi.org/10.1103/PhysRevD.78.114024}{\emph{Phys. Rev. D}
  {\bfseries 78} (2008) 114024}
  [\href{https://arxiv.org/abs/0805.3355}{{\ttfamily 0805.3355}}].

\bibitem{Miller:2007ae}
G.~A. Miller, \emph{{Densities, parton distributions, and measuring the
  nonspherical shape of the nucleon}},
  \href{https://doi.org/10.1103/PhysRevC.76.065209}{\emph{Phys. Rev. C}
  {\bfseries 76} (2007) 065209}
  [\href{https://arxiv.org/abs/0708.2297}{{\ttfamily 0708.2297}}].

\bibitem{Burkardt:2007rv}
M.~Burkardt, \emph{\textit{\textsc{spin-orbit correlations and single-spin
  asymmetries}}},  in \emph{{Exclusive Reactions at High Momentum Transfer}},
  A.~Radyushkin and P.~Stoler, eds., pp.~78--86, World Scientific, 2008
  [\href{https://arxiv.org/abs/0709.2966}{{\ttfamily 0709.2966}}].

\bibitem{Kotzinian:1994dv}
A.~Kotzinian, \emph{{New quark distributions and semi-inclusive
  electroproduction on polarized nucleons}},
  \href{https://doi.org/10.1016/0550-3213(95)00098-D}{\emph{Nucl. Phys.}
  {\bfseries B441} (1995) 234}
  [\href{https://arxiv.org/abs/hep-ph/9412283}{{\ttfamily hep-ph/9412283}}].

\bibitem{Zhang:2013dow}
{\scshape Jefferson Lab Hall A} collaboration, \emph{{Measurement of
  ``pretzelosity'' asymmetry of charged pion production in semi-inclusive deep
  inelastic scattering on a polarized $^3$He target}},
  \href{https://doi.org/10.1103/PhysRevC.90.055209}{\emph{Phys. Rev. C}
  {\bfseries 90} (2014) 055209}
  [\href{https://arxiv.org/abs/1312.3047}{{\ttfamily 1312.3047}}].

\bibitem{Ralston:1979ys}
J.~P. Ralston and D.~E. Soper, \emph{Production of dimuons from high-energy
  polarized proton-proton collisions},
  \href{https://doi.org/10.1016/0550-3213(79)90082-8}{\emph{Nucl. Phys.}
  {\bfseries B152} (1979) 109}.

\bibitem{Kotzinian:1995cz}
A.~M. Kotzinian and P.~J. Mulders, \emph{Longitudinal quark polarization in
  transversely polarized nucleons},
  \href{https://doi.org/10.1103/PhysRevD.54.1229}{\emph{Phys. Rev. D}
  {\bfseries 54} (1996) 1229}
  [\href{https://arxiv.org/abs/hep-ph/9511420}{{\ttfamily hep-ph/9511420}}].

\bibitem{Diehl:2005jf}
M.~Diehl and P.~H\"agler, \emph{Spin densities in the transverse plane and
  generalized transversity distributions},
  \href{https://doi.org/10.1140/epjc/s2005-02342-6}{\emph{Eur. Phys. J. C}
  {\bfseries 44} (2005) 87}
  [\href{https://arxiv.org/abs/hep-ph/0504175}{{\ttfamily hep-ph/0504175}}].

\bibitem{Boffi:2009sh}
S.~Boffi, A.~V. Efremov, B.~Pasquini and P.~Schweitzer, \emph{{Azimuthal spin
  asymmetries in light-cone constituent quark models}},
  \href{https://doi.org/10.1103/PhysRevD.79.094012}{\emph{Phys. Rev. D}
  {\bfseries 79} (2009) 094012}
  [\href{https://arxiv.org/abs/0903.1271}{{\ttfamily 0903.1271}}].

\bibitem{Bacchetta:2013pqa}
A.~Bacchetta and A.~Prokudin, \emph{{Evolution of the helicity and
  transversity. Transverse-momentum-dependent parton distributions}},
  \href{https://doi.org/10.1016/j.nuclphysb.2013.07.013}{\emph{Nucl. Phys.}
  {\bfseries B875} (2013) 536}
  [\href{https://arxiv.org/abs/1303.2129}{{\ttfamily 1303.2129}}].

\bibitem{Avakian:2007mv}
H.~Avakian, A.~V. Efremov, K.~Goeke, A.~Metz, P.~Schweitzer and T.~Teckentrup,
  \emph{Are there approximate relations among transverse momentum dependent
  distribution functions?},
  \href{https://doi.org/10.1103/PhysRevD.77.014023}{\emph{Phys. Rev. D}
  {\bfseries 77} (2008) 014023}
  [\href{https://arxiv.org/abs/0709.3253}{{\ttfamily 0709.3253}}].

\bibitem{Wandzura:1977qf}
S.~Wandzura and F.~Wilczek, \emph{{Sum rules for spin-dependent
  electroproduction: test of relativistic constituent quarks}},
  \href{https://doi.org/10.1016/0370-2693(77)90700-6}{\emph{Phys. Lett. B}
  {\bfseries 72} (1977) 195}.

\bibitem{Bastami:2018xqd}
S.~Bastami et~al., \emph{{Semi-Inclusive deep-inelastic scattering in
  Wandzura-Wilczek-type approximation}},
  \href{https://doi.org/10.1007/JHEP06(2019)007}{\emph{JHEP} {\bfseries 06}
  (2019) 007} [\href{https://arxiv.org/abs/1807.10606}{{\ttfamily
  1807.10606}}].

\bibitem{Accardi:2009au}
A.~Accardi, A.~Bacchetta, W.~Melnitchouk and M.~Schlegel, \emph{{What can break
  the Wandzura-Wilczek relation?}},
  \href{https://doi.org/10.1088/1126-6708/2009/11/093}{\emph{JHEP} {\bfseries
  11} (2009) 093} [\href{https://arxiv.org/abs/0907.2942}{{\ttfamily
  0907.2942}}].

\bibitem{Huang:2011bc}
{\scshape Jefferson Lab Hall A} collaboration, \emph{{Beam-Target Double-Spin
  Asymmetry \(A_{LT}\) in Charged Pion Production from Deep Inelastic
  Scattering on a Transversely Polarized \(^{3}\)He Target at 1.4\(<Q^2<\)2.7
  GeV\(^{2}\)}},
  \href{https://doi.org/10.1103/PhysRevLett.108.052001}{\emph{Phys. Rev. Lett.}
  {\bfseries 108} (2012) 052001}
  [\href{https://arxiv.org/abs/1108.0489}{{\ttfamily 1108.0489}}].

\bibitem{Airapetian:1999tv}
{\scshape \textsc{HERMES}} collaboration, \emph{{Evidence for a Single-Spin
  Azimuthal Asymmetry in Semi-inclusive Pion Electroproduction}},
  \href{https://doi.org/10.1103/PhysRevLett.84.4047}{\emph{Phys. Rev. Lett.}
  {\bfseries 84} (2000) 4047}
  [\href{https://arxiv.org/abs/hep-ex/9910062}{{\ttfamily hep-ex/9910062}}].

\bibitem{Airapetian:2001eg}
{\scshape \textsc{HERMES}} collaboration, \emph{Single-spin azimuthal
  asymmetries in electroproduction of neutral pions in semi-inclusive
  deep-inelastic scattering},
  \href{https://doi.org/10.1103/PhysRevD.64.097101}{\emph{Phys. Rev. D}
  {\bfseries 64} (2001) 097101}
  [\href{https://arxiv.org/abs/hep-ex/0104005}{{\ttfamily hep-ex/0104005}}].

\bibitem{Airapetian:2002mf}
{\scshape \textsc{HERMES}} collaboration, \emph{Measurement of single-spin
  azimuthal asymmetries in semi-inclusive electroproduction of pions and kaons
  on a longitudinally polarised deuterium target},
  \href{https://doi.org/10.1016/S0370-2693(03)00566-5}{\emph{Phys. Lett. B}
  {\bfseries 562} (2003) 182}
  [\href{https://arxiv.org/abs/hep-ex/0212039}{{\ttfamily hep-ex/0212039}}].

\bibitem{Airapetian:2005jc}
{\scshape \textsc{HERMES}} collaboration, \emph{{Subleading-twist effects in
  single-spin asymmetries in semi-inclusive deep-inelastic scattering on a
  longitudinally polarized hydrogen target}},
  \href{https://doi.org/10.1016/j.physletb.2005.06.067}{\emph{Phys. Lett. B}
  {\bfseries 622} (2005) 14}
  [\href{https://arxiv.org/abs/hep-ex/0505042}{{\ttfamily hep-ex/0505042}}].

\bibitem{Airapetian:2006rx}
{\scshape \textsc{HERMES}} collaboration, \emph{Beam-spin asymmetries in the
  azimuthal distribution of pion electroproduction},
  \href{https://doi.org/10.1016/j.physletb.2007.03.015}{\emph{Phys. Lett. B}
  {\bfseries 648} (2007) 164}
  [\href{https://arxiv.org/abs/hep-ex/0612059}{{\ttfamily hep-ex/0612059}}].

\bibitem{Airapetian:2019mov}
{\scshape HERMES} collaboration, \emph{{Beam-helicity asymmetries for
  single-hadron production in semi-inclusive deep-inelastic scattering from
  unpolarized hydrogen and deuterium targets}},
  \href{https://doi.org/10.1016/j.physletb.2019.134886}{\emph{Phys. Lett. B}
  {\bfseries 797} (2019) 134886}
  [\href{https://arxiv.org/abs/1903.08544}{{\ttfamily 1903.08544}}].

\bibitem{Mulders:1995dh}
P.~J. Mulders and R.~D. Tangerman, \emph{{The complete tree-level result up to
  order \(1/Q\) for polarized deep-inelastic leptoproduction}},
  \href{https://doi.org/10.1016/0550-3213(95)00632-X}{\emph{Nucl. Phys.}
  {\bfseries B461} (1996) 197}
  [\href{https://arxiv.org/abs/hep-ph/9510301}{{\ttfamily hep-ph/9510301}}],
  [\textit{Erratum ibid:}
  \href{https://doi.org/10.1016/S0550-3213(96)00648-7}{\textbf{B484} (1997)
  538}].

\bibitem{Christ:1966zz}
N.~Christ and T.~D. Lee, \emph{{Possible Tests of \( C_{\text{st}} \) and \(
  T_{\text{st}} \) Invariances in \( l^{\pm} + N \to l^{\pm} + \Gamma \) and \(
  A \to B + e^{+} + e^{-} \)}},
  \href{https://doi.org/10.1103/PhysRev.143.1310}{\emph{Phys. Rev.} {\bfseries
  143} (1966) 1310}.

\bibitem{Airapetian:2009ab}
{\scshape HERMES} collaboration, \emph{{Search for a two-photon exchange
  contribution to inclusive deep-inelastic scattering}},
  \href{https://doi.org/10.1016/j.physletb.2009.11.041}{\emph{Phys. Lett. B}
  {\bfseries 682} (2010) 351}
  [\href{https://arxiv.org/abs/0907.5369}{{\ttfamily 0907.5369}}].

\bibitem{Gamberg:2017gle}
L.~Gamberg, Z.-B. Kang, D.~Pitonyak and A.~Prokudin, \emph{{Phenomenological
  constraints on $A_N$ in $p^\uparrow p\to \pi\, X$ from Lorentz invariance
  relations}},
  \href{https://doi.org/10.1016/j.physletb.2017.04.061}{\emph{Phys. Lett. B}
  {\bfseries 770} (2017) 242}
  [\href{https://arxiv.org/abs/1701.09170}{{\ttfamily 1701.09170}}].

\bibitem{Aschenauer:2015ndk}
E.~C. Aschenauer, U.~D'Alesio and F.~Murgia, \emph{{TMDs and SSAs in hadronic
  interactions}}, \href{https://doi.org/10.1140/epja/i2016-16156-4}{\emph{Eur.
  Phys. J. A} {\bfseries 52} (2016) 156}
  [\href{https://arxiv.org/abs/1512.05379}{{\ttfamily 1512.05379}}].

\bibitem{Airapetian:2013bim}
{\scshape HERMES} collaboration, \emph{{Transverse target single-spin asymmetry
  in inclusive electroproduction of charged pions and kaons}},
  \href{https://doi.org/10.1016/j.physletb.2013.11.021}{\emph{Phys. Lett. B}
  {\bfseries 728} (2014) 183}
  [\href{https://arxiv.org/abs/1310.5070}{{\ttfamily 1310.5070}}].

\bibitem{Jaffe:1993xb}
R.~L. Jaffe and X.~Ji, \emph{{Novel quark fragmentation functions and the
  nucleon's transversity distribution}},
  \href{https://doi.org/10.1103/PhysRevLett.71.2547}{\emph{Phys. Rev. Lett.}
  {\bfseries 71} (1993) 2547}
  [\href{https://arxiv.org/abs/hep-ph/9307329}{{\ttfamily hep-ph/9307329}}].

\bibitem{Efremov:1992pe}
A.~V. Efremov, L.~Mankiewicz and N.~A. T{\"o}rnqvist, \emph{{Jet handedness as
  a measure of quark and gluon polarization}},
  \href{https://doi.org/10.1016/0370-2693(92)90451-9}{\emph{Phys. Lett. B}
  {\bfseries 284} (1992) 394}.

\bibitem{Collins:1993kq}
J.~C. Collins, S.~F. Heppelmann and G.~A. Ladinsky, \emph{{Measuring
  transversity densities in singly polarized hadron-hadron and lepton-hadron
  collisions}}, \href{https://doi.org/10.1016/0550-3213(94)90078-7}{\emph{Nucl.
  Phys.} {\bfseries B420} (1994) 565}
  [\href{https://arxiv.org/abs/hep-ph/9305309}{{\ttfamily hep-ph/9305309}}].

\bibitem{Airapetian:2011wu}
{\scshape \textsc{HERMES}} collaboration, \emph{{Measurement of the
  virtual-photon asymmetry $A_2$ and the spin-structure function $g_2$ of the
  proton}}, \href{https://doi.org/10.1140/epjc/s10052-012-1921-5}{\emph{Eur.
  Phys. J. C} {\bfseries 72} (2012) 1921}
  [\href{https://arxiv.org/abs/1112.5584}{{\ttfamily 1112.5584}}].

\bibitem{Adams:1997tq}
{\scshape Spin Muon} collaboration, \emph{{Spin structure of the proton from
  polarized inclusive deep-inelastic muon-proton scattering}},
  \href{https://doi.org/10.1103/PhysRevD.56.5330}{\emph{Phys. Rev. D}
  {\bfseries 56} (1997) 5330}
  [\href{https://arxiv.org/abs/hep-ex/9702005}{{\ttfamily hep-ex/9702005}}].

\bibitem{Abe:1998wq}
{\scshape E143} collaboration, \emph{{Measurements of the proton and deuteron
  spin structure functions $g_1$ and $g_2$}},
  \href{https://doi.org/10.1103/PhysRevD.58.112003}{\emph{Phys. Rev. D}
  {\bfseries 58} (1998) 112003}
  [\href{https://arxiv.org/abs/hep-ph/9802357}{{\ttfamily hep-ph/9802357}}].

\bibitem{Anthony:2002hy}
{\scshape E155} collaboration, \emph{{Precision measurement of the proton and
  deuteron spin structure functions $g_2$ and asymmetries $A_2$}},
  \href{https://doi.org/10.1016/S0370-2693(02)03015-0}{\emph{Phys. Lett. B}
  {\bfseries 553} (2003) 18}
  [\href{https://arxiv.org/abs/hep-ex/0204028}{{\ttfamily hep-ex/0204028}}].

\bibitem{Armstrong:2018xgk}
{\scshape SANE} collaboration, \emph{{Revealing Color Forces with Transverse
  Polarized Electron Scattering}},
  \href{https://doi.org/10.1103/PhysRevLett.122.022002}{\emph{Phys. Rev. Lett.}
  {\bfseries 122} (2019) 022002}
  [\href{https://arxiv.org/abs/1805.08835}{{\ttfamily 1805.08835}}].

\bibitem{Burkardt:2008ps}
M.~Burkardt, \emph{{Transverse force on quarks in deep-inelastic scattering}},
  \href{https://doi.org/10.1103/PhysRevD.88.114502}{\emph{Phys. Rev. D}
  {\bfseries 88} (2013) 114502}
  [\href{https://arxiv.org/abs/0810.3589}{{\ttfamily 0810.3589}}].

\bibitem{Aslan:2019jis}
F.~P. Aslan, M.~Burkardt and M.~Schlegel, \emph{{Transverse force tomography}},
  \href{https://doi.org/10.1103/PhysRevD.100.096021}{\emph{Phys. Rev. D}
  {\bfseries 100} (2019) 096021}
  [\href{https://arxiv.org/abs/1904.03494}{{\ttfamily 1904.03494}}].

\bibitem{Ackerstaff:1998av}
{\scshape \textsc{HERMES}} collaboration, \emph{{The \hermes Spectrometer}},
  \href{https://doi.org/10.1016/S0168-9002(98)00769-4}{\emph{Nucl. Instr. and
  Meth. A} {\bfseries 417} (1998) 230}
  [\href{https://arxiv.org/abs/hep-ex/9806008}{{\ttfamily hep-ex/9806008}}].

\bibitem{Airapetian:2004yf}
{\scshape \textsc{HERMES}} collaboration, \emph{The \hermes polarized hydrogen
  and deuterium gas target in the \hera electron storage ring},
  \href{https://doi.org/10.1016/j.nima.2004.11.020}{\emph{Nucl. Instr. and
  Meth. A} {\bfseries 540} (2005) 68}
  [\href{https://arxiv.org/abs/physics/0408137}{{\ttfamily physics/0408137}}].

\bibitem{Sokolov:1963zn}
A.~A. Sokolov and I.~M. Ternov, \emph{On polarization and spin effects in the
  theory of synchrotron radiation}, {\emph{Sov. Phys. Dokl.} {\bfseries 8}
  (1964) 1203} [Phys. Dokl.8,1203(1964)].

\bibitem{Akopov:2000qi}
N.~Akopov et~al., \emph{{The \hermes dual-radiator ring imaging Cherenkov
  detector}}, \href{https://doi.org/10.1016/S0168-9002(01)00932-9}{\emph{Nucl.
  Instr. and Meth. A} {\bfseries 479} (2002) 511}
  [\href{https://arxiv.org/abs/physics/0104033}{{\ttfamily physics/0104033}}].

\bibitem{Sjostrand:2000wi}
T.~Sj\"ostrand, P.~Eden, C.~Friberg, L.~L\"onnblad, G.~Miu, S.~Mrenna et~al.,
  \emph{{High-energy-physics event generation with \textit{\textsc{Pythia}}
  6.1}}, \href{https://doi.org/10.1016/S0010-4655(00)00236-8}{\emph{Comput.
  Phys. Commun.} {\bfseries 135} (2001) 238}
  [\href{https://arxiv.org/abs/hep-ph/0010017}{{\ttfamily hep-ph/0010017}}].

\bibitem{Sjostrand:2001yu}
T.~Sj\"ostrand, L.~L\"onnblad and S.~Mrenna, \emph{{\textit{\textsc{Pythia}}
  6.2: Physics and Manual}}, 2001.

\bibitem{Airapetian:2012ki}
{\scshape HERMES} collaboration, \emph{{Multiplicities of charged pions and
  kaons from semi-inclusive deep-inelastic scattering by the proton and the
  deuteron}}, \href{https://doi.org/10.1103/PhysRevD.87.074029}{\emph{Phys.
  Rev. D} {\bfseries 87} (2013) 074029}
  [\href{https://arxiv.org/abs/1212.5407}{{\ttfamily 1212.5407}}].

\bibitem{Boglione:2016bph}
M.~Boglione, J.~Collins, L.~Gamberg, J.~O. Gonzalez-Hernandez, T.~C. Rogers and
  N.~Sato, \emph{{Kinematics of current region fragmentation in semi-inclusive
  deeply inelastic scattering}},
  \href{https://doi.org/10.1016/j.physletb.2017.01.021}{\emph{Phys. Lett. B}
  {\bfseries 766} (2017) 245}
  [\href{https://arxiv.org/abs/1611.10329}{{\ttfamily 1611.10329}}].

\bibitem{Berger:1987zu}
E.~L. Berger, \emph{Semi-inclusive inelastic electron scattering from nuclei},
  in \emph{Proceedings of \textsc{NPAS} Workshop on Electronuclear Physics with
  Internal Targets, SLAC, January 5-8, 1987}, pp.~82--91, 1987.

\bibitem{Mulders:2000jt}
P.~J. Mulders, \emph{{Current fragmentation in semiinclusive leptoproduction}},
  \href{https://doi.org/10.1063/1.1413147}{\emph{AIP Conf. Proc.} {\bfseries
  588} (2001) 75} [\href{https://arxiv.org/abs/hep-ph/0010199}{{\ttfamily
  hep-ph/0010199}}].

\bibitem{Diehl:2005pc}
M.~Diehl and S.~Sapeta, \emph{On the analysis of lepton scattering on
  longitudinally or transversely polarized protons},
  \href{https://doi.org/10.1140/epjc/s2005-02242-9}{\emph{Eur. Phys. J. C}
  {\bfseries 41} (2005) 515}
  [\href{https://arxiv.org/abs/hep-ph/0503023}{{\ttfamily hep-ph/0503023}}].

\bibitem{Solmitz:1964xw}
F.~T. Solmitz, \emph{\textit{\textsc{analysis of experiments in particle
  physics}}},
  \href{https://doi.org/10.1146/annurev.ns.14.120164.002111}{\emph{Annu. Rev.
  Nucl. Sci.} {\bfseries 14} (1964) 375}.

\bibitem{Weibull:1951aa}
W.~Weibull, \emph{{A Statistical Distribution Function of Wide Applicability}},
  {\emph{J. Appl. Mech.} {\bfseries 18} (1951) 293}.

\bibitem{Cahn:1978se}
R.~N. Cahn, \emph{{Azimuthal dependence in leptoproduction: A simple parton
  model calculation}},
  \href{https://doi.org/10.1016/0370-2693(78)90020-5}{\emph{Phys. Lett. B}
  {\bfseries 78} (1978) 269}.

\bibitem{Schnell:2015gaa}
G.~Schnell, \emph{{Monte Carlo methods for \textsc{TMD} analyses}},
  \href{https://doi.org/10.1051/epjconf/20158502024}{\emph{EPJ Web Conf.}
  {\bfseries 85} (2015) 02024}.

\bibitem{Akushevich:1998ft}
I.~Akushevich, H.~B\"ottcher and D.~Ryckbosch, \emph{{\radgen 1.0: Monte Carlo
  Generator for Radiative Events in DIS on Polarized and Unpolarized Targets}},
   in \emph{{Monte Carlo generators for HERA physics. Proceedings, Workshop,
  Hamburg, Germany, 1998-1999}}, pp.~554--565, 1998
  [\href{https://arxiv.org/abs/hep-ph/9906408}{{\ttfamily hep-ph/9906408}}].

\bibitem{Brun:1978fy}
R.~Brun, R.~Hagelberg, M.~Hansroul and J.~C. Lassalle, \emph{{Geant: Simulation
  Program for Particle Physics Experiments. User Guide and Reference Manual}},
  1978.

\bibitem{Brun:1987ma}
R.~Brun, F.~Bruyant, M.~Maire, A.~C. McPherson and P.~Zanarini,
  \emph{{GEANT3}}, 1987.

\bibitem{supplemental}
{\scshape HERMES} collaboration, A.~Airapetian et~al., \emph{Azimuthal single-
  and double-spin asymmetries in semi-inclusive deep-inelastic lepton
  scattering by transversely polarized protons}, Supplemental Material (2020).

\bibitem{Anselmino:2007fs}
M.~Anselmino, M.~Boglione, U.~D'Alesio, A.~Kotzinian, F.~Murgia, A.~Prokudin
  et~al., \emph{{Transversity and Collins functions from \textsc{SIDIS} and
  \(e^+e^-\) data}},
  \href{https://doi.org/10.1103/PhysRevD.75.054032}{\emph{Phys. Rev. D}
  {\bfseries 75} (2007) 054032}
  [\href{https://arxiv.org/abs/hep-ph/0701006}{{\ttfamily hep-ph/0701006}}].

\bibitem{Anselmino:2013vqa}
M.~Anselmino, M.~Boglione, U.~D'Alesio, S.~Melis, F.~Murgia and A.~Prokudin,
  \emph{{Simultaneous extraction of transversity and Collins functions from new
  semi-inclusive deep inelastic scattering and \( e^{+} e^{-} \) data}},
  \href{https://doi.org/10.1103/PhysRevD.87.094019}{\emph{Phys. Rev. D}
  {\bfseries 87} (2013) 094019}
  [\href{https://arxiv.org/abs/1303.3822}{{\ttfamily 1303.3822}}].

\bibitem{Anselmino:2015sxa}
M.~Anselmino, M.~Boglione, U.~D'Alesio, J.~O. Gonzalez~Hernandez, S.~Melis,
  F.~Murgia et~al., \emph{{Collins functions for pions from SIDIS and new
  $e^+e^-$ data: A first glance at their transverse momentum dependence}},
  \href{https://doi.org/10.1103/PhysRevD.92.114023}{\emph{Phys. Rev. D}
  {\bfseries 92} (2015) 114023}
  [\href{https://arxiv.org/abs/1510.05389}{{\ttfamily 1510.05389}}].

\bibitem{Lin:2017stx}
H.-W. Lin, W.~Melnitchouk, A.~Prokudin, N.~Sato and H.~Shows, \emph{{First
  Monte Carlo Global Analysis of Nucleon Transversity with Lattice QCD
  Constraints}},
  \href{https://doi.org/10.1103/PhysRevLett.120.152502}{\emph{Phys. Rev. Lett.}
  {\bfseries 120} (2018) 152502}
  [\href{https://arxiv.org/abs/1710.09858}{{\ttfamily 1710.09858}}].

\bibitem{Barone:2019yvn}
V.~Barone, F.~Bradamante, A.~Bressan, A.~Kerbizi, A.~Martin, A.~Moretti et~al.,
  \emph{{Transversity distributions from difference asymmetries in
  semi-inclusive DIS}},
  \href{https://doi.org/10.1103/PhysRevD.99.114004}{\emph{Phys. Rev. D}
  {\bfseries 99} (2019) 114004}
  [\href{https://arxiv.org/abs/1902.08445}{{\ttfamily 1902.08445}}].

\bibitem{DAlesio:2020vtw}
U.~D'Alesio, C.~Flore and A.~Prokudin, \emph{{Role of the Soffer bound in
  determination of transversity and the tensor charge}},
  \href{https://doi.org/10.1016/j.physletb.2020.135347}{\emph{Phys. Lett. B}
  {\bfseries 803} (2020) 135347}
  [\href{https://arxiv.org/abs/2001.01573}{{\ttfamily 2001.01573}}].

\bibitem{Cammarota:2020qcw}
{\scshape Jefferson Lab Angular Momentum (JAM)} collaboration, \emph{{The
  origin of single transverse-spin asymmetries in high-energy collisions}},
  \href{https://arxiv.org/abs/2002.08384}{{\ttfamily 2002.08384}}.

\bibitem{Alexakhin:2005iw}
{\scshape \textsc{COMPASS}} collaboration, \emph{{First Measurement of the
  Transverse Spin Asymmetries of the Deuteron in Semi-inclusive Deep Inelastic
  Scattering}},
  \href{https://doi.org/10.1103/PhysRevLett.94.202002}{\emph{Phys. Rev. Lett.}
  {\bfseries 94} (2005) 202002}
  [\href{https://arxiv.org/abs/hep-ex/0503002}{{\ttfamily hep-ex/0503002}}].

\bibitem{Ageev:2006da}
{\scshape \textsc{COMPASS}} collaboration, \emph{{A new measurement of the
  Collins and Sivers asymmetries on a transversely polarised deuteron target}},
  \href{https://doi.org/10.1016/j.nuclphysb.2006.10.027}{\emph{Nucl. Phys.}
  {\bfseries B765} (2007) 31}
  [\href{https://arxiv.org/abs/hep-ex/0610068}{{\ttfamily hep-ex/0610068}}].

\bibitem{Adolph:2012sn}
{\scshape COMPASS} collaboration, \emph{{I -- Experimental investigation of
  transverse spin asymmetries in \(\mu\)-\(p\) \textsc{SIDIS} processes:
  Collins asymmetries}},
  \href{https://doi.org/10.1016/j.physletb.2012.09.055}{\emph{Phys. Lett. B}
  {\bfseries 717} (2012) 376}
  [\href{https://arxiv.org/abs/1205.5121}{{\ttfamily 1205.5121}}].

\bibitem{Alekseev:2008aa}
{\scshape \textsc{COMPASS}} collaboration, \emph{{Collins and Sivers
  asymmetries for pions and kaons in muon--deuteron \textsc{DIS}}},
  \href{https://doi.org/10.1016/j.physletb.2009.01.060}{\emph{Phys. Lett. B}
  {\bfseries 673} (2009) 127}
  [\href{https://arxiv.org/abs/0802.2160}{{\ttfamily 0802.2160}}].

\bibitem{Adolph:2014zba}
{\scshape COMPASS} collaboration, \emph{{Collins and Sivers asymmetries in
  muonproduction of pions and kaons off transversely polarised protons}},
  \href{https://doi.org/10.1016/j.physletb.2015.03.056}{\emph{Phys. Lett. B}
  {\bfseries 744} (2015) 250}
  [\href{https://arxiv.org/abs/1408.4405}{{\ttfamily 1408.4405}}].

\bibitem{Qian:2011py}
{\scshape Jefferson Lab Hall A} collaboration, \emph{{Single Spin Asymmetries
  in Charged Pion Production from Semi-Inclusive Deep Inelastic Scattering on a
  Transversely Polarized $^3$He Target at $Q^{2}=$1.4--2.7 GeV$^{2}$}},
  \href{https://doi.org/10.1103/PhysRevLett.107.072003}{\emph{Phys. Rev. Lett.}
  {\bfseries 107} (2011) 072003}
  [\href{https://arxiv.org/abs/1106.0363}{{\ttfamily 1106.0363}}].

\bibitem{Seidl:2006}
{\scshape Belle} collaboration, \emph{{Measurement of Azimuthal Asymmetries in
  Inclusive Production of Hadron Pairs in \(e^+e^-\) Annihilation at Belle}},
  \href{https://doi.org/10.1103/PhysRevLett.96.232002}{\emph{Phys. Rev. Lett.}
  {\bfseries 96} (2006) 232002}
  [\href{https://arxiv.org/abs/hep-ex/0507063}{{\ttfamily hep-ex/0507063}}].

\bibitem{Seidl:2008xc}
{\scshape Belle} collaboration, \emph{{Measurement of azimuthal asymmetries in
  inclusive production of hadron pairs in \(e^{+}e^{-}\) annihilation at
  \(\sqrt{s} =\) 10.58 GeV}},
  \href{https://doi.org/10.1103/PhysRevD.78.032011}{\emph{Phys. Rev. D}
  {\bfseries 78} (2008) 032011}
  [\href{https://arxiv.org/abs/0805.2975}{{\ttfamily 0805.2975}}],
  [\textit{Erratum ibid:}
  \href{https://doi.org/10.1103/PhysRevD.86.039905}{\textbf{86} (2012)
  039905}].

\bibitem{TheBABAR:2013yha}
{\scshape BaBar} collaboration, \emph{{Measurement of Collins asymmetries in
  inclusive production of charged pion pairs in $e^+e^-$ annihilation at
  BABAR}}, \href{https://doi.org/10.1103/PhysRevD.90.052003}{\emph{Phys. Rev.
  D} {\bfseries 90} (2014) 052003}
  [\href{https://arxiv.org/abs/1309.5278}{{\ttfamily 1309.5278}}].

\bibitem{Ablikim:2015pta}
{\scshape BESIII} collaboration, \emph{{Measurement of Azimuthal Asymmetries in
  Inclusive Charged Dipion Production in $e^+e^-$ Annihilations at $\sqrt{s}$ =
  3.65 GeV}}, \href{https://doi.org/10.1103/PhysRevLett.116.042001}{\emph{Phys.
  Rev. Lett.} {\bfseries 116} (2016) 042001}
  [\href{https://arxiv.org/abs/1507.06824}{{\ttfamily 1507.06824}}].

\bibitem{Airapetian:2008sk}
{\scshape \textsc{HERMES}} collaboration, \emph{{Evidence for a transverse
  single-spin asymmetry in leptoproduction of \(\pi^{+}\pi^{-}\) pairs}},
  \href{https://doi.org/10.1088/1126-6708/2008/06/017}{\emph{JHEP} {\bfseries
  06} (2008) 017} [\href{https://arxiv.org/abs/0803.2367}{{\ttfamily
  0803.2367}}].

\bibitem{Adolph:2012nw}
{\scshape COMPASS} collaboration, \emph{{Transverse spin effects in hadron-pair
  production from semi-inclusive deep inelastic scattering}},
  \href{https://doi.org/10.1016/j.physletb.2012.05.015}{\emph{Phys. Lett. B}
  {\bfseries 713} (2012) 10} [\href{https://arxiv.org/abs/1202.6150}{{\ttfamily
  1202.6150}}].

\bibitem{Adolph:2014fjw}
{\scshape COMPASS} collaboration, \emph{{A high-statistics measurement of
  transverse spin effects in dihadron production from muon-proton
  semi-inclusive deep-inelastic scattering}},
  \href{https://doi.org/10.1016/j.physletb.2014.06.080}{\emph{Phys. Lett. B}
  {\bfseries 736} (2014) 124}
  [\href{https://arxiv.org/abs/1401.7873}{{\ttfamily 1401.7873}}].

\bibitem{Vossen:2011fk}
{\scshape Belle} collaboration, \emph{{Observation of Transverse Polarization
  Asymmetries of Charged Pion Pairs in $e^+e^-$ Annihilation near
  $\sqrt{s}=10.58$ GeV}},
  \href{https://doi.org/10.1103/PhysRevLett.107.072004}{\emph{Phys. Rev. Lett.}
  {\bfseries 107} (2011) 072004}
  [\href{https://arxiv.org/abs/1104.2425}{{\ttfamily 1104.2425}}].

\bibitem{Adamczyk:2015hri}
{\scshape STAR} collaboration, \emph{{Observation of Transverse Spin-Dependent
  Azimuthal Correlations of Charged Pion Pairs in $p^\uparrow+p$ at
  $\sqrt{s}=200$ GeV}},
  \href{https://doi.org/10.1103/PhysRevLett.115.242501}{\emph{Phys. Rev. Lett.}
  {\bfseries 115} (2015) 242501}
  [\href{https://arxiv.org/abs/1504.00415}{{\ttfamily 1504.00415}}].

\bibitem{Bacchetta:2012ty}
A.~Bacchetta, A.~Courtoy and M.~Radici, \emph{{First extraction of valence
  transversities in a collinear framework}},
  \href{https://doi.org/10.1007/JHEP03(2013)119}{\emph{JHEP} {\bfseries 03}
  (2013) 119} [\href{https://arxiv.org/abs/1212.3568}{{\ttfamily 1212.3568}}].

\bibitem{Radici:2015mwa}
M.~Radici, A.~Courtoy, A.~Bacchetta and M.~Guagnelli, \emph{{Improved
  extraction of valence transversity distributions from inclusive dihadron
  production}}, \href{https://doi.org/10.1007/JHEP05(2015)123}{\emph{JHEP}
  {\bfseries 05} (2015) 123}
  [\href{https://arxiv.org/abs/1503.03495}{{\ttfamily 1503.03495}}].

\bibitem{Radici:2018iag}
M.~Radici and A.~Bacchetta, \emph{{First Extraction of Transversity from a
  Global Analysis of Electron-Proton and Proton-Proton Data}},
  \href{https://doi.org/10.1103/PhysRevLett.120.192001}{\emph{Phys. Rev. Lett.}
  {\bfseries 120} (2018) 192001}
  [\href{https://arxiv.org/abs/1802.05212}{{\ttfamily 1802.05212}}].

\bibitem{Benel:2019mcq}
J.~Benel, A.~Courtoy and R.~Ferro-Hernandez, \emph{{A constrained fit of the
  valence transversity distributions from dihadron production}},
  \href{https://doi.org/10.1140/epjc/s10052-020-8039-y}{\emph{Eur. Phys. J. C}
  {\bfseries 80} (2020) 465}
  [\href{https://arxiv.org/abs/1912.03289}{{\ttfamily 1912.03289}}].

\bibitem{Yamanaka:2018uud}
{\scshape JLQCD} collaboration, \emph{{Nucleon charges with dynamical overlap
  fermions}}, \href{https://doi.org/10.1103/PhysRevD.98.054516}{\emph{Phys.
  Rev. D} {\bfseries 98} (2018) 054516}
  [\href{https://arxiv.org/abs/1805.10507}{{\ttfamily 1805.10507}}].

\bibitem{Gupta:2018qil}
R.~Gupta, Y.-C. Jang, B.~Yoon, H.-W. Lin, V.~Cirigliano and T.~Bhattacharya,
  \emph{{Isovector charges of the nucleon from $2+1+1$-flavor lattice QCD}},
  \href{https://doi.org/10.1103/PhysRevD.98.034503}{\emph{Phys. Rev. D}
  {\bfseries 98} (2018) 034503}
  [\href{https://arxiv.org/abs/1806.09006}{{\ttfamily 1806.09006}}].

\bibitem{Alexandrou:2018eet}
C.~Alexandrou, K.~Cichy, M.~Constantinou, K.~Jansen, A.~Scapellato and
  F.~Steffens, \emph{{Transversity parton distribution functions from lattice
  QCD}}, \href{https://doi.org/10.1103/PhysRevD.98.091503}{\emph{Phys. Rev. D}
  {\bfseries 98} (2018) 091503(R)}
  [\href{https://arxiv.org/abs/1807.00232}{{\ttfamily 1807.00232}}].

\bibitem{Aubert:2015hha}
{\scshape BaBar} collaboration, \emph{{Collins asymmetries in inclusive charged
  $KK$ and $K\pi$ pairs produced in $e^+e^-$ annihilation}},
  \href{https://doi.org/10.1103/PhysRevD.92.111101}{\emph{Phys. Rev. D}
  {\bfseries 92} (2015) 111101}
  [\href{https://arxiv.org/abs/1506.05864}{{\ttfamily 1506.05864}}].

\bibitem{Anselmino:2015fty}
M.~Anselmino, M.~Boglione, U.~D'Alesio, J.~O. Gonzalez~Hernandez, S.~Melis,
  F.~Murgia et~al., \emph{{Extracting the kaon Collins function from $e^+e^-$
  hadron pair production data}},
  \href{https://doi.org/10.1103/PhysRevD.93.034025}{\emph{Phys. Rev. D}
  {\bfseries 93} (2016) 034025}
  [\href{https://arxiv.org/abs/1512.02252}{{\ttfamily 1512.02252}}].

\bibitem{Efremov:2001ia}
A.~V. Efremov, K.~Goeke and P.~Schweitzer, \emph{{Predictions for azimuthal
  asymmetries in pion and kaon production in SIDIS off a longitudinally
  polarized deuterium target at HERMES}},
  \href{https://doi.org/10.1007/s100520200918}{\emph{Eur. Phys. J. C}
  {\bfseries 24} (2002) 407}
  [\href{https://arxiv.org/abs/hep-ph/0112166}{{\ttfamily hep-ph/0112166}}].

\bibitem{Zhao:2014qvx}
{\scshape Jefferson Lab Hall A} collaboration, \emph{{Single spin asymmetries
  in charged kaon production from semi-inclusive deep inelastic scattering on a
  transversely polarized $^3He$ target}},
  \href{https://doi.org/10.1103/PhysRevC.90.055201}{\emph{Phys. Rev. C}
  {\bfseries 90} (2014) 055201}
  [\href{https://arxiv.org/abs/1404.7204}{{\ttfamily 1404.7204}}].

\bibitem{Artru:1995bh}
X.~Artru, J.~Czy\.{z}ewski and H.~Yabuki, \emph{{Single spin asymmetry in
  inclusive pion production, Collins effect and the string model}},
  \href{https://doi.org/10.1007/s002880050342}{\emph{Z. Phys. C} {\bfseries 73}
  (1997) 527} [\href{https://arxiv.org/abs/hep-ph/9508239}{{\ttfamily
  hep-ph/9508239}}].

\bibitem{Andersson:1983ia}
B.~Andersson, G.~Gustafson, G.~Ingelman and T.~Sj\"ostrand, \emph{Parton
  fragmentation and string dynamics},
  \href{https://doi.org/10.1016/0370-1573(83)90080-7}{\emph{Phys. Rept.}
  {\bfseries 97} (1983) 31}.

\bibitem{Wang:2018wqo}
X.~Wang, Y.~Yang and Z.~Lu, \emph{{Double Collins effect in $e^+ e^- \to
  \Lambda \bar\Lambda X$ and $e^+ e^- \to \Lambda \pi X$ processes in a diquark
  spectator model}},
  \href{https://doi.org/10.1103/PhysRevD.97.114015}{\emph{Phys. Rev. D}
  {\bfseries 97} (2018) 114015}
  [\href{https://arxiv.org/abs/1802.01843}{{\ttfamily 1802.01843}}].

\bibitem{Efremov:2004tp}
A.~V. Efremov, K.~Goeke, S.~Menzel, A.~Metz and P.~Schweitzer, \emph{{Sivers
  effect in semi-inclusive DIS and in the Drell--Yan process}},
  \href{https://doi.org/10.1016/j.physletb.2005.03.010}{\emph{Phys. Lett. B}
  {\bfseries 612} (2005) 233}
  [\href{https://arxiv.org/abs/hep-ph/0412353}{{\ttfamily hep-ph/0412353}}].

\bibitem{Vogelsang:2005cs}
W.~Vogelsang and F.~Yuan, \emph{Single-transverse spin asymmetries: From deep
  inelastic scattering to hadronic collisions},
  \href{https://doi.org/10.1103/PhysRevD.72.054028}{\emph{Phys. Rev. D}
  {\bfseries 72} (2005) 054028}
  [\href{https://arxiv.org/abs/hep-ph/0507266}{{\ttfamily hep-ph/0507266}}].

\bibitem{Anselmino:2005ea}
M.~Anselmino, M.~Boglione, U.~D'Alesio, A.~Kotzinian, F.~Murgia and
  A.~Prokudin, \emph{{Extracting the Sivers function from polarized
  semi-inclusive deep inelastic scattering data and making predictions}},
  \href{https://doi.org/10.1103/PhysRevD.72.094007}{\emph{Phys. Rev. D}
  {\bfseries 72} (2005) 094007}
  [\href{https://arxiv.org/abs/hep-ph/0507181}{{\ttfamily hep-ph/0507181}}],
  [\textit{Erratum ibid:}
  \href{https://doi.org/10.1103/PhysRevD.72.099903}{\textbf{72} (2005)
  099903}].

\bibitem{Anselmino:2005nn}
M.~Anselmino, M.~Boglione, U.~D'Alesio, A.~Kotzinian, F.~Murgia and
  A.~Prokudin, \emph{{Role of Cahn and Sivers effects in deep inelastic
  scattering}}, \href{https://doi.org/10.1103/PhysRevD.71.074006}{\emph{Phys.
  Rev. D} {\bfseries 71} (2005) 074006}
  [\href{https://arxiv.org/abs/hep-ph/0501196}{{\ttfamily hep-ph/0501196}}].

\bibitem{Collins:2005ie}
J.~C. Collins, A.~V. Efremov, K.~Goeke, M.~Grosse-Perdekamp, S.~Menzel, A.~Metz
  et~al., \emph{Sivers effect in semiinclusive deeply inelastic scattering},
  \href{https://doi.org/10.1103/PhysRevD.73.014021}{\emph{Phys. Rev. D}
  {\bfseries 73} (2006) 014021}
  [\href{https://arxiv.org/abs/hep-ph/0509076}{{\ttfamily hep-ph/0509076}}].

\bibitem{Anselmino:2008sga}
M.~Anselmino, M.~Boglione, U.~D'Alesio, A.~Kotzinian, S.~Melis, F.~Murgia
  et~al., \emph{Sivers effect for pion and kaon production in semi-inclusive
  deep inelastic scattering},
  \href{https://doi.org/10.1140/epja/i2008-10697-y}{\emph{Eur. Phys. J. A}
  {\bfseries 39} (2009) 89} [\href{https://arxiv.org/abs/0805.2677}{{\ttfamily
  0805.2677}}].

\bibitem{Anselmino:2012aa}
M.~Anselmino, M.~Boglione and S.~Melis, \emph{{Strategy towards the extraction
  of the Sivers function with transverse momentum dependent evolution}},
  \href{https://doi.org/10.1103/PhysRevD.86.014028}{\emph{Phys. Rev. D}
  {\bfseries 86} (2012) 014028}
  [\href{https://arxiv.org/abs/1204.1239}{{\ttfamily 1204.1239}}].

\bibitem{Sun:2013dya}
P.~Sun and F.~Yuan, \emph{{Energy evolution for the Sivers asymmetries in hard
  processes}}, \href{https://doi.org/10.1103/PhysRevD.88.034016}{\emph{Phys.
  Rev. D} {\bfseries 88} (2013) 034016}
  [\href{https://arxiv.org/abs/1304.5037}{{\ttfamily 1304.5037}}].

\bibitem{Gamberg:2013kla}
L.~Gamberg, Z.-B. Kang and A.~Prokudin, \emph{{Indication on the Process
  Dependence of the Sivers Effect}},
  \href{https://doi.org/10.1103/PhysRevLett.110.232301}{\emph{Phys. Rev. Lett.}
  {\bfseries 110} (2013) 232301}
  [\href{https://arxiv.org/abs/1302.3218}{{\ttfamily 1302.3218}}].

\bibitem{Echevarria:2014xaa}
M.~G. Echevarria, A.~Idilbi, Z.-B. Kang and I.~Vitev, \emph{{QCD evolution of
  the Sivers asymmetry}},
  \href{https://doi.org/10.1103/PhysRevD.89.074013}{\emph{Phys. Rev. D}
  {\bfseries 89} (2014) 074013}
  [\href{https://arxiv.org/abs/1401.5078}{{\ttfamily 1401.5078}}].

\bibitem{Anselmino:2016uie}
M.~Anselmino, M.~Boglione, U.~D'Alesio, F.~Murgia and A.~Prokudin, \emph{{Study
  of the sign change of the Sivers function from STAR Collaboration $W/Z$
  production data}}, \href{https://doi.org/10.1007/JHEP04(2017)046}{\emph{JHEP}
  {\bfseries 04} (2017) 046}
  [\href{https://arxiv.org/abs/1612.06413}{{\ttfamily 1612.06413}}].

\bibitem{Boglione:2018dqd}
M.~Boglione, U.~D'Alesio, C.~Flore and J.~O. Gonzalez-Hernandez,
  \emph{{Assessing signals of TMD physics in SIDIS azimuthal asymmetries and in
  the extraction of the Sivers function}},
  \href{https://doi.org/10.1007/JHEP07(2018)148}{\emph{JHEP} {\bfseries 07}
  (2018) 148} [\href{https://arxiv.org/abs/1806.10645}{{\ttfamily
  1806.10645}}].

\bibitem{Adolph:2012sp}
{\scshape COMPASS} collaboration, \emph{{II -- Experimental investigation of
  transverse spin asymmetries in \(\mu\)-\(p\) \textsc{SIDIS} processes: Sivers
  asymmetries}},
  \href{https://doi.org/10.1016/j.physletb.2012.09.056}{\emph{Phys. Lett. B}
  {\bfseries 717} (2012) 383}
  [\href{https://arxiv.org/abs/1205.5122}{{\ttfamily 1205.5122}}].

\bibitem{Adolph:2016dvl}
{\scshape COMPASS} collaboration, \emph{{Sivers asymmetry extracted in SIDIS at
  the hard scales of the Drell--Yan process at COMPASS}},
  \href{https://doi.org/10.1016/j.physletb.2017.04.042}{\emph{Phys. Lett. B}
  {\bfseries 770} (2017) 138}
  [\href{https://arxiv.org/abs/1609.07374}{{\ttfamily 1609.07374}}].

\bibitem{Burkardt:2003yg}
M.~Burkardt, \emph{{Quark correlations and single spin asymmetries}},
  \href{https://doi.org/10.1103/PhysRevD.69.057501}{\emph{Phys. Rev. D}
  {\bfseries 69} (2004) 057501}
  [\href{https://arxiv.org/abs/hep-ph/0311013}{{\ttfamily hep-ph/0311013}}].

\bibitem{Burkardt:2004ur}
M.~Burkardt, \emph{{Sivers mechanism for gluons}},
  \href{https://doi.org/10.1103/PhysRevD.69.091501}{\emph{Phys. Rev. D}
  {\bfseries 69} (2004) 091501}
  [\href{https://arxiv.org/abs/hep-ph/0402014}{{\ttfamily hep-ph/0402014}}].

\bibitem{Signori:2013mda}
A.~Signori, A.~Bacchetta, M.~Radici and G.~Schnell, \emph{{Investigations into
  the flavor dependence of partonic transverse momentum}},
  \href{https://doi.org/10.1007/JHEP11(2013)194}{\emph{JHEP} {\bfseries 11}
  (2013) 194} [\href{https://arxiv.org/abs/1309.3507}{{\ttfamily 1309.3507}}].

\bibitem{Airapetian:2004zf}
{\scshape \textsc{HERMES}} collaboration, \emph{Quark helicity distributions in
  the nucleon for up, down, and strange quarks from semi-inclusive
  deep-inelastic scattering},
  \href{https://doi.org/10.1103/PhysRevD.71.012003}{\emph{Phys. Rev. D}
  {\bfseries 71} (2005) 012003}
  [\href{https://arxiv.org/abs/hep-ex/0407032}{{\ttfamily hep-ex/0407032}}].

\bibitem{deFlorian:2007hc}
D.~de~Florian, R.~Sassot and M.~Stratmann, \emph{Global analysis of
  fragmentation functions for protons and charged hadrons},
  \href{https://doi.org/10.1103/PhysRevD.76.074033}{\emph{Phys. Rev. D}
  {\bfseries 76} (2007) 074033}
  [\href{https://arxiv.org/abs/0707.1506}{{\ttfamily 0707.1506}}].

\bibitem{Echevarria2018}
M.~Echevarria and G.~Schnell, \emph{Sivers asymmetries for electroproduction of
  protons and antiprotons},  (to be published).

\bibitem{Lefky:2014eia}
C.~Lefky and A.~Prokudin, \emph{{Extraction of the distribution function
  $h^{\perp}_{1T}$ from experimental data}},
  \href{https://doi.org/10.1103/PhysRevD.91.034010}{\emph{Phys. Rev. D}
  {\bfseries 91} (2015) 034010}
  [\href{https://arxiv.org/abs/1411.0580}{{\ttfamily 1411.0580}}].

\bibitem{Musch:2010ka}
B.~U. Musch, P.~H{\"a}gler, J.~W. Negele and A.~Sch{\"a}fer, \emph{{Exploring
  quark transverse momentum distributions with lattice QCD}},
  \href{https://doi.org/10.1103/PhysRevD.83.094507}{\emph{Phys. Rev. D}
  {\bfseries 83} (2011) 094507}
  [\href{https://arxiv.org/abs/1011.1213}{{\ttfamily 1011.1213}}].

\bibitem{Yoon:2017qzo}
B.~Yoon, M.~Engelhardt, R.~Gupta, T.~Bhattacharya, J.~R. Green, B.~U. Musch
  et~al., \emph{{Nucleon transverse momentum-dependent parton distributions in
  lattice QCD: Renormalization patterns and discretization effects}},
  \href{https://doi.org/10.1103/PhysRevD.96.094508}{\emph{Phys. Rev. D}
  {\bfseries 96} (2017) 094508}
  [\href{https://arxiv.org/abs/1706.03406}{{\ttfamily 1706.03406}}].

\bibitem{Mao:2014aoa}
W.~Mao, Z.~Lu and B.-Q. Ma, \emph{{Transverse single-spin asymmetries of pion
  production in semi-inclusive DIS at subleading twist}},
  \href{https://doi.org/10.1103/PhysRevD.90.014048}{\emph{Phys. Rev. D}
  {\bfseries 90} (2014) 014048}
  [\href{https://arxiv.org/abs/1405.3876}{{\ttfamily 1405.3876}}].

\bibitem{Parsamyan:2018ovx}
B.~Parsamyan, \emph{{Measurement of longitudinal-target-polarization dependent
  azimuthal asymmetries in SIDIS at COMPASS experiment}},
  \href{https://doi.org/10.22323/1.297.0259}{\emph{PoS} {\bfseries DIS2017}
  (2018) 259} [\href{https://arxiv.org/abs/1801.01488}{{\ttfamily
  1801.01488}}].

\bibitem{Avakian:2010ae}
{\scshape \textsc{CLAS}} collaboration, \emph{{Measurement of Single- and
  Double-Spin Asymmetries in Deep Inelastic Pion Electroproduction with a
  Longitudinally Polarized Target}},
  \href{https://doi.org/10.1103/PhysRevLett.105.262002}{\emph{Phys. Rev. Lett.}
  {\bfseries 105} (2010) 262002}
  [\href{https://arxiv.org/abs/1003.4549}{{\ttfamily 1003.4549}}].

\bibitem{Airapetian:2009ac}
{\scshape HERMES} collaboration, \emph{{Single-spin azimuthal asymmetry in
  exclusive electroproduction of \(\pi^{+}\) mesons on transversely polarized
  protons}}, \href{https://doi.org/10.1016/j.physletb.2009.11.039}{\emph{Phys.
  Lett. B} {\bfseries 682} (2010) 345}
  [\href{https://arxiv.org/abs/0907.2596}{{\ttfamily 0907.2596}}].

\bibitem{Airapetian:2018rlq}
{\scshape HERMES} collaboration, \emph{{Longitudinal double-spin asymmetries in
  semi-inclusive deep-inelastic scattering of electrons and positrons by
  protons and deuterons}},
  \href{https://doi.org/10.1103/PhysRevD.99.112001}{\emph{Phys. Rev. D}
  {\bfseries 99} (2019) 112001}
  [\href{https://arxiv.org/abs/1810.07054}{{\ttfamily 1810.07054}}].

\bibitem{Airapetian:2001iy}
{\scshape HERMES} collaboration, \emph{{Single-spin azimuthal asymmetry in
  exclusive electroproduction of \(\pi^{+}\) mesons}},
  \href{https://doi.org/10.1016/S0370-2693(02)01780-X}{\emph{Phys. Lett. B}
  {\bfseries 535} (2002) 85}
  [\href{https://arxiv.org/abs/hep-ex/0112022}{{\ttfamily hep-ex/0112022}}].

\bibitem{Oganessyan:2002pc}
K.~A. Oganessyan, P.~J. Mulders and E.~De~Sanctis, \emph{{Double-spin
  \(\cos\phi\) asymmetry in semi-inclusive electroproduction}},
  \href{https://doi.org/10.1016/S0370-2693(02)01532-0}{\emph{Phys. Lett. B}
  {\bfseries 532} (2002) 87}
  [\href{https://arxiv.org/abs/hep-ph/0201061}{{\ttfamily hep-ph/0201061}}].

\bibitem{Anselmino:2006yc}
M.~Anselmino, A.~Efremov, A.~Kotzinian and B.~Parsamyan, \emph{{Transverse
  momentum dependence of the quark helicity distributions and the Cahn effect
  in double-spin asymmetry \(A_{LL}\) in semiinclusive DIS}},
  \href{https://doi.org/10.1103/PhysRevD.74.074015}{\emph{Phys. Rev. D}
  {\bfseries 74} (2006) 074015}
  [\href{https://arxiv.org/abs/hep-ph/0608048}{{\ttfamily hep-ph/0608048}}].

\bibitem{Alekseev:2010dm}
{\scshape COMPASS} collaboration, \emph{{Azimuthal asymmetries of charged
  hadrons produced by high-energy muons scattered off longitudinally polarised
  deuterons}}, \href{https://doi.org/10.1140/epjc/s10052-010-1461-9}{\emph{Eur.
  Phys. J. C} {\bfseries 70} (2010) 39}
  [\href{https://arxiv.org/abs/1007.1562}{{\ttfamily 1007.1562}}].

\bibitem{Trentadue:1993ka}
L.~Trentadue and G.~Veneziano, \emph{{Fracture functions. An improved
  description of inclusive hard processes in QCD}},
  \href{https://doi.org/10.1016/0370-2693(94)90292-5}{\emph{Phys. Lett. B}
  {\bfseries 323} (1994) 201}.

\bibitem{Anselmino:2011ss}
M.~Anselmino, V.~Barone and A.~Kotzinian, \emph{{SIDIS in the target
  fragmentation region: Polarized and transverse momentum dependent fracture
  functions}},
  \href{https://doi.org/10.1016/j.physletb.2011.03.067}{\emph{Phys. Lett. B}
  {\bfseries 699} (2011) 108}
  [\href{https://arxiv.org/abs/1102.4214}{{\ttfamily 1102.4214}}].

\bibitem{Bacchetta:2008xw}
A.~Bacchetta, D.~Boer, M.~Diehl and P.~J. Mulders, \emph{{Matches and
  mismatches in the descriptions of semi-inclusive processes at low and high
  transverse momentum}},
  \href{https://doi.org/10.1088/1126-6708/2008/08/023}{\emph{JHEP} {\bfseries
  08} (2008) 023} [\href{https://arxiv.org/abs/0803.0227}{{\ttfamily
  0803.0227}}].

\bibitem{Pappalardo:2008zza}
L.~L. Pappalardo, \emph{{Transverse spin effects in polarized semi inclusive
  deep inelastic scattering}}, Ph.D. thesis, University of Ferrara, 2008.

\bibitem{Diefenthaler:2010zz}
M.~Diefenthaler, \emph{{Signals for transversity and transverse momentum
  dependent quark distribution functions studied at the HERMES experiment}},
  Ph.D. thesis, University Erlangen-Nuremberg, 2010.

\bibitem{James:1975dr}
F.~James and M.~Roos, \emph{{Minuit - a system for function minimization and
  analysis of the parameter errors and correlations}},
  \href{https://doi.org/10.1016/0010-4655(75)90039-9}{\emph{Comput. Phys.
  Commun.} {\bfseries 10} (1975) 343}.

\end{thebibliography}\endgroup
